\documentclass[iop,numberedappendix,appendixfloats]{emulateapj}
\pdfoutput=1
\usepackage{graphicx}
\usepackage{amsmath}
\usepackage{longtable}
\usepackage[flushleft]{threeparttable}
\usepackage{txfonts}
\defcitealias{buri12}{GB12}
\defcitealias{lero12}{L12}
\defcitealias{lero13}{L13}
\defcitealias{krum05}{KMK05}
\defcitealias{krum07}{KT07}

\newcommand{\hcn}{\mbox{HCN(1--0)}}

\newcommand{\coone}{\mbox{CO(1--0)}}
\newcommand{\ihcn}{I_\mathrm{HCN10}}
\newcommand{\eihcn}{\sigma_\mathrm{HCN10}}
\newcommand{\ico}{I_\mathrm{CO}}

\newcommand{\itir}{I_\mathrm{TIR}}
\newcommand{\lhcn}{L_\mathrm{HCN}}
\newcommand{\lco}{L_\mathrm{CO}}
\newcommand{\ltir}{L_\mathrm{TIR}}
\newcommand{\lfir}{L_\mathrm{FIR}}
\newcommand{\smol}{\Sigma_\mathrm{mol}}

\newcommand{\sdens}{\Sigma_\mathrm{dense}}
\newcommand{\satom}{\Sigma_\mathrm{atom}}
\newcommand{\sgas}{\Sigma_\mathrm{gas}}
\newcommand{\sstar}{\Sigma_\mathrm{star}}
\newcommand{\ssfr}{\Sigma_\mathrm{SFR}}
\newcommand{\ssfrler}{\Sigma_\mathrm{SFR}(\mathrm{L12})}
\newcommand{\sfemol}{SFE_\mathrm{mol}}
\newcommand{\sfedens}{SFE_\mathrm{dense}}
\newcommand{\fdens}{f_\mathrm{dense}}
\newcommand{\rank}{r_\mathrm{s}}

\newcommand{\aco}{\alpha_\mathrm{CO}}
\newcommand{\ahcn}{\alpha_\mathrm{HCN}}
\newcommand{\gco}{\gamma_\mathrm{CO}}
\newcommand{\ghcn}{\gamma_\mathrm{HCN}}
\newcommand{\neff}{n_\mathrm{eff}}

\newcommand{\si}[1]{#1$^\star$}
\newcommand{\no}[1]{#1\phantom{$^{\star}$}}
\newcommand{\tellzero}[1]{}

\newlength{\qq} 
\newcommand{\mybox}[1]{\makebox[0.42\textwidth][c]{ 
\begin{minipage}[b]{0.42\textwidth}{ 
\vspace{0.5\qq}\hspace{\stretch{1}}#1\hspace{\stretch{1}}\vspace{0.5\qq}} 
\end{minipage} 
}}

\begin{document} 

\title{Variations in the Star Formation Efficiency of the Dense Molecular Gas across the Disks of Star-Forming Galaxies
\footnote{Based on observations with the IRAM 30-m telescope. IRAM is supported by CNRS/INSU (France), the MPG (Germany), and the IGN (Spain).}}
\author{
Antonio Usero\altaffilmark{1}, 
Adam K. Leroy\altaffilmark{2,3}, 
Fabian Walter\altaffilmark{4},
Andreas Schruba\altaffilmark{5},
Santiago Garc\'ia-Burillo\altaffilmark{1},
Karin Sandstrom\altaffilmark{6},
Frank Bigiel\altaffilmark{7},
Elias Brinks\altaffilmark{8},
Carsten Kramer\altaffilmark{9},
Erik Rosolowsky\altaffilmark{10},
Karl-Friedrich Schuster\altaffilmark{11},
W.~J.~G. de Blok\altaffilmark{12,13,14}
}
\altaffiltext{1}{Observatorio Astron\'omico Nacional (IGN), C/ Alfonso XII 3, Madrid 28014, Spain; \email{a.usero@oan.es}}
\altaffiltext{2}{National Radio Astronomy Observatory, 520 Edgemont Road, Charlottesville, VA 22903, USA}
\altaffiltext{3}{The Ohio State University, 140 W 18$^{\rm th}$ St, Columbus, OH 43210, USA}
\altaffiltext{4}{Max-Planck Institut f\"ur Astronomie, K\"onigstuhl 17, D-69117 Heidelberg, Germany}
\altaffiltext{5}{Max-Planck-Institut f\"ur extra-terrestrische Physik, Giessenbachstrasse 1, D-85748 Garching, Germany}
\altaffiltext{6}{Steward Observatory, University of Arizona, 933 N. Cherry Ave, Tucson, AZ 85721, USA}
\altaffiltext{7}{Institut f\"ur theoretische Astrophysik, Zentrum f\"ur Astronomie der Universit\"at Heidelberg, Albert-Ueberle Str. 2, D-69120 Heidelberg, Germany}
\altaffiltext{8}{Centre for Astrophysics Research, University of Hertfordshire, Hatfield AL10 9AB, UK}
\altaffiltext{9}{Instituto de Radioastronom\'ia Milim\'etrica (IRAM), Av. Divina Pastora 7, Nucleo Central, 18012 Granada, Spain}
\altaffiltext{10}{Department of Physics, University of Alberta, 4-183 CCIS, Edmonton, AB, T6G 2E1, Canada}
\altaffiltext{11}{Institut de RadioAstronomie Millim\'etrique (IRAM), 300 rue de la Piscine, 38406 St. Martin d'H\`eres, France}
\altaffiltext{12}{Netherlands Institute for Radio Astronomy (ASTRON), 
Postbus 2, 7990  AA Dwingeloo, the Netherlands}
\altaffiltext{13}{Astrophysics, Cosmology and Gravity Centre, Department 
of Astronomy, University of Cape Town, Private Bag X3, Rondebosch 7701, 
South Africa}
\altaffiltext{14}{Kapteyn Astronomical Institute, University of
  Groningen, PO Box 800, 9700 AV Groningen, the Netherlands}

\slugcomment{Submitted}
 
 \begin{abstract}
We present a new survey of HCN~(1--0) emission, a tracer of dense molecular gas, focused on the little-explored regime of normal star-forming galaxy disks.
Combining HCN, CO, and infrared (IR) emission, we investigate the role of dense gas in star formation, finding systematic variations in both 
the apparent dense gas fraction (traced by the HCN-to-CO ratio) and the apparent star formation efficiency of dense gas (traced by the IR-to-HCN ratio). The latter may be unexpected, given the recent popularity of gas density threshold models to explain star formation scaling relations. Our survey used the IRAM 30-m telescope to 
observe \hcn, CO~(1--0), and several other emission lines across 29 nearby disk galaxies
whose CO (2--1) emission has previously been mapped by the HERACLES survey. 
We detected HCN in 48 out of 62 observed positions. Because 
our observations achieve a typical resolution of $\sim 1.5$~kpc and span a range of galaxies and galactocentric radii ($56\%$ lie at $r_{\rm gal}~>~1$~kpc), we are 
able to investigate the properties of the dense gas as a function of local conditions in a galaxy disk. We focus on how the ratios IR-to-CO, HCN-to-CO, and IR-to-HCN 
(observational cognates of the star formation efficiency, dense gas fraction, and dense gas star formation efficiency) 
depend on the stellar surface density, $\sstar$, and the molecular-to-atomic gas ratio, $\smol / \satom$. The HCN-to-CO ratio is low, often $\sim1/30$, and correlates tightly with 
both the molecular-to-atomic ratio and the stellar mass surface density across a range of $2.1$~dex (factor of $\approx125$) in both parameters. Thus for the assumption of
fixed CO-to-H$_2$ and HCN-to-dense gas conversion factors, the dense gas fraction depends strongly on location in the disk, being higher in the high surface density, highly molecular 
parts of galaxies. At the same time, the IR-to-HCN ratio (closely related to the star formation efficiency of dense molecular gas) decreases systematically with these same parameters 
and is $\sim6-8$ times lower near galaxy  centers than in the outer regions of the galaxy disks. For fixed conversion factors, these results are incompatible with a simple 
model in which star formation depends only on the amount of gas mass above some density threshold. Moreover, only a very specific set of environment-dependent conversion 
factors can render our observations compatible with such a model. Whole cloud models, such as the theory of turbulence regulated star formation, do a
better job of matching our observations. We explore one such model in which variations in the Mach number driving many of the trends within galaxy disks while density 
contrasts drive the differences between disk and merging galaxies.
\end{abstract}
\maketitle

%
%________________________________________________________________

\section{Introduction}
\label{s-intro}

It is not yet fully understood how the properties of molecular gas clouds affect their ability to form stars. A strong relation between 
gas density\footnote{Throughout this paper ``density'' refers to volume density of H$_2$ molecules unless otherwise 
specified.} ($n$) and star formation is expected, because denser gas pockets will be more prone to collapse and will collapse
more quickly. A key method to test this expectation has been to assemble observations of molecular lines with different 
critical densities and to compare these observations to tracers of recent star formation. Because such observations do not
need to resolve individual clouds or star-forming cores, this approach may be the most practical way to systematically 
study the effect of gas density on star formation over large parts of the universe. In this paper, we present new observations
of the HCN (1--0) transition that cover large ($\sim$~kpc) parts of nearby galaxy disks and compare them to estimates of the recent star 
formation rate and CO emission, a tracer for the molecular gas content (H$_2$). \hcn, with an effective excitation density  $\gtrsim10^4-10^5$~cm$^{-3}$, has (along with HCO$^{+}$ (1--0))
become the most common tracer of dense gas over large parts of galaxies. By contrasting it with the bulk of the molecular gas traced by CO 
emission and the rate of recent star formation, we aim to gain insight into how physical conditions internal to the molecular gas, especially 
density, affect the star formation process.

Starting with the groundbreaking papers by \citet{gao04a,gao04b}, several studies \citep[e.g.][]{grac06,grac08,buss08,june09,buri12} have shown that the total star formation rate (SFR) of galaxies scales more directly (meaning more linearly)
with the luminosity of molecular lines that have high excitation densities (e.g., HCN) than with bulk tracers of the molecular gas (e.g., CO). While
low-$J$ CO emission tracks star formation nearly linearly in the disks of local star forming galaxies (e.g., \citealt{bigi08}, \citealt{schr11}, and \citealt{lero13}, hereafter \citetalias{lero13}),  the relationship becomes nonlinear (multivalued or steeper than linear) when starburst regions and merging galaxies are considered \citep[e.g.,][]{genz10,daddi10}. Considering
spirals and (ultra)luminous infrared galaxies ($\ltir\geq10^{11}~L_\odot$; hereafter (U)LIRGs) as a single population, the SFR is a superlinear function of the 
bulk gas tracer CO. By contrast, \citet{gao04b} showed that total SFR scales almost linearly with the \hcn{} luminosity  across the entire range of galaxies. 
In a follow-up study, \citet{wu05}  concluded that the large parts of galaxies sampled by \citet{gao04b} and Galactic dense cores align along the same 
HCN(1--0)--IR relation. On the assumption that line luminosities are proportional to the mass of emitting gas, these results imply that the SFR per unit 
{\em total} molecular mass is higher in the brighter (higher SFR) systems, while the SFR per unit {\em dense} molecular mass is nearly constant.  

Explanations of the distinct relationships between CO, recent SFR, and HCN can broadly be grouped into two classes. One group of
explanations, which we refer to as ``density threshold models'', broadly posit that the SFR is determined by the mass of molecular gas above
a certain {\em threshold} density, with gas above this density forming stars at an approximately fixed rate everywhere (i.e., SFR/$M_{\rm dense}$ is $\sim$ constant).
\citet{gao04b} and \citet{wu05} advocate such a view, as do \citet{lada10,lada12} and \citet{evans14} based on studies of clouds in the Solar Neighborhood.
A competing class of models, which we refer to as ``whole cloud'' or ``turbulence regulated models,'' posit that the properties of the whole star-forming cloud affect the efficiency of 
star formation. That is, the average density, Mach number, and other properties of the entire cloud set the local threshold for star formation, the fraction of gas above this 
threshold, and the speed with which star-forming gas collapses. \citet{fedd12} synthesize several such models; in this paper, we focus on comparisons to 
\citet[][hereafter \citetalias{krum05}]{krum05} and \citet[][hereafter \citetalias{krum07}]{krum07}, which considered such a ``whole cloud'' picture in the context of observations of
molecular lines with varying critical density. In these models, both the SFR  per unit of total molecular mass and the fraction of dense gas increase with the 
average density within the clouds ($\bar{n}$). The Mach ($\mathcal{M}$) number also affects the density distribution and criteria for star formation,
with a high $\mathcal{M}$ broadening the probability distribution function of gas densities and increasing high density condensations, though its numerical 
effect varies somewhat with the details of the model used \citep{fedd12}. 

Characterizing the relation between dense gas and star formation is critical to distinguish between these models, and, in turn,
to understand how larger scale conditions in galaxies affect star formation. For example, several recent results have highlighted the
potential importance of dynamical effects in explaining observed variations in the SFR per unit gas \citep{genz10,daddi10,buri12,meidt13}.
For a full picture of how such large scale phenomena propagate into cloud-scale variations in the star formation process, we must first understand
how physical conditions within the gas \citep[which may be affected by large-scale conditions, e.g., see][]{hughes13} affect star formation in the cloud.

Observations of a diverse sample of normal galaxy disks can help distinguish these models but such a data set has remained lacking. The pioneering surveys of 
dense gas in galaxies mostly offer rough averages over the entire molecular/star-forming disks or values at bright galaxy centers 
\citep[e.g.,][]{gao04a,grac06,grac08,buss08,june09,krip08,crock12,buri12}. The existing handful of observations targeting dense gas in galaxy 
disks at cloud scales  have been time-consuming and provide only small-size samples (e.g., \citealt{brou05} and \citealt{buch13} in M~31 and M~33, respectively).
Observations comparing extragalactic observations to individual clouds in the Milky Way \citep{wu05,heid10,lada12} have proved enlightening,
but the Galactic clouds studied sample a more restricted range of conditions than nearby galaxies and also suffer from small sample sizes and narrow perspective.

To bridge these studies, we undertook a new HCN survey  that targeted large parts of a significant sample of galaxy disks.
From 2008 until 2011, we used the IRAM 30-m telescope to observe HCN (1--0) emission from the disks of 29 nearby star-forming galaxies. We drew our targets
from the HERACLES survey, which constructed maps of CO(2--1) line emission from a set of 48 nearby galaxies  
(\citetalias{lero13}; first maps and survey presented by \citealt{lero09}). With a
typical resolution of $\sim 1.5$~kpc, each of our HCN observations blends together a large population of individual molecular clouds. We expect this averaging to attenuate the 
intrinsic variability and evolutionary effects visible when studying individual star-forming complexes \citep[e.g.,][]{schruba10}. However this resolution is still fine enough
that we can isolate many local variables related to the physical state of the interstellar medium (ISM), for example stellar and gas surface density, interstellar radiation field, and so 
on. A particular emphasis of our survey is to study the disks (not just the centers) of star-forming galaxies, and so our observations span a wide range of galactocentric radii, 
from galaxy centers out to $\sim75\%$ of the {\em optical radius} (defined as the radius of the  25~mag~arcsec$^{-2}$ isophote in the B-band). This allows us to probe a wide range of local conditions and, because HERACLES builds on many surveys 
at other wavelengths (see Sect. \ref{s-data}), the data exist to readily estimate these local conditions. In the end, our HCN survey provides context for cloud-based studies in the Milky Way and Local Group galaxies, while contrasting the starbursts, whole-galaxy integrals, and bright galaxy centers targeted in previous extragalactic studies  
(see also Bigiel et al, subm.).

This paper is structured as follows. We describe our observations and the ancillary data sets in Sect.~\ref{s-data}.
In Sect.~\ref{s-dg-gb12}, we put our data in context by comparing them to star formation scaling relations obtained from unresolved observations of galaxies. 
In Sect.~\ref{s-env}, we focus specifically on our data and investigate the systematic variations across galaxy disks of the star formation efficiencies and the dense gas fraction in the molecular clouds. 
In Sect.~\ref{s-models}, we compare these trends with the predictions from density-threshold and turbulence-regulated models of star formation.  
In Sect.~\ref{s-cf}, we study how the compatibility between our observations and these models depends on the assumed CO and HCN conversion factors.
In Sect.~\ref{s-discuss}, we try to define a common scenario for resolved and unresolved observations.  
Sect.~\ref{s-sum} summarizes the paper. 

%__________________________________________________________________

\section{Data and Physical Parameters}
\label{s-data}

\subsection{Data}

\subsubsection{A New Survey of HCN (1--0) Emission in Galaxy Disks}
\label{sss-30m-obs}

\begin{table*}
\begin{center}
\begin{threeparttable}
\caption{Galaxies Observed in the HCN (1--0) Survey}
\label{t-gal}
                                                                                                                                                                                                                                                                                                                                                                                                       
\begin{tabular}{rrrrrrrr}                                                                                                                                                                                                                                                                                                                                                                                  
\hline\hline                                                                                                                                                                                                                                                                                                                                                                                              
\noalign{\smallskip}                                                                                                                                                                                                                                                                                                                                                                                      
\hspace*{\fill} Name \hspace*{\fill} & \hspace*{\fill} $RA_\mathrm{J2000}$ \hspace*{\fill} & \hspace*{\fill} $Dec_\mathrm{J2000}$ \hspace*{\fill} & \hspace*{\fill} $D$ \hspace*{\fill} & \hspace*{\fill} $R_{25}$ \hspace*{\fill} & \hspace*{\fill} $i$ \hspace*{\fill} & \hspace*{\fill} $PA$ \hspace*{\fill} & \hspace*{\fill} res. \hspace*{\fill} \\               
\hspace*{\fill} NGC \hspace*{\fill} & \hspace*{\fill} $(h:m:s)$ \hspace*{\fill} & \hspace*{\fill} $(\degr:\arcmin:\arcsec)$ \hspace*{\fill} & \hspace*{\fill} (Mpc) \hspace*{\fill} & \hspace*{\fill} $(\arcmin)$ \hspace*{\fill} & \hspace*{\fill} $(\degr)$ \hspace*{\fill} & \hspace*{\fill} $(\degr)$ \hspace*{\fill} & \hspace*{\fill} (kpc) \hspace*{\fill} \\
\noalign{\smallskip}                                                                                                                                                                                                                                                                                                                                                                                      
\hline                                                                                                                                                                                                                                                                                                                                                                                                     
\noalign{\smallskip}                                                                                                                                                                                                                                                                                                                                                                                      
0628 & 01:36:41.8 & $+$15:47:00 &  7.2 &  4.9 &  7 &  20 & 0.98\\
2146 & 06:18:37.7 & $+$78:21:25 & 12.8 &  2.7 & 54 & 123 & 1.74\\
2403 & 07:36:51.1 & $+$65:36:03 &  3.2 &  7.9 & 63 & 124 & 0.43\\
2798 & 09:17:22.8 & $+$41:59:59 & 24.7 &  1.2 & 85 & 152 & 3.35\\
2903 & 09:32:10.1 & $+$21:30:03 &  8.9 &  5.9 & 65 & 204 & 1.21\\
2976 & 09:47:15.3 & $+$67:55:00 &  3.6 &  3.6 & 65 & 335 & 0.49\\
3034 & 09:55:52.7 & $+$69:40:46 &  3.9 &  5.5 & 77 &  68 & 0.53\\
3049 & 09:54:49.5 & $+$09:16:15 & 19.2 &  1.0 & 58 &  28 & 2.61\\
3077 & 10:03:19.1 & $+$68:44:02 &  3.8 &  2.7 & 46 &  45 & 0.52\\
3184 & 10:18:17.0 & $+$41:25:28 & 11.8 &  3.7 & 16 & 179 & 1.60\\
3198 & 10:19:55.0 & $+$45:32:59 & 14.1 &  3.2 & 72 & 215 & 1.91\\
3351 & 10:43:57.7 & $+$11:42:14 &  9.3 &  3.6 & 41 & 192 & 1.26\\
3521 & 11:05:48.6 & $-$00:02:09 & 11.2 &  4.2 & 73 & 340 & 1.52\\
3627 & 11:20:15.0 & $+$12:59:30 &  9.4 &  5.1 & 62 & 173 & 1.28\\
3938 & 11:52:49.4 & $+$44:07:15 & 17.9 &  1.8 & 14 &  15 & 2.43\\
4254 & 12:18:49.6 & $+$14:24:59 & 14.4 &  2.5 & 32 &  55 & 1.95\\
4321 & 12:22:54.9 & $+$15:49:21 & 14.3 &  3.0 & 30 & 153 & 1.94\\
4536 & 12:34:27.0 & $+$02:11:17 & 14.5 &  3.5 & 59 & 299 & 1.97\\
4569 & 12:36:49.8 & $+$13:09:47 &  9.9 &  4.6 & 66 &  23 & 1.34\\
4579 & 12:37:43.5 & $+$11:49:05 & 16.4 &  2.5 & 39 & 100 & 2.23\\
4631 & 12:42:08.0 & $+$32:32:29 &  9.0 &  7.2 & 86 &  86 & 1.22\\
4725 & 12:50:26.6 & $+$25:30:03 & 11.9 &  4.9 & 54 &  36 & 1.62\\
4736 & 12:50:53.0 & $+$41:07:14 &  4.7 &  3.9 & 41 & 296 & 0.64\\
5055 & 13:15:49.2 & $+$42:01:45 &  7.9 &  5.9 & 59 & 102 & 1.07\\
5194 & 13:29:52.7 & $+$47:11:43 &  7.9 &  3.9 & 20 & 172 & 1.07\\
5457 & 14:03:12.6 & $+$54:20:57 &  6.7 & 12.0 & 18 &  39 & 0.91\\
5713 & 14:40:11.5 & $-$00:17:20 & 21.4 &  1.2 & 48 &  11 & 2.91\\
6946 & 20:34:52.2 & $+$60:09:14 &  6.8 &  5.7 & 33 & 243 & 0.92\\
7331 & 22:37:04.1 & $+$34:24:57 & 14.5 &  4.6 & 76 & 168 & 1.97\\
\noalign{\smallskip}
\hline               
\end{tabular}

\begin{tablenotes}
\item{\sc Note. ---} Center coordinates, distance, size, and orientation of the observed sources.  All values are taken from \citetalias{lero12}, when available. The remainder objects are NGC~2146 and 
NGC~2798 \citep[all data taken from][]{schr11}, NGC~3034 (all from HyperLeda, \citealt{patu03}, except distance, from \citealt{walt02}), 
NGC~3077 \citep[from][]{walt08}, and NGC~4631 \citep[from][]{irwi11}.
\end{tablenotes}
\end{threeparttable}
\end{center}
\end{table*}

From 2008 through 2011, we used the IRAM 30-m telescope to observe HCN(1--0) line emission at 62 positions in 29 galaxies drawn from the HERACLES survey (Table~\ref{t-gal}).
For comparison to the overall molecular reservoir, we also observed the CO(1--0) line at 58 of those positions (recall that HERACLES observed CO (2--1)). All of our targets
are star-forming galaxies ($\ltir<10^{11}~L_\odot$; hereafter SF galaxies); most of our targets are disk galaxies, though several are known starburst galaxies (e.g., we include M~82) and several of our central pointings
cover nuclear starbursts. The pointings were chosen to cover  a wide range in environment properties and star formation activity, while also maximizing the likelihood of detection.
To this end, we targeted regions with relatively bright CO(2--1) emission in the HERACLES maps, picking a set of such regions that span a wide range in galactic radius (up to $75\%$ of the optical radius; see maps in Appendix~\ref{a-tabl}).  
Our final set of HCN(1--0) detections spans a range of  $2.1$~dex (factor of $\approx125$) in both stellar surface density and molecular--to--atomic mass ratio, a range of  1.9~dex  (factor of $\approx80$) in molecular gas surface density,  and a range of 1.7~dex (factor of $\approx50$) in SFR surface density. 

The 
angular resolution of the IRAM 30-m at the frequency of \hcn\ is $\sim28\arcsec$, which sets our working resolution in this paper. The corresponding spatial resolution at the distances of our
targets ranges from $0.4$ to $3.4$~kpc (rightmost column in Table~\ref{t-gal}) with an average value of $\approx 1.5$~kpc.

For observations in January and December 2008 we used the old AB receivers at the 30-m telescope. In August 2009 and June 2011 we used the EMIR receiver (\citealt{cart12}; see Table~\ref{t-febe}).
Whenever possible with regard to the extent of the emission and the observing schedule, we observed in wobbler-switching mode with a total ON--OFF throw of $\pm240''$ in azimuth and a wobbling frequency of 0.5~Hz. The orientation 
of the wobbler throw direction is fixed in azimuth, so the throw in equatorial coordinates depends on the hour angle of the target. We used the \mbox{HERACLES} CO(2--1) maps
to ensure that every position was observed only when the OFF positions were free of molecular line emission. Due to this constraint, wobbler-switching mode was not viable 
for a few positions because the throw would have intersected the galaxy. In those cases we adopted a position-switching scheme and selected a suitable OFF position with coordinates chosen based on the \mbox{HERACLES} maps.

We checked the focus of the telescope on planets or bright quasars at the beginning of each session and then every few hours and, if relevant, at sunset and sunrise.
Every $\sim 1$--$1.5$~hours, we corrected the telescope pointing using a point-like source close to the target galaxy. The magnitude of these corrections typically agreed with
the nominal pointing accuracy of the telescope ($\sim2\arcsec$ rms).  Every 15 minutes, we obtained a standard chopper wheel calibration, which we used to place the data on the
antenna temperature scale ($T_\mathrm{A}^*$). In order to convert to main beam temperature ($T_\mathrm{MB}$), we adopted forward and beam efficiencies from the 
IRAM documentation\footnote{http://www.iram.es/IRAMES/mainWiki/Iram30mEfficiencies} that were up-to-date at the time of the observations (see Table~\ref{t-lin}). The expected uncertainties in the resulting gain calibration
are $\lesssim10\%$. We checked this by observing well-known line-calibrators on a non-systematic basis. Comparing these to the 30-m telescope spectra collected by \citet{maue89},
we confirm this accuracy. As a crosscheck, we also compared our HCN(1--0) line intensities at five galaxy centers to those measured  by \citet{krip08}, also with the IRAM 30-m telescope. The
measured intensities agree within $<10\%$.

We reduced the data using the CLASS package of the GILDAS software library\footnote{http://www.iram.fr/IRAMFR/GILDAS}. The raw spectra taken 
with EMIR (since 2009 onwards) are $\sim$4~GHz wide. For each line at each position, we fitted a baseline to a $0.5$~GHz-wide part of the spectrum 
centered on the line frequency Doppler shifted by using the systemic velocity of the observed galaxy. For observations taken with the AB 
receivers (in 2008), we used the entire 0.5~GHz bandwidth when fitting the baseline. Depending on the quality of the baseline, we fitted a polynomial
baseline of degree one or three. In each case we avoided the velocity range of molecular emission from the galaxy known from the HERACLES maps.
When necessary, we removed spurious spikes from the spectra. In the positions observed in 2009 ($\sim20\%$ of the total), we also suppressed a systematic small-amplitude
ripple in the HCN(1--0) spectra via flagging and linear interpolation in the Fourier transformed spectrum. For each line and position, we averaged all the spectra except those 
that showed egregious artifacts, poor baselines or abnormal noise levels to produce a final spectrum. 

From these reduced, averaged spectra, we calculate velocity-integrated intensities, expressed throughout the paper on the main beam temperature 
($T_\mathrm{MB}$) scale in units of K~km~s$^{-1}$. We derive these intensities summing over the velocity range known to contain molecular line 
emission from the CO (2--1) maps (the same region that we avoided in baseline fitting). We estimate the statistical error, $\sigma_\mathrm{line}$, via

\begin{equation}
\sigma_\mathrm{line}=rms_\mathrm{channel}\sqrt{\delta V\Delta V},
\end{equation}

\noindent
where $rms_\mathrm{channel}$ is the rms noise, in Kelvin, for each channel, $\delta V$ is the channel width and $\Delta V$ is the width of the line window. 
Throughout this paper, we 
disregard the fine structure of HCN, which is not apparent in our observations, and quote a single integrated intensity for the $J = 1 \rightarrow 0$ transition. 
The two brightest fine-structure levels of the $J=1 \rightarrow 0$ transition lie only $\sim5$~km~s$^{-1}$ apart. This is comparable to our typical spectral 
resolution ($3.4-6.8$~km~s$^{-1}$) and three times smaller than the smallest line width (FWHM) of any detected line in our survey.
 
We acquired and reduced the CO (1--0) data in a similar way, although the integration times  were much shorter than for HCN (1--0) because
CO (1--0) is a much brighter line. In most of this paper we derive molecular surface densities from the CO(2--1) HERACLES maps (Sect.~\ref{ss-data-param}), 
because, unlike the CO(1--0) data, they are available at all observed positions. In Appendix~\ref{ss-check-mol}, we use the \coone\ data where available to
demonstrate that the CO excitation conditions do not bias our main results.
 
Appendix~\ref{a-tabl} summarizes the results of our 30-m survey. In Figs.~\ref{f-map-1}--\ref{f-map-5}, we show our 62 observed positions on the HERACLES CO(2--1) maps
of the target galaxies. Figs.~\ref{f-spec-1}--\ref{f-spec-14} show the final \hcn\ and \coone\ spectra. Table~\ref{t-pos} reports the velocity integrated intensities, $\ihcn$ and 
$I_\mathrm{CO10}$, at each position, indicating statistical errors, $\eihcn$ and $\sigma_\mathrm{CO10}$, in brackets.
The quoted errors do not include the $\approx 10\%$ uncertainty that we expect in the overall flux calibration. 
 For the analysis in this paper,
we consider all spectra with $SNR_\mathrm{line}\equiv(I_\mathrm{line}/\sigma_\mathrm{line})\geq4$ to be detections. As a check, we verified that in all cases 
the line shapes of HCN were consistent with the shape of the CO(2--1) line at matched spatial and spectral resolution (Figs.~\ref{f-spec-1}--\ref{f-spec-14}). We applied the
same check to spectra with lower signal-to-noise ratio and, based on this criterion, we promoted three HCN spectra with $3.4\leq SNR_\mathrm{HCN}\leq3.7$ 
to be detections. In summary, we detect \hcn\ at 48 positions with a median $SNR_\mathrm{HCN}$ of $\sim9.5$. The HCN detections span $\sim2.5$~dex in $\ihcn$ 
(a factor of $\simeq330$) and extend from galaxy centers up to $r_{25}=0.75$. The detection at the largest radius occurs in NGC~6946 at a galactocentric radius of $8.5$~kpc. We detect the CO(1--0) line in 57 out of 58 positions with a median $SNR_\mathrm{CO}$ of $\sim35$.  When reporting non-detections, we give $4\sigma$ upper limits.

\begin{table}
\begin{center}
\begin{threeparttable}
\caption{Receiver-Backend Combinations.}
\begin{tabular}{l l l }
\hline\hline
\noalign{\smallskip}				
 Month/year & Receiver(s) & Backend  \\
\noalign{\smallskip}				
\hline
\noalign{\smallskip}							
 01/2008   & A100$+$B100 & 1~MHz filterbank       \\							            	             
 12/2008   & A100$+$B100 & 1~MHz filterbank       \\
 08/2009   & E090      & WILMA (2 MHz resolution) \\
 06/2011   & E090      & WILMA (2 MHz resolution) \\
\noalign{\smallskip}								
\hline
\end{tabular}
\label{t-febe}
\end{threeparttable}
\end{center}
\end{table}

\begin{table}
\begin{center}
\begin{threeparttable}
\caption{Observed Lines}
\label{t-lin}
\begin{tabular}{r r r r r }
\hline\hline
\noalign{\smallskip}
Line 			& Frequency & {Beam} & \multicolumn{2}{c}{$T_\mathrm{MB}/T_\mathrm{A}^*$}\\
				&	(GHz)			& ($\arcsec$) & [2008] & [2009/2011]\\
\noalign{\smallskip}
\hline
\noalign{\smallskip}
HCN(1--0)		&  88.6			& 27.8 & 1.23 & 1.17\\
CO(1--0)			& 115.3			& 21.3 & 1.28 & 1.22 \\
\noalign{\smallskip}
\hline
\end{tabular}
\begin{tablenotes}
\item{\sc Note. ---} Line rest frequencies. Reported beam size corresponds to the FWHM of the 30-m beam. 
Separate $T_\mathrm{MB}/T_\mathrm{A}^*$ ratios are given for the 2008 and 2009/2011 campaigns, which used different receivers (see text and Table~\ref{t-febe} for details).
\end{tablenotes}
\end{threeparttable}
\end{center}
\end{table}

\subsubsection{Ancillary Data}
\label{sss-anc-data}

Our target galaxies belong to the HERACLES sample, which means that we have CO(2--1) emission mapped at $\sim13\arcsec$ across their disks. 
HERACLES targets many individually well-studied nearby galaxies and was designed to overlap with other surveys that provide data across the electromagnetic spectrum.
  The data from these other
surveys and from numerous studies of individual galaxies, provide us with an excellent characterization of the local properties of the ISM and of the star formation rates
at each of our pointings. A detailed account can be found in \citet[][hereafter L12; see also \citetalias{lero13}]{lero12}. In short, we make use of the following data:

\begin{itemize}
\item CO(2--1) line emission, from HERACLES \citepalias{lero13}. 
\item IR emission in the $3.6-160$~$\mu$m range, from the {\sl Spitzer} Infrared Galaxies Survey \citep[SINGS,][]{kenn03} and the Local Volume Legacy Survey \citep[LVL,][]{dale09}.
\item Continuum-subtracted H$\alpha$ emission, from SINGS, LVL, GoldMine \citep{gavazzi03}, and a compilation of other, smaller studies
\citep{hoopes01,boselli02,knapen04}.
\item {\sc Hi} line emission, from The {\sc Hi} Nearby Galaxy Survey \citep[THINGS,][]{walt08} and a collection of new and archival VLA data (\citetalias{lero13} and Leroy et al., in prep.). 
\end{itemize}
	
Except in the few cases noted below, we convolved all the available maps to our working resolution of $28\arcsec$, i.e., the resolution of our HCN data. We  closely followed \citetalias{lero12} to estimate physical parameters from them.
 
\subsection{Physical Parameter Estimates}
\label{ss-data-param}

All surface densities in this paper are {\em face-on} values. For this, the equations to convert intensities measured along the line of sight to surface densities include a ${\cos(i)}$ factor, where ${i}$ is the adopted inclination of the galaxy disks (Table~\ref{t-gal}). The cosine factor corrects for the observational bias that makes measured intensities to increase with ${i}$ as the beam intersects a larger volume of the disk.

In all cases we have tried to keep the physical parameters in this paper closely linked to observed intensities. That is, total molecular surface density is a recasting
of CO intensity, dense gas surface density is a linear translation of HCN intensity, and so on for {\sc Hi} (21-cm intensity), star formation (IR intensity from 24$\mu$m),
and stellar surface density (near-IR intensity). We explore additional effects related to parameter estimation in the appendix but the results of this paper can mostly be 
straightforwardly read in terms of observables and we plot these as an alternative axis whenever possible.

\subsubsection{Molecular Surface Density}

The mass surface density of molecular gas ($\smol$, including helium) is commonly derived from the velocity-integrated intensity of the CO~(1--0) line.  For 
practical reasons, in this paper we derive $\smol$ from the HERACLES observations of the CO~(2--1) line, $I_\mathrm{CO21}$; we can easily convolve our
full maps at higher resolution to match the HCN(1--0) beam and derive a CO(2--1) intensity for each point (while the CO 1--0 line is observed at higher resolution
than the HCN 1--0 line because of its rest frequency). 
We then define $\ico$ as the CO(1--0) intensity inferred from the HERACLES observations on the assumption of a fixed  CO(2-1)--to--CO(1--0) line ratio, $R_{21}$:

\begin{equation}
\ico=\frac{I_\mathrm{CO21}}{R_{21}}.
\label{e-ico}
 \end{equation}

We adopt $R_{21}=0.7$, which is the average ratio found by \citetalias{lero13} by comparing their HERACLES CO(2--1) data with \coone\ data available in the 
literature (including the data presented in this study) at matched angular resolution. Our new observations of CO(1--0) support this characteristic ratio as appropriate
for our data. 
In Appendix~\ref{ss-check-mol}, we redo large portions of our analysis with the available CO (1--0) data and confirm that the assumption of a fixed $R_{21}$ 
does not bias our results.

From $\ico$, we derive 

\begin{equation}
\label{e-smol}
\smol=\aco\ico\cos(i),
\end{equation}

\noindent where $\aco$ is the  CO--to--molecular mass conversion factor, here phrased in terms of CO (1--0). By default, we assume throughout our sample a fixed, Galaxy-like conversion factor 
$\aco=\aco^0\equiv4.4$~$M_\odot$~pc$^{-2}$~(K~km~s$^{-1}$)$^{-1}$ (derived from a H$_2$ column density conversion of 
$X_\mathrm{CO}=2\times10^{20}$~cm$^{-2}$~(K~km~s$^{-1}$)$^{-1}$ and a 1.36 helium-correction factor). We explore the impact of varying this assumption in
Sect.~\ref{s-cf}. 

Note that our approach to $\aco$ could simply be recast as adopting a fixed conversion factor for the CO (2--1) intensity. For an in-depth investigation into the
CO-to-H$_2$ conversion factor in HERACLES galaxies, we refer the reader to \citet{sand13}.
 
\subsubsection{Dense Molecular Gas Surface Density}

Molecular clouds host cores and dense filaments comprised of gas with densities several orders of magnitude higher than the cloud 
average ($\bar{n}$). The emission lines of molecules with high electric dipoles (e.g., HCN, HCO$^+$, CS) trace this dense component, 
in which collisions can thermally excite their energy levels. A conversion factor ($\alpha_\mathrm{HCN}$) can be defined to calculate the 
mass surface density of dense molecular gas (including helium), $\sdens$, from the HCN(1--0) intensity:

\begin{equation}
\label{e-dens}
\sdens=\alpha_\mathrm{HCN}\ihcn\cos(i).
\end{equation}

We initially assume a fixed $\alpha_\mathrm{HCN}=\ahcn^0\equiv10$~$M_\odot$~pc$^{-2}$~(K~km~s$^{-1}$)$^{-1}$, which  \citet{gao04b}
argued to be reasonable for the disks of normal, star-forming galaxies. In discussing $\sdens$, we consider this quantity to represent the surface density of 
gas at densities above a fixed cutoff, $n_\mathrm{dense}$, where $n_\mathrm{dense}$ is of the order of the star formation threshold expected 
from some models and observations, i.e., $\sim10^4-10^5$~cm$^{-3}$. The exact value of $n_\mathrm{dense}$ does not matter for
most of our analysis, as long as this density is associated with the dense, often star-forming substructures within clouds.

A potentially important issue is how  $n_\mathrm{dense}$ compares with the density required to effectively excite the HCN(1--0) line 
($\neff$). For the HCN(1--0) line, the critical density at which the de-excitation rates by collisions and by spontaneous 
emission are equal is $n_  \mathrm{crit}=1.25\times10^6$~cm$^{-3}$ at $T_\mathrm{K}=20$~K \citep[taking the rate coefficients listed 
in the LAMDA database\footnote{{http://home.strw.leidenuniv.nl/$\sim$moldata/}};][]{scho05}. However, it is well-known from radiative transfer 
models \citep[e.g.,][]{scov74} that a line can be effectively excited at densities lower than $n_\mathrm{crit}$ thanks to opacity effects.
The excitation threshold is thus sensitive to variations in the gas temperature, the velocity gradient, and the HCN abundance within the clouds.
Based on a grid of models with the RADEX radiative transfer code \citep{vtak07}, we find  
$\neff$ could be $\sim10^4-10^5$~cm$^{-3}$ for typical cloud conditions.
With our definitions, any variation in this excitation threshold could imply a change in $\ahcn$. 
We explore how an $\ahcn$ that varies with environment would impact our conclusions in Sect.~\ref{s-cf}. 

\subsubsection{Other Local Conditions Within Our Beams}

{\sl Atomic gas.} We assume that the 21~cm {\sc Hi} emission is optically thin and derive the mass surface density of atomic gas 
($\satom$, including helium), from the velocity-integrated intensity of the line, $I_\textsc{Hi}$:

\begin{equation}
\frac{\satom}{M_\odot~\mathrm{pc}^{-2}}=1.98\times10^{-2}\frac{I_\textsc{Hi}}{\mathrm{K~km~s^{-1}}}\cos(i). 
\end{equation}

The total mass surface density of gas, $\sgas$, is defined as $\sgas=\smol+\satom$ and the molecular-to-atomic ratio is $\smol/\satom$.
In the observed positions where HCN is detected, the molecular gas makes up $\sim50-100\%$ (median $88\%$) of the total gas surface density. 

{\sl Stars.} We estimate the mass surface density of old stars from the $3.6$~$\mu$m intensity, $I_{3.6}$ via:

\begin{equation}
\frac{\sstar}{M_\odot~\mathrm{pc}^{-2}}=280\frac{I_{3.6}}{\mathrm{MJy~sr^{-1}}}\cos(i),
\end{equation}

\noindent which we consider uncertain by $\sim 50\%$ due to contamination from young stars and dust emission \citep{meid12}
and uncertainties in the mass-to-light ratio \citep{zibetti09,meidt14}.

{\sl Dust Properties and IR Spectral Energy Distribution.} We convolve the LVL and SINGS {\em Spitzer} images to a 
common resolution ($\sim40\arcsec$), construct radial profiles for each galaxy, and fit the average spectral 
energy distribution in each ring using the dust models of \citet{drai07}; for details of this specific processing see \citet{lero12}.
From these models we derive several dust properties within radial bins, of which the $24~\mu$m--to--TIR  ratio, the ``cirrus'' 
emissivity,  and the dust--to--gas mass ratio (after comparing with CO and {\sc Hi} data also convolved to $40\arcsec$) are 
used in this paper to derive star formation rates (see below and Appendix~\ref{ss-check-sfr}). These quantities are all ratios (rather than absolute intensities) and we
assume that they hold, at least on average, at the somewhat higher working resolution of our HCN observations.

{\sl Star Formation Rate (SFR).} To easily compare our results with most studies of molecular lines in  galaxies, we estimate the 
SFR surface density ($\ssfr$) from the total IR intensity ($\itir$):

 \begin{equation}
 \label{e-sfrir}
 \frac{\ssfr}{M_\odot~\mathrm{Myr^{-1}~pc^{-2}}} = 1.87\times10^{-3}\frac{\itir}{L_\odot~\mathrm{pc}^{-2}~\mathrm{sr^{-1}}}\cos(i).
 \end{equation} 

Here, $\itir$ is calculated from the measured $24~\mu$m intensity ($I_{24}$) and the dust models, which produce a 24-to-TIR ratio that rigorously
holds at slightly larger scales than our beam. This can alternatively be thought of as approximately using the 24$\mu$m emission to make an aperture correction
that matches the lower resolution TIR, estimated from $24~\mu$m$, 70~\mu$m, and $160~\mu$m data, to our $28\arcsec$ HCN beam. 

Equation~\ref{e-sfrir} corresponds to  the calibration by \citet[][see also \citealp{kenn12}]{murp11} and we prefer this approach because it
links us cleanly to previous work comparing HCN and CO, which has focused mostly on the integrated IR luminosity as a star formation rate tracer.
This allows our results to be seamlessly read in terms of IR luminosity surface density, even though we label them ``SFR.'' 
In Appendix~\ref{ss-check-sfr}, we take advantage of the ancillary data set and build alternative SFR tracers (H$\alpha$, $24\mu$m, and a linear 
combination of both) to confirm that the choice of star formation rate tracer has a minimal impact on our results. All the SFR calibrations in this paper 
assume a \citet{krou01} IMF. 

\subsection{Extragalactic Reference Sample}
\label{ss-gb12}

We compare our data to observations of molecular lines in 106 SF galaxies and (U)LIRGs compiled by \citet[][hereafter GB12]{buri12}. GB12
combined published data with new observations using the IRAM 30-m telescope. GB12 present CO(1--0) and HCN(1--0) luminosities for 
all of their targets, estimating these from higher-order transitions in a few high-redshift objects. Their sample includes both upper limits (non-detections)
and lower limits for objects without full mapping. In principle, we derive the total SFR (in $M_\odot$~Myr$^{-1}$) for these targets from the total IR luminosity ($\ltir$) again
using Equation~\ref{e-sfrir}. However, following \citet{grac08}, we rather use $1.3\times \lfir$ instead of $\ltir$, where $\lfir$ is the far IR luminosity. This allows us to avoid AGN--driven 
contamination of $\ltir$ in the brightest (U)LIRGs, while it hardly changes the estimated SFR in objects powered by star formation. In particular, once the most conspicuous AGN sources are removed, the $\ltir/\lfir$ ratios of the GB12 sample show a small scatter of 0.05~dex rms ($12\%$) around a median value of $\simeq1.3$. 
Since we lack TIR data for some GB12 objects, we also use ${(1.3\times\lfir)}$ rather than ${\ltir}$ to  classify them as either  SF galaxies or (U)LIRGs (Sect.~\ref{s-intro}). In principle, this affects the classification of only five galaxies.
 
For simplicity and in order to present the data as directly as possible, we assume the constant conversion factors quoted in Sect.~\ref{ss-data-param} to 
derive (dense) molecular masses from the measured line luminosities. This is a non-trivial assumption, given the ample evidence that these factors are a 
few times lower in (U)LIRGs than in SF galaxies \citep[e.g., see][for a review]{bola13}. The main influence of these factors on our first-order analysis is likely 
to be some offset between (U)LIRGs and SF galaxies. We return to the topic of conversion factors in Sect.~\ref{s-cf}. An additional concern is the 
unobscured SFR, which is not perfectly accounted for by the IR luminosity and could make up a significant fraction of the total SFR in SF galaxies, thus 
causing additional offsets \citepalias{buri12}.  

For the 83 objects that have measured sizes for the star-forming/molecular disk, we convert luminosity/flux measurements to estimates of the average 
$\sdens$, $\smol$, and $\ssfr$. We divide the total molecular (dense) mass or SFR in each galaxy by its disk area. The measured sizes are typically
given as FWHM values of the light distribution, so that we quote $\left< \Sigma \right> = M / (\pi\mathrm{FWHM^2_{maj}}/4)$. Here, $\mathrm{FWHM_{maj}}$ would be the Full Width at Half Maximum along the the major axis of the star-forming/molecular disk. This definition is equivalent to that of the {\em face-on} surface densities derived from our data (Sect.~\ref{ss-data-param}).
When calculating ${\left<\Sigma\right>}$, we implicitly neglect any aperture correction to account for the
fraction of the mass inside the FWHM, which would be a factor of $0.5$ for a two-dimensional Gaussian and potentially as low as $\sim 0.15$ for an exponential 
disk. It is also driven by the total mass and so may be viewed as somewhat closer to the peak $\Sigma$ value than 
the value likely to be found at any given locale in the galaxy (in the same way that the ratio between the total area below a Gaussian curve and its FWHM 
approximately gives its peak value). For the example of a face-on exponential disk, the surface density $\left< \Sigma \right>$ is, in fact, $\sim 4$ times
the peak value of the exponential because the aperture correction (i.e., the mass outside the FWHM) is so large.  For a Gaussian, the number approaches
$\sim 1.4$. Most SF galaxies likely fall somewhere in this range \citep[e.g.,][]{youn95}. We return to this source of bias when we compare our sampling of
galaxy disks to these unresolved data (Sect.~\ref{s-dg-gb12}). For now, we caution the reader that $\left< \Sigma \right>$ derived in this way is biased somewhat high.
 
\subsection{Analysis tools}
\label{ss-data-ana}

We parameterize the relations between pairs of variables using power-law fits, which we use to capture general trends in the data. We derive the 
best fit power law from total-least-square minimization, typically in the log-log space, excluding non-detections and assuming equal uncertainties along both axes.  
Quoted uncertainties in the fit outputs are $1\sigma$ levels analytically derived from the equations of error propagation. Our lower/upper limits are typically compatible 
with the solutions that we find, so we do not expect that their exclusion biases the fit outputs. Given the irregular sampling of our targets, we caution that 
care should be taken before extrapolating our fits to other regimes or data sets. 

We also find it useful to measure the strength of the correlations between pairs of variables in a non-parametric way, which we do using 
Spearman's rank correlation coefficient ($\rank$; hereafter rank coefficient). Throughout this paper, rank coefficients do not  consider limits (i.e., non-detections). 
We measure the significance of $\rank$ by its p--value, i.e., the probability that the absolute value of the rank coefficient for uncorrelated data 
could be equal or higher than the measured value by chance.  Throughout this paper, we consider that a correlation is significant when the 
p-value of its rank coefficient is lower than $2.5\%$.
   
We performed most of the data analysis within the environment of statistical computing R \citep{rref}. 

\section{Comparison with Unresolved Observations: Star Formation Scaling Relations}
\label{s-dg-gb12}

\begin{figure*}
\centering
\includegraphics[width=0.49\textwidth]{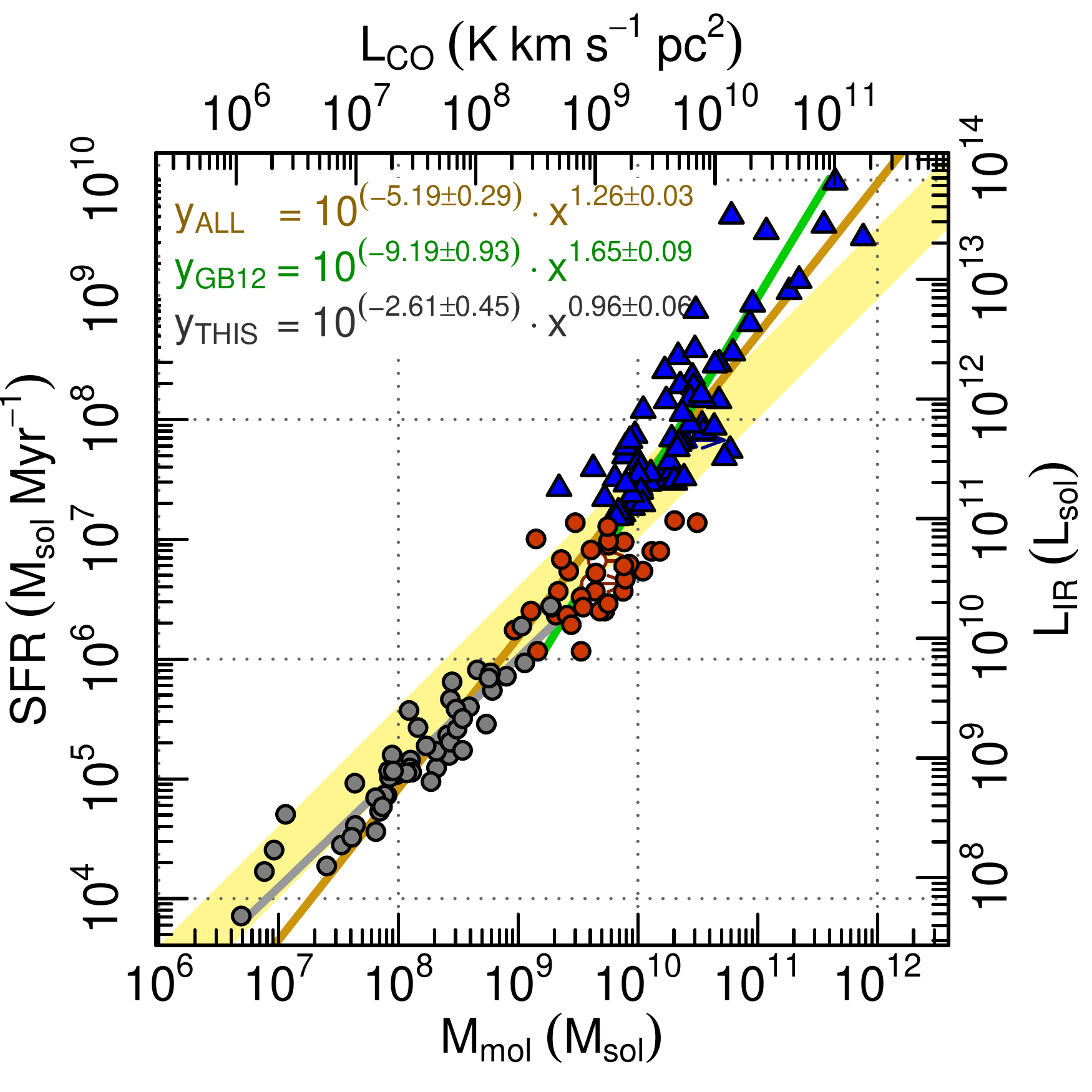}\hspace{0.25cm}
\includegraphics[width=0.49\textwidth]{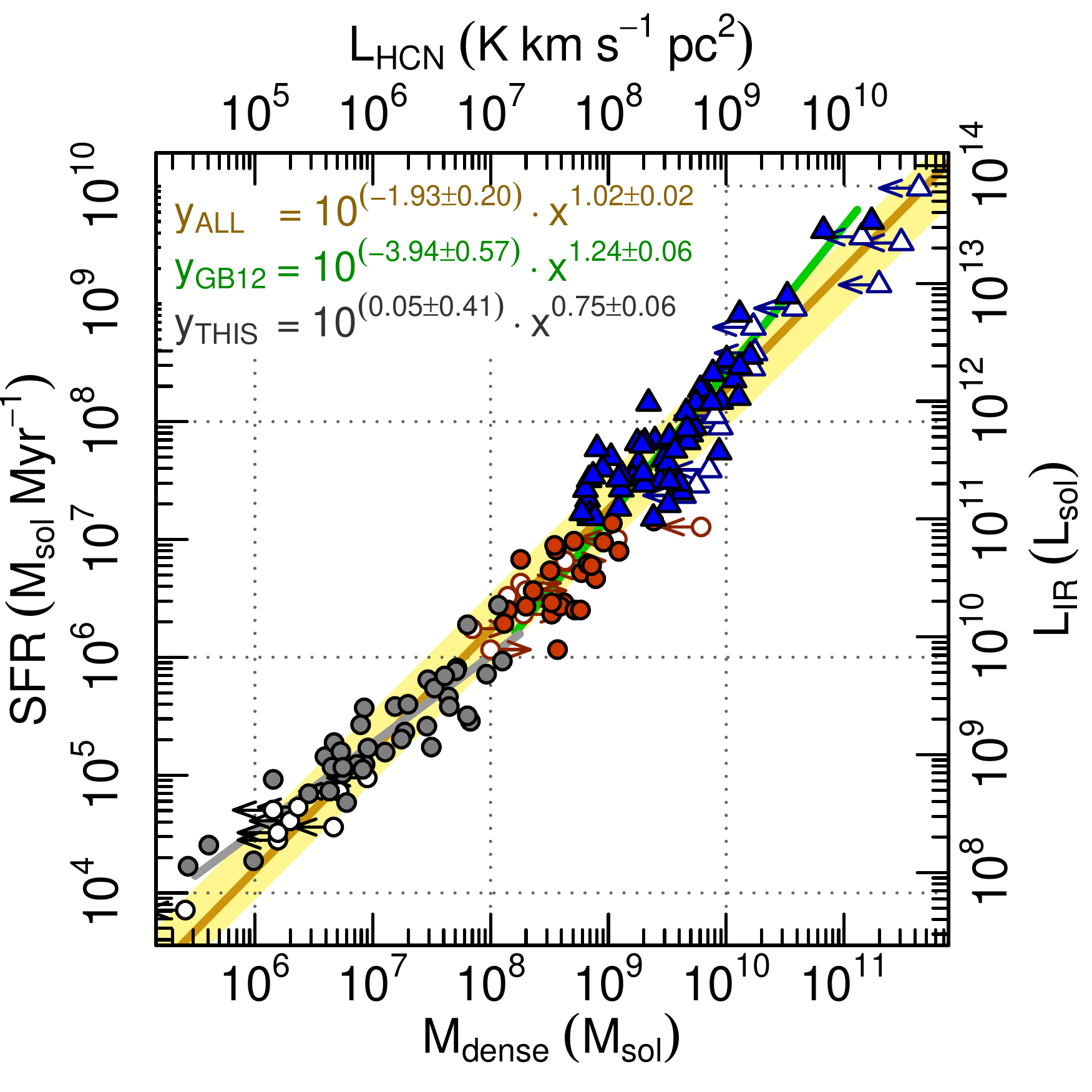}\\
\includegraphics[width=0.49\textwidth]{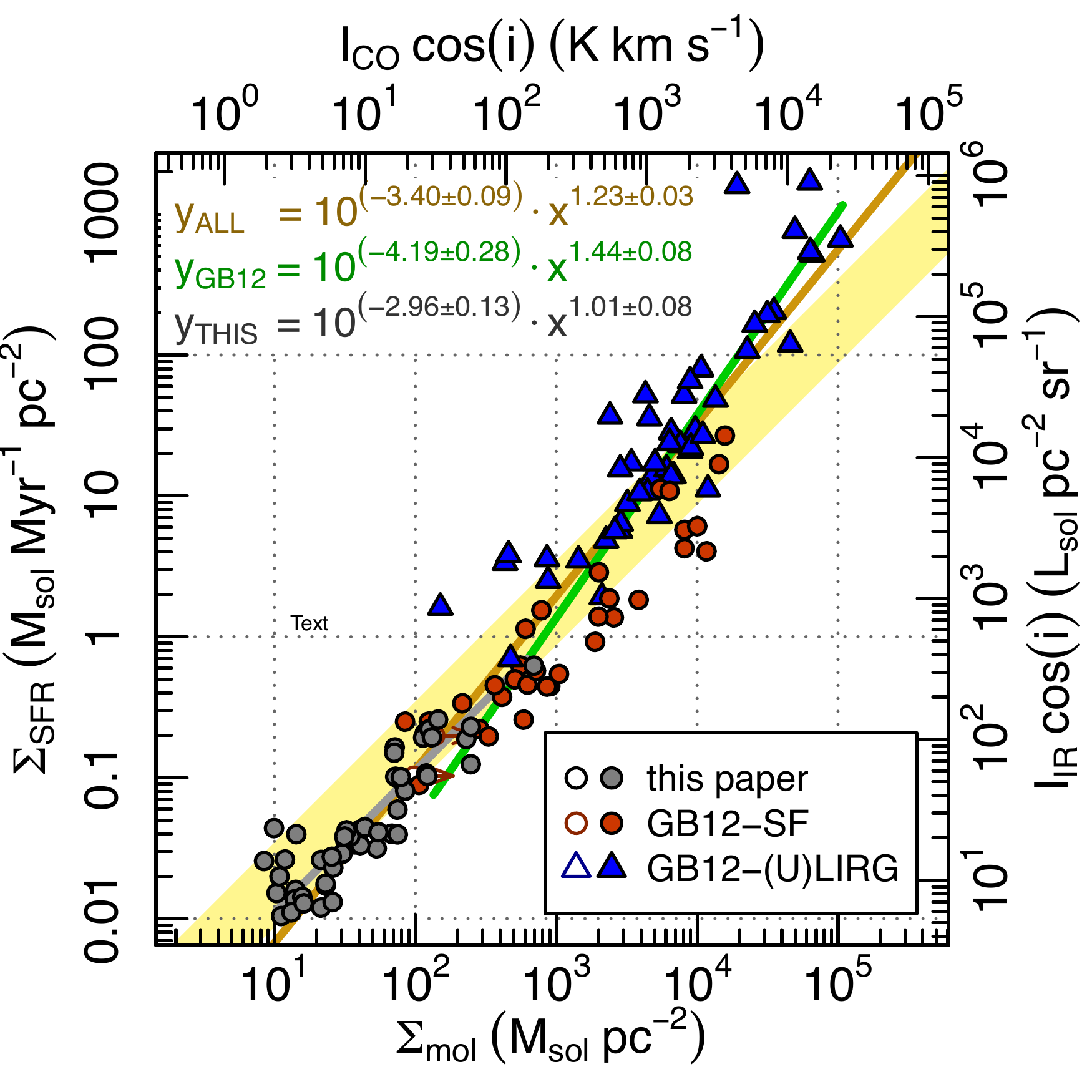}\hspace{0.25cm}
\includegraphics[width=0.49\textwidth]{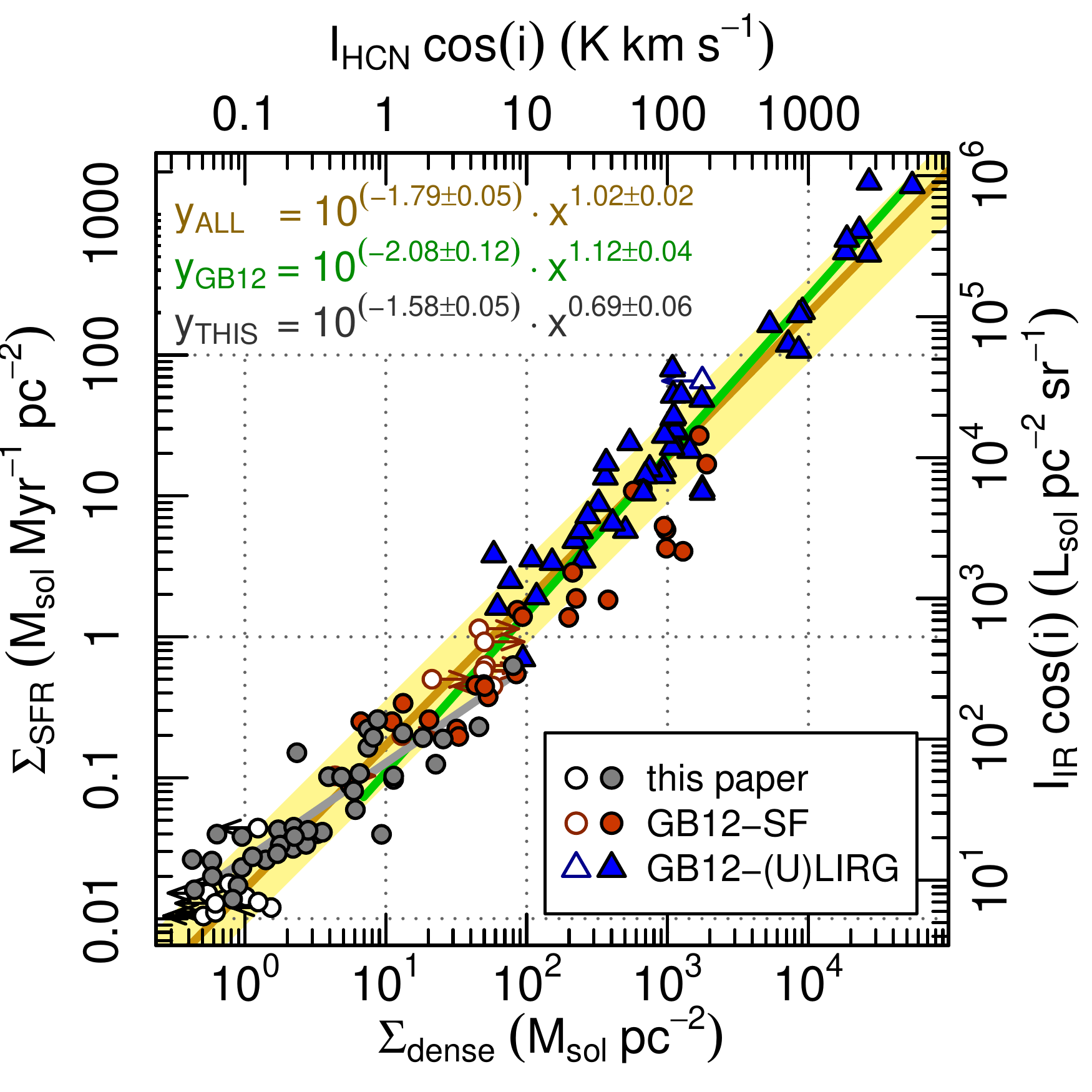}
\caption{
{\em Top row:}
Total SFR ($\ssfr$) as a function of the mass of molecular ({\em left}) and
dense ({\em right}) molecular gas for pointings in galaxy disks (this paper, gray points) and
unresolved star-forming galaxies (red) and (U)LIRGs (blue) from GB12. 
{\em Bottom row:}
Surface density of recent SFR ($\ssfr$) as a function of the surface density of total ({\em left}) and
dense ({\em right}) molecular gas for the same  data sets. 
The top and right-hand axes of each panel display the data in terms of observed quantities. 
Open symbols indicate limits in the direction 
of the attached arrows. The equations in the figures report  power-law fits to our data (gray line), the GB12 sample (green line), 
and all the data (brown line). 
Errors at $1\sigma$-level  in the fit parameters are indicated.
For comparison, the yellow area shows a fixed ratio (power law index $1$) with a factor of 2 scatter.}
\label{f-ks1}%
\end{figure*}

Our HCN detections span a range in HCN (1--0) luminosity (within the beam) of $\sim3\times10^4-2\times10^7$~K~km~s$^{-1}$~pc$^2$. Thus, 
they span the luminosity gap between individual Galactic clouds \citep[up to $\sim1\times10^4$~K~km~s$^{-1}$~pc$^2$ in][]{wu05} and entire galaxy disks 
\citepalias[down to $\sim1\times10^7$~K~km~s$^{-1}$~pc$^2$ in][]{buri12}. This reflects that, at our working resolution of $\sim 1.5$~kpc, we probe ensembles of 
clouds at sub-galactic scales.  Because this scale has been barely explored in previous studies, we begin our analysis by comparing in Fig.~\ref{f-ks1} our data (gray circles) to unresolved observations of galaxies drawn from GB12 (colored symbols; see Sect.~\ref{ss-gb12}).

The top row of Fig.~\ref{f-ks1} plots the total SFR in our pointings (within the $28''$ beam) and in the GB12 galaxies as a function of the (dense) molecular mass traced by HCN ($M_\mathrm{dense}$; left-hand column) and CO ($M_\mathrm{mol}$; right-hand column).  The bottom row shows the same information, but in terms of surface densities of SFR and (dense) gas mass. These surface density scaling relations are often called Kennicutt-Schmidt (KS) plots, and the accompanying power-law fits (``KS laws'') are a common tool for empirical studies of the relations between star formation and the ISM.  We remind that, for sake of simplicity, we adopt fixed CO and HCN conversion factors for both SF galaxies and (U)LIRG. Thus, these are also luminosity-luminosity scaling relations and we include axes showing the corresponding IR and line luminosities.

As expected from previous studies, in all four panels of Fig.~\ref{f-ks1} the SFR shows a good correspondence with the (dense) molecular gas traced by the HCN and CO emission. We also see this in Table~\ref{t-cor1}, which lists the rank coefficients for the plots in Fig.~\ref{f-ks1} considering various groupings
of our data and the GB12 data. All of these correlations are significant  (p-value $<2.5\%$) and high ($\gtrsim 0.8$ for total SFR/masses, except for the subset of SF galaxies from GB12; $\gtrsim 0.9$ for surface densities), thus reflecting the tight correlations seen in the 
figures. Unexpectedly, the rank coefficients of the SFR-vs-$M_\mathrm{dense}$ and SFR-vs-$M_\mathrm{mol}$ relations, on the one hand, and of the  $\ssfr$-vs-$\smol$ and $\ssfr$-vs-$\sdens$ relations, on the other, are very similar for every data set. Although the slopes of the relations vary,
implying differences in the physics at play for different molecular tracers, the CO and HCN lines appears to be almost equally good at predicting the SFR at these scales. We caution the reader that this result may not hold at different scales and/or different data samples.

\begin{table}
\begin{center}
\begin{threeparttable}
\caption{Rank Correlations for (Dense) Molecular -- SFR Scalings}

\begin{tabular}{rrrrr}
\hline\hline
\noalign{\smallskip}
& \multicolumn{1}{l}{M$_\mathrm{dense}$ vs.} & \multicolumn{1}{l}{M$_\mathrm{mol}$ vs.} & \multicolumn{1}{l}{$\sdens$ vs.} & \multicolumn{1}{l}{$\smol$ vs.} \\
 & \multicolumn{1}{l}{SFR} & \multicolumn{1}{l}{SFR} & \multicolumn{1}{l}{$\ssfr$} & \multicolumn{1}{l}{$\ssfr$} \\
\noalign{\smallskip}
\hline
\noalign{\smallskip}
{\em this paper}+GB12 & \si{0.96} & \si{0.96} & \si{0.97} & \si{0.97}\\
{\em this paper} & \si{0.89} & \si{0.95} & \si{0.88} & \si{0.87}\\
GB12 & \si{0.89} & \si{0.86} & \si{0.92} & \si{0.89}\\
GB12 -- SF only & \si{0.54} & \si{0.55} & \si{0.95} & \si{0.92}\\
GB12 -- (U)LIRG only & \si{0.82} & \si{0.83} & \si{0.91} & \si{0.89}\\
\noalign{\smallskip}
\hline               
\end{tabular}

\begin{tablenotes}
\item{\sc Note. ---} Rank correlation coefficients relating star formation and gas surface densities (Fig.~\ref{f-ks1})
for various groupings of our data and the GB12 data. 
Star symbols indicate significant correlations (p-value $< 2.5\%$). 
\end{tablenotes}
\label{t-cor1}
\end{threeparttable}
\end{center}
\end{table}

Fig.~\ref{f-ks1} shows that our data extend the range of gas masses and surface densities spanned by unresolved observations. In terms of total gas masses (top row), 
our brightest points, mostly vigorously star-forming galaxy centers, lie right below the least active SF galaxies  (dark red circles). 
In terms of surface densities (bottom row), there is a partial overlap between them. 
Pointings at larger
galactocentric radii, where the lines are typically fainter (Sect.~\ref{s-env}), have masses and surface densities as much as two and one order of magnitude lower than the galaxies in the GB12 sample, respectively.  Recall that we expect the surface densities estimated from the unresolved (GB12) observations to be biased somewhat high by the lack of an aperture correction (Sect.~\ref{ss-gb12}). 
Correcting by these factors, which should affect both axes, would
tend to increase the GB12 overlap with our data set in terms of surface density.

The bottom-row panels of Fig.~\ref{f-ks1} might support the idea that (U)LIRGs and SF galaxies form distinct star formation sequences
(\citealt{genz10}, \citealt{daddi10}, \citetalias{buri12}). With or without aperture correction applied to the unresolved observations, 
our data and the SF galaxies from GB12 align along a continuous and monotonous sequence. In contrast, the (U)LIRG sample lies parallel,
but seemingly offset to higher $\ssfr$ for the same (dense) gas surface density. The offset is particularly noticeable in the $\ssfr-\smol$ plane, and  would be even larger if we assumed lower-than-Galactic CO and HCN conversion factors for the (U)LIRG subsample of \citetalias{buri12}. 

In the four panels of Fig.~\ref{f-ks1}, adding our data to the GB12 sample has the effect of lowering the best-fit power law indices. However, the reason for this change differs between the left-hand and right-hand panels. In the  $\ssfr-\smol$ plane (similarly in the $SFR-M_\mathrm{mol}$ plane), the global power-law index drops from $\sim1.4$ to $\sim1.2$ because our data (power-law index of $1.01\pm0.08$) reinforce the linear trend shown by the SF galaxies in the GB12 sample (index of $0.97\pm0.07$; fit not shown 
in the plots). By contrast, in the  $\ssfr-\sdens$ plane (similarly in the $SFR-M_\mathrm{dense}$ plane) the global index changes from $\sim1.1$ to $\sim1.0$ when our data are included because our data by themselves follow a significantly sublinear  relation (index of $0.69\pm0.06$).  In this plot, the SF galaxies alone are fitted 
by a higher index of $0.88\pm0.06$. This highlights a main conclusion of this paper: that, perhaps surprisingly, {\em the apparent efficiency with which
dense gas forms stars seems to vary systematically across our sample of galaxy disk pointings}. The sense of the variation is that regions with 
lower dense gas fractions and lower dense gas surface densities tend to have a higher apparent rate of star formation per unit dense gas.

In summary, despite the continuity between our data and SF galaxies, Fig.~\ref{f-ks1} suggests that the star formation laws for the dense molecular gas 
are not the same at galactic and sub-galactic scales. This is not illogical, given that many of the new conclusions in this paper come from data points
with low surface density. These typically correspond to off-center positions and, because they are relatively faint in line emission, they will not dominate the
total galaxy luminosity. Our results thus suggest that the average star formation properties of molecular clouds vary across galaxy disks. We investigate this topic in detail
in the following sections, focusing on our new data set. Doing so, we take advantage of the fact that our data set resolves disk galaxies into discrete regions and
targets the well-characterized HERACLES (i.e., THINGS and SINGS) galaxies.

\section{Dense Gas Fraction and Star Formation Efficiency Across Galaxy Disks}
\label{s-env}

\subsection{Normalized Quantities: $f_{\rm dense}$, $SFE_{\rm mol}$, and $SFE_{\rm dense}$}

In this section, we combine our HCN observations with ancillary data of the observed pointings (Sect.~\ref{sss-anc-data}). We explore how the beam-averaged properties of the dense gas in 
unresolved molecular clouds depend on local conditions within a galaxy. Doing so, it is crucial to bear in mind
that, because the molecular interstellar medium in galaxy disks can be sparse and clumpy at our typical spatial resolution, 
neither beam-integrated masses (``luminosities") nor
beam-averaged surface densities (``intensities") can be straightforwardly translated into 
intrinsic cloud properties \citep[see extended discussion in][]{lero13b}. For example, in the simplistic assumption 
that the beam had an area $A$ and probed $\mathcal{N}$ identical spherical clouds of mass $m_\mathrm{cloud}$ and projected area $a_\mathrm{cloud}$, we would find that

\begin{equation}
M=m_\mathrm{cloud}\times\mathcal{N},
\end{equation}

\begin{equation}
\Sigma=\Sigma_\mathrm{cloud}\times a_\mathrm{cloud}\times\frac{\mathcal{N}}{A},
\end{equation}

\noindent where $\Sigma_\mathrm{cloud}\equiv m_\mathrm{cloud}/a_\mathrm{cloud}$ is the intrinsic surface density in a cloud and $M$ and $\Sigma$ are the mass and surface density 
measured at  beam scales\footnote{As usual in extragalactic studies, we assume that the cloud-cloud velocity 
dispersion prevents clouds in the foreground from absorbing the emission of those behind them.}.

Because we work at a fixed angular resolution, beam-integrated masses ($M$) are strongly dependent on the source distance, that determines the number of clouds captured by the beam ($\mathcal{N}$) to a significant extent. This observational bias is partially corrected when working with surface densities. However, 
because our observations span a large range of galactocentric radii, we expect  large variations in the $\mathcal{N}/A$ ratio, 
which measures the ``filling factor'' of clouds per unit area. These filling factor variations, which may largely affect $\ssfr$, $\smol$, and $\sdens$
in the same way, can drive surface density correlations (e.g., see Sect.~\ref{s-dg-gb12} and Table~\ref{t-cor1}),
masking more subtle changes in intrinsic cloud properties.

The simplest way to eliminate the $\mathcal{N}/A$ factor, assuming that it acts approximately equally on all surface density terms,
is to represent our data in terms of three surface density {\em ratios} that are also frequently used in the literature: the dense gas fraction ($\fdens\equiv\sdens/\smol$), 
the star formation efficiency of the molecular gas ($\sfemol \equiv \ssfr/\smol$, i.e., the inverse of the molecular gas depletion time, 
with dimension of time$^{-1}$), and the star formation efficiency of the dense molecular gas ($\sfedens \equiv \ssfr/\sdens$).  
These three parameters are related through the following equation:

\begin{equation}
\label{e-sfemol-fdens}
\sfemol=\sfedens\times\fdens.
\end{equation}
 
 \noindent If all the star formation and molecular emission arises from the same set of clouds inside a beam,
 then these ratios capture the average intrinsic cloud properties over the population within the beam, independent of any filling factor variations. 

Our simple approach to physical parameter estimation means that for most of this paper these rations have direct observational analogs. The ratio
of HCN-to-CO emission maps to the dense gas fraction, the ratio of infrared to HCN emission maps directly to the dense gas star formation efficiency,
and the ratio of infrared to CO emission maps to the overall molecular gas star formation efficiency.
 
\subsection{Environmental Trends}
 
\begin{figure*}
\centering
\par
\vspace{-1cm}
\includegraphics[width=0.45\textwidth]{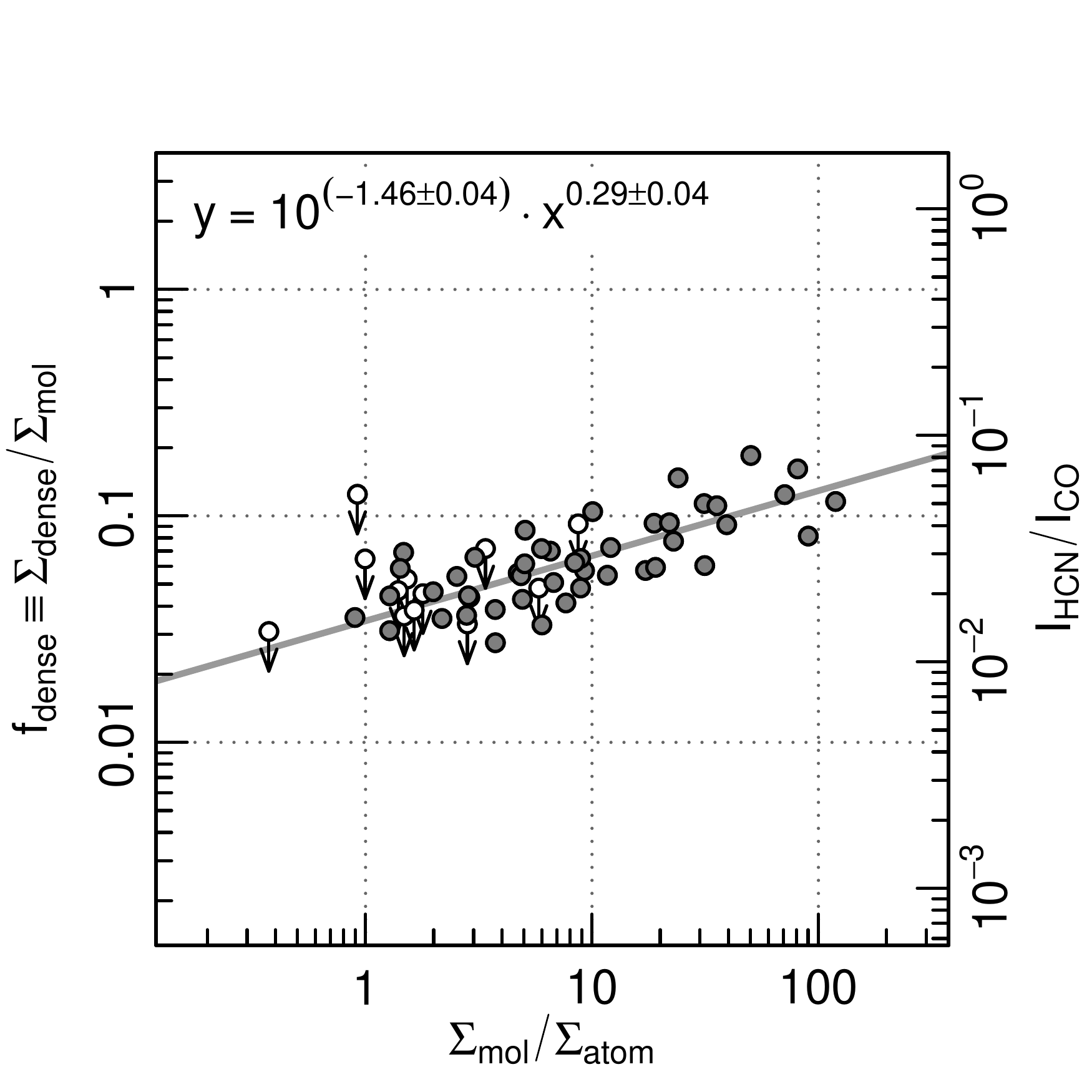}\hspace{1cm}
\includegraphics[width=0.45\textwidth]{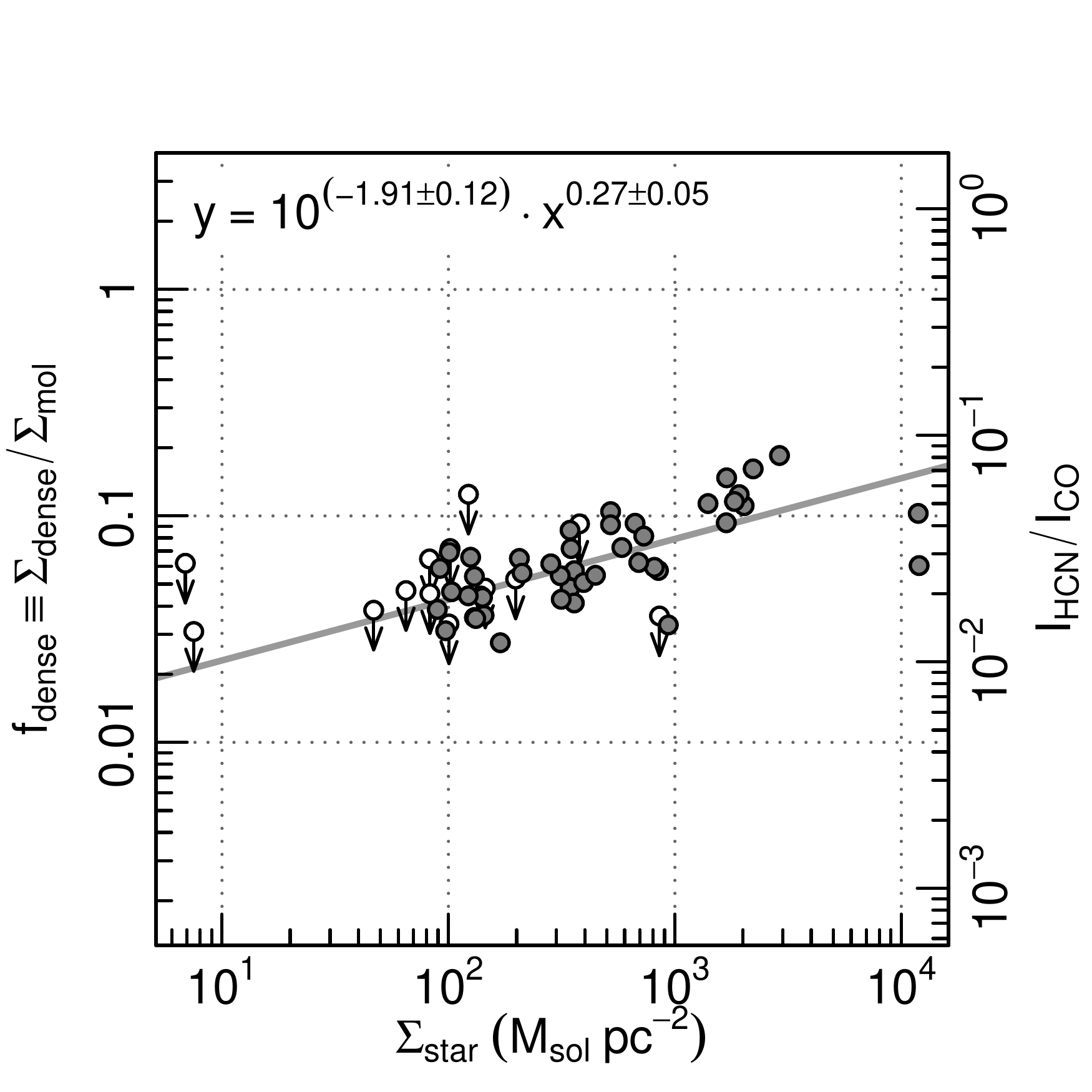}\\
\vspace{-1cm}
\includegraphics[width=0.45\textwidth]{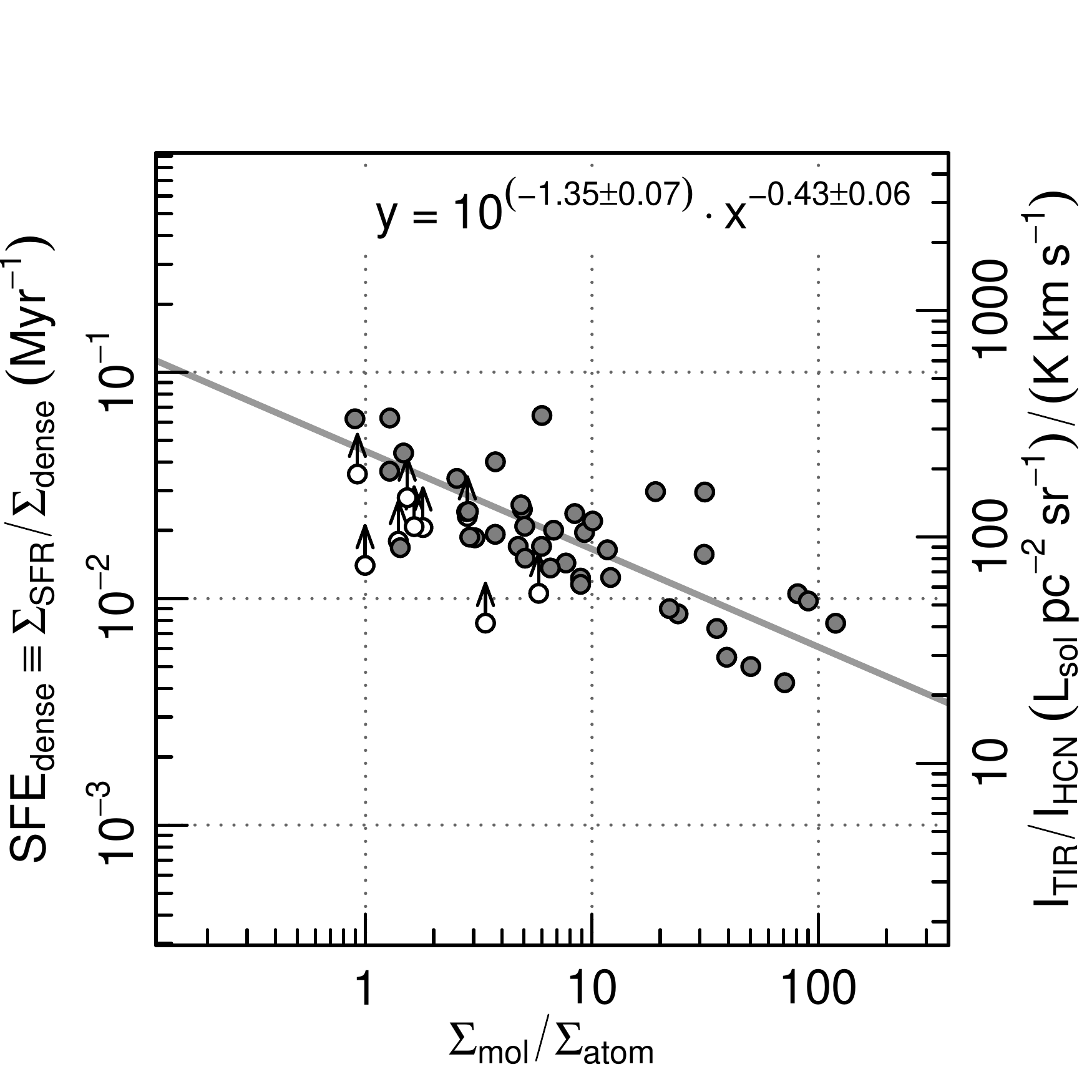}\hspace{1cm}
\includegraphics[width=0.45\textwidth]{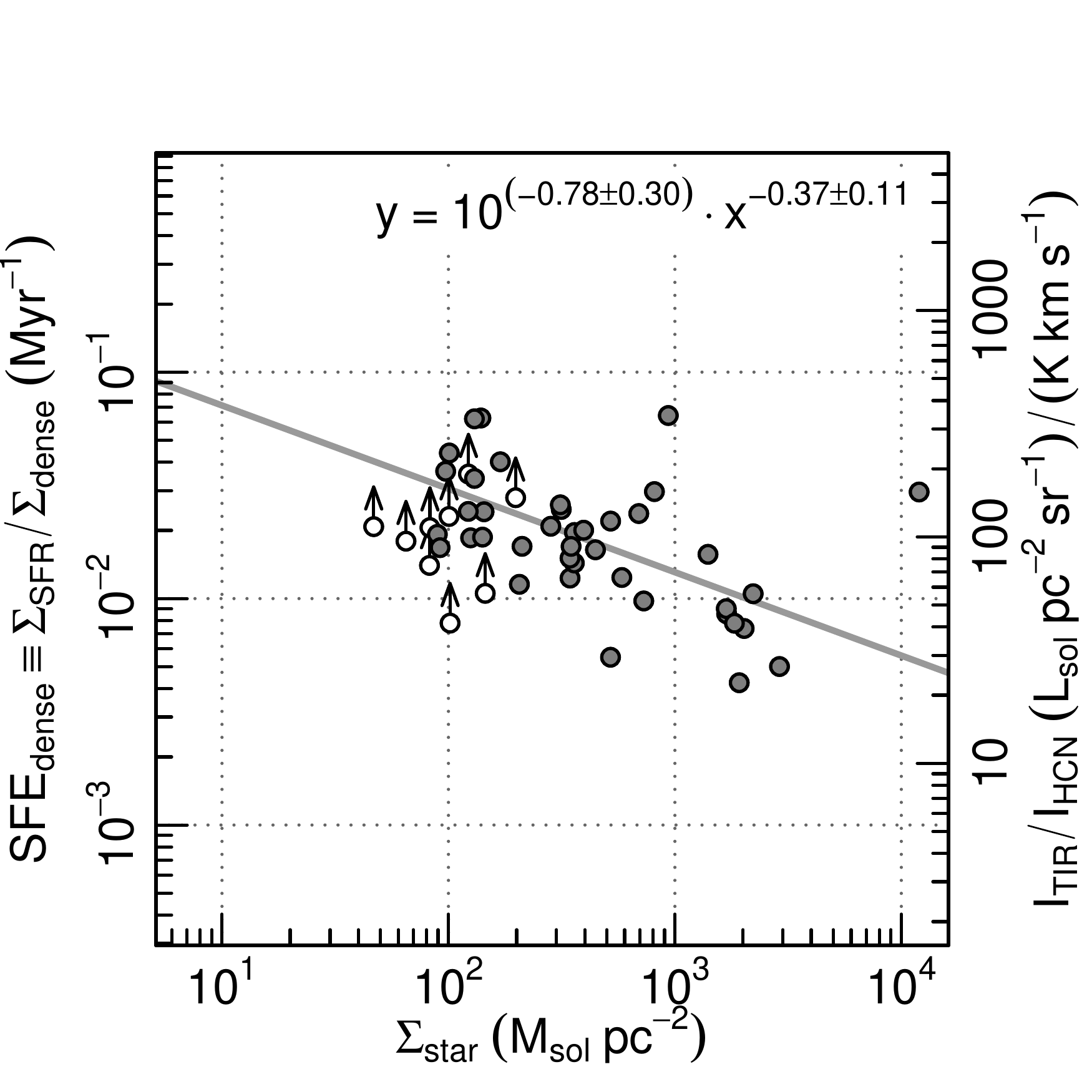}\\
\vspace{-1cm}
\includegraphics[width=0.45\textwidth]{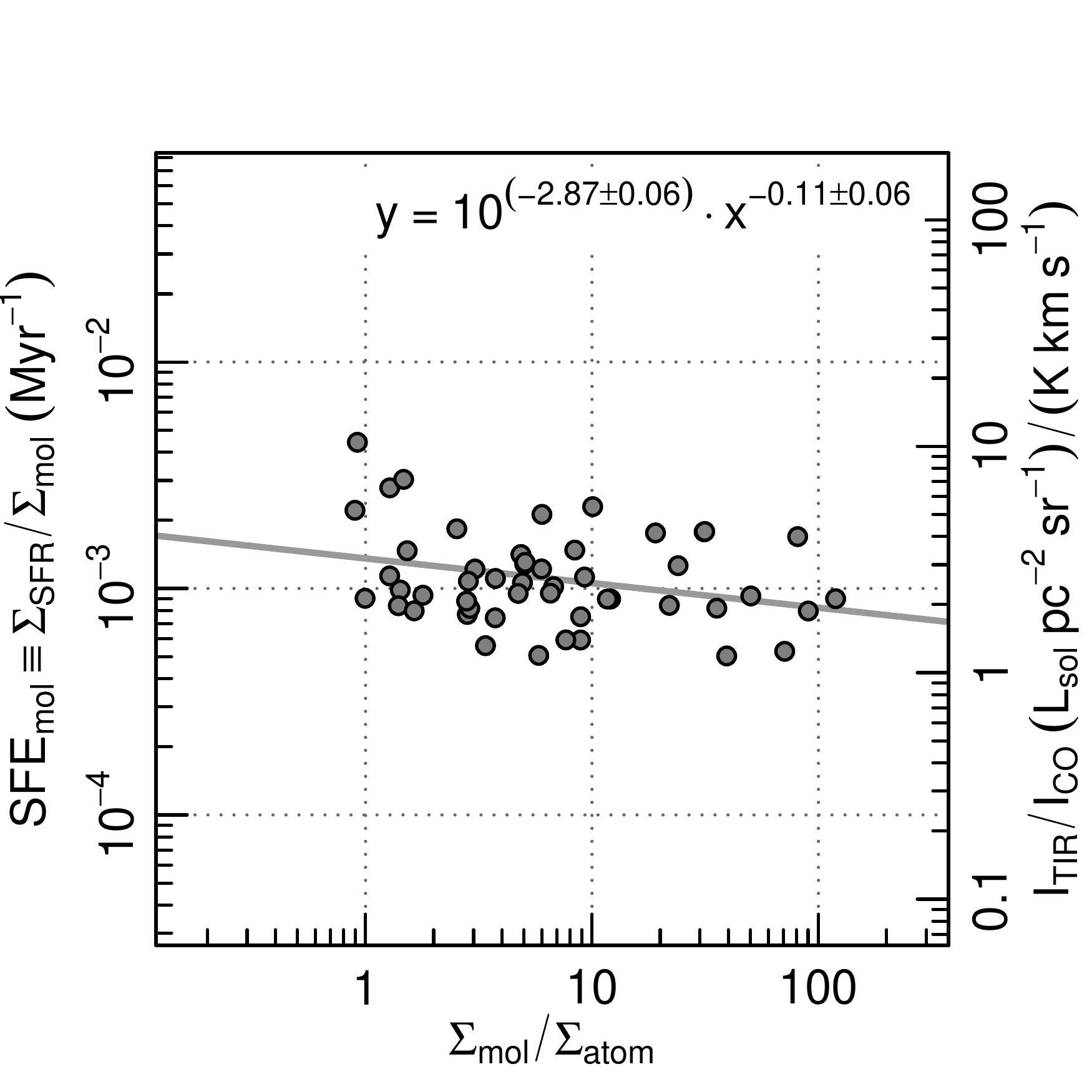}\hspace{1cm}
\includegraphics[width=0.45\textwidth]{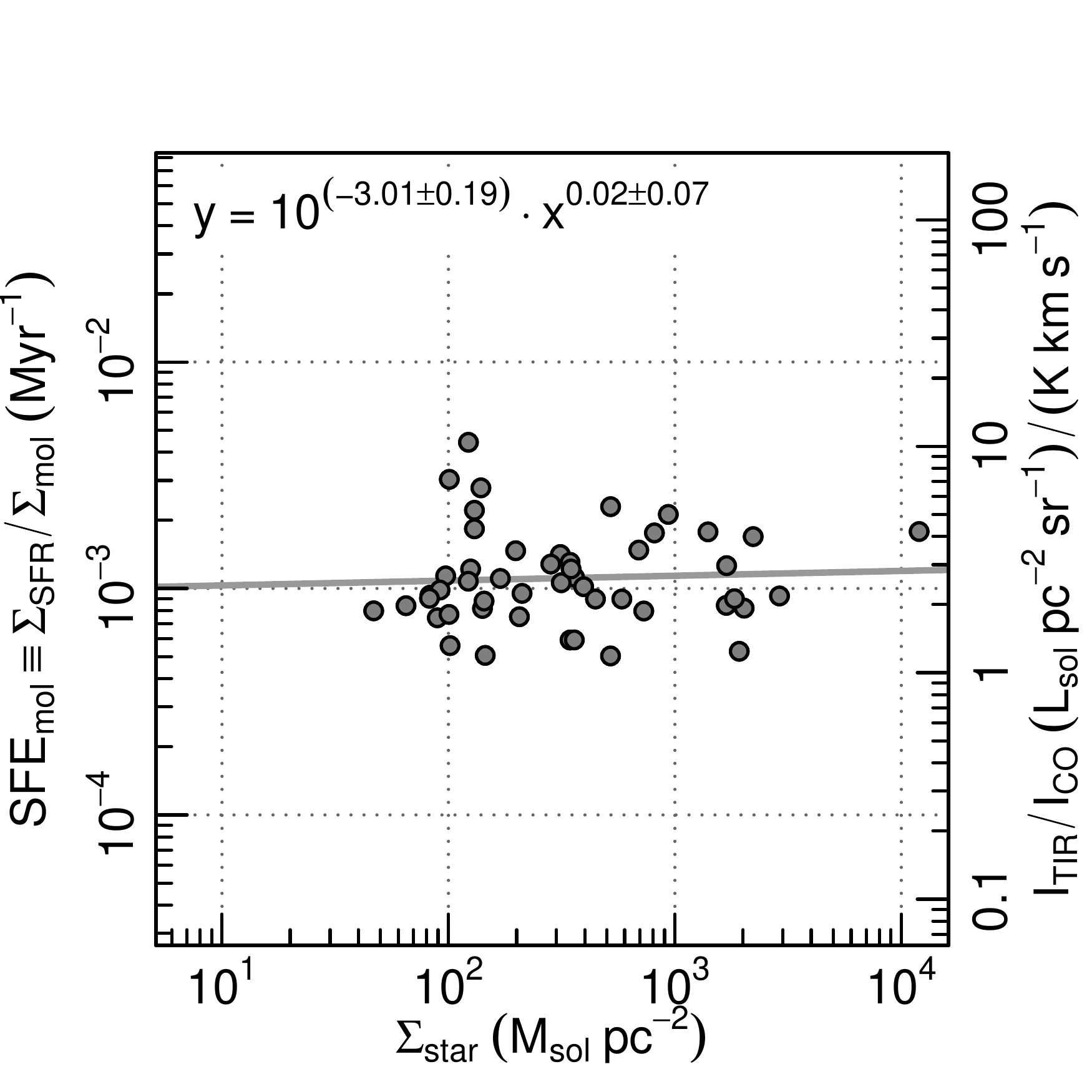}

\caption{Dependence on environment in our observations. From top to bottom, the dense gas fraction $\fdens$ (top row), the star formation efficiency of dense gas $\sfedens$ (middle row), and 
the star formation efficiency of molecular gas $\sfemol$ (bottom row) as a function of: the molecular-to-atomic ratio ($\smol/\satom$, left-hand column) and the stellar mass surface density ($\sstar$, right-hand column).  The right-hand axis of each panel display the data in terms of observed quantities.
Filled and empty symbols correspond to detections and non-detections, respectively. The solid line shows our best fit to the data (Sect.~\ref{ss-data-ana}). Its equation is reported in the top right corner. }
\label{f-env1}%
\end{figure*}

After exploring a variety of ISM parameters, we found that the systematic variations of $\fdens$, $\sfedens$, and $\sfemol$ in 
our sample are best captured by correlations with the molecular-to-atomic ratio, $\smol/\satom$, and the stellar mass surface density, $\sstar$.
These two quantities also have the advantage of being independent of one another and representing extremely simple functions of the available data,
which allows us to confirm with little analysis that those correlations are not spurious.  We stress, however, that many other ISM parameters that are relevant to our study are highly covariant with $\smol/\satom$ and $\sstar$ (e.g., metallicity, midplane pressure),
so that we cannot rigorously ascertain what ultimately drives the trends discussed below. The reader should
keep in mind that both $\smol/\satom$ and $\sstar$ decline with increasing galactocentric radius, so that the lower values in Fig.~\ref{f-env1} typically correspond to outer positions in the disks. 
  
Fig.~\ref{f-env1} shows $\fdens$ ($\propto I_{\rm HCN}/I_{\rm CO}$; top row), $\sfedens$ ($\propto I_{\rm IR}/I_{\rm HCN}$; middle row) and $\sfemol$ ($\propto I_{\rm IR}/I_{\rm CO}$; bottom row) as a function of  $\smol/\satom$ (left column) and  $\sstar$ (right column).
The Figure shows several well-defined, systematic trends across the $\sim2.1$~dex (a factor of $\approx125$) spanned by both ISM parameters. 
As either $\smol/\satom$ or $\sstar$ increases, 
$\fdens$ increases by a factor of $\gtrsim4$ (fitted power-law indices $\sim0.3$; see equations in the top corners of the panels), whereas  $\sfedens$ decreases by a 
factor of $\sim8$ (power-law indices $\sim-0.4$). In contrast, $\sfemol$ appears virtually independent of $\smol/\satom$ and  $\sstar$ within the scatter, in rough agreement 
with more detailed studies in HERACLES galaxies \citepalias{lero13}. These trends are confirmed in Table~\ref{t-env1}, where we list the corresponding rank coefficients. 
Clearly, $\smol/\satom$ and $\sstar$ show strong (anti)correlations with  $\fdens$ and $\sfedens$ but 
no significant correlations with $\sfemol$.

\begin{table}
\begin{center}
\begin{threeparttable}
\caption{Rank coefficients for $\fdens$, $\sfedens$, and $\sfemol$ as a function of local conditions in the galaxy}
\label{t-env1}
                                         
\begin{tabular}{rrrr}                       
\hline\hline                               
\noalign{\smallskip}                       
& $\fdens$ & $\sfedens$ & $\sfemol$ \\  
                                             
\noalign{\smallskip}                       
\hline                                      
\noalign{\smallskip}                       
$r_{25}$ & \si{-0.59} & \si{ 0.43} & \no{-0.07}\\
$\smol/\satom$ & \si{ 0.75} & \si{-0.72} & \no{-0.22}\\
$\sstar$ & \si{ 0.67} & \si{-0.53} & \no{ 0.09}\\
\noalign{\smallskip}
\hline               
\end{tabular}

\begin{tablenotes}
\item{\sc Note. ---} 
Stars symbols indicate significant correlations (p-value $< 2.5\%$)
For a typical number of 45 fully detected data points, 
p-values are $\lesssim50\%$, $10\%$, $5\%$, $2.5\%$, $1\%$, and $0.1\%$ for $|\rank|\gtrsim0.10$, $0.25$, $0.30$, $0.34$, $0.39$, and $0.49$, respectively.
\end{tablenotes}
\end{threeparttable}
\end{center}
\end{table}

Fig.~\ref{f-env1} argues strongly that some intrinsic properties of molecular clouds vary systematically across galaxy disks. For our default assumption of fixed CO and HCN conversion factors, 
we find that $\fdens$ and $\sfedens$ show opposite trends, in the sense that the $\sfedens$ decreases and $\fdens$ increases as we move from low-$\sstar$, low $\smol/\satom$ disks
towards high $\sstar$, high $\smol/\satom$ galaxy centers. From Equation~\ref{e-sfemol-fdens}, this renders $\sfemol$ almost uncorrelated with the local conditions considered here.  In 
Appendix~\ref{s-crosscheck}, we demonstrate that these results are robust against changes in the CO transition used to trace molecular gas (i.e., $J=1-0$ instead of $J=2-1$) or in the 
chosen SFR tracer.

Cast in observational terms, our results show that the ratio of HCN-to-CO intensity ($\propto\sdens/\smol$) increases systematically from a typical value of $\approx1/30$ (in K) in the disks of normal star-forming
galaxies to a higher ratio of $\approx1/10$ in actively star-forming galaxy centers and starbursts, i.e., a factor of $\approx 3.5$ increase over the two orders of magnitude in $\sstar$
and $\smol/\satom$ that we study. At the same time, the IR-to-HCN ratio ($\propto\ssfr/\sdens$) drops, by a factor of $\approx 6$ over the same range, so that active galaxy centers appear to display systematically less 
infrared emission (thus, lower star formation rate) per unit dense gas emission. These two trends cancel, though not entirely, so that the ratio of IR-to-CO ($\propto\ssfr/\smol$) varies more weakly over the same range, systematically changing by 
less than a factor of $2$.

Broadly, these results have two straightforward interpretations: either $\sfedens$ and $\fdens$ indeed change as described, or physical conditions within the molecular clouds change
systematically in such a way that the conversions of CO and HCN intensity to total and dense molecular gas produce the trends that we observe. In either case, the observations imply
that the physical conditions in the molecular ISM, and in molecular clouds specifically, change systematically across galaxy disks, but in a way that conspires to yield only weak apparent
variations in the SFR per unit molecular gas. We explore these two interpretations in the next sections.

\section{Comparison to Star Formation Models}
\label{s-models}

Because dense gas is believed to play a pivotal role in star formation, our results on $\fdens$ and $\sfedens$ have implications for some of the most common models of star formation 
applied to galaxies. In this section, we compare our observations to examples of the two broad classes of current models: 
first (Sect.~\ref{ss-thres}), the straightforward density-threshold model by \citet{gao04a,gao04b} and \citet{wu05}; 
second (Sect.~\ref{s-krumholz}), turbulent whole cloud models (\citetalias{krum05}; \citetalias{krum07}; \citealt{fedd12}). 
We specifically focus on the theory of turbulence-regulated star formation discussed by
\citetalias{krum05}  and \citetalias{krum07}, but refer the reader to a wider summary of whole-cloud models by \citet{fedd12}. The alternative model by \citet{nara08} 
is based on hydrodynamical simulations of entire galaxies,  but shares a similar framework and results in similar predictions to the \citetalias{krum05}.

\subsection{Density Threshold Models}
\label{ss-thres}

In a density threshold model, the star formation rate in a region depends chiefly on the available mass of molecular gas above a certain volume density. 
The simplest form of such a model, and the most commonly adopted in extragalactic research, assumes an environment-independent, fixed star formation efficiency of the dense gas traced by, e.g., the \hcn\ line (our $\sfedens$). Such a view offers a simple explanation for the almost linear correlation between the HCN(1--0) and IR luminosities of 
galaxies found in the early studies by \citet{gao04a,gao04b}. That relation contrasts with the (globally) superlinear relation between the CO(1--0) and IR luminosities when (U)LIRGs are included. 
Within this paradigm, the IR-molecular relations in  galaxies emerge because of a higher $\fdens$ in the IR-brighter objects. In support of a density threshold model
for star formation, \citet{wu05} concluded that  galaxies and  Galactic dense cores with $\ltir>10^{4.5}~L_\odot$ (as required to get reliable SFR estimates from the IR continuum) 
align along the same HCN(1--0)--IR relation. Other studies of Galactic clouds based on different observational methods have been found in qualitative agreement \citep{lada12,heid10,evans14},  
favoring this simple picture in which the dense gas reservoir is the main determinant of the star formation rate.

At face value, our results in galaxy disks are clearly at odds with these simple threshold models. We find that the $\sfedens$ at kiloparsec scales systematically varies 
by a significant factor of $\approx 6$ across $\sim2$~dex in $\smol/\satom$ and $\sstar$. We do find that $\fdens$  systematically increases by a similar factor 
moving from galaxy disks to galaxy centers,  but the interplay of the two quantities means that $\sfemol$ remains constant across the disks within the scatter. The
apparently varying $\sfedens$ in our observations means that the ability of dense gas to form stars must depend on environment and thus, presumably, 
on conditions in the host molecular cloud. This is at direct odds with the basic assumption of threshold models.
 
One can envisage more complex density-threshold models that could fit our resolved observations in galaxy disks. In principle, a model that allowed for variations in the 
density threshold and/or in the star formation timescale of the dense gas could solve the discrepancies \citep[e.g., see][]{fedd12}. However, making such adjustments quickly
moves one into a gray area: if a density threshold depends sensitively on conditions in the parent molecular cloud and host galaxy, then it increasingly resembles the whole-cloud models.
Alternatively, systematic variations in how CO and HCN trace the molecular and dense molecular medium could partly account for the observed trends, salvaging a more universal 
density-threshold model. In Sect.~\ref{ss-cf-thres}, we explore whether our observations could be consistent with a  simple density--threshold model given plausible variations in the 
conversion factors. 

\subsection{The Turbulence Regulated Whole-Cloud Model}
\label{s-krumholz}

``Whole cloud models'' offer an alternative to the universal density threshold view. Here we use this term to refer to models in which conditions throughout a molecular cloud
ultimately affect the efficiency with which dense gas forms stars. These models have been contrasted with threshold models in the literature several times, with
recent Milky Way results invoked to support a simple density threshold \citep{evans14}, while arguments using low resolution extragalactic data used to argue for
whole cloud models \citep[e.g.,][]{krum13}. Even considering only analytic models, a wide suite of whole cloud models exist in the literature \citep{fedd12}. These almost all
consider the star formation efficiency per free fall time, $SFE_{\rm ff}$, as the fiducial quantity and so encode a fundamental dependence on density \citep[e.g.,][]{mckee07}. 
$SFE_{\rm ff}$ further depends in most models on the virial parameter, Mach number, magnetic field, and turbulent character within the clouds.

\subsubsection{KMK05 Model Overview}
\label{ss-krumholz-model}

With such a wide set of free parameters, it is outside the scope of our paper to compare our data to the full set of models or possible cloud conditions. Indeed, because in our survey we have
very limited knowledge of the cloud population at sub-beam scales, we lack many of the measurements needed for such a comparison. Instead, the main thrust of our
treatment of whole cloud models is to ask: can such models plausibly create the kind of correlated variations in $\sfedens$ and $\fdens$ that we see in our data? To do so,
we focus on one of the earliest whole cloud models, the theory of turbulence-regulated star formation put forward by \citetalias{krum05}. In a later paper, \citetalias{krum07} 
combined their theoretical treatment with a simple radiation transfer model\footnote{Code available at http://www.ucolick.org/~krumholz/codes/} that implicitly provides the conversion factors between gas 
masses and the observed molecular line luminosities (e.g., HCN(1--0)). To simplify the comparison, we refer here to the original equations by \citetalias{krum05} and treat conversion factors
as fixed.

In the \citetalias{krum05} theory, the star formation within a virialized molecular cloud with a lognormal distribution of densities depends primarily on the average gas density ($\bar{n}$) and the 
\mbox{intracloud} turbulence. The latter is parameterized by the Mach number, $\mathcal{M}$.  The sense of the dependence is that $\bar{n}$ sets the characteristic timescale for star formation, the free-fall time, 
so that a high density implies more efficient star formation per absolute time interval. Meanwhile in the \citetalias{krum05} model, a higher degree of intracloud turbulence, i.e., a higher $\mathcal{M}$, tends to inhibit star formation\footnote{Note that this feature is not shared by some other models, in which higher Mach numbers tend to drive higher rates of star 
formation rates; see \citet{fedd12}.}.

\begin{figure*}
\centering
\includegraphics[width=0.49\textwidth]{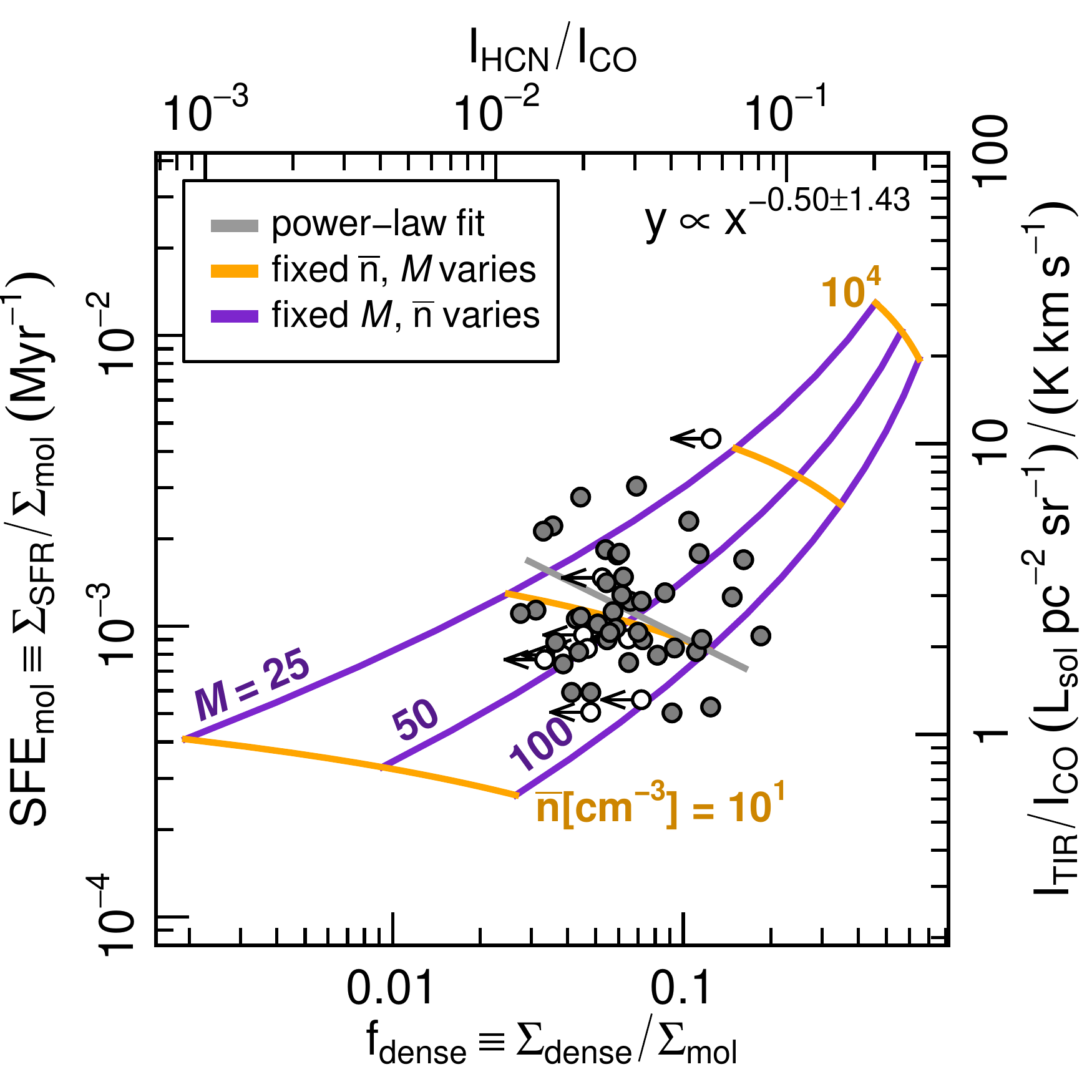}\hfill
\includegraphics[width=0.49\textwidth]{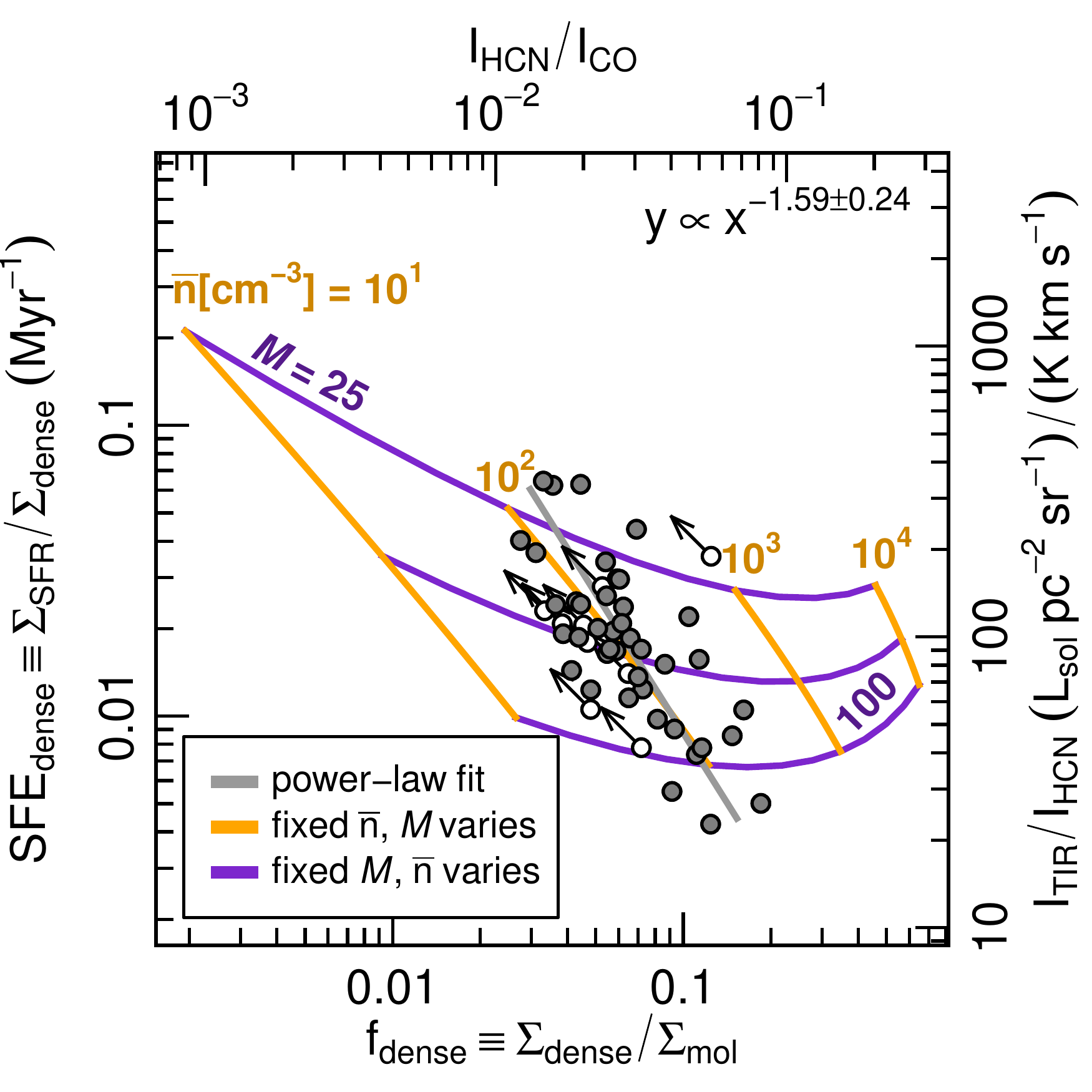}
\caption{Comparison between the $\sfemol$-$\fdens$ (left-hand panel) and $\sfedens$-$\fdens$ (right-hand panel) relations in our data 
and in the \citetalias{krum05} model.  
The top and right-hand axes of each panel display the data in terms of observed quantities.
For the data, symbols are the same as in Fig.~\ref{f-ks1}. The power-law fit equation (gray line) is shown in the top-right corner. The 
model predictions are represented by colored lines. 
To obtain the purple curves, labelled by the assumed $\mathcal{M}$  in the left-hand panel, we fix the Mach number and vary the 
average gas density. To obtain the orange lines,  labelled by the assumed $\bar{n}$ in the right-hand panel, we fix the average density 
and vary the Mach number.
}
\label{f-k07-1}%
\end{figure*}

For our purposes, the \citetalias{krum05} model boils down to the following Equations (complemented by Equation~\ref{e-sfemol-fdens}):

\begin{eqnarray}
\label{e-k05-2}
\sfemol=\epsilon_\mathrm{SF}\frac{(\mathcal{M}/100)^{-0.32}}{\tau_\mathrm{ff}(\bar{n})},\\
\label{e-k05-1}
\fdens=
\frac{1}{2}\left[1+\mathrm{erf}\left(\frac{\sigma_\mathrm{PDF}^2-\log{(n_\mathrm{dense}/\bar{n})}}{2^{3/2}\sigma_\mathrm{PDF}}\right)\right],\\
\sigma^2_\mathrm{PDF}\approx\log\left(1+\frac{3\mathcal{M}^2}{4}\right).
\label{e-k05-3}
\end{eqnarray}

\noindent Here, $\tau_\mathrm{ff}$ is the free-fall timescale evaluated at the average volume density ($\bar{n}$), which is proportional to $1/\sqrt{\bar{n}}$; $\epsilon_\mathrm{SF}$ is an 
efficiency parameter of the order of $\sim1.4\%$ that depends on the virial state of the clouds; $\sigma_{\rm PDF}$ is the width of the lognormal probability distribution function of 
gas density within a cloud, determined by the Mach number ($\mathcal{M}$). 

The \citetalias{krum05} model has two key features. First, unlike a simple threshold model, whole cloud models predict both $\fdens$ and $\sfemol$, and by extension $\sfedens$. Although these
models have more free parameters, they also attempt to explain more of the star formation process; by contrast, the density threshold models push the topic of regulating star formation
into the formation of dense gas, which they do not address. Second, the \citetalias{krum05} model has two different mechanisms to alter $\fdens$: the fraction of dense gas increases along with both the average density, $\bar{n}$, and the turbulent Mach number, $\mathcal{M}$. The latter dependence occurs because
a higher Mach number broadens the density PDF (Eq.~\ref{e-k05-3}).
Consequently, for a fixed average density, it also increases the mass of gas with densities above the cutoff for the HCN-traced gas, $n_\mathrm{dense}$ (this happens as long as $n_\mathrm{dense}>\bar{n}$, which is always expected for plausible conditions; see Sect.~\ref{ss-data-param}). In the \citetalias{krum05} model, the variations in $\bar{n}$ and $\mathcal{M}$  affect the star formation efficiencies in different ways, however:
\begin{itemize}  
\item  If $\bar{n}$ increases, $\tau_\mathrm{ff}$ decreases, enhancing $\sfemol$. In contrast, the product  $\fdens\times\tau_\mathrm{ff}$  changes 
little across a plausible range in $\bar{n}$, which would render $\sfedens$ almost constant.
\item  If $\mathcal{M}$ increases, $\fdens$ increases, while $\sfemol$ would moderately decrease. Combining these, $\sfedens$ would experience 
a more significant drop (due to the additional $1/\fdens$ factor).
\end{itemize}

Based on this model, \citetalias{krum07} (see also Sect.~\ref{s-discuss}) posit that an increase by 2-3 orders of magnitude in $\bar{n}$ along the IR sequence of galaxies 
could mostly explain the different power-law indices for the IR-CO(1--0) and IR-HCN(1--0) relations in galaxies. While a high Mach number in (U)LIRGs compared to 
normal SF galaxies could also come into play, \citetalias{krum07} argue that this parameter would likely play a secondary role.

\subsubsection{Results}
\label{ss-krumholz-fixx} 

Because we do not know the detailed structure of molecular clouds within our beams, the agreement between our data and Equations~\ref{e-k05-2}-\ref{e-k05-3} depends on some assumptions about four parameters that are not strongly constrained  {\em a priori}: the CO and HCN conversion factors, the density cutoff for  HCN emission ($n_\mathrm{dense}$), and 
$\epsilon_\mathrm{SF}$. To match the region of $\sfemol$--$\sfedens$--$\fdens$ parameter space spanned by our observations, we
set $n_\mathrm{dense}$ to $2.8\times10^5$~cm$^{-3}$, as assumed by \citetalias{krum07}, but adopt a $\epsilon_\mathrm{SF}$ efficiency parameter that is 1/5 times the value preferred by them. The same mismatch in $\epsilon_\mathrm{SF}$ was noted by \citetalias{buri12}. 

Figure \ref{f-k07-1} displays our data in the $\sfemol$-$\fdens$ (left) and $\sfedens$-$\fdens$ (right) planes, with power-law fits shown as grey lines. We overlay 
the curves predicted for realistic conditions when we vary either $\bar{n}$ (fixed $\mathcal{M}=$ 25, 50, and 100, from top to bottom; purple) or $\mathcal{M}$ 
(fixed $\bar{n}=$ $10$, $10^2$, $10^3$, and $10^4$~cm$^{3}$, from left to right; orange). Thus the purple and orange grid shows a plausible parameter space from
the \citetalias{krum05} overlaid on our observations.

Figure \ref{f-k07-1} shows that our data are better described by the orange tracks obtained by varying the Mach number ($\mathcal{M}$) while holding the average density constant at $\approx 100$~cm$^{-3}$.  
The mismatch between our data and the curves for constant $\mathcal{M}$ (purple) is particularly apparent in the right-hand panel, which shows that the
spread in $\sfedens$ that we measure is hard to reproduce only by varying the average density of otherwise identical clouds. As a simple way to quantify this (dis)agreement, 
we fitted the data in each panel with curves predicted by the \citetalias{krum05} model, fixing either $\bar{n}$ or $\mathcal{M}$ while leaving the other parameter free.
In the $\sfedens-\fdens$ space, the best-fit curve for fixed $\mathcal{M}$ has a four times higher reduced-$\chi^2$ than the best-fit curve for fixed $\bar{n}$ (in the
$\sfemol-\fdens$ plane the fits have more comparable quality).

Thus a simple comparison to the whole-cloud models of \citetalias{krum05} shows that, with reasonable inputs (and after adjusting the $\epsilon_\mathrm{SF}$ parameter), the models span a range of $\sfemol$--$\sfedens$--$\fdens$ 
parameter space that can match our data well. For the specific functional forms of the \citetalias{krum05} models, our data are well fit by an implementation of the models
in which clouds are significantly more turbulent, but not much denser, near the galaxy centers than in the disks. If this were the case, the trends that we find across galaxy 
disks would be different from those between SF galaxies and (U)LIRG, which appear to be chiefly driven by an enhancement in $\bar{n}$ (see Sect.~\ref{s-discuss}). However,
we caution that the specific numerical dependencies of the \citetalias{krum05} model do differ from other, more recent, analytic models, including those that otherwise adopt
the \citetalias{krum05} approach but treat the free fall time on multiple scales. Broadly, the wider parameter space spanned by whole cloud models can accommodate our
data in a way that a simple threshold cannot.

\section{The Effect of Conversion Factors}
\label{s-cf}

So far, we have assumed fixed, Milky Way-like conversion factors  for CO and HCN. Under this assumption, our data strongly favored whole-cloud models over a universal
density threshold. An alternative hypothesis would be that the changing physical conditions within our clouds mainly change the conversion between CO and total molecular
gas mass or HCN and dense molecular gas mass. Though an exhaustive study of the conversion factor is well beyond the scope of this paper, here we test a suite of plausible
conversion factor variations and investigate how they could alter the simple picture discussed above. Specifically, we build a grid of plausible conversion factors for CO and HCN and, at every point of the grid, 
we check whether our data, combined with these conversion factors, exhibit basic agreement with either of our two models.

\subsection{Background}

Significant deviations in $\aco$ from its Milky Way value have been predicted and observed in galaxies, both at global and smaller scales \citep[see the review by][]{bola13}. Broadly, 
$\aco$ appears to exhibit two main behaviors. First, $\aco$ seems to increase with decreasing metallicity (or dust--to--gas ratio), as CO emission 
becomes confined to a smaller fraction of a molecular cloud. Second, $\aco$ tends to drop in extreme star forming environments (e.g., galaxy centers, (U)LIRGs, and starbursts) 
where the gas is more turbulent and more excited. \citet{bola13} suggested that this second trend could be usefully parameterized as a decrease in $\aco$ with increasing 
total mass surface density ($\Sigma_\mathrm{tot}=\sstar+\sgas$), although $\Sigma_\mathrm{tot}$ may not be its ultimate physical driver. 

Compared with $\aco$, the HCN conversion factor, $\ahcn$,  has been much more poorly characterized. There are, however, practical reasons to expect 
that the two conversion factors follow each other to some extent. First, $\ahcn$ must be lower in some (U)LIRGs than in SF galaxies to ensure that $\fdens<1$.
More physically, if $\ahcn$ emerges from a turbulent, opaque medium --- even if it makes up only a portion of a typical cloud --- we expect similar dependences
on the degree of turbulence and excitation. 
 Second, the density above which the HCN(1--0) line is effectively emitted could vary as certain cloud properties change (e.g., average density, temperature, turbulent velocity gradient). This would also make $\ahcn$, that converts the line luminosity into  gas mass above a fixed density cutoff, vary (Sect.~\ref{ss-data-param}).

We adopt a simple, {\em ad hoc} approach to gauge the interplay of conversion factors, models of star formation, and our observations. To do so,
we simply assume that, to some very coarse approximation, both $\aco$ and $\ahcn$ may vary as some power law function of $\sstar$, that is:

\begin{eqnarray}
\label{e-xhcn}
\ahcn(\sstar,\ghcn)=\ahcn^0\left(\frac{\sstar}{100~M_\odot~\mathrm{pc}^{-2}}\right)^{\ghcn},\\
\label{e-xco}
\aco(\sstar,\gco)=\aco^0\left(\frac{\sstar}{100~M_\odot~\mathrm{pc}^{-2}}\right)^{\gco}.
\end{eqnarray}

\noindent 
Here, the prefactors are set to the default Galaxy-like HCN and CO conversion factors defined in Sect.~\ref{ss-data-param}.
Equations~\ref{e-xhcn} and \ref{e-xco} are reminiscent of the empirical law proposed by \citet{bola13} to match a set of $\aco$ values culled from the literature, including 
those estimated by \citet{sand13} across the disks of HERACLES galaxies. \citet{bola13} specifically proposed $\gco=-0.5$ above a total mass surface density of 
$100~M_\odot$~pc$^{-2}$ in addition to an exponential metallicity dependent term. We neglect the metallicity term because our survey focuses heavily
on the metal-rich parts of the HERACLES sample (we checked that this term would hardly affect our results if included). We approximate 
$\Sigma_\mathrm{tot} \sim \sstar$ because, in our sample, the total surface density is almost always dominated by $\sstar$ (usually by a factor of $\gtrsim 3$).
Finally, we have formally dropped the $\Sigma_{\rm tot}$ threshold advocated by \citet{bola13} because essentially all of our detections have
$\sstar$ above this value (see Figure \ref{f-env1}).

We emphasize that  Equations~\ref{e-xhcn} and \ref{e-xco} represent a coarse approach to the variability of conversion factors. 
The detailed study of HERACLES galaxies by \citet{sand13} suggests that $\aco$ in an individual galaxy can be better described by a flat profile
over most of the disk and an abrupt drop near its center and we do not expect disk surface density to be the only (or even main) physical driver
of conversion factor variations. Nevertheless, power laws offer a useful tool for the kind of exploratory calculations that
we carry out in this section. These are already highly approximate due to the lack of observational and theoretical constraints on $\ahcn$ and a
power law formalism allows us to treat $\ahcn$ and $\aco$ in a similar way with only a minimal set of free parameters ($\gco$ and $\ghcn$). 
Moreover, we are specifically interested in how conversion factor variations could ``rescue'' a density threshold model.
Fig.~\ref{f-env1} shows that, at least in the specific case of our set of pointings, $\ahcn$ should roughly behave as a power-law function of $\sstar$ in order to 
cancel the systematic gradients in IR-to-HCN that we observe.

We can consider a grid of plausible $\gco$ and $\ghcn$, with ``plausible'' defined according to the following rules:
\begin{itemize}
\item[i)]  Most evidence in the literature point to lower conversion factors in regions of higher surface density (galaxy centers and (U)LIRGs), so we assume $\gco \leq 0$.
\item[ii)]  We expect the fraction of gas in the molecular phase, traced by $\smol/\satom$, to increase with the depth of the gravitational potential well, traced by $\sstar$. We find that we require $\gco>-0.82$ to ensure that this ratio and $\sstar$ are positively and significantly correlated and we consider only values above this value.
\item[iii)] We expect that $\fdens$ increases along with the HCN/CO line ratio. We find this happens when $\ghcn-\gco>-0.42$. That is, the $\ahcn/\aco$ ratio cannot be so low at galaxy centers  that
we actually have {\em lower} dense gas fractions for higher HCN/CO ratios.
\item[iv)] We require that less than $5\%$ of our positions have a nominal $\fdens>1$, which is unphysical (even widespread $\fdens \sim 1$ is implausible). For this, we find that 
the condition $\ghcn-\gco<0.58$ is necessary. That is, the $\ahcn/\aco$ ratio cannot be so high at galaxy centers that all of the molecular gas is in the dense phase.  
\end{itemize} 

The four conditions stated above define a rectangular region in the ($\gco$, $\ghcn-\gco$) plane that encapsulates 
plausible variations in the CO and HCN conversion factors. The changes in $\gco$ and $\ghcn$ will modify the mutual relations between
$\fdens$, $\sfemol$, and $\sfedens$ inferred from our observations.  
 Thus, exploring this parameter space, we study the compatibility between our observations and models as a function of possible
conversion factor variations.
In this analysis, we keep $\aco^0$ and $\ahcn^0$  in Equations~\ref{e-xhcn} and \ref{e-xco} fixed. Any change in these parameters would globally shift the data points, which is irrelevant for the comparison between observed and predicted trends that we carry out below.

  \begin{figure}
   \centering
\includegraphics[width=0.49\textwidth]{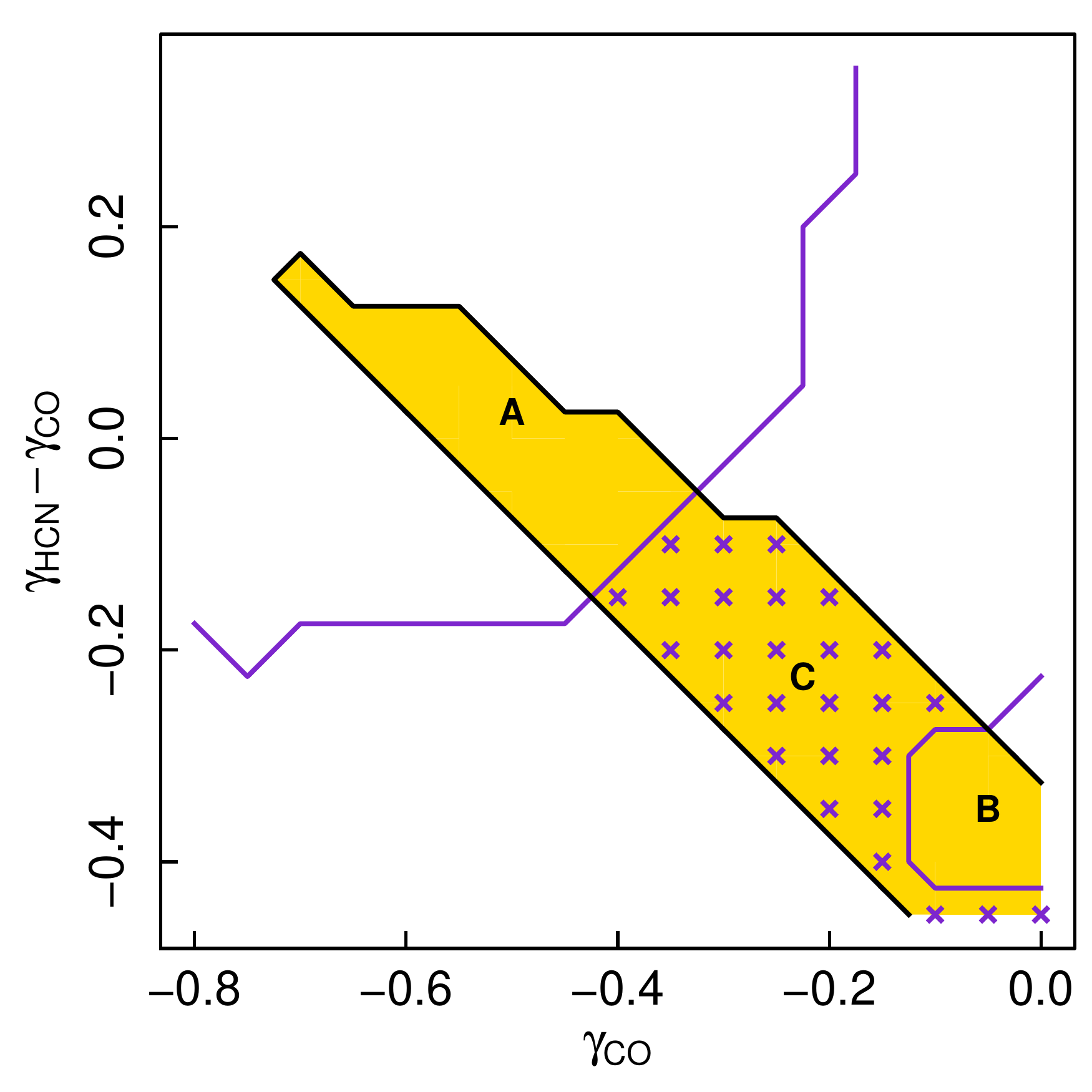}

\caption{Compatibility between our data and the density-threshold models across the  ($\gco$, $\ghcn-\gco$) region defined by our boundary 
conditions (see text; we only represent $\ghcn-\gco<0.35$ to keep the plotting box square). 
Within the orange area, the observations and the density-threshold model are compatible ($\sfedens$ is neither significantly correlated with $\sstar$ nor with $\smol/\satom$; see text for details). The purple contours enclose the regime 
there are systematic variations in either $\sfemol$ and $\fdens$, but both variables are either anticorrelated or not significantly correlated at all. The purple crosses indicate the intersection between this regime and the region where  $\sfedens$ shows no systematic variations. This defines section "C", while sections "A" and "B" are located at either side of it. }
\label{f-thres1}%
\end{figure}

 \begin{figure*}[!t]
   \centering
    \begin{tabular}{c@{\extracolsep{3pt}}c}
     \large fixed $\bar{n}$, $\mathcal{M}$ varies & \large fixed $\mathcal{M}$, $\bar{n}$ varies \\
\includegraphics[width=0.49\textwidth]{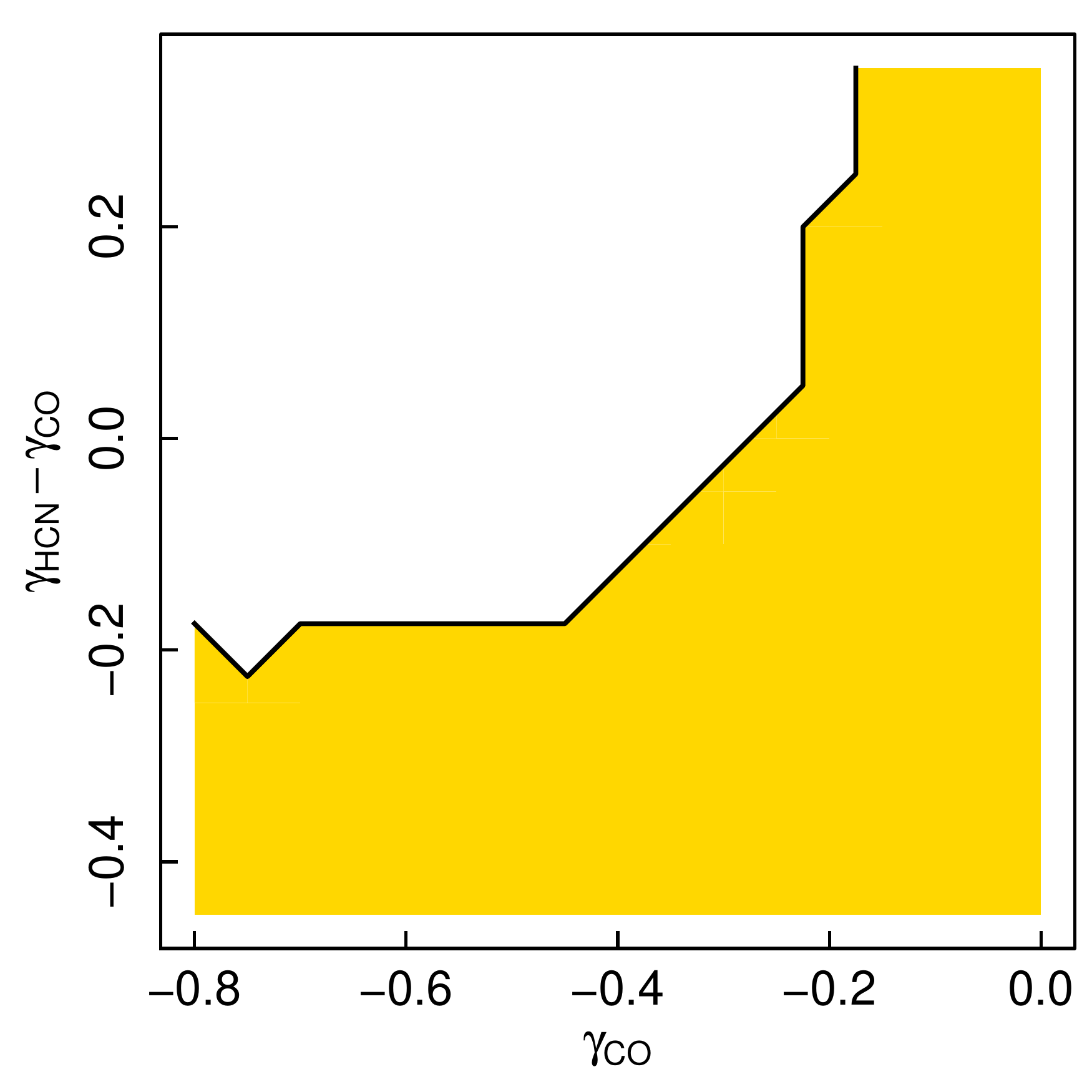}&
\includegraphics[width=0.49\textwidth]{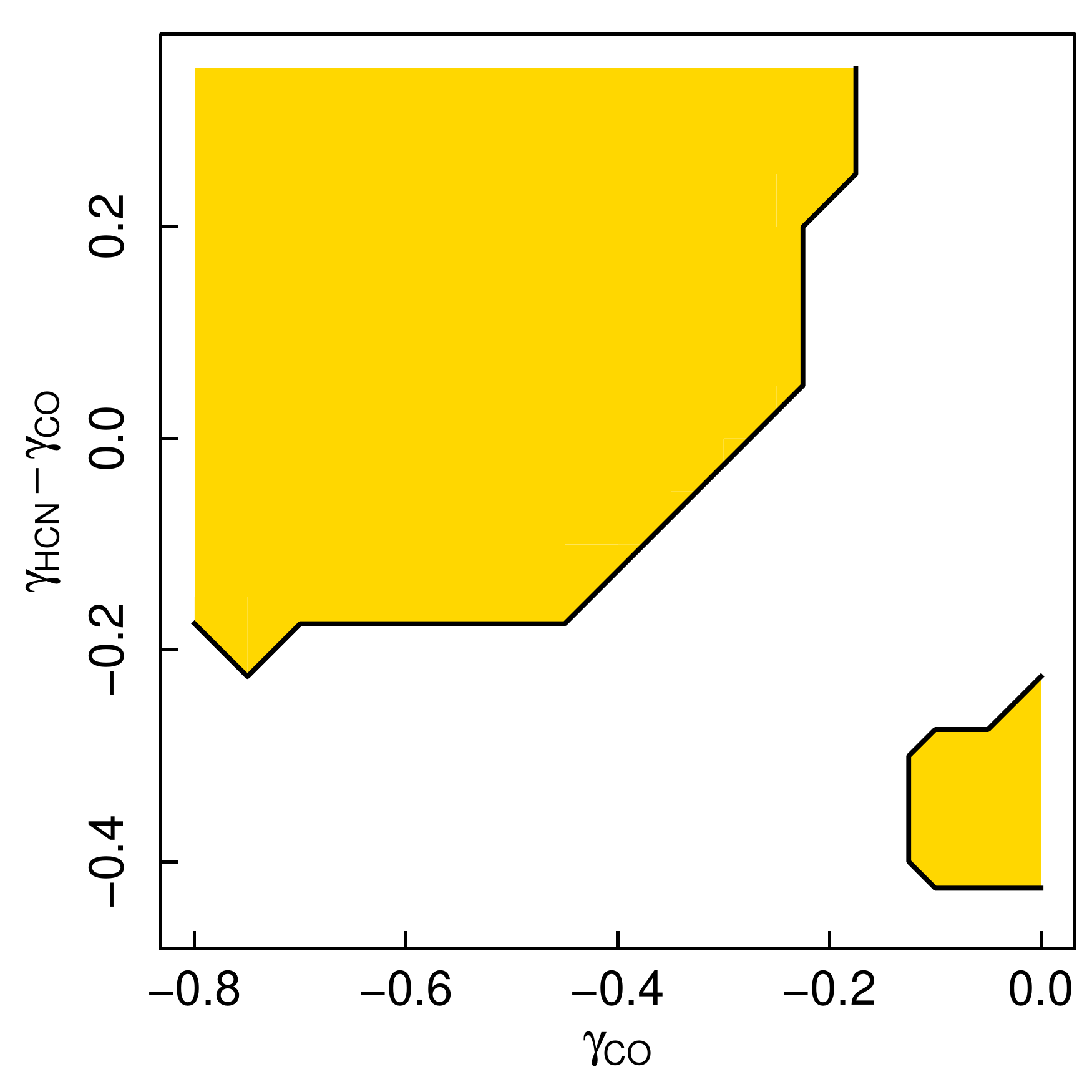}\\
\end{tabular}
\caption{Compatibility between our data and the \citetalias{krum05} model for variation in the Mach number (left-hand panel) and in the 
average density (right-hand panel). Each panel covers the  ($\gco$, $\ghcn-\gco$) region defined by our boundary conditions. The orange area 
indicates the ($\gco$, $\ghcn-\gco$) values for which  observations and theory are compatible (see text for details). 
}
\label{f-k07-2}%
\end{figure*}

\subsection{Density Threshold Models}
\label{ss-cf-thres}

The key assumption of the simple density-threshold hypothesis supported by, e.g., \citet{gao04b} and \citet{wu05} is that $\sfedens$ is fixed within the scatter.  In Figure~\ref{f-thres1}, the orange-colored area within the 
thick black contour is the region of the $(\gco,\ghcn-\gco)$ parameter space where this could be fulfilled: inside it, $\sfedens$ (derived from our data and
the conversion factors at each point) would not significantly correlate with $\sstar$, nor with $\smol/\satom$. Essentially, this region stretches along the 
straight lines for which $-0.55\lesssim\ghcn\lesssim-0.35$, which is the $\ahcn$ required to cancel out the gradients in $\sfedens$ found in Sect.~\ref{s-env}. 
We remark that, to cancel these gradients, $\ahcn$ should roughly behave as such a power law  when restricted to our sample of pointings, regardless its true functional form.

In more detail, the region without systematic variations in $\sfedens$ consists of three different sections where the threshold model has a decreasing predictive power for our sample of pointings. In the first section, labeled ``A" in the figure,  $\sfemol$ and $\fdens$ are mutually correlated and show systematic variations with respect to either $\smol/\satom$ or $\sstar$. Thus, {\em a priori},  systematic increases in  $\sfemol$ could be ascribed to increases in $\fdens$, as expected for a threshold model. Section ``A'' corresponds to values of $-0.7\lesssim\gco\lesssim-0.3$, i.e., compatible with the $\gco=-0.5$ value in \citet{bola13}, 
and to variations in $\ahcn$ that are largely in lockstep with those in $\aco$ ($|\ghcn-\gco|\lesssim0.2$), 
perhaps in response to the same physical mechanisms. The significant correlation
between $\fdens$ and $\sfemol$ arises  because the $\sstar$ dependent term in $\ahcn$ flattens
$\sfedens$, whereas the steep trend in $\fdens$ is barely affected as variations in $\ahcn$ and $\aco$ cancel each other out.

In a second section, labeled ``B'',  the systematic gradients in $\fdens$, $\sfedens$, 
and $\sfemol$ across galaxy disks all cancel out. This leaves the density threshold theory plausible, but essentially unprobed within the scatter of our data.
Section ``B" corresponds to 
 only weak variations in $\aco$ across 
galaxy disks, but much stronger gradients in $\ahcn$. Essentially, $\ahcn$ varies in such a way that it cancels the
observed variations in the IR-to-HCN and HCN-to-CO ratios, whereas the trend in the IR-to-CO ratio remains unchanged.

Finally, there is a third region, labelled ``C", where there are systematic variations in either $\sfemol$ and $\fdens$, but both variables are either anticorrelated or not significantly correlated at all. In this case, $\fdens$ could not be considered a major driver of $\sfemol$, which would rather be regulated by an hypotethical {\em hidden} variable.  Therefore, even though $\sfedens$ would remain constant, the density threshold model would not be not predictive of $\sfemol$.

\subsection{The Turbulence-Regulated Model}
\label{ss-cf-krumholz}

To illustrate the impact of changing $\ghcn$ and $\gco$ on the  \citetalias{krum05} model, we consider two extreme cases: one in which we 
keep $\bar{n}$ fixed while varying $\mathcal{M}$ and one in which we fix $\mathcal{M}$ and vary $\bar{n}$ (see Sect.~\ref{ss-krumholz-model} and 
Fig.~\ref{f-k07-1} for an illustration of such models). Of course, in a real system both quantities may vary and,
as discussed above, the \citetalias{krum05} model represents only one implementation of a turbulence-regulated whole-cloud model.
The essential features of the models, which our data should reproduce to be compatible with the theory in a region of ($\gco$, $\ghcn-\gco$) space, are:

\begin{itemize}
\item For the fixed $\mathcal{M}$ case, both $\fdens$ and $\sfemol$ should increase when $\bar{n}$ increases; that is, 
we require a significant positive correlation between $\fdens$ and $\sfemol$.
\item For the fixed $\bar{n}$ case, as Figure \ref{f-k07-1} shows,  $\fdens$ increases  when $\mathcal{M}$ increases, 
whereas $\sfemol$ decreases more mildly; as a result, we expect either an anticorrelation between $\fdens$ and $\sfemol$, or even no significant 
correlation given the observational uncertainties.
\end{itemize}

 We implement these qualitative behaviors as requirements on the p-value of the rank 
correlation coefficient between $\fdens$ and $\sfemol$ and show the valid regions in ($\gco$, $\ghcn-\gco$) space in Fig.~\ref{f-k07-2}. 
Fig.~\ref{f-k07-2} shows that almost all of the plausible ($\gco$, $\ghcn$) parameter space can be covered by some variant
of the turbulence-regulated model. Under the assumption of fixed $\bar{n}$ (left-hand panel), our observations agree with this theory
when $\gco\gtrsim-0.2$ or $\ghcn-\gco\lesssim-0.2$. This includes the case of fixed conversion factors ($\gco=\ghcn=0$) adopted throughout
the first part of the paper. If $\mathcal{M}$ remains fixed while $\bar{n}$ varies (right-hand panel), observations and models agree within the complementary region, 
but also in a small patch  at $\gco\simeq0$ and $\ghcn-\gco\simeq-0.4$. This is the patch where the CO and HCN conversion factors conspire to eliminate all gradients
($\fdens$, $\sfemol$, and $\sfedens$) from our sample. Of course, inasmuch as this conspiracy yields a homogenous 
population of star-forming clouds throughout the sample, our observations could be compatible with any cloud-scale star formation theory. 
 
 \subsection{Conclusions}

Our exploration of plausible $\aco$ and $\ahcn$ shows that compatibility with a simple density-threshold model of star formation places specific,
non-trivial requirements on the behavior of the CO and HCN conversion factors. Essentially, $\ahcn$ must vary in a way that cancels out the observed
variation in the IR-to-HCN ratio and its value must be fairly tightly coupled to $\aco$. By contrast, the turbulence-regulated whole cloud model of
\citetalias{krum05} can be compatible with almost any region of $\aco$-$\ahcn$ space that we explore. To some degree, this reflects the 
larger suite of free parameters in the whole cloud models, but this does not entirely invalidate the contrast: it may very well be that more physics,
and so more free parameters, are needed to explain the observations.

\section{Discussion}
\label{s-discuss}

We interpret our results to offer a challenge to simple density threshold models, though not to conclusively rule them out. Whole-cloud
models like \citetalias{krum05} have more success explaining our observations because they can span a wider part of $\sfemol$-$\fdens$-$\sfedens$
space. We discuss shortly the implications of combining each of these models with our data and the unresolved observations from \citetalias{buri12}. 

\subsection{Implications for Density Threshold Models}

Density-threshold models found support in the remarkable linear IR-HCN(1--0) relation  found by \citet{gao04a} and \citet{wu05}, which 
encompassed systems covering $\sim7-8$~dex in $\ltir$. This correlation represents almost the only extragalactic observational support for
such a model. In plots that span such an enormous range in luminosity, our data do broadly ``fill in'' the previously missing part of this correlation.
However, in detail our data also exhibit clear, systematic variations in the IR-to-HCN luminosity ratio. That is, luminosity-luminosity
plots or power law fits spanning many decades would miss significant, systematic environmental trends that appear as ``scatter'' around
the luminosity-luminosity scaling. We suspect that along with intrinsic cloud-cloud scatter \citep{ma13} and 
observational uncertainties, the wide range in measured $SFR/\lhcn$ ratios \citep[$>2$~dex and $\sim1.4$~dex in Galactic cores and 
galaxy averages, respectively;][GB12]{wu05} may be driven by the same systematic variations that we see in this survey. 

Other studies have highlighted variable $\sfedens$: recent studies at whole-galaxy scales have pointed out that the average $\sfedens$ of (U)LIRGs 
is at least 3-4 times higher than that of SF galaxies \citepalias[][see also \citealt{gao07}]{buri12}, unlike what \citet{gao04a} concluded. Our own fit to 
the GB12 data implies that  the average $\sfedens$ would increase as $\ltir^{0.27\pm0.04}$. Moreover, studying the center of our own Milky Way
galaxy using ammonia lines, \citet{long13} demonstrated this high surface density region of a normal disk galaxy to show a systematically low
$\sfedens$, the same sense of the trend ($\sfedens$ anticorrelates with $\sstar$) that we observe in this paper.

For a fixed HCN-to-dense gas conversion factor, these results are clearly at odds with the assumption of a fixed $\sfedens$ that underlies threshold 
models. We show that our resolved observations could be compatible with a fixed (within the scatter) $\sfedens$ if $\aco$ and $\ahcn$ varied in some plausible way.
However, we emphasize that doing so would remove the major extragalactic argument for a simple density threshold: the simple correlation of
HCN and IR luminosity. Moreover, an {\em ad hoc} tuning of $\ahcn$ cannot address the higher apparent $\sfedens$ in (U)LIRGs, since the basic
expectation is that their $\ahcn$ should be lower than that of SF galaxies. Applying such a correction would exacerbate, rather than
alleviate, the difference in $\sfedens$ between both kinds of galaxies \citepalias[for details, see][]{buri12}; but see \citet{papa12} for an alternative
discussion of conversion factors in (U)LIRGs.

\subsection{Turbulence-Regulated Models and a Simple Observational View}

 \begin{figure*}
   \centering
\includegraphics[width=0.48\textwidth]{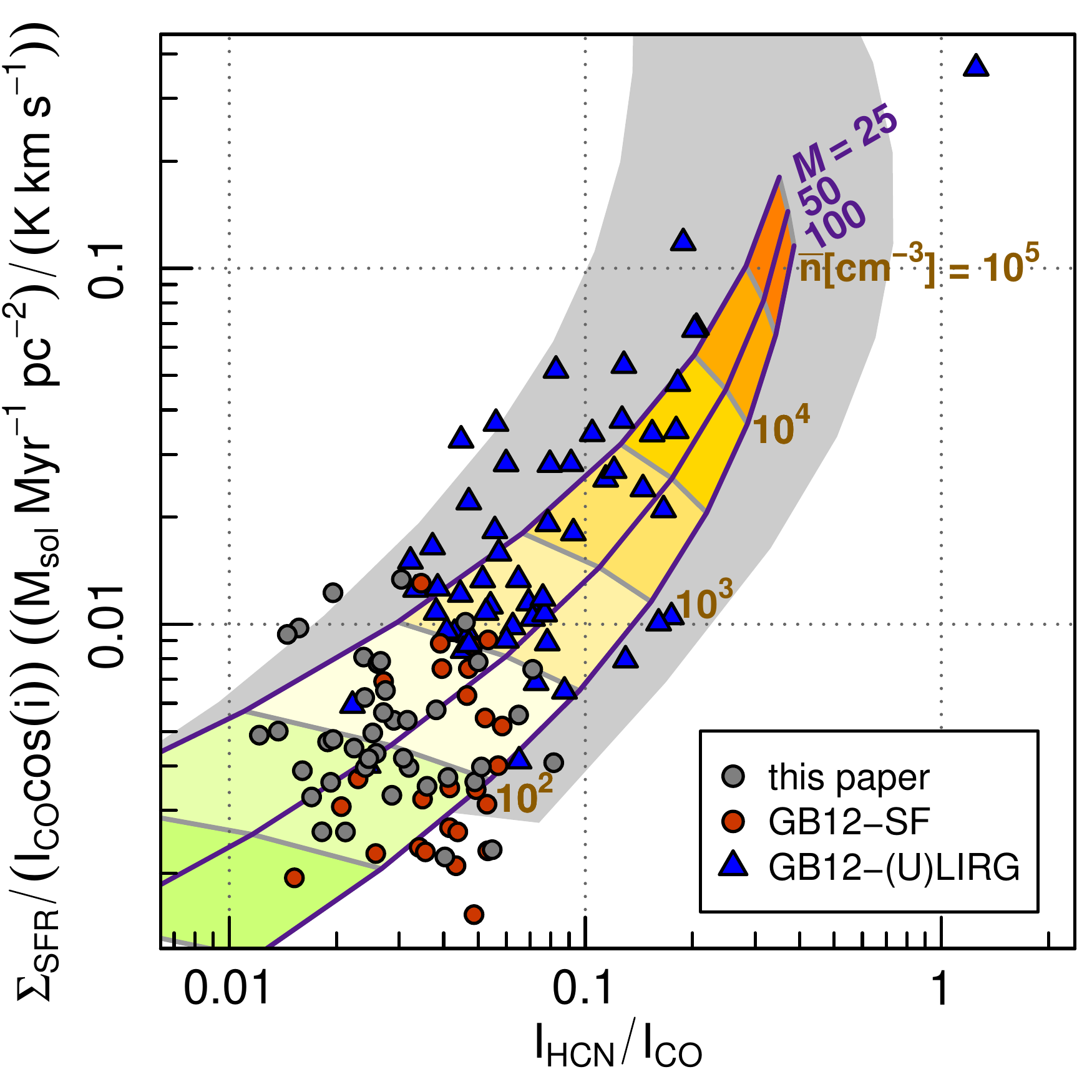}\hspace{0.65cm}
\includegraphics[width=0.48\textwidth]{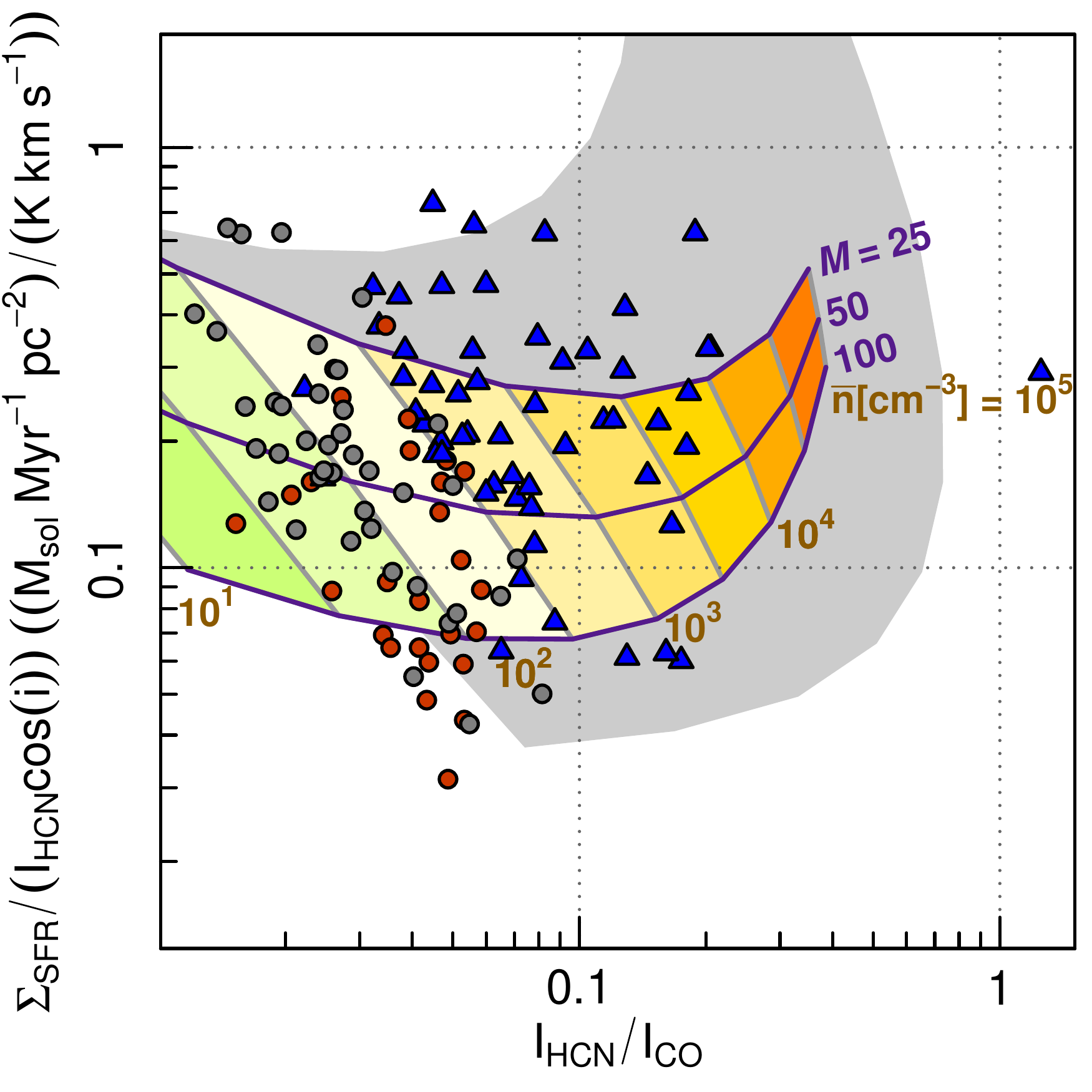}\\

\caption{Comparison between observations and the \citetalias{krum05} model: $\ssfr/\ico$ (left-hand panel) and $\ssfr/\ihcn$ (right-hand panel) as functions of $\ihcn/\ico$ in our data and the GB12 extragalactic sample (note that intensity and luminosity ratios are formally identical, which allows us to include in this figure those galaxies of the GB12 sample that lack an estimate of the molecular disk area). Symbols as in Fig.~\ref{f-ks1}. For the sake of visualization, limits are not displayed. The purple lines correspond to lines of constant Mach number $\mathcal{M}=25$, 50 and 100, as indicated by the labels of the same color. They are obtained by increasing the average gas density ($\bar{n}$) from $10$~cm$^{-3}$ to $10^5$~cm$^{-3}$ in steps of 0.5~dex. The gray lines correspond to constant $\bar{n}$ conditions, as indicated by the brown labels (for clarity, only one in two lines is labeled). The space between them is colored in green-orange-red hues, with reddish ones corresponding to higher densities. 
To convert (dense) gas masses into line luminosities, we assume the default HCN and CO conversion factors of this paper.
The gray area in the background indicates the range of observables that can be reproduced by the  \citetalias{krum07} model, where non-fixed HCN and CO conversion factors are derived from a simple radiative transfer code coupled to the \citetalias{krum05} cloud model. The upper and lower envelopes of this area correspond to the {\em normal} and {\em starburst} set of parameters defined by  \citetalias{krum07}. They assume Mach numbers of 30 and 80, respectively. 
}
\label{f-k07}%
\end{figure*}

Whole cloud, turbulence-regulated models can produce a wide range of points in the $\sfemol$--$\fdens$--$\sfedens$ space. The specific model
of \citetalias{krum05} therefore appears able to reproduce our data for almost all assumptions about conversion factors, though recall that, as in
\citetalias{buri12}, we had to rescale the absolute scale of the star formation rate in the model (i.e., the efficiency parameter $\epsilon_\mathrm{SF}$). We saw that, for the case
of fixed conversion factors, our data agree particularly well with an implementation of the \citetalias{krum05} model in which the average cloud
density remains approximately fixed while the turbulent Mach number increases with increasing stellar surface density.

This primary dependence on the Mach number may seem at odds with the normal explanation for increased $\sfemol$ in galaxy centers, which
would attribute the contrast between galaxy disks and (U)LIRGs mainly to gas density. We address this in Fig.~\ref{f-k07}, where we combine our data 
with the \citetalias{buri12} data in the mostly observational spaces $\ssfr/\ico$ vs. $\ihcn/\ico$ (left-hand panel) and $\ssfr/\ihcn$ vs. $\ihcn/\ico$ (right-hand panel),
where $\ssfr$ is estimated linearly from the IR. This figure encapsulates our best current observational understanding of HCN, CO, and star formation in galaxies. In both panels, the combined data sets span a contiguous region of the parameter space\footnote{
The only outlier is APM~08279+5255, a luminous BAL quasar at redshift $z\simeq3.9$ with an exceptionally high $\lhcn/\lco$ ratio. Multitransition studies of this source has shown that its HCN excitation is driven by IR pumping to a significant extent \citep{weiss07}, in which case the line luminosity is not a reliable tracer of gas mass.}.
Remarkably, there is no obvious separation between (U)LIRGs and SF galaxies (i.e., no hint of bimodality), unlike what Fig.~\ref{f-ks1} suggested.  
Fig.~\ref{f-k07} shows that, for SF galaxies in both our sample and \citetalias{buri12}, there is no strong correlation between $\ssfr/\ihcn$ and $\ihcn/\ico$ (right-hand panel). 
Such a correlation exists, but only in the contrast between SF galaxies and (U)LIRGs and more weakly within the (U)LIRG sample. At the same time, both sets
of SF galaxies (ours and \citetalias{buri12}) show a wide range in $\ssfr/\ihcn$, including an anti-correlation with $\ihcn/\ico$ that becomes
much clearer and unambigous when we consider the $\sstar$ as the independent variable in our own sample.

Colored regions in the background compare all data to the predictions of the \citetalias{krum05} model. These are the same shown in Fig.~\ref{f-k07-1}, except that predicted surface densities of (dense) molecular are converted to line intensities via our default fixed conversion factors. The regions with constant $\bar{n}$ are colored from green to red, with red hues indicating higher densities, for the sake of visualization. 

In the $\ihcn/\ico$ vs. $\ssfr/\ico$ plot (left-hand panel), the collection of data from SF disks through (U)LIRGs tend to follow the purple lines (constant Mach number), 
with most but not all of them spreading between $\mathcal{M}=25$ and $\mathcal{M}=100$. Therefore, as we move from SF galaxies to (U)LIRGs,  $\ihcn/\ico$ and $\ssfr/\ico$ would
mostly increase  because $\bar{n}$ changes from $100$~cm$^{-3}$ (solid line near the bottom-left corner) to  $\sim10^{4}$~cm$^{-3}$. In this plot, changes in $\mathcal{M}$ would be secondary and would mostly introduce scatter around the main trend.

The  $\ihcn/\ico$ vs. $\ssfr/\ihcn$ plot in right-hand panel provides us with a complementary view. Here, the most relevant trend comes from considering only our 
resolved observations of SF disks (gray dots), which spread out roughly along lines of fixed $\bar{n}$, mostly between $\bar{n}=10^{1.5}$~cm$^{-3}$ and $\bar{n}=10^{2.5}$~cm$^{-3}$ (green-yellowish areas). 
Therefore, as we move across galaxy disks towards their centers, $\ihcn/\ico$ would increase and $\ssfr/\ihcn$ would decrease mostly because $\mathcal{M}$ 
increases from $25$ to $>50$. Changes in $\bar{n}$ would play a secondary role here.
A less prominent but significant trend is found when SF galaxies and (U)LIRGs are globally compared. In this case, the average difference in $\sfedens$ between the two families would mostly stem from a difference in the Mach number.

Thus the \citetalias{krum05} model provides a qualitative picture that can reasonably account for the observations. We caution that the quantitative details of this
picture are far from settled, however. Firstly, we had to rescale their model to produce the comparison grids, 
and the exact numbers remain imperfect matches to typical average cloud 
properties: e.g., typical Milky Way clouds have $\mathcal{M} \approx 10$ and a volume averaged density $\bar{n} \sim 100$~cm$^{-3}$ \citep[e.g.,][]{heyer09}, not 
perfect matches to the models.  Secondly, a significant fraction of data points lie outside the  generous $25-100$ range in Mach numbers that we have assumed to generate the comparison grids. 
Finally, the model requires the clouds to be less turbulent in (U)LIRGs than in SF galaxies, which is not the sense
seen in observations. Still the sense of the trends seems reasonable: several studies suggest that clouds in galaxy centers are denser but also more turbulent than those in the galactic disks 
(e.g., \citealt{oka01}; \citealt{ros05}; Leroy et al., submitted).

These inconsistencies between the \citetalias{krum05} model and the extragalactic data might stem from our crude assumption of fixed CO and HCN conversion factors for all data points. 
They might be partly alleviated by solving the radiative transfer independently in each pointing/source to calculate more accurate conversion factors.
 To illustrate this, we plot in Fig.~\ref{f-k07} as a grey background the predictions from the  \citetalias{krum07} model, which combines the  \citetalias{krum05} model with a basic radiative transfer model.  The gray area is limited by two sets of cloud conditions defined by \citetalias{krum07}: one tries to reproduce typical conditions in SF galaxies, and assumes, in particular, $\mathcal{M}=30$ (upper envelope); the second represents conditions in starburst galaxies, and assumes 
 $\mathcal{M}=80$ (lower envelope). In both cases the average gas density ($\bar{n}$) is left as the only free parameter.
 We globally shifted the predictions in the vertical direction to better match the data clouds. The most apparent result is that, once the radiative transfer is taken into account, the model can explain the observed range in $\ssfr/\ico$ and $\ssfr/\ihcn$ ratios with a narrower range of Mach numbers (30--80). 
 
 In spite of this relative improvement, the \citetalias{krum07} model neither reproduces the data to a satisfactory extent, though. E.g., it still requires a too broad range of Mach numbers and it predicts clouds in (U)LIRGs to be less turbulent that those in SF galaxies. These two flaws might stem from the simplifications in the radiative transfer code use by \citetalias{krum07} (see the paper for details). They may also be solved by combining radiative transfer models with the more recent turbulent models in \citet{fedd12} and following. In general, the assumption of virialized clouds with lognormal density PDF implicit in the   \citetalias{krum05} model does not necessarily hold in the disturbed ISM of many (U)LIRGs.

\subsection{Simple Predictions and Logical Next Steps}

The contrast of our data with both models indicates logical next steps. First, while the CO-to-H$_2$ conversion factor has received
substantial attention, the conversion between HCN and the dense gas mass remains far more uncertain. Especially the dependence of this
value on environment is very uncertain. We expect HCN (1--0) 
and similar high critical density, low excitation lines (e.g., HCO$^{+}$ (1--0), CS (2--1), or HNC(1--0)) 
to be the most accessible tracers of dense gas in external galaxies for the foreseeable
future. A quantitative understanding of how $\ahcn$ and similar quantities depend on environment will be essential. 
New observations, e.g., of optically thin isotopologues over a wide range of environments, are clearly needed \citep[e.g.,][]{papa12}.

Along similar lines, the success of the ``whole cloud'' models hinges on the properties of individual turbulent molecular clouds. These properties, including
average density and Mach number, can be measured and compared among environments. In the same way that knowing $\ahcn$ will provide a
rigorous constraint on threshold models, folding cloud properties into the comparison to whole-cloud models will vastly constrain the area of viable
parameter space. A number of current mm-wave facilities are able to resolve individual clouds at the distances of many of our targets 
\citep[e.g.][]{col14}.

\section{Summary}
\label{s-sum}

We have used the IRAM 30-m telescope to observe the \hcn\ line in $62$ positions (48 detections) across the disks of 29 nearby galaxies
also targeted by the HERACLES CO(2--1) survey. Combining these data with previous multiwavelength observations, we study the 
relations between the dense molecular gas traced by the HCN emission ($n\gtrsim10^4-10^5$~cm$^{-3}$), the bulk of the molecular gas 
(traced by the CO data), and the star formation rate (SFR; traced by infrared emission). We highlight that:

\begin{itemize}
\item Unlike most previous studies of HCN in galaxies, our observations span a wide range of galactocentric radii (from galaxy centers up to $\sim80\%$ of the optical radius)
at linear resolution ($\sim1.5$~kpc on average) that isolates local properties of a galaxy disk while still averaging over a large population of clouds. This allows
us to probe a wide range in environmental parameters.

\item We compare our data set to a compilation of unresolved observations of star-forming (SF) galaxies and (U)LIRGs 
\citepalias[from][]{buri12}. Our resolved observations seamlessly expand the range of surface densities sampled by HCN studies 
of extreme systems to lower values. They also fill in an intermediate, so far largely unexplored, luminosity regime in the
HCN-IR luminosity correlation found by \citet{gao04b,wu05}.

\end{itemize}

We focus our analysis on three parameters: the dense gas fraction ($\fdens\propto$~HCN/CO), the star formation rate 
per unit of dense molecular gas ($\sfedens\propto$~SFR/HCN) and the star formation rate per unit of molecular gas ($\sfedens\propto$~SFR/CO). 
The last two are often referred to as the ``star formation efficiency'' of the dense and total molecular gas. We study
how these quantities relate to conditions in the disk. We focus on the molecular-to-atomic mass ratio ($\smol/\satom$) 
and the stellar mass surface density ($\sstar$) as useful indicators of the local conditions in the disk. Both quantities are
highly covariate with galactocentric radius. We find that:

\begin{itemize}

\item For the simplest assumption of fixed luminosity-to-mass conversion factors for CO and HCN, we find that 
$\fdens$ ($\propto {\rm HCN/CO}$) tightly correlates with both $\smol/\satom$ and $\sstar$ (Spearman's rank coefficients $0.67$ and $0.75$). 
Across $\sim2.1$~dex in both disk structure parameters, $\fdens$ systematically varies by a factor of $\sim4$. 

\item In contrast, $\sfedens \propto {\rm IR/HCN}$ {\em anti}-correlates with  $\smol/\satom$ and $\sstar$ (rank coefficients $-0.72$ and $-0.53$), systematically 
decreasing by a factor of $\sim6-8$ as we move from disk positions towards galaxy centers. The two trends cancel 
each other out, so that $\sfemol$ shows no systematic variation with either $\smol/\satom$ nor with $\sstar$. 
This implies that, to some degree, the apparent constancy of $\sfemol$ known from previous studies arises from a conspiracy,
as the sub-resolution cloud populations must vary systematically to produce the changing HCN-to-IR and HCN-to-CO ratios.

\end{itemize}

We compare our results to two families of standard models of star formation in galaxies:

\begin{itemize}

\item First, we consider a simple density-threshold model, which, in its simplest version,  assumes that the SFR is proportional 
only to the mass of dense gas above a certain density threshold traced by the HCN emission (thus, a fixed $\sfedens$). Such
studies have been motivated by local studies of Galactic clouds and the apparent linear correlation of HCN and IR
across many decades in luminosity. Such a model is clearly at immediate odds with the apparent systematic variation in $\sfedens$ across 
galaxy disks. Further, it predicts a strong, direct dependence of $\sfemol$ on $\fdens$ that our data do not immediately show. For
fixed conversion factors, our data do not appear compatible with a simple threshold model and contradict one
of the main observational pieces of evidence used to argue for such data.

\item In order for our observations to agree with a simple density threshold model, the HCN-to-dense gas mass conversion factor, 
$\ahcn$, must vary in such a way as to cancel out the observed variations in the HCN-to-IR ratio. We explore the topic
more quantitatively and show that only a very restricted range of possible combinations of CO and HCN conversion factors can satisfy such a model.

\item We contrast the density threshold model with ``whole cloud models,'' focusing on the theory of turbulence-regulated star formation 
formulated by \citetalias{krum05} as an example 
\citep[see also \citetalias{krum07} and, for a wider view of such models,][]{fedd12}.  
 In the \citetalias{krum05} model, $\sfemol$ increases with the average density of the molecular gas ($\bar{n}$), 
whereas it is inhibited to some degree by increased turbulence (high Mach number, $\mathcal{M}$). Meanwhile, $\fdens$ increases with both 
$\bar{n}$ and $\mathcal{M}$. We find that, for fixed conversion factors, the \citetalias{krum05} model can reproduce our data well,
provided that the increase in $\fdens$ towards galaxy centers is driven by variations in $\mathcal{M}$ rather 
than in $\bar{n}$. This good agreement exists for almost the full range of plausible conversion factors that we consider, reflecting
the flexibility of the models.

\end{itemize}

We interpret these results to represent a challenge to simple density threshold models, though they do not rule them out.
The simple correlation between IR and HCN luminosity spanning from starbursts to Galactic cores has represented an important piece
of support for such models, our survey of galaxy disks shows that the main sites of star formation in the local universe do not obey
such a simple relation (see also a dedicated study of M~51 by Bigiel et al, in prep.).

The appendices report more details of the observations and investigate the effects of varying how we trace the SFR and molecular 
gas on our conclusions, showing that they remain robust to substitution of CO transition or SFR tracer.

\begin{acknowledgements} 
We thank the IRAM staff for their help during the observations. 
Antonio Usero acknowledges support from Spanish grants AYA2012-32295 and FIS2012-32096. 
Santiago Garc\'ia-Burillo acknowledges support from Spanish grants 
AYA2010-15169 and AYA2012-32295 and from the Junta de Andaluc\'ia through TIC-114 and the Excellence Project P08-TIC-03531. 
Frank Bigiel acknowledges support from DFG grant BI~1546/1-1.
The work of W.J.G. de Blok was supported by the European
Commission (grant \mbox{FP7-PEOPLE-2012-CIG \#333939}).
We acknowledge the 
usage of the HyperLeda database (http://leda.univ-lyon1.fr). The National Radio Astronomy Observatory is a facility of the National Science Foundation operated 
under cooperative agreement by Associated Universities, Inc.
\end{acknowledgements}

\bibliographystyle{apj} 
\bibliography{densegalbib-v02}

%%%%%%%%%%%%%%%%%%%%%%%%%%%%%%
\appendix

\section{Alternative CO and SFR tracers}
\label{s-crosscheck}

\subsection{Checking $\smol$}
\label{ss-check-mol}

By default in this paper, we assume a fixed CO(2--1)--to--CO(1--0) line ratio $R_{21}=0.7$ to estimate $\smol$ directly from the intensity of the CO(2--1) line (Equation~\ref{e-smol}). There is evidence that $R_{21}$ could systematically vary across galaxy disks to some extent, however (see \citealt{lero09} for the case of HERACLES galaxies; also references therein).  In particular, $R_{21}$ could be up to $\sim1.5$ times higher near some galaxy centers, as the gas excitation tends to be more efficient. By neglecting these variations, we could be overestimating $\smol$ near galaxy centers. Thanks to our complementary observations of CO(1--0) (Sect.~\ref{sss-30m-obs}), available for most of the observed positions, we can confirm that the assumption of a fixed  $R_{21}$ has a minimal effect on our results. To do this, we derive an alternative  $\smol$ after substituting  $R_{21}=0.7$ in Equation~\ref{e-ico} with the line ratio derived from the CO(1--0) and CO(2--1) observations in each observed position. The CO(2--1) spectral cubes were previously convolved to the angular resolution of the CO(1--0) observations ($22\arcsec$) to account for aperture effects.  This implicitly assumes that the  $R_{21}$ ratio is similar at $22\arcsec$ and $28\arcsec$ resolution.

\begin{table*}
\begin{center}
\begin{threeparttable}
\caption{Rank coefficients for $\fdens$, $\sfedens$ and $\sfemol$ as a function of local parameters, when either the CO(2--1) or the CO(1--0) transitions are used to estimate $\smol$}
\label{t-env3}
                                                                       
\begin{tabular}{rrrr@{\extracolsep{6pt}}rrr}                                                         
\hline\hline                                                                    
\noalign{\smallskip} 
& \multicolumn{3}{c}{$\smol$ from CO(2-1)} &   \multicolumn{3}{c}{$\smol$ from CO(1-0)} \\
 \noalign{\smallskip}
\cline{2-4}\cline{5-7}
\noalign{\smallskip}                                                           
& $\fdens$ & $\sfedens$ & $\sfemol$ & $\fdens$ & $\sfedens$ & $\sfemol$ \\
                                                                                  
\noalign{\smallskip}                                                            
\hline                                                                           
\noalign{\smallskip}                                                            
$r_{25}$ & \si{-0.59} & \si{ 0.43} & \no{-0.07} & \si{-0.55} & \si{ 0.43} & \no{-0.10}\\
$\smol/\satom$ & \si{ 0.75} & \si{-0.72} & \no{-0.22} & \si{ 0.77} & \si{-0.68} & \no{ 0.00}\\
$\sstar$ & \si{ 0.67} & \si{-0.53} & \no{ 0.09} & \si{ 0.76} & \si{-0.53} & \si{ 0.34}\\
\noalign{\smallskip}
\hline               
\end{tabular}

\begin{tablenotes}
\item{\sc Note. ---}  Star symbols indicate significant correlations. 
\end{tablenotes}
\end{threeparttable}
\end{center}
\end{table*}

\begin{figure}
   \centering
\includegraphics[width=0.33\hsize]{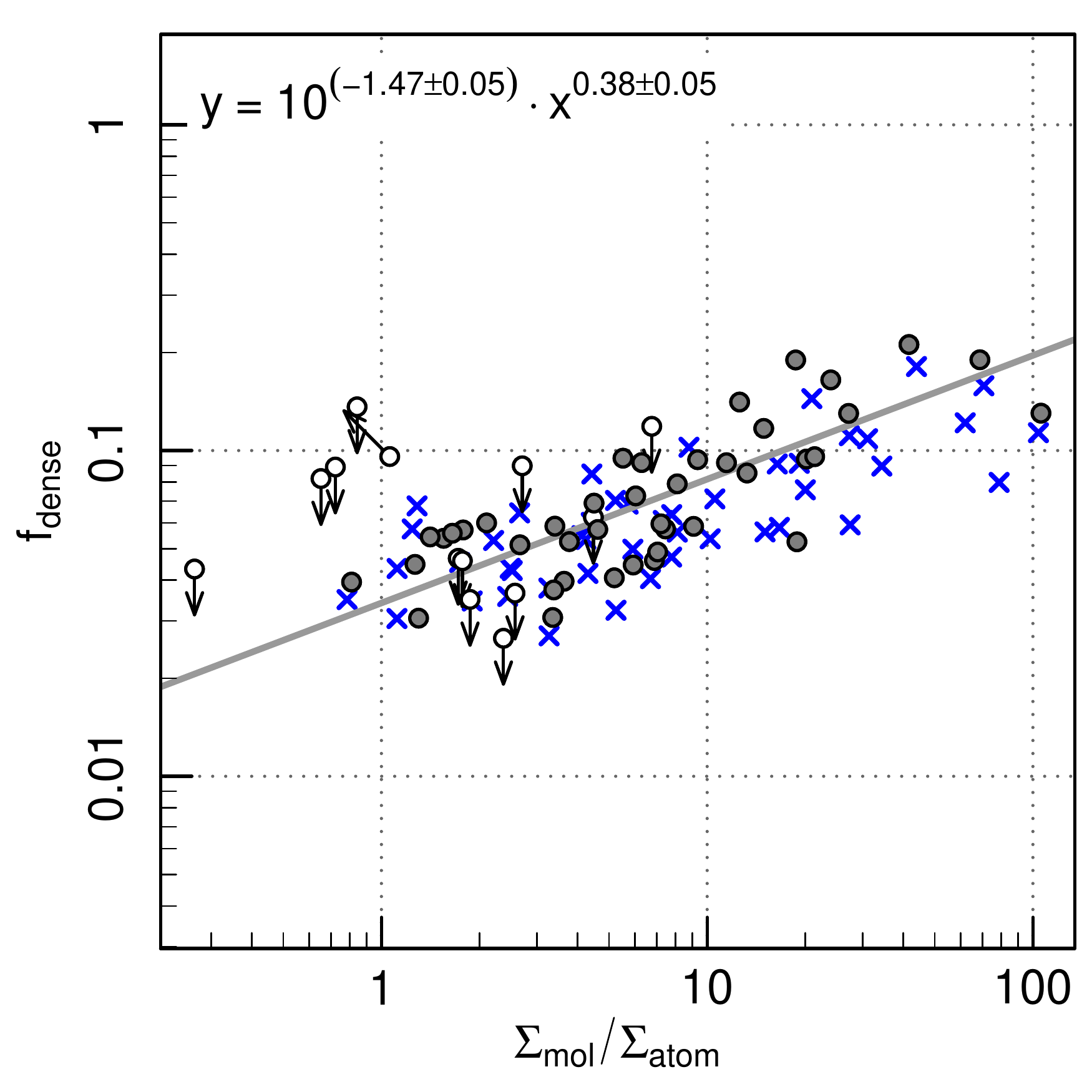}
\includegraphics[width=0.33\hsize]{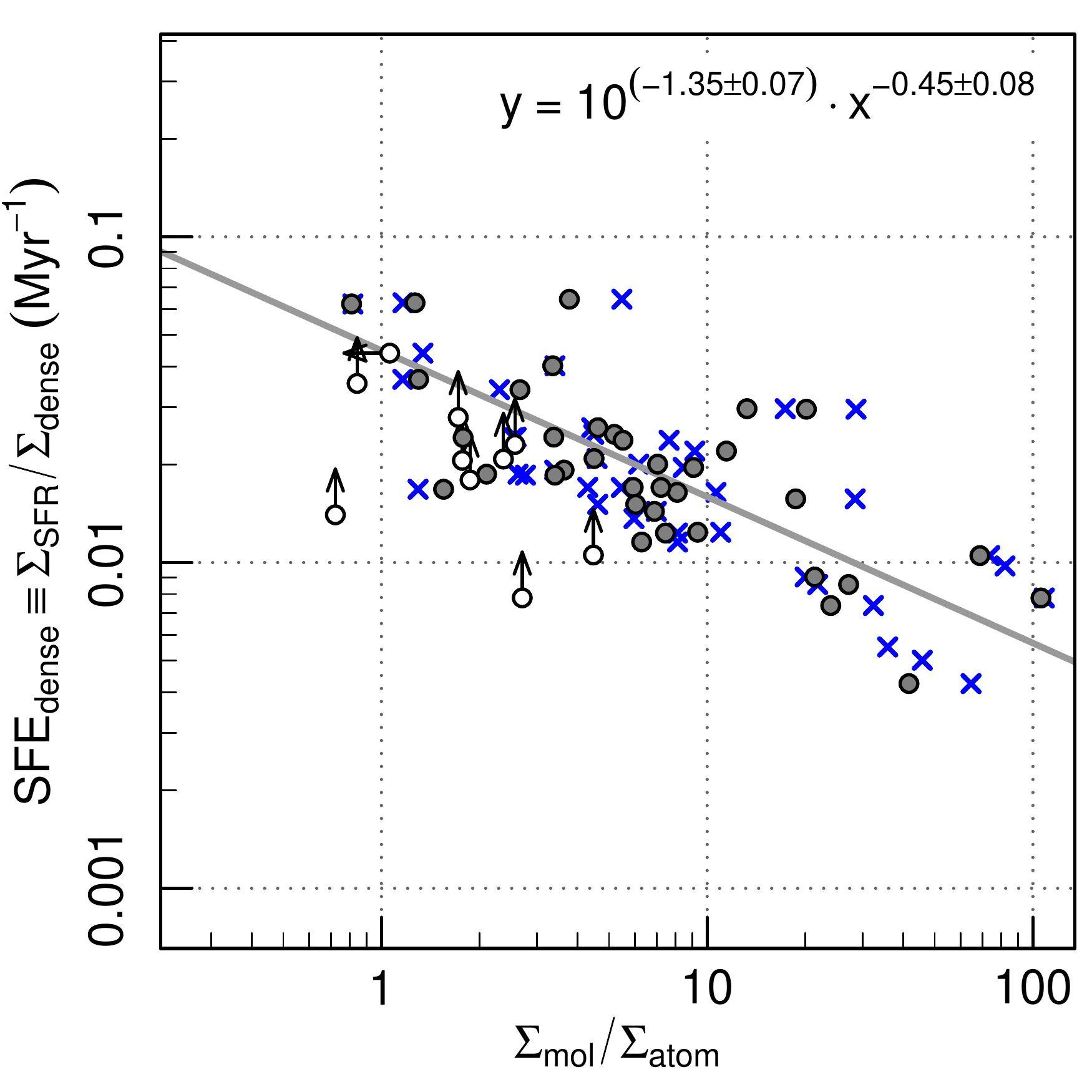}
\includegraphics[width=0.33\hsize]{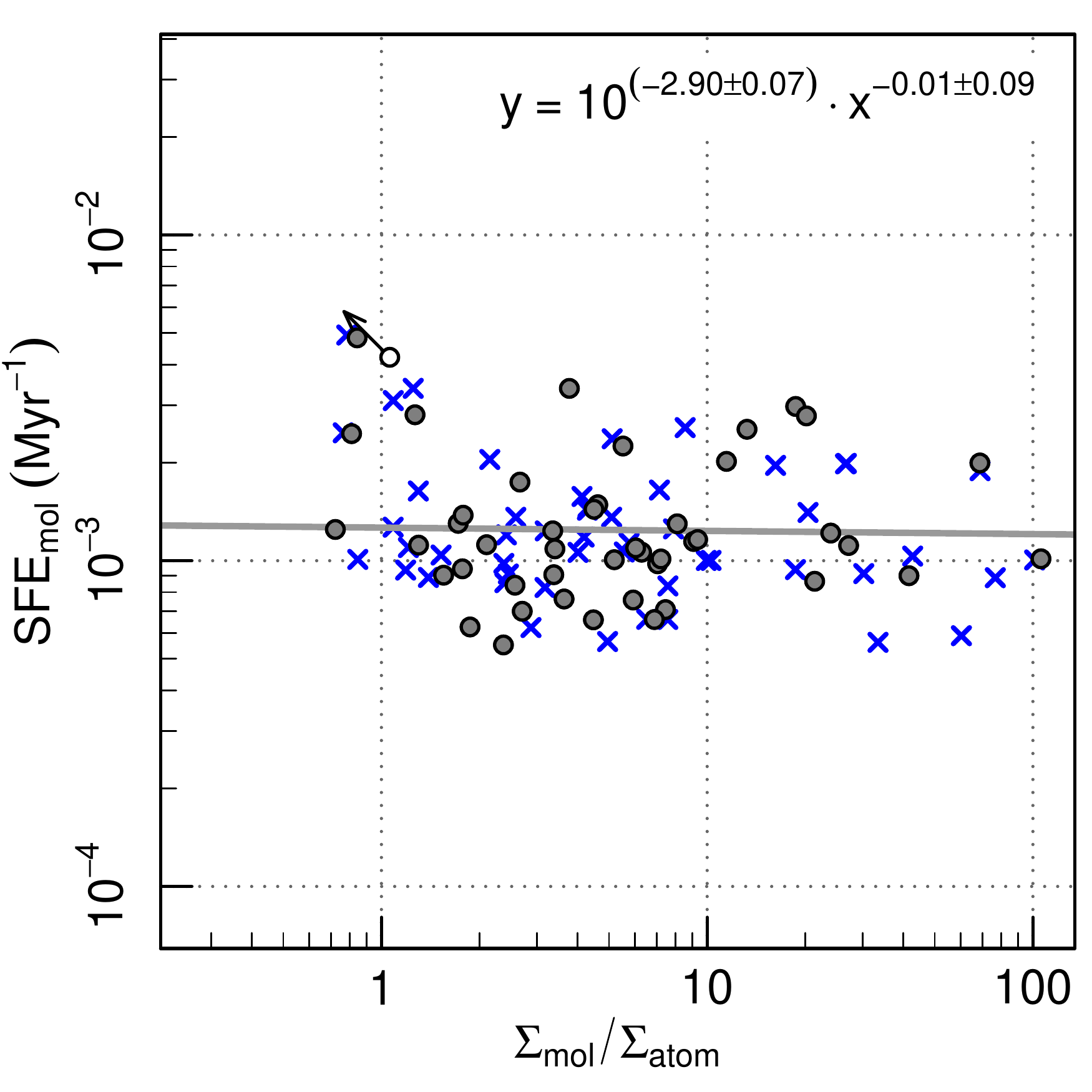}
\caption{$\fdens$ (left-hand panel), $\sfedens$ (middle panel) and $\sfemol$ (right-hand panel) as a function of the molecular-to-atomic ratio.
For grey dots, $\smol$ is derived from CO(1--0) data as indicated in the text. For blue crosses, it is derived from the HERACLES CO(2--1) spectra.  
}
\label{f-xco1}%
\end{figure}

Fig.~\ref{f-xco1} shows $\fdens$, $\sfedens$ and $\sfemol$ as a function of $\smol/\satom$ when $\smol$ is derived from CO(1--0) (grey dots) and CO(2--1) (blue crosses).  The plots where the variables are plotted against $\sstar$ are similar to these, so we do not show them here. In each panel, the cloud of blue crosses was slightly adjusted so that the two data clouds have the same median along both axes. This allows us to better compare the trends that we find for both CO transitions. It is apparent in Fig.~\ref{f-xco1} that those trends are nearly identical in each case.  The fitted power laws are the same as those in Fig.~\ref{f-env1} within the errors, although they tend to be slightly steeper when $\smol$ is derived from the CO(1-0) intensity (i.e., by $\lesssim0.1$). This is consistent with a small increase in $R_{21}$ towards the galaxy centers.  Table~\ref{t-env3} shows that the rank coefficients discussed in Sect.~\ref{s-env} are marginally sensitive to the chosen CO transition. The only noteworthy difference is that the correlation between $\sfemol$ and $\sstar$ is more significant for CO(1--0).

\subsection{Checking $\ssfr$}
\label{ss-check-sfr}

Thanks to the our ample ancillary data set, we can confirm that the trends in $\sfedens$ and $\sfemol$  found in Sect.~\ref{s-env} are robust against the choice of the SFR tracer. To illustrate this, we obtained simple SFR estimates from the  H$\alpha$ ($I_\mathrm{H\alpha}$) and $24~\mu$m  intensities, adopting the calibrations quoted by \citet{kenn12}. These were respectively determined by \citet{murp11} and \citet{riek09}.  In addition, we followed  \citetalias{lero12} to combine  H$\alpha$  and $24~\mu$m  data into a hybrid SFR estimate that takes into account the unobscured and obscured star formation they respectively trace: 

\begin{equation}
\label{e-sfr}
\frac{\ssfrler}{M_\odot~\mathrm{Myr^{-1}~pc^{-2}}}=
634\frac{I_\mathrm{H\alpha}}{\mathrm{erg~s^{-1}~sr^{-1}~cm^{-2}}}\cos(i)+
1.3\frac{I_{24}^\mathrm{cirrus-corrected}}{400~\mathrm{MJy~sr^{-1}}}\cos(i).
\end{equation}

In Equation~\ref{e-sfr}, the measured $24~\mu$m intensity is corrected for the contribution from dust heated by pervasive radiation fields ("cirrus") unrelated to recent star formation. The cirrus component is proportional to the mass surface density of dust and a certain emissivity. Both parameters are derived from the dust models mentioned in Sect.~\ref{ss-data-param}. A detailed comparison of different SFR tracers in the context of the HERACLES  survey can be found in  \citetalias{lero12} and \citetalias{lero13}.

\begin{figure*}
   \centering
   \begin{tabular}{c@{\extracolsep{3pt}}c@{\extracolsep{3pt}}c}
 \small $SFR=24~\mu$m & \small $SFR=$~L12 & \small $SFR=$~H$\alpha$ \\  
\includegraphics[width=0.33\textwidth]{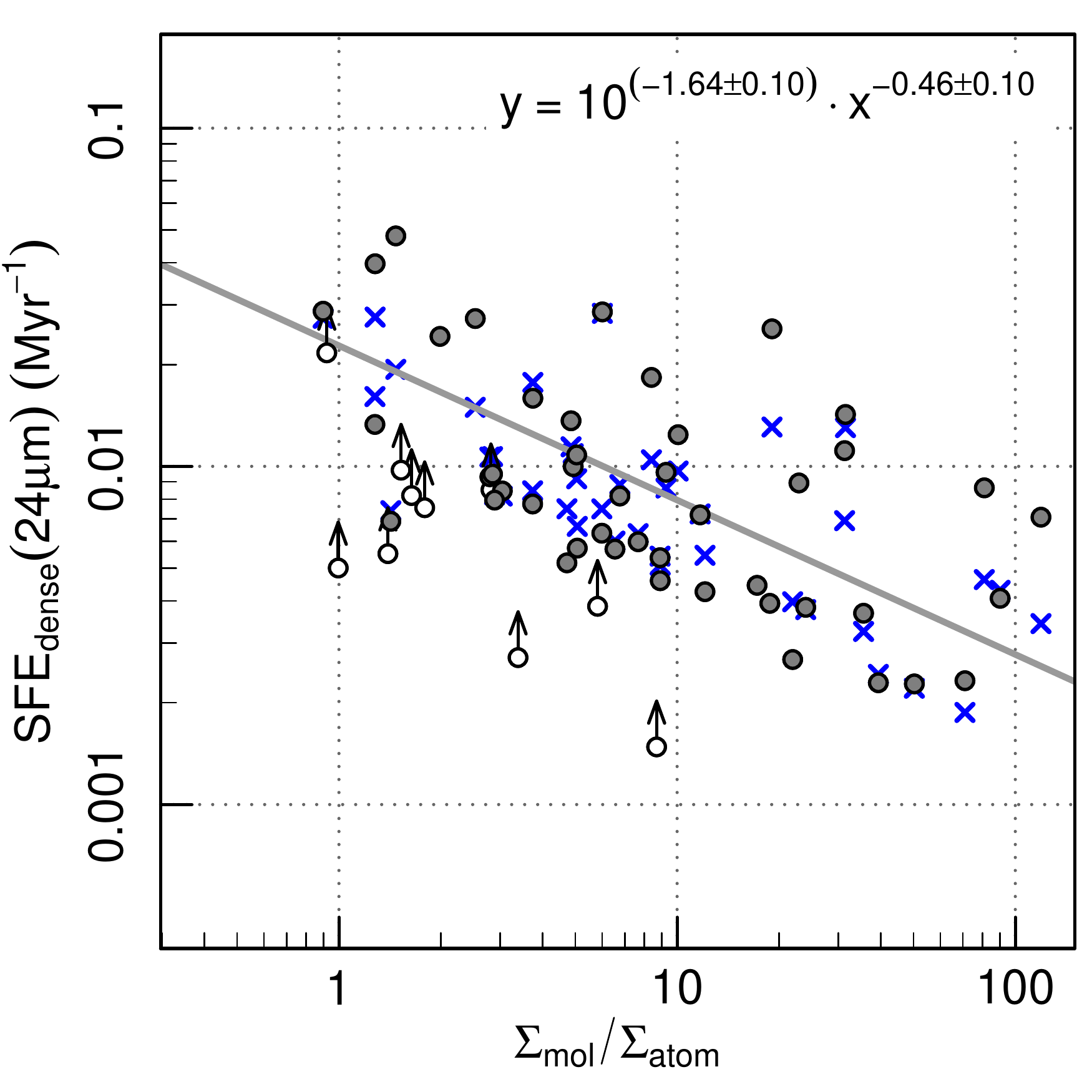}&
\includegraphics[width=0.33\textwidth]{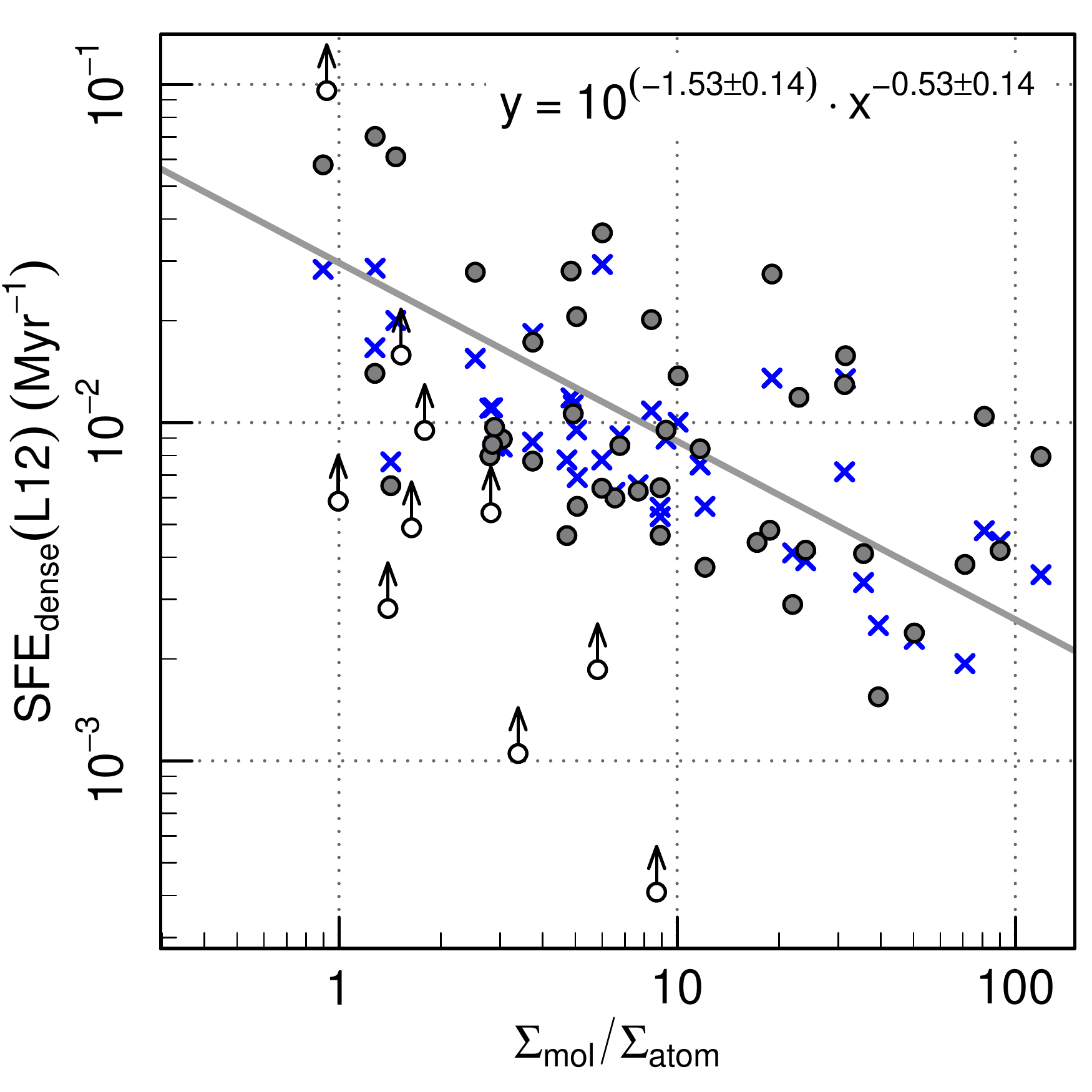}&
\includegraphics[width=0.33\textwidth]{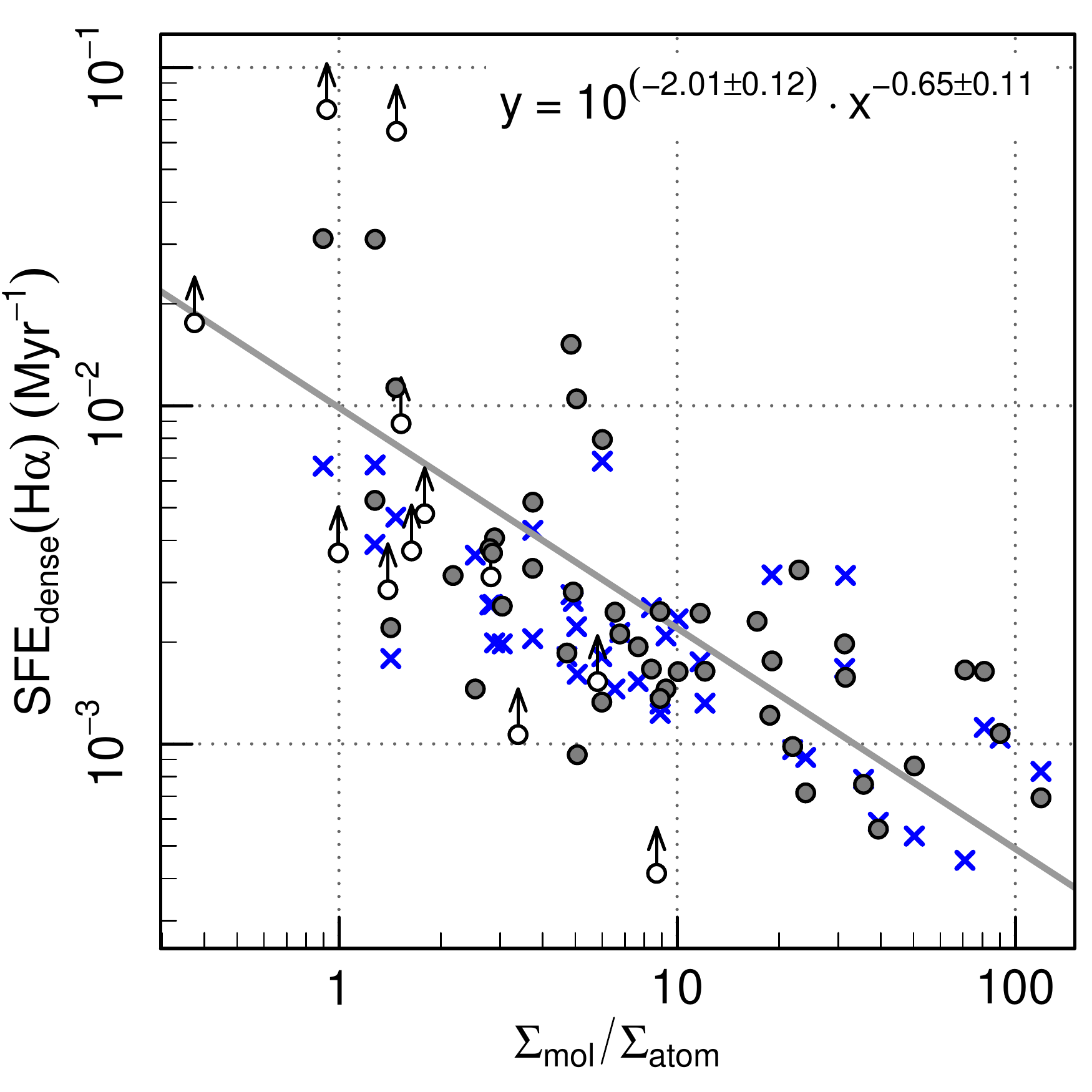}\\
\includegraphics[width=0.33\textwidth]{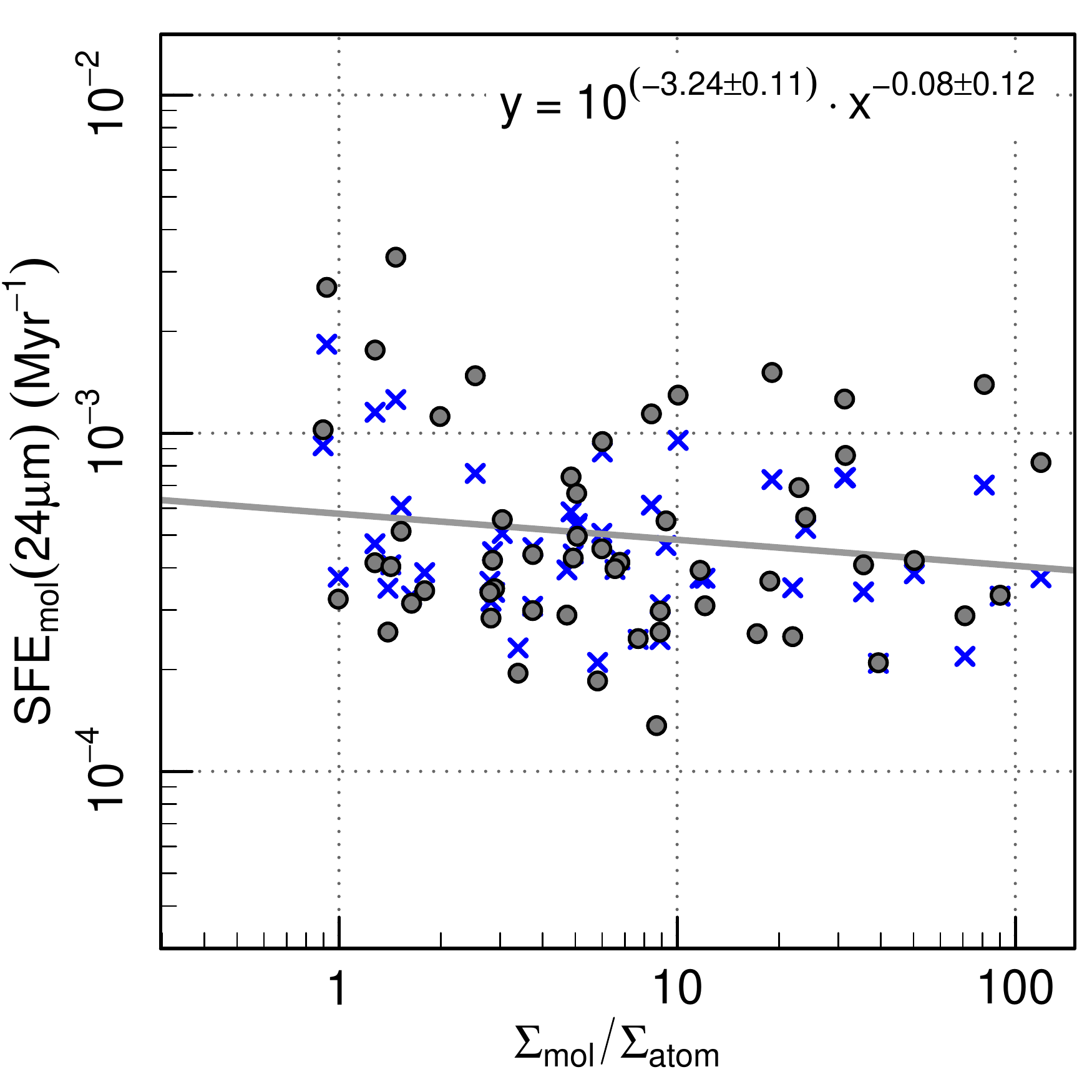}&
\includegraphics[width=0.33\textwidth]{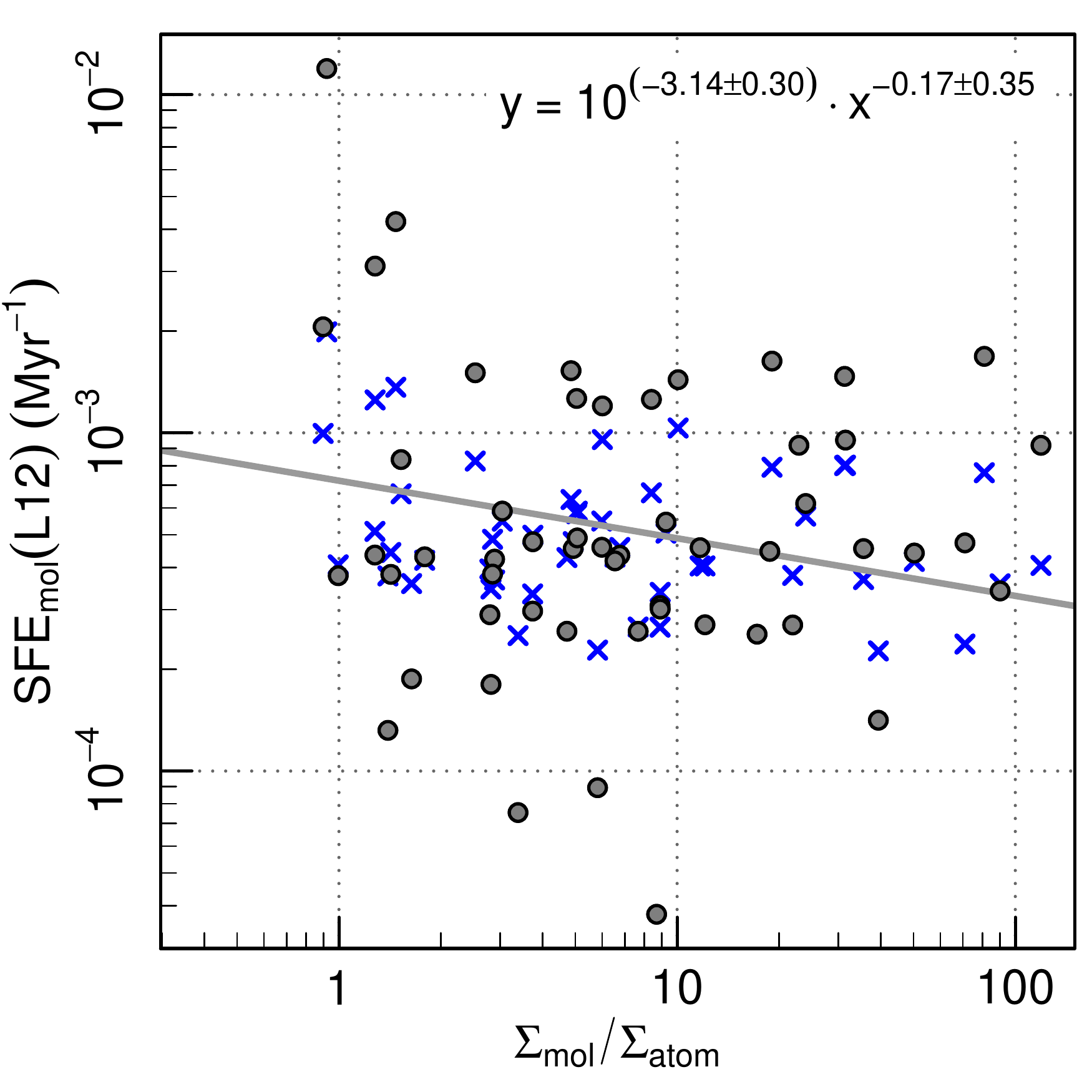}&
\includegraphics[width=0.33\textwidth]{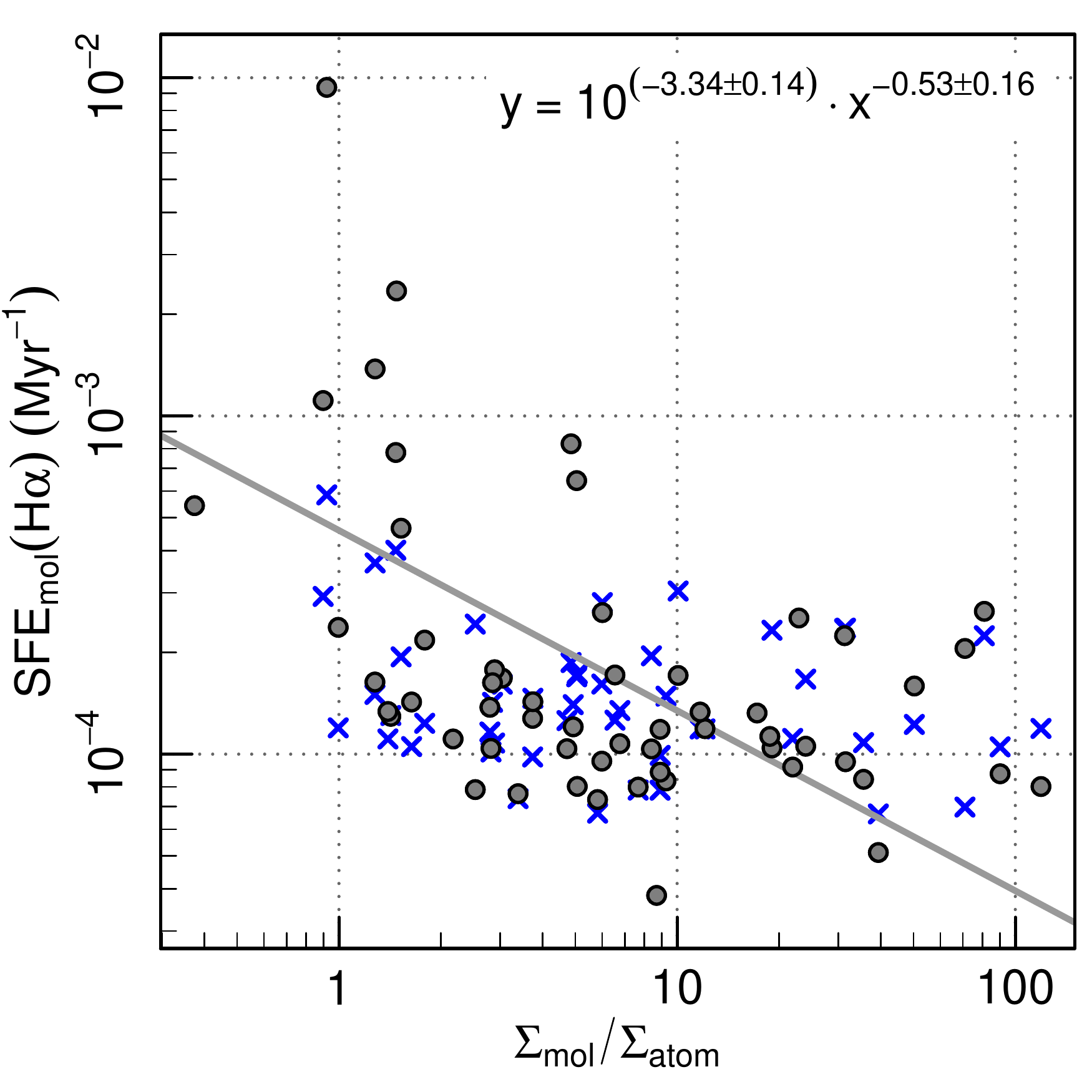}
\end{tabular}

\caption{$\sfedens$ (top row) and $\sfemol$ (bottom row) as a function of the molecular-to-atomic ratio. From left to right, the SFR is derived from the $24~\mu$m emission, a linear combination of $24~\mu$m and  H$\alpha$ intensities (\citetalias{lero12}), and the H$\alpha$ emission. 
}  
\label{f-sfrtracers1}%
\end{figure*}

We plot in Fig.~\ref{f-sfrtracers1} the two star formation efficiencies against the molecular-to-atomic ratio, but now using the three alternative SFR tracers: the  $24~\mu$m continuum (left-hand column),  the hybrid \citetalias{lero12} calibration (middle column), and the H$\alpha$ line (right-hand column). The plots against $\sstar$, not shown here, are similar to these. For the sake of comparison,  we represent the results for our default SFR tracer (TIR) by means of blue crosses.  These data points have been vertically shifted in each panel so that their median $\ssfr$ is the same as that of the grey dots. This allows us to focus on trends, letting aside the relative calibration between the different tracers.

 The SFR calibrators have different levels of intrinsic scatter and are sensitive to different components of star formation. Some mild changes in the power-law fits can be expected, since the fraction of unobscured (i.e., H$\alpha$-traced) star formation is known to increase towards fainter regions \citepalias{lero12}. In spite of all this, blue crosses and grey dots show similar trends (e.g., as captured by the fitted power-law indices) in most cases.   We find only two noteworthy features, in the $\sfemol$(L12) and  in the $\sfemol$(H$\alpha$) panels:\begin{itemize}
  
 \item  A few data points increase the dispersion in $\sfemol$(L12) at relatively low $\smol/\satom$.  On the one hand, there are three positions whose  $\sfemol$(L12) is well below the mean. They have an abnormally high cirrus correction\footnote{Their cirrus emissivity is normal, but the $\Sigma_\mathrm{dust}/I_\mathrm{24\mu m}$ ratio is high.} in Equation~\ref{e-sfr} and their  $\ssfr$(L12) is also the lowest in our sample, $\lesssim 2\times10^{-3}~M_\odot$~Myr$^{-1}$~pc$^{-2}$. These SFR levels are similar to the  $\lesssim 1\times10^{-3}~M_\odot$~Myr$^{-1}$~pc$^{-2}$ threshold below which hybrid SFR estimators may become unreliable \citepalias{lero12}.  On the other hand, there are a few positions  whose  $\sfemol$(L12) is well above the mean. They have some of the highest H$\alpha$--to--24~$\mu$m intensity ratios in our sample. We discuss them in the next point. 
  
 \item The H$\alpha$-bright positions mentioned above tilt the otherwise flat fit to $\sfemol$(H$\alpha$).  Our sample does not include enough  H$\alpha$-bright positions to assess a possible difference in $\sfemol$ between the two SFR regimes dominated by obscured and unobscured star formation.  In particular, the unobscured star formation only amounts to $\sim5\%-55\%$ (median 30\%) of the total SFR(L12) in positions detected in HCN. In our case,  pure IR and the L12 values likely provide more reliable estimates of the total SFR. Furthermore, in their thorough study of HERACLES galaxies \citetalias{lero13} found that the environmental trends in the molecular depletion time (i.e., the inverse of $\sfemol$)  were not significantly sensitive to the chosen SFR tracer.   
 \end{itemize}

  Something similar is found when paying attention to the rank coefficients between the star formation efficiencies and the ISM parameters,  which are listed in Table~\ref{t-env2}.
  On the one hand, $\sfedens$ remains strongly and significantly correlated with $\smol/\satom$ and $\sstar$ in all cases
  On the other hand, $\sfemol$ remains uncorrelated with the ISM parameters, except when H$\alpha$ is chosen as SFR tracer.

\begin{table*}
\begin{center}
\begin{threeparttable}
\caption{Rank coefficients for $\sfedens$ and $\sfemol$ as a function of local parameters, when different SFR tracers are used to estimate $\ssfr$}
\label{t-env2}

\begin{tabular}{rrr@{\extracolsep{3pt}}rr@{\extracolsep{3pt}}rr@{\extracolsep{3pt}}rr}                                                                                                                                                                                                   
                                                                                 
\hline\hline                                                                                                  
\noalign{\smallskip}  
 & \multicolumn{2}{c}{TIR (default)} &
  \multicolumn{2}{c}{24$\mu$m} &  
  \multicolumn{2}{c}{L12} &
  \multicolumn{2}{c}{H$\alpha$} \\
  \noalign{\smallskip}
\cline{2-3}\cline{4-5}\cline{6-7}\cline{8-9} 
\noalign{\smallskip}   
                                                                                       
& $\sfedens$ & $\sfemol$ & $\sfedens$ & $\sfemol$ & $\sfedens$ & $\sfemol$ & $\sfedens$ & $\sfemol$ \\
                                                                                                                
\noalign{\smallskip}                                                                                          
\hline                                                                                                         
\noalign{\smallskip}                                                                                          
$r_{25}$ & \si{ 0.43} & \no{-0.07} & \si{ 0.36} & \no{-0.05} & \si{ 0.35} & \no{-0.13} & \si{ 0.68} & \si{ 0.34}\\
$\smol/\satom$ & \si{-0.72} & \no{-0.22} & \si{-0.56} & \no{-0.08} & \si{-0.51} & \no{ 0.01} & \si{-0.72} & \si{-0.43}\\
$\sstar$ & \si{-0.53} & \no{ 0.09} & \si{-0.41} & \no{ 0.12} & \si{-0.37} & \no{ 0.22} & \si{-0.58} & \no{-0.20}\\
\noalign{\smallskip}
\hline               
\end{tabular}

\end{threeparttable}
\end{center}
\end{table*}

\newpage
\section{Data}
\label{a-tabl}

\begin{figure}
\begin{center}
\includegraphics[height=0.3\textheight]{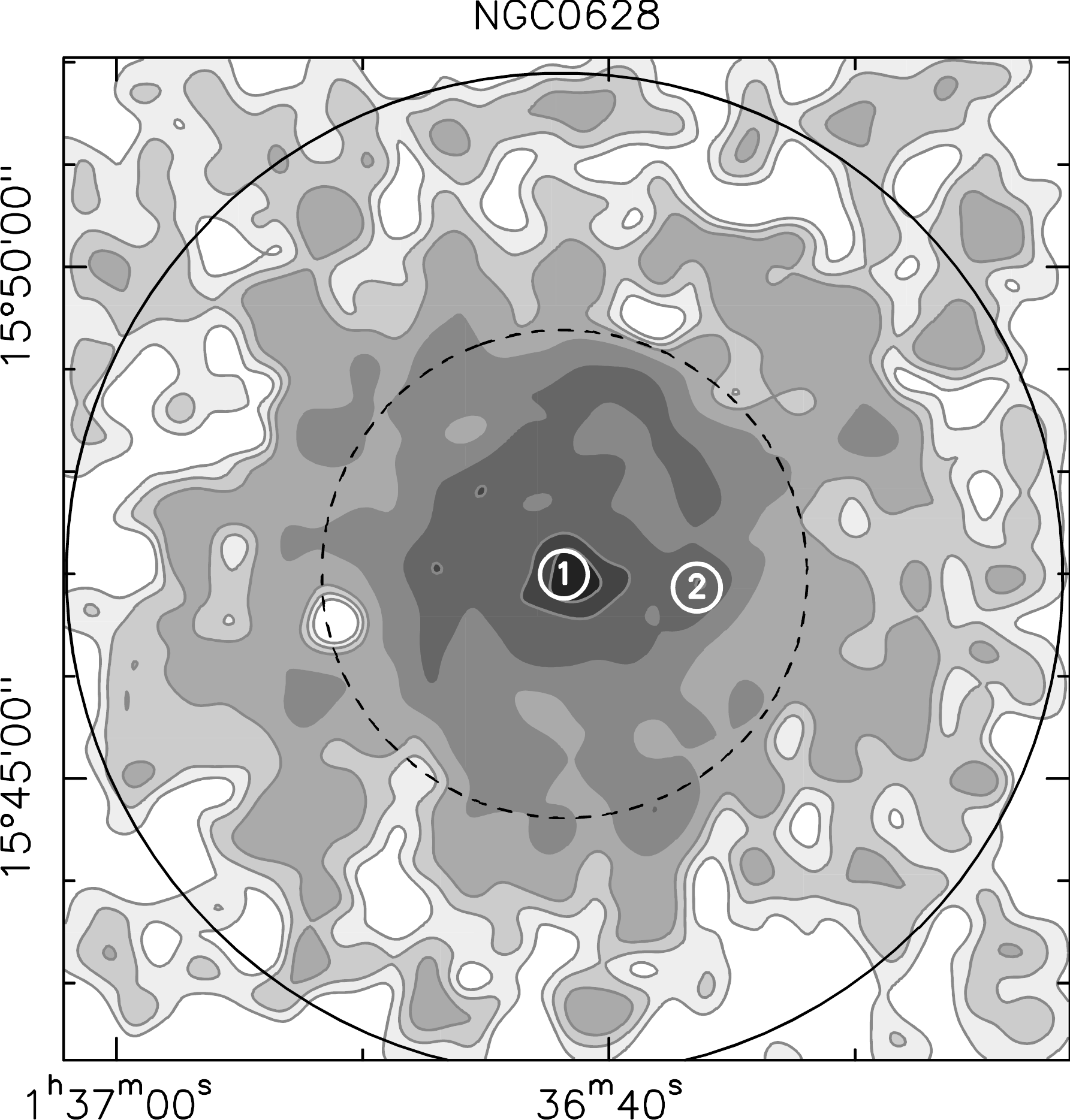}
\includegraphics[height=0.3\textheight]{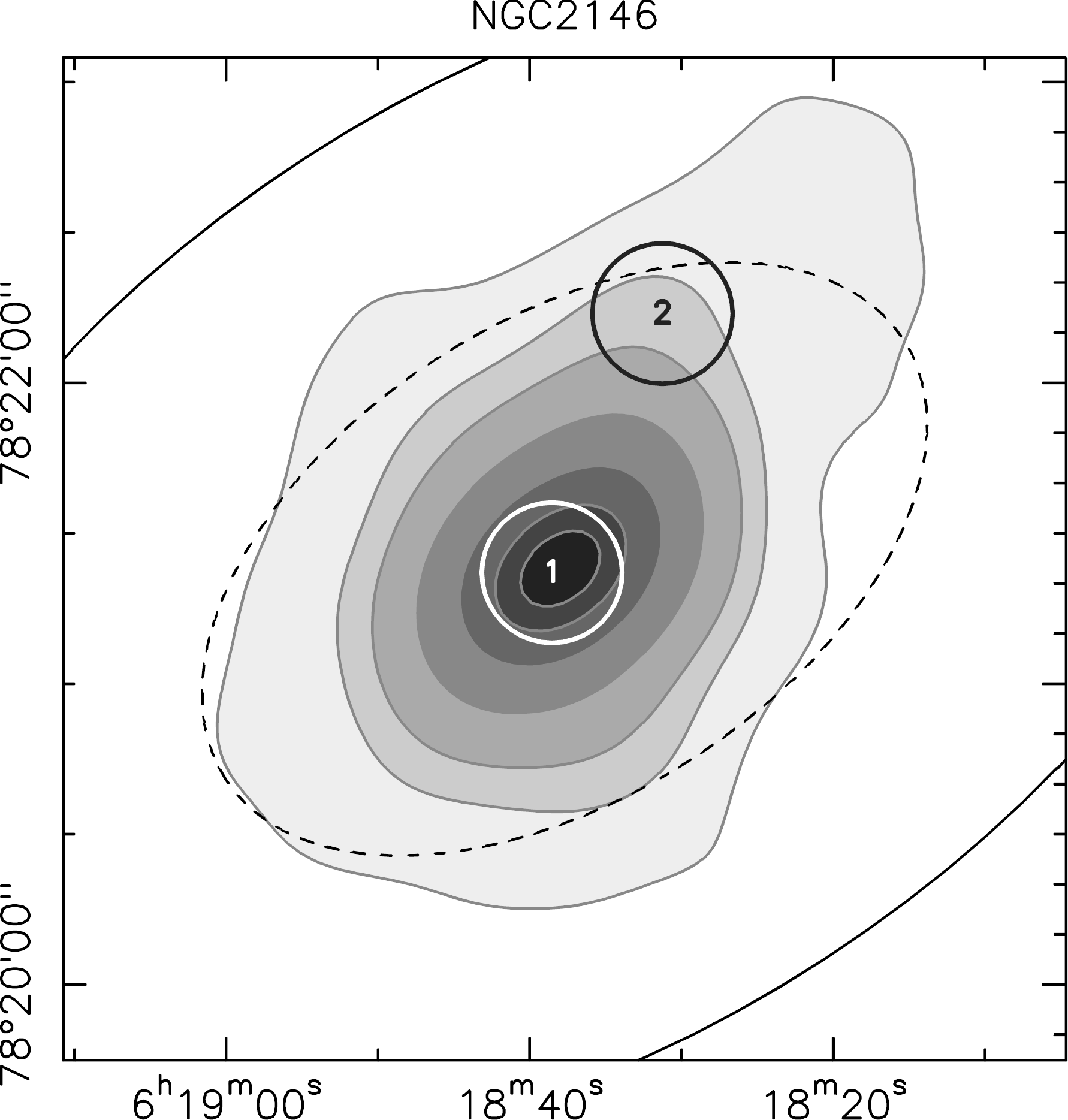}\\
\includegraphics[height=0.3\textheight]{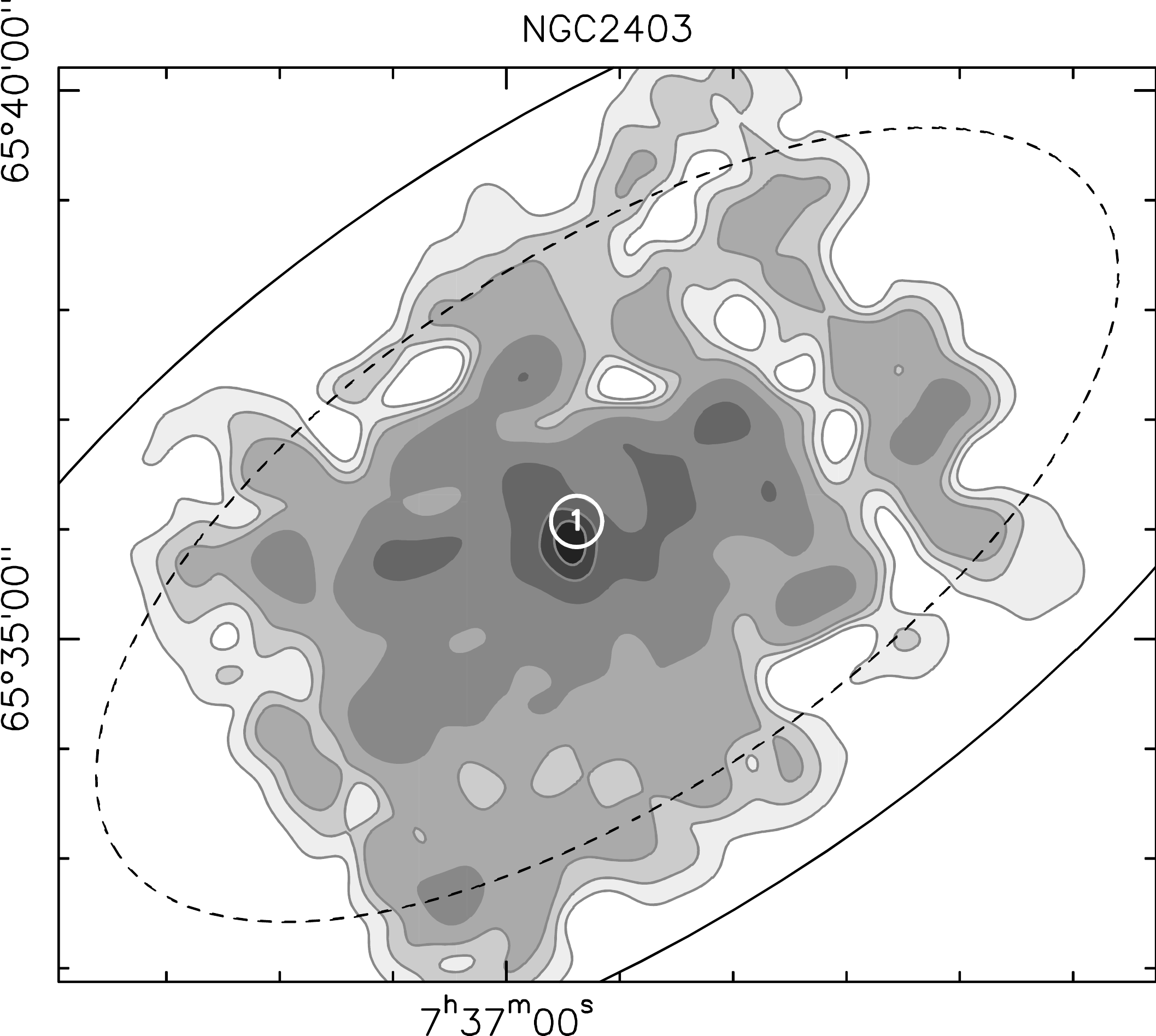}
\includegraphics[height=0.3\textheight]{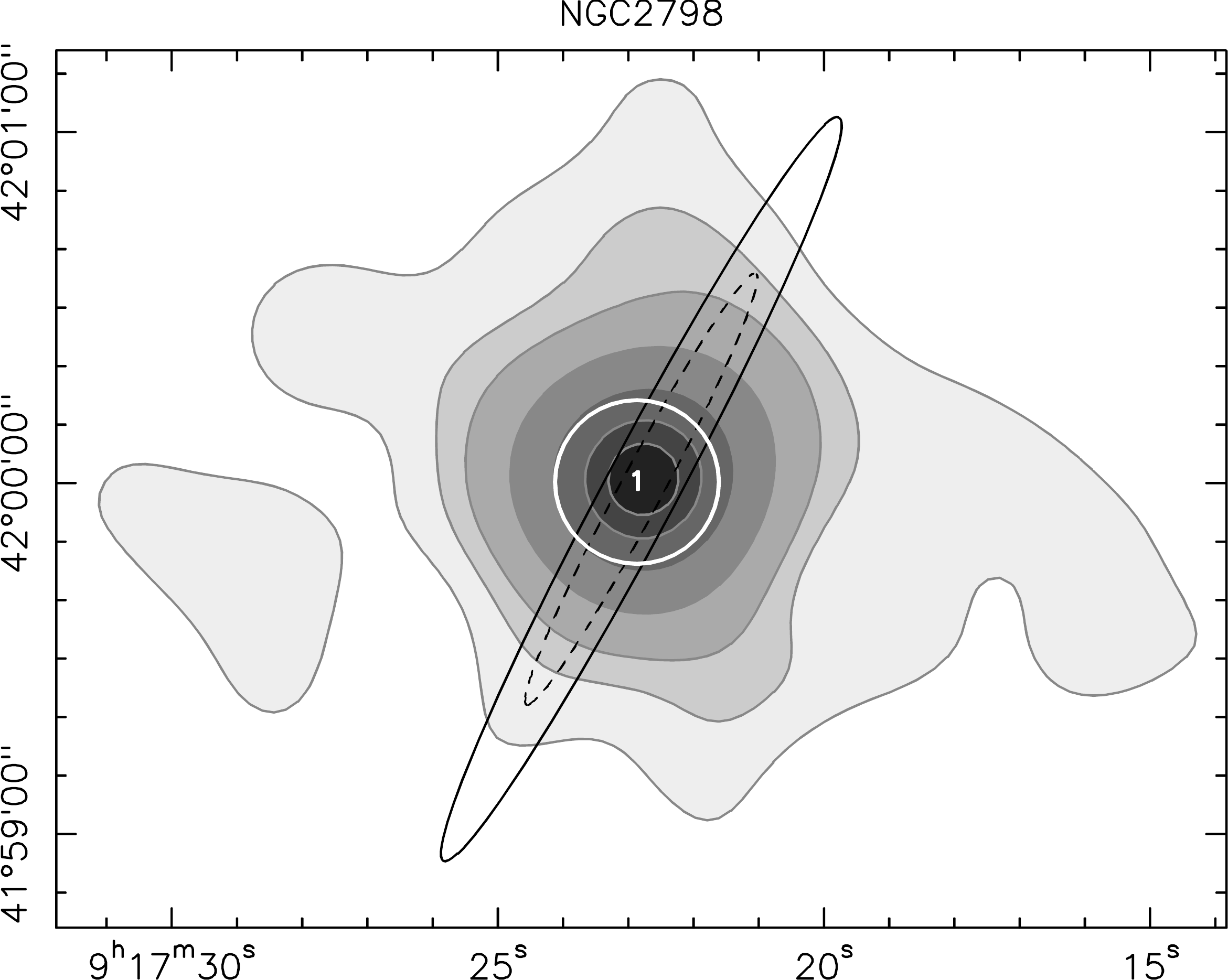}\\
\includegraphics[height=0.3\textheight]{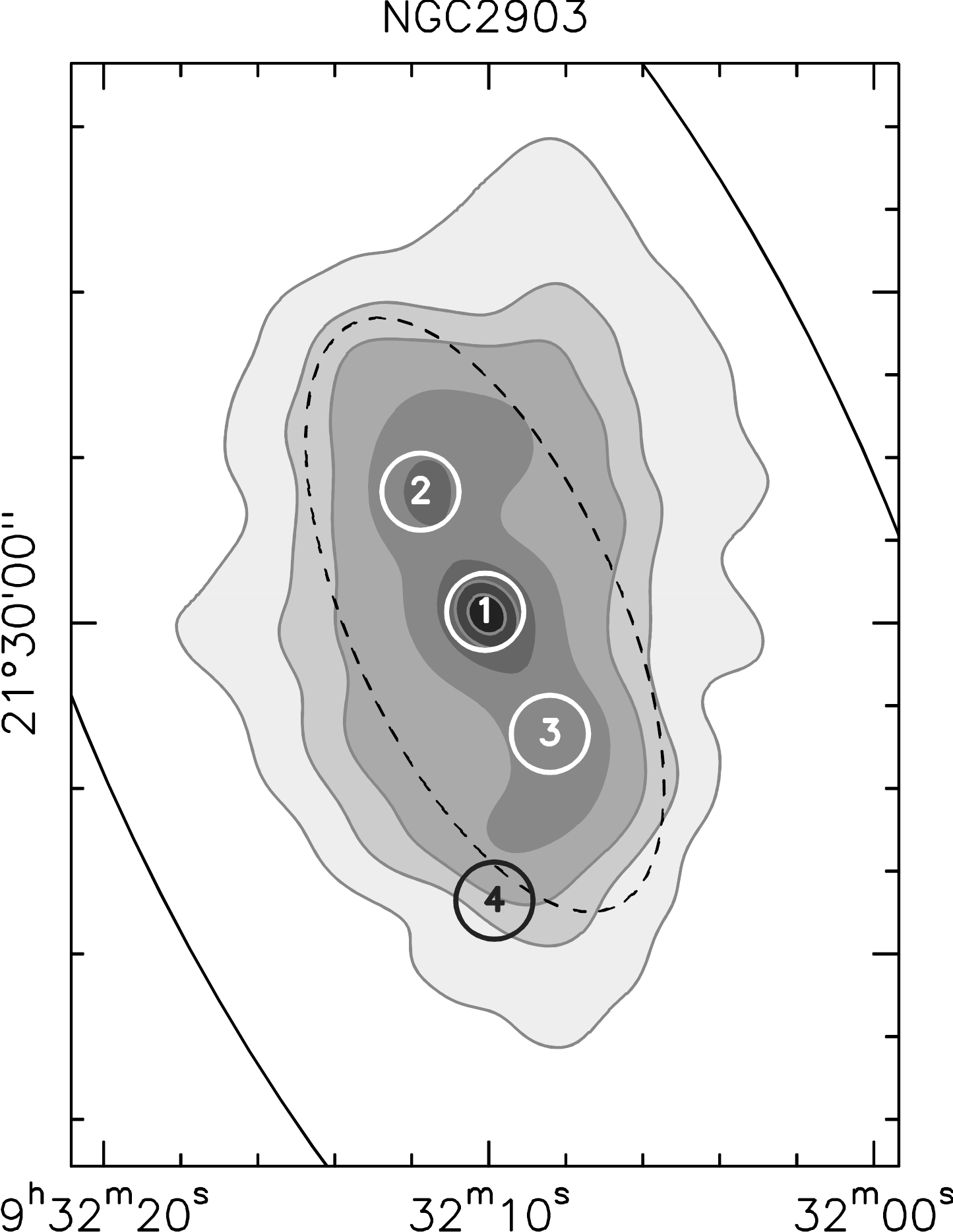}
\includegraphics[height=0.3\textheight]{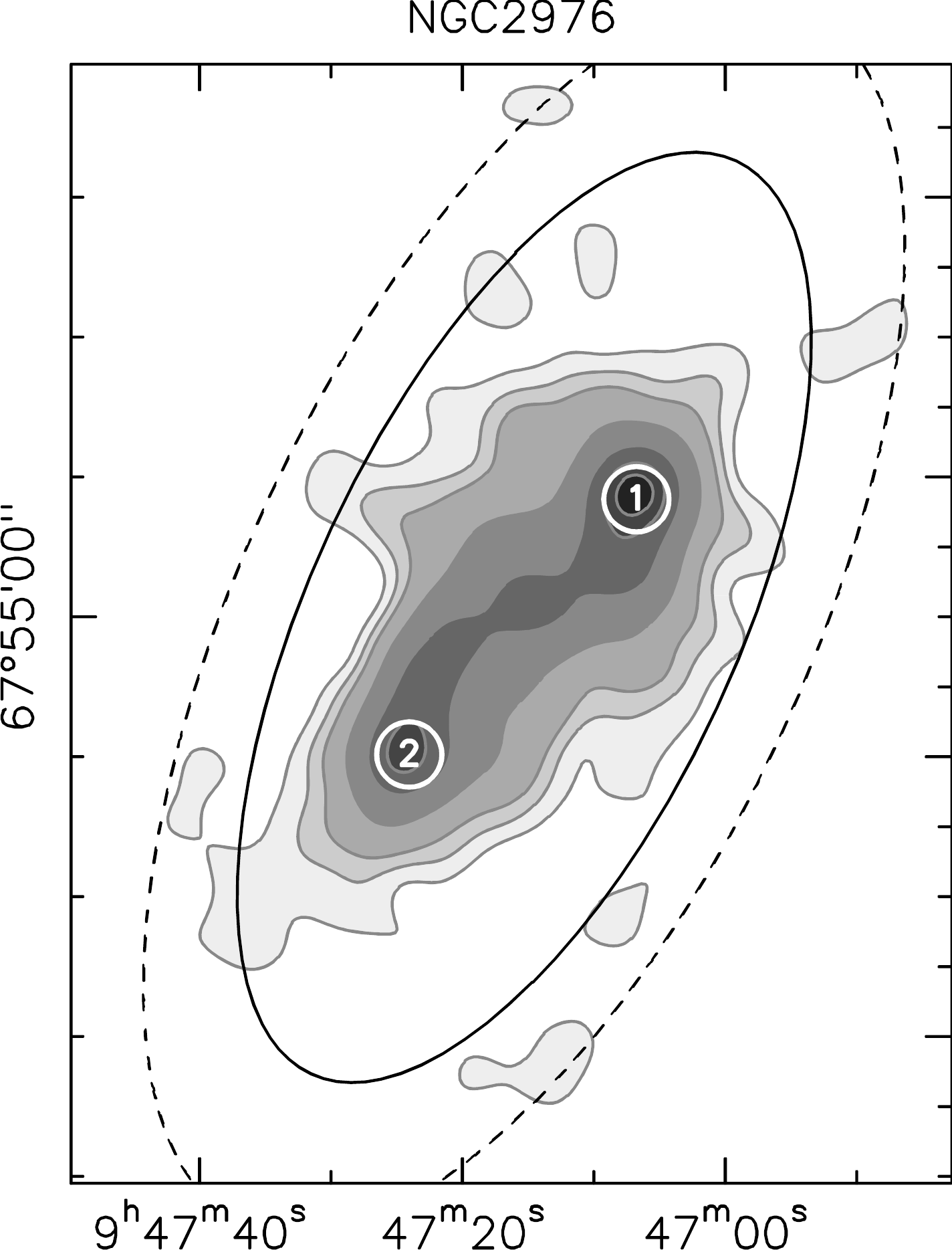}
\end{center}
\caption{Map of the HERACLES CO(2--1) velocity-integrated intensity in NGC~0628, NGC~2146, NGC~2403, NGC~2798, NGC~2903, and NGC~2976 at $28\arcsec$ resolution. These illustrative maps where obtained by a crude integration over all velocity channels, so they might not be accurate at low levels. The contours correspond to 1\%, 5\%, 10\%, 25\%, 50\%, 70\%, and 90\% of the maximum map value.  The observed positions are indicated by numbers surrounded by $28\arcsec$--wide circles. The solid and dashed ellipses correspond to $R_{25}$ and 5~kpc galactocentric radii, respectively.} 
\label{f-map-1}
\end{figure}

\begin{figure}
\begin{center}
\includegraphics[height=0.3\textheight]{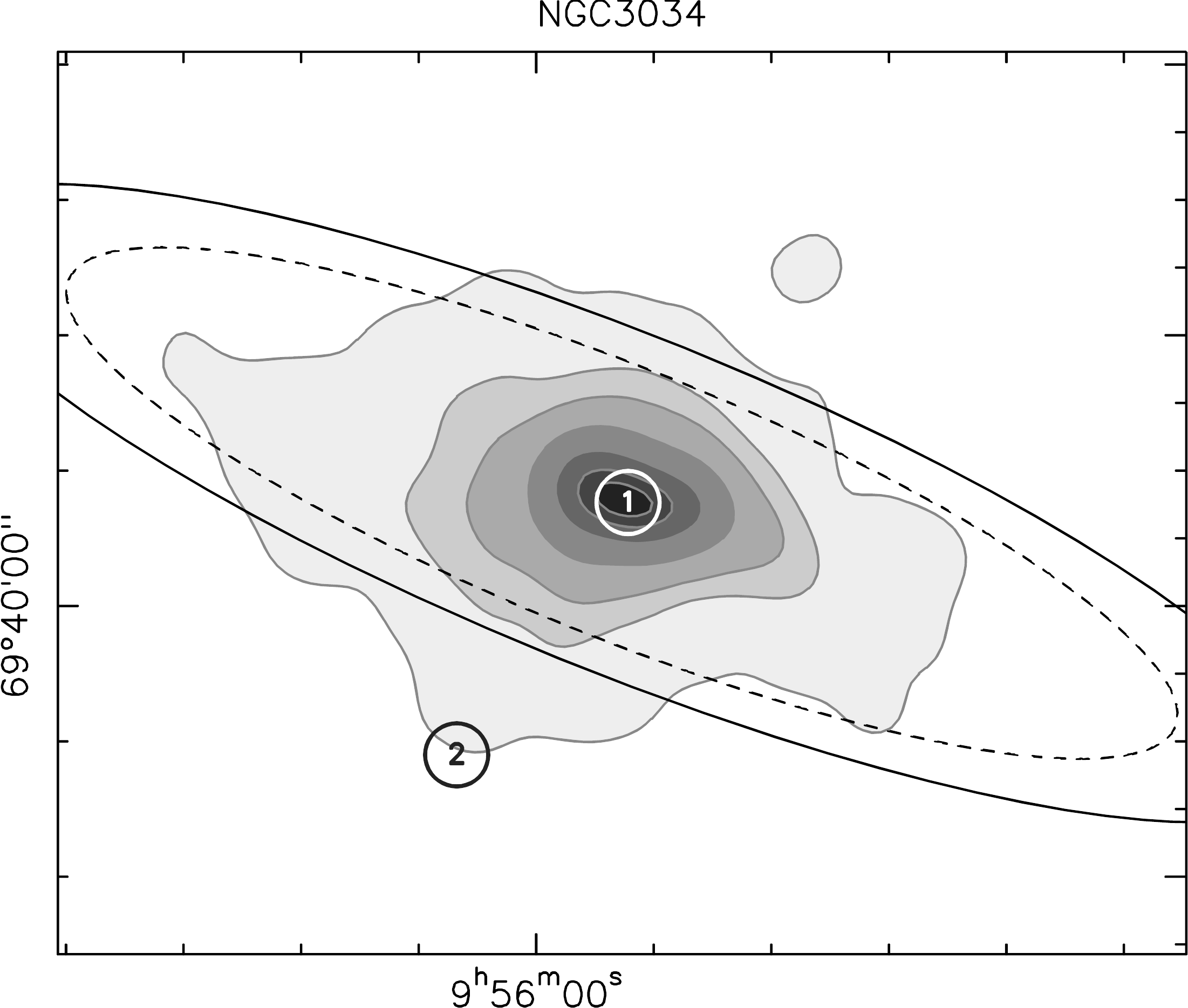}
\includegraphics[height=0.3\textheight]{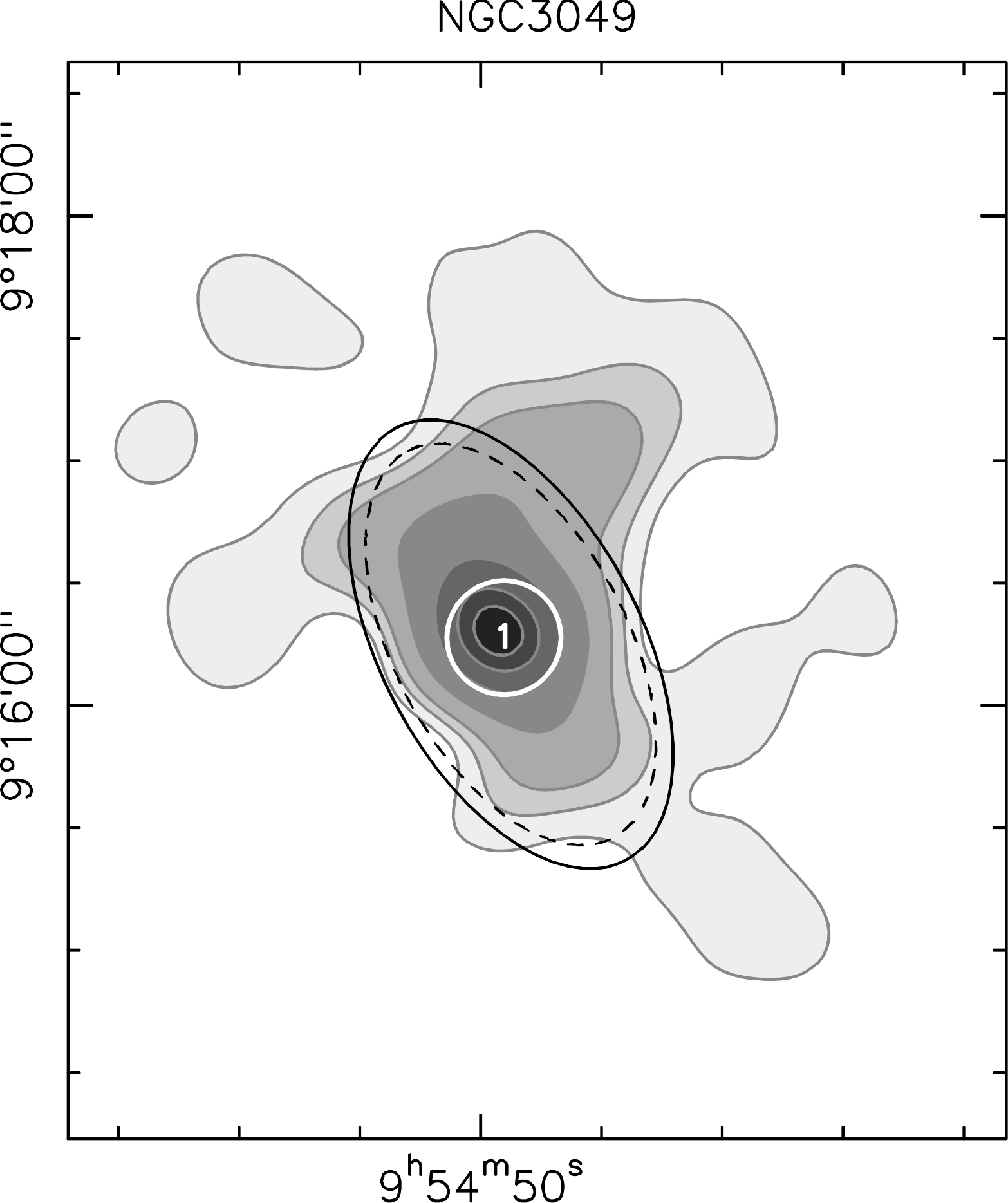}\\
\includegraphics[height=0.3\textheight]{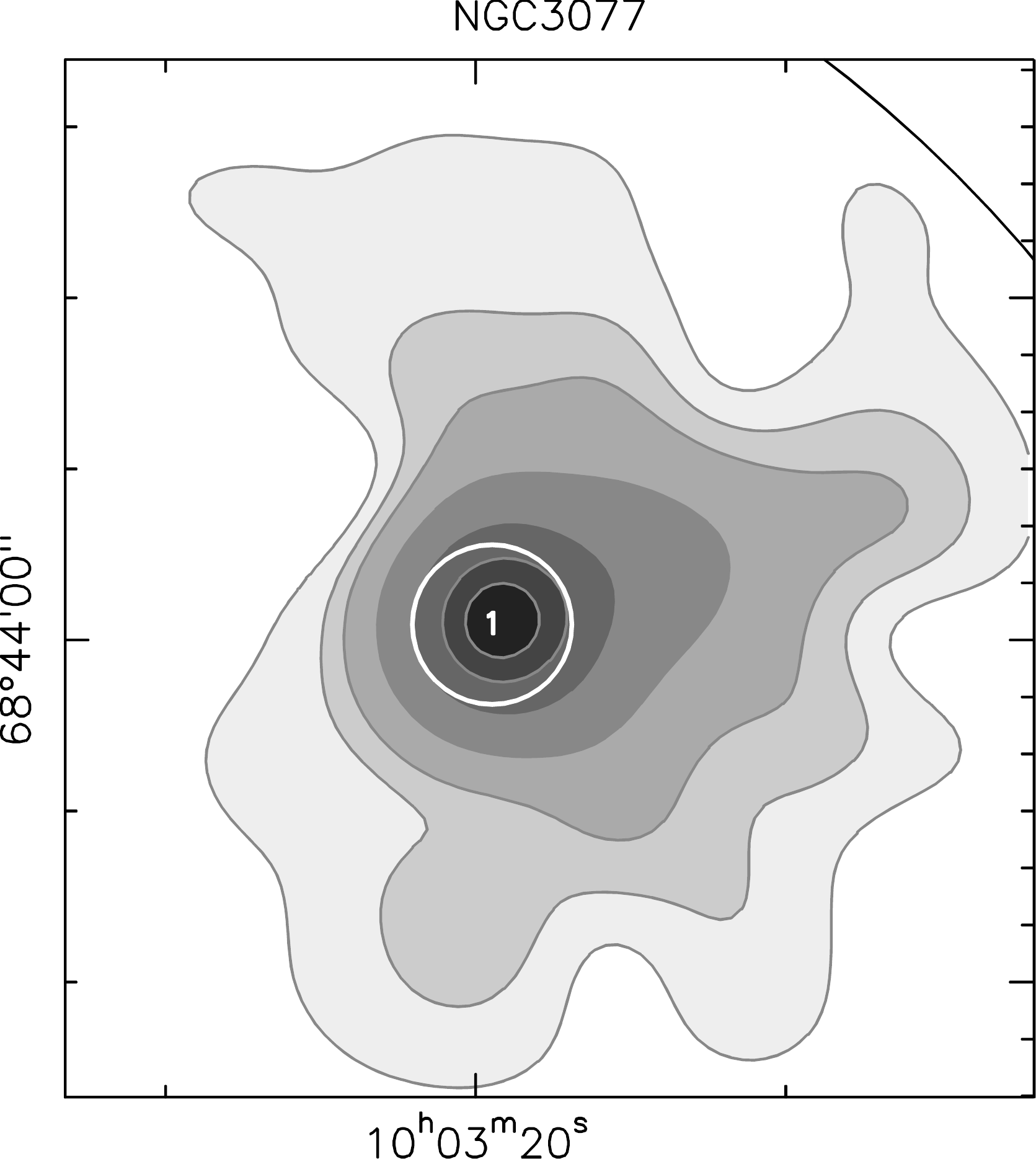}
\includegraphics[height=0.3\textheight]{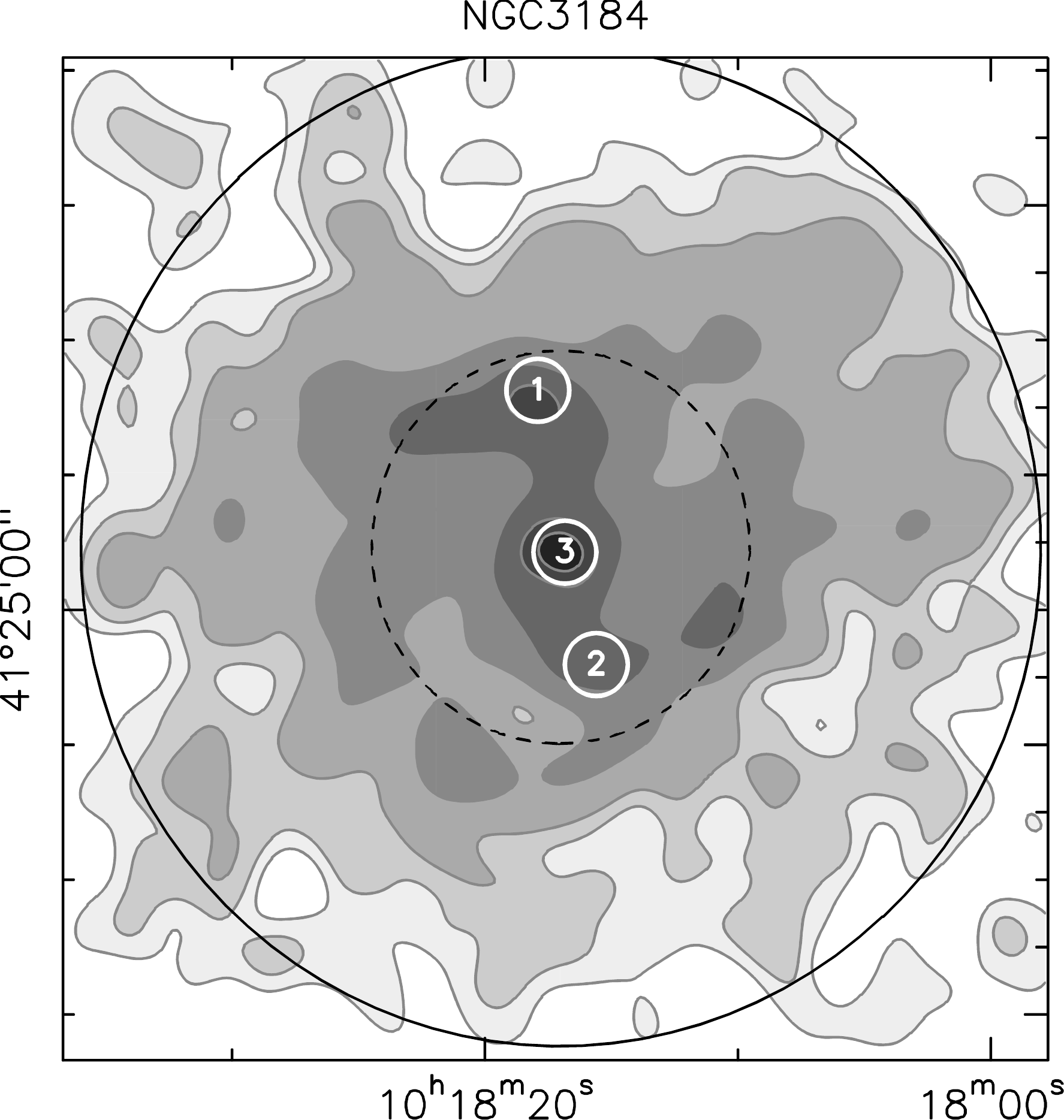}\\
\includegraphics[height=0.3\textheight]{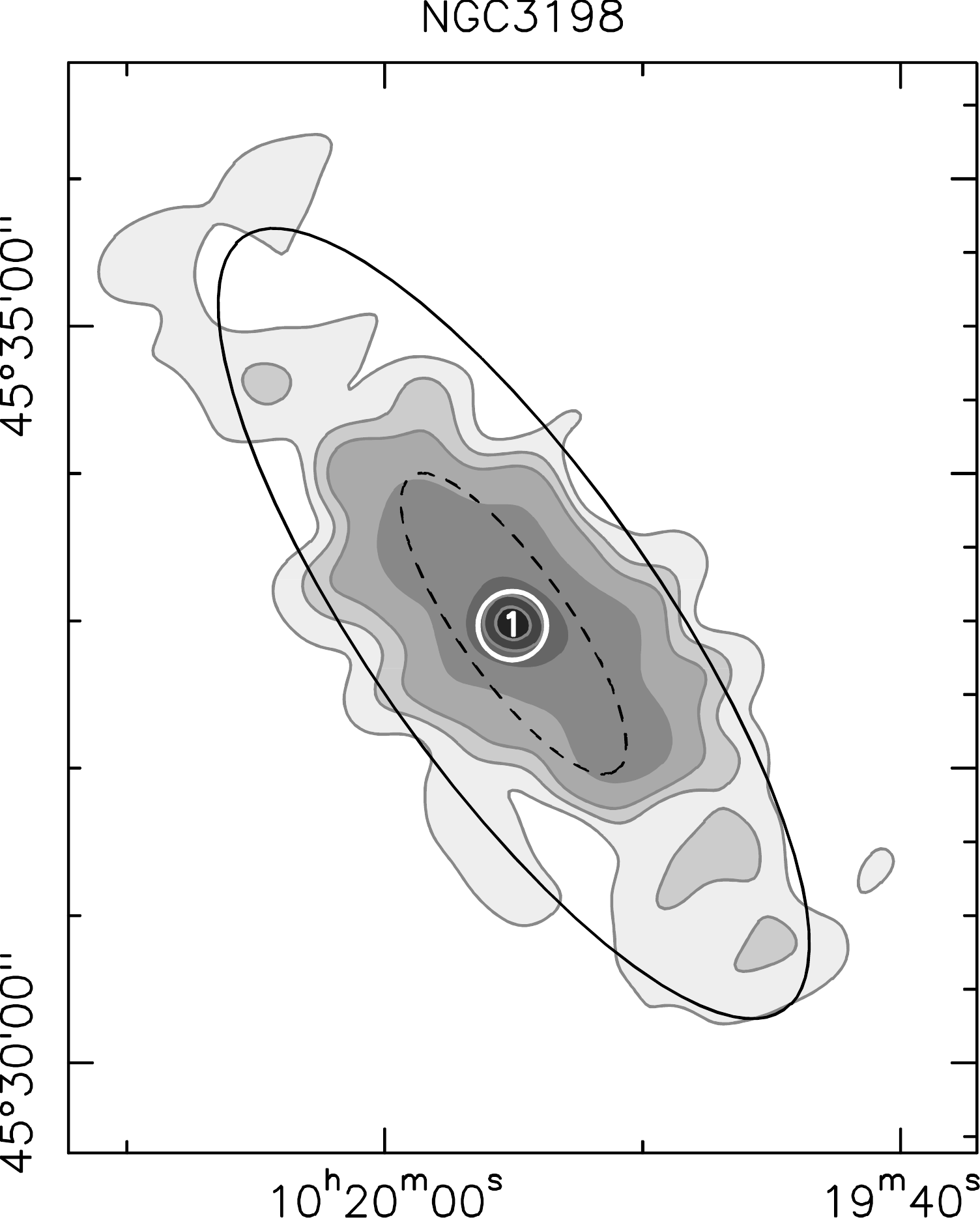}
\includegraphics[height=0.3\textheight]{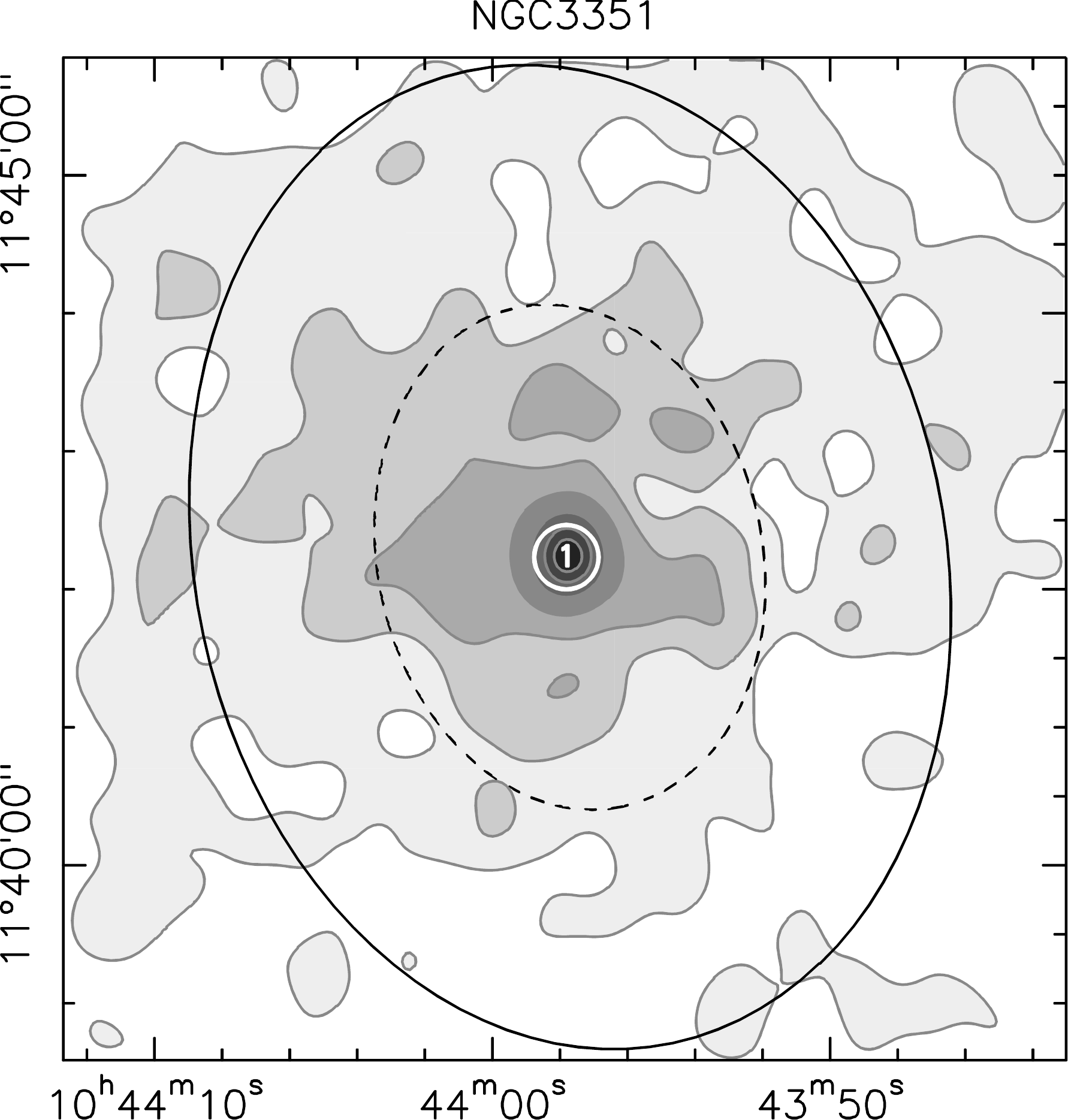}
\end{center}
\caption{Same as Fig.~\ref{f-map-1} for NGC~3034, NGC~3049, NGC~3077, NGC~3184, NGC~3198, and NGC~3351.} 
\label{f-map-2}
\end{figure}

\begin{figure}
\begin{center}
\includegraphics[height=0.3\textheight]{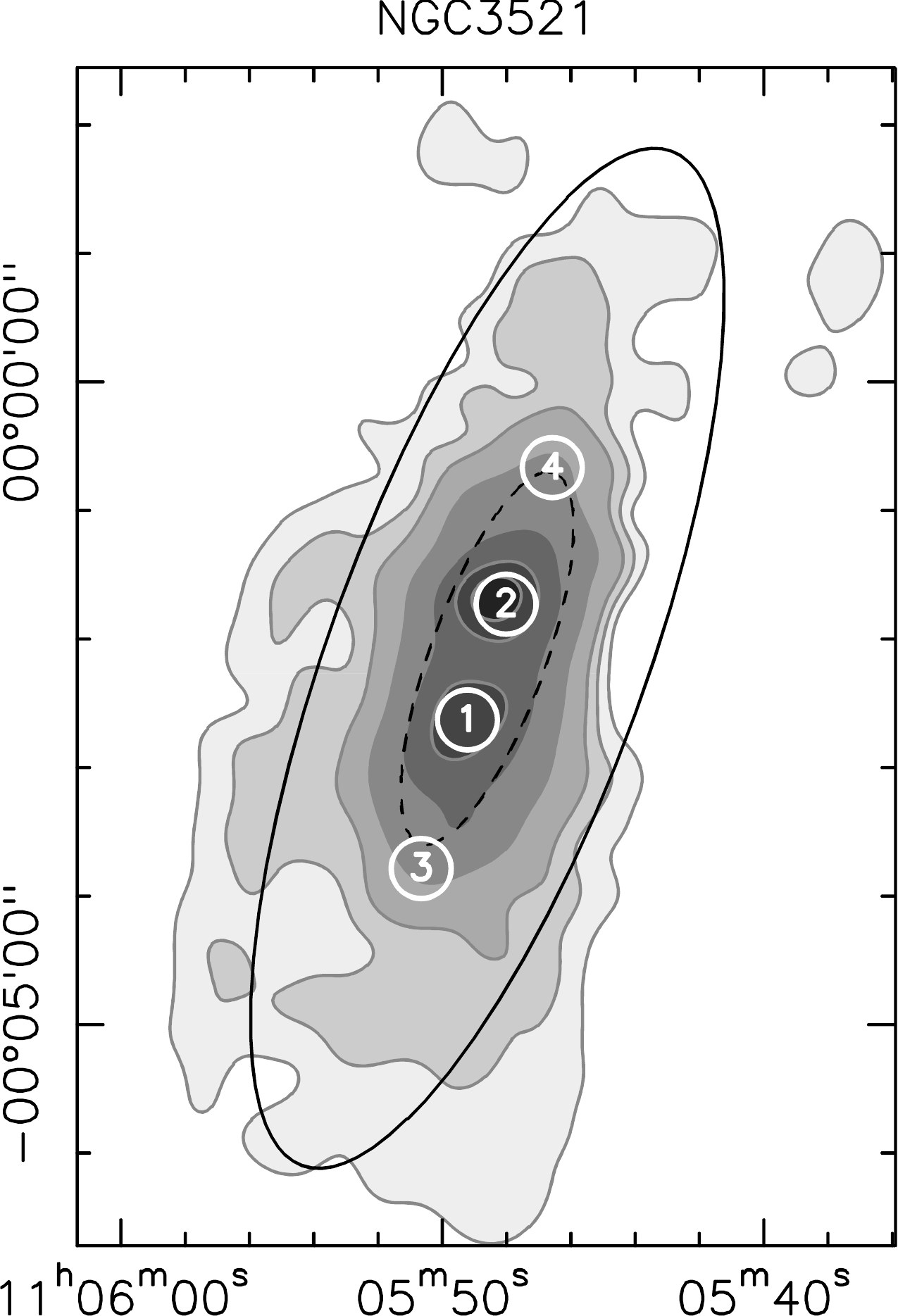}
\includegraphics[height=0.3\textheight]{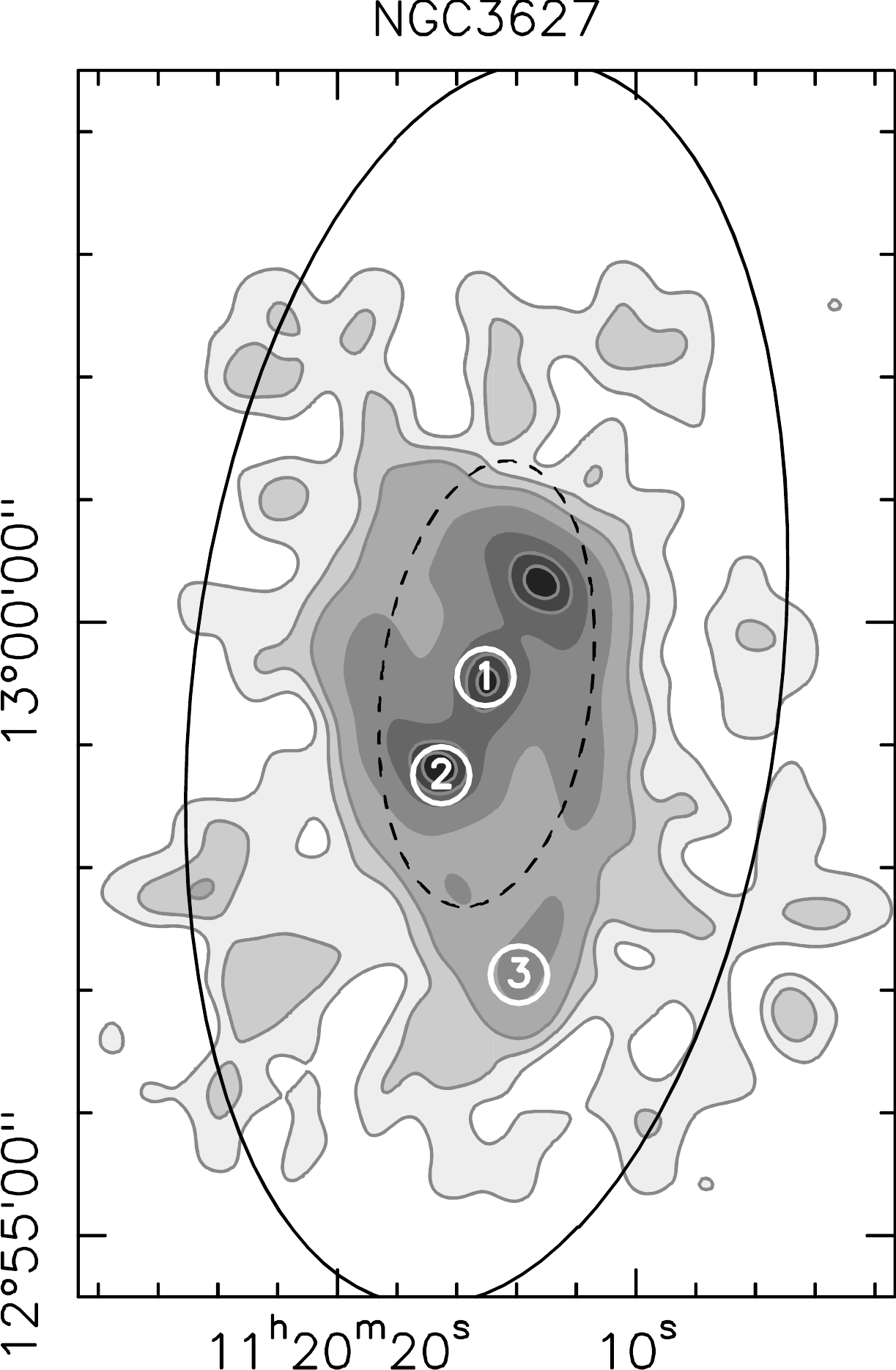}\\
\includegraphics[height=0.3\textheight]{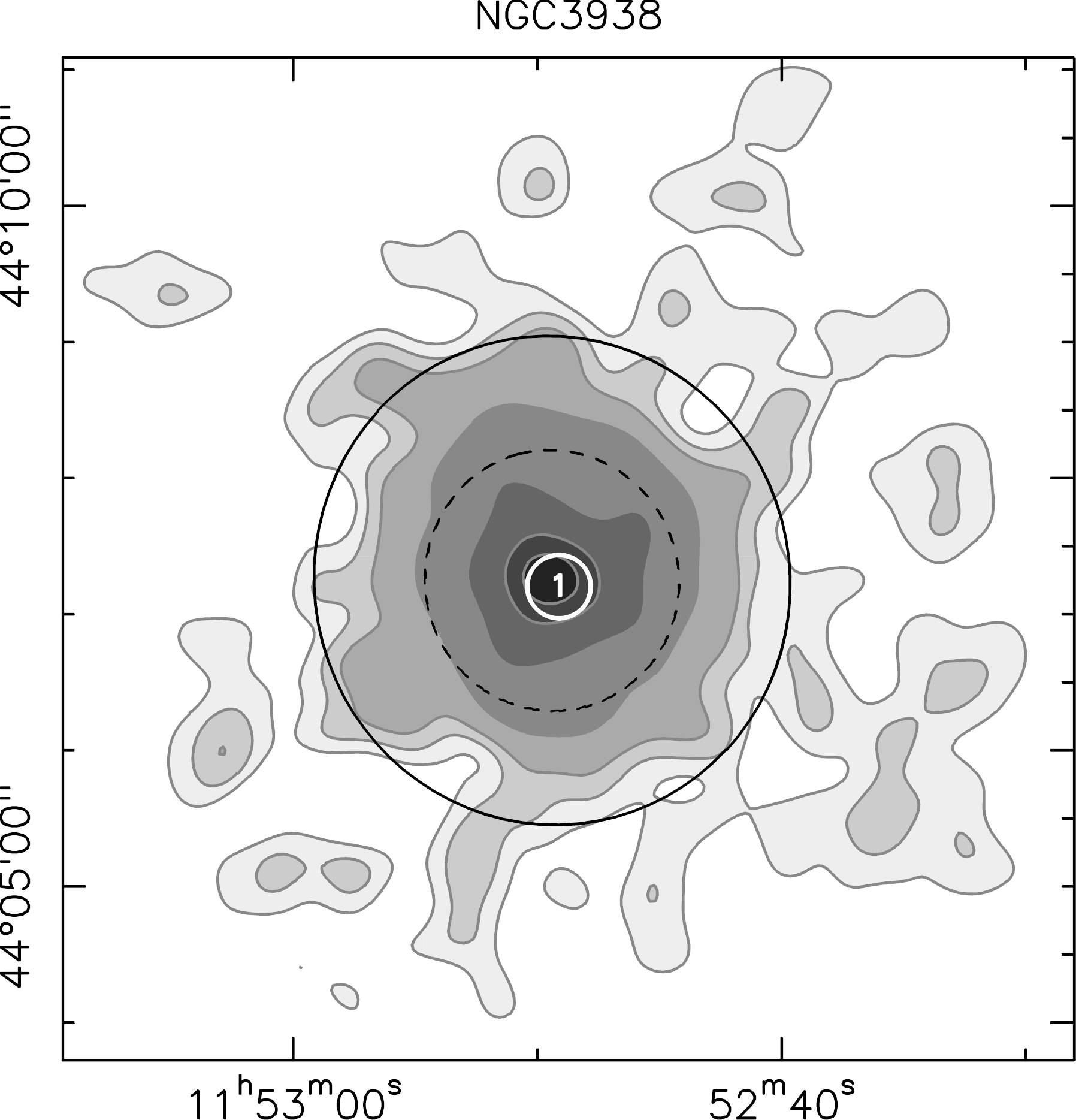}
\includegraphics[height=0.3\textheight]{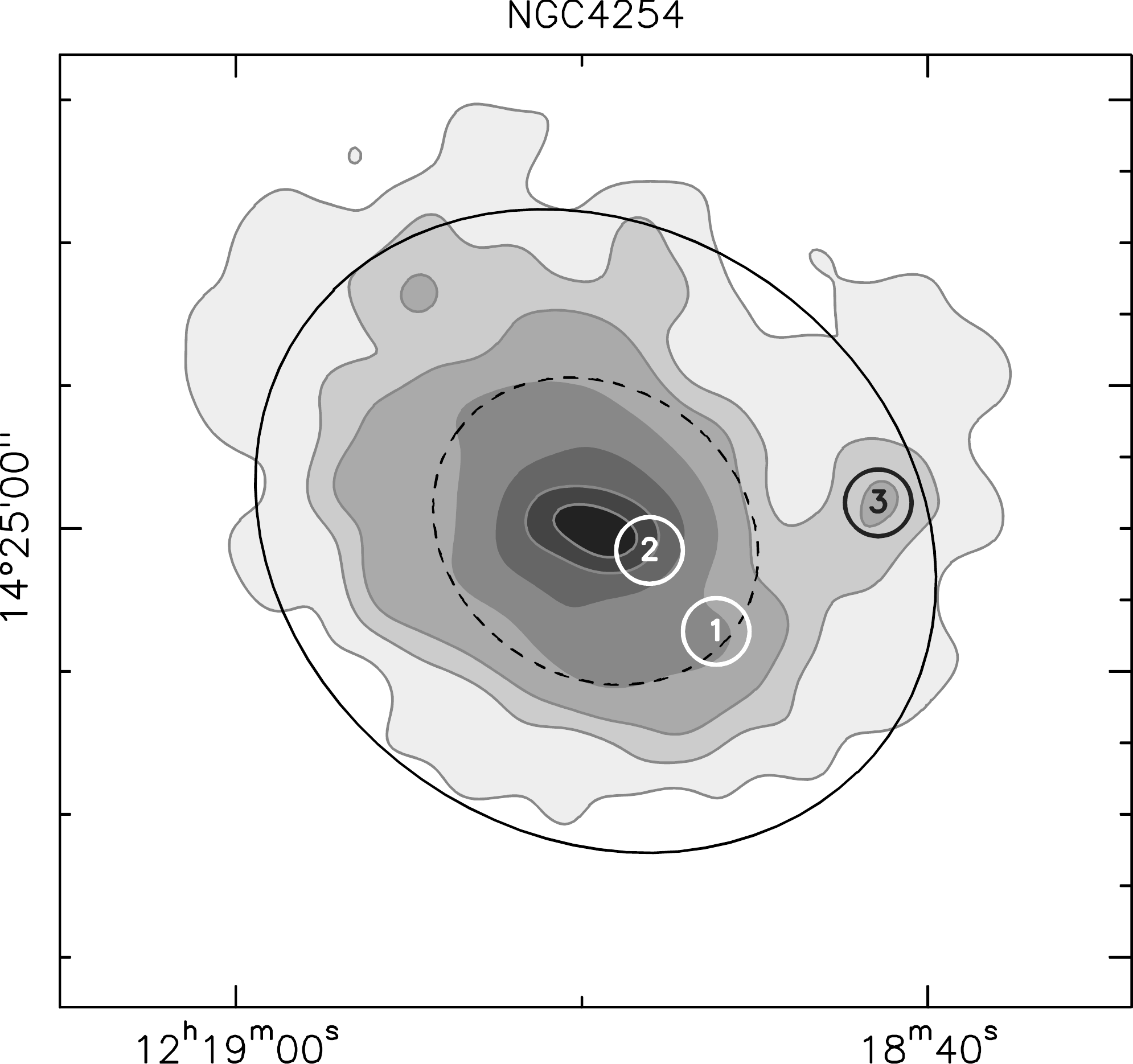}\\
\includegraphics[height=0.3\textheight]{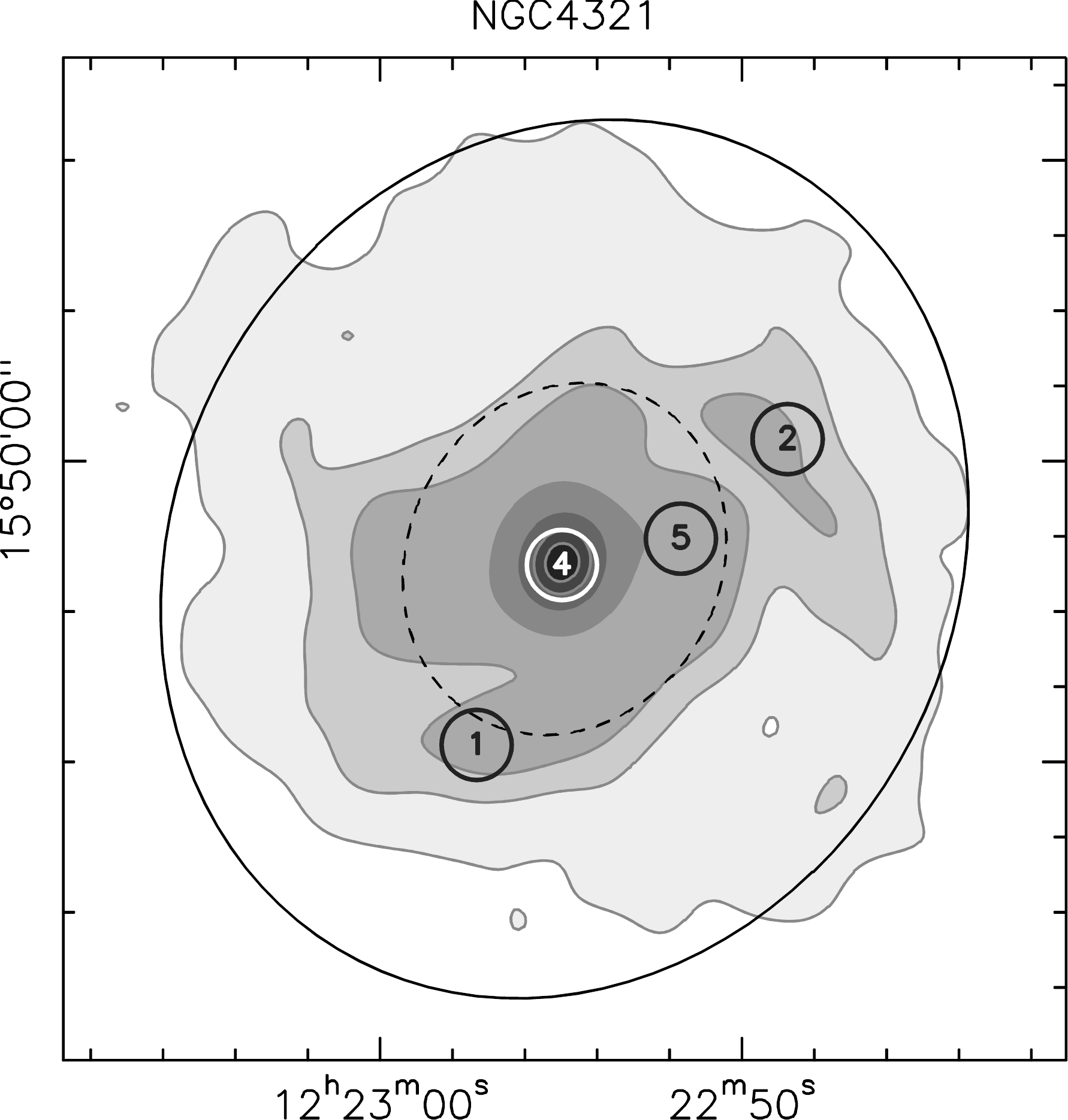}
\includegraphics[height=0.3\textheight]{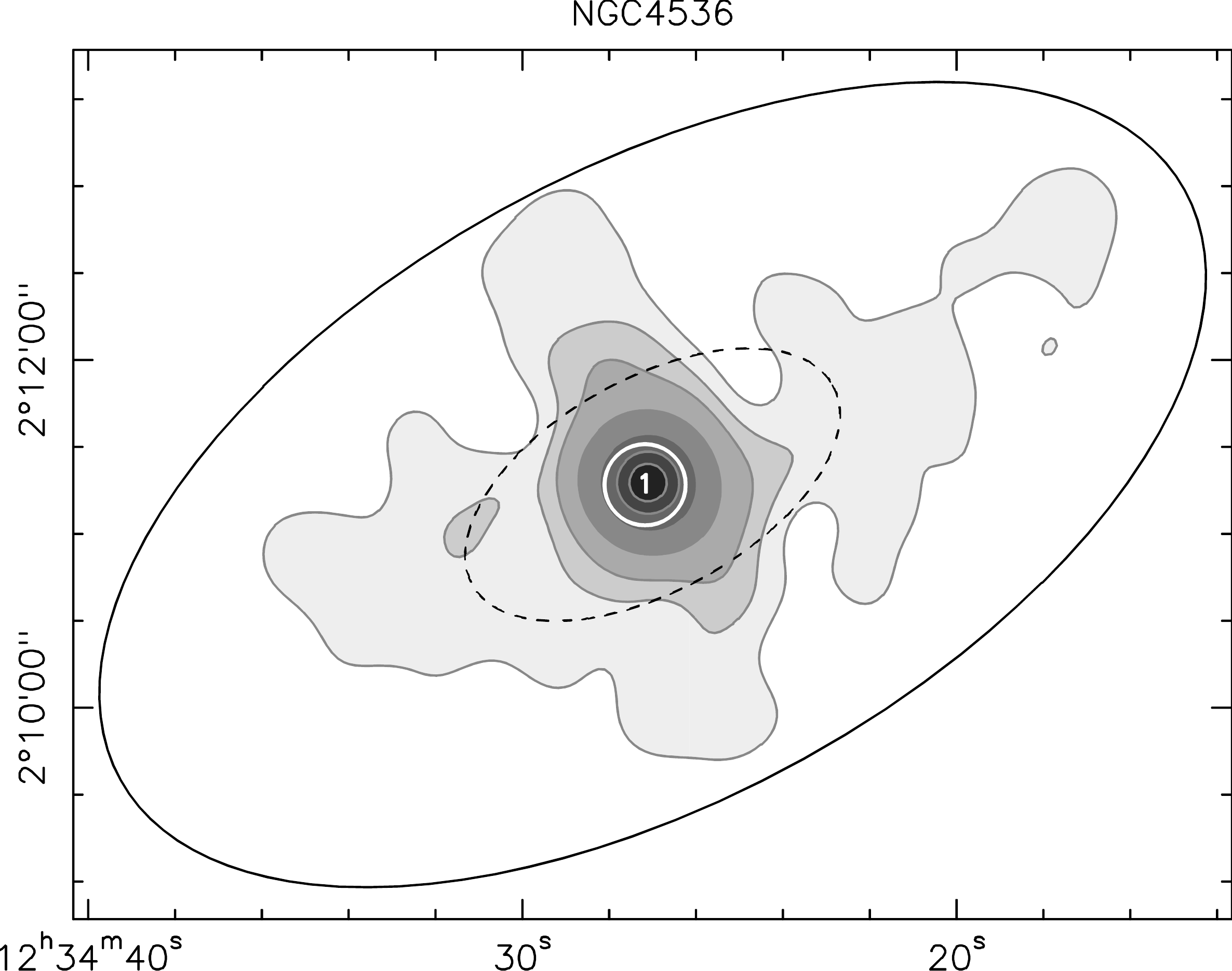}
\end{center}
\caption{Same as Fig.~\ref{f-map-1} for NGC~3521, NGC~3627, NGC~3938, NGC~4254, NGC~4321, and NGC~4536.} 
\label{f-map-3}
\end{figure}

\begin{figure}
\begin{center}
\includegraphics[height=0.3\textheight]{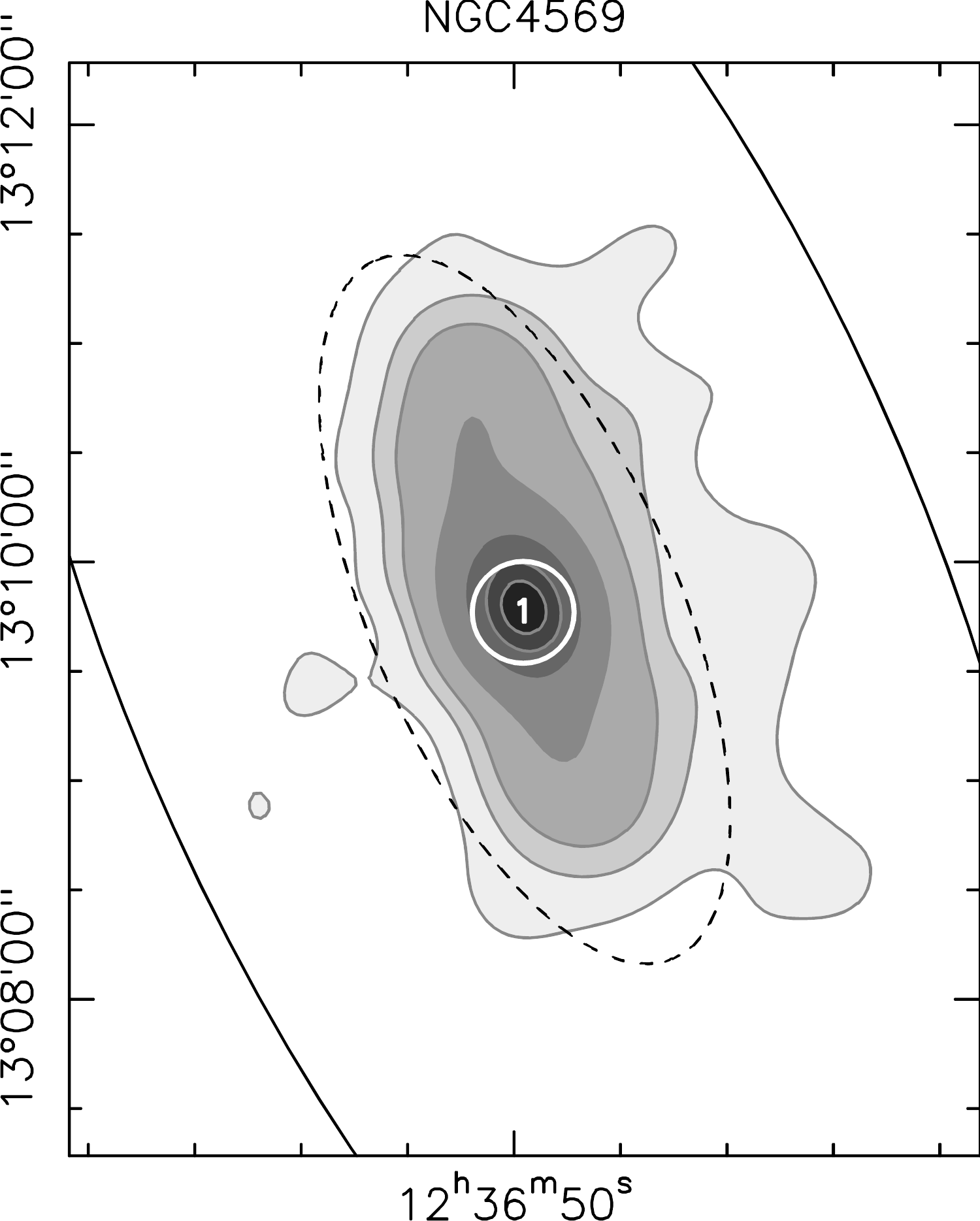}
\includegraphics[height=0.3\textheight]{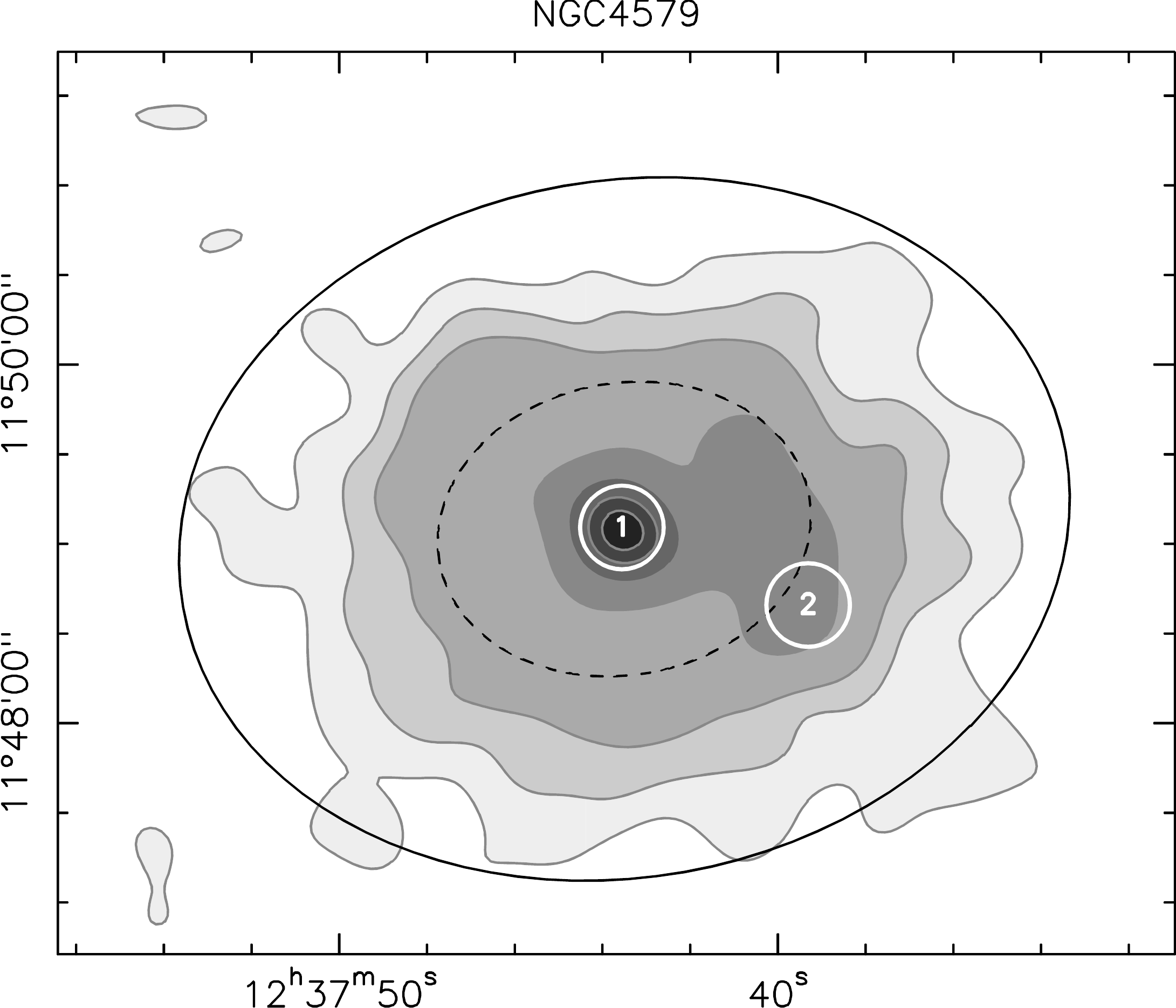}\\
\includegraphics[height=0.2\textheight]{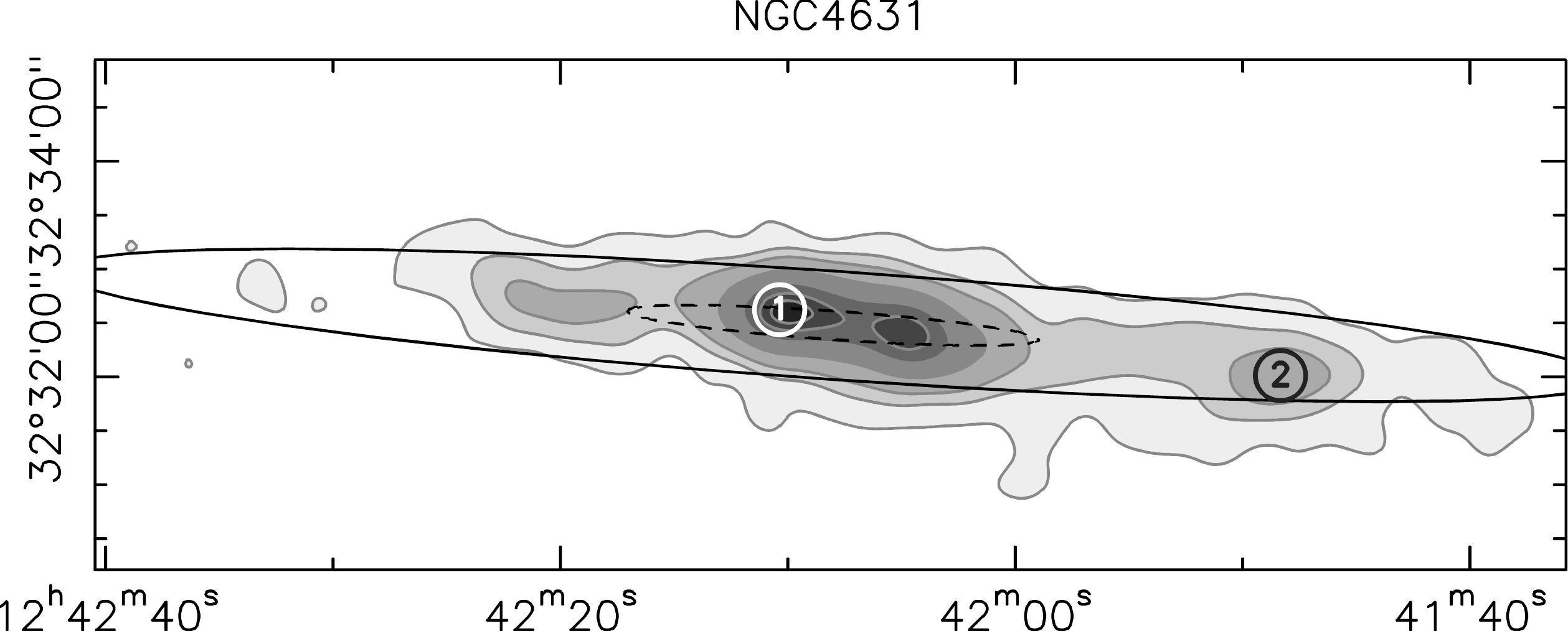}\\
\includegraphics[height=0.3\textheight]{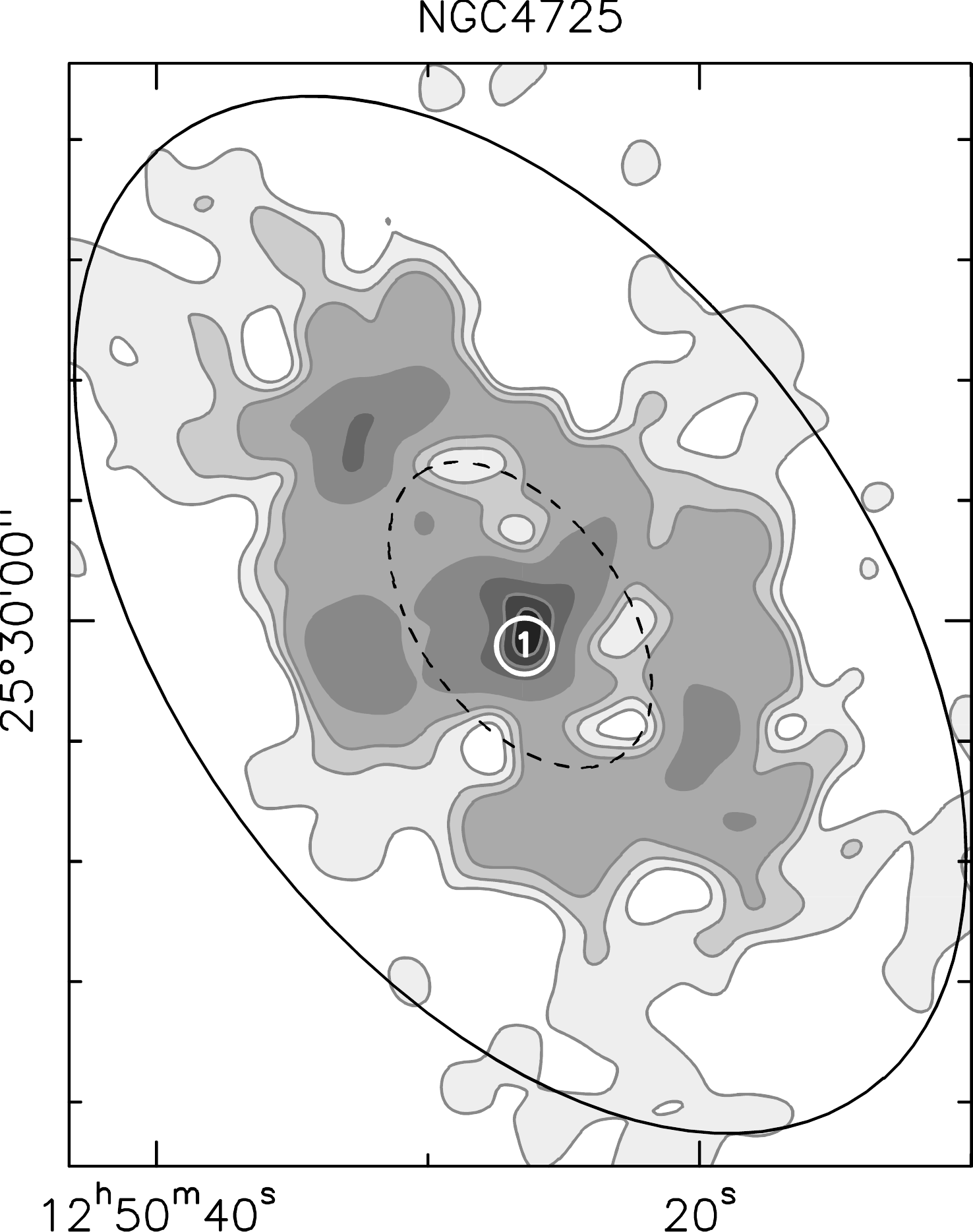}
\includegraphics[height=0.25\textheight]{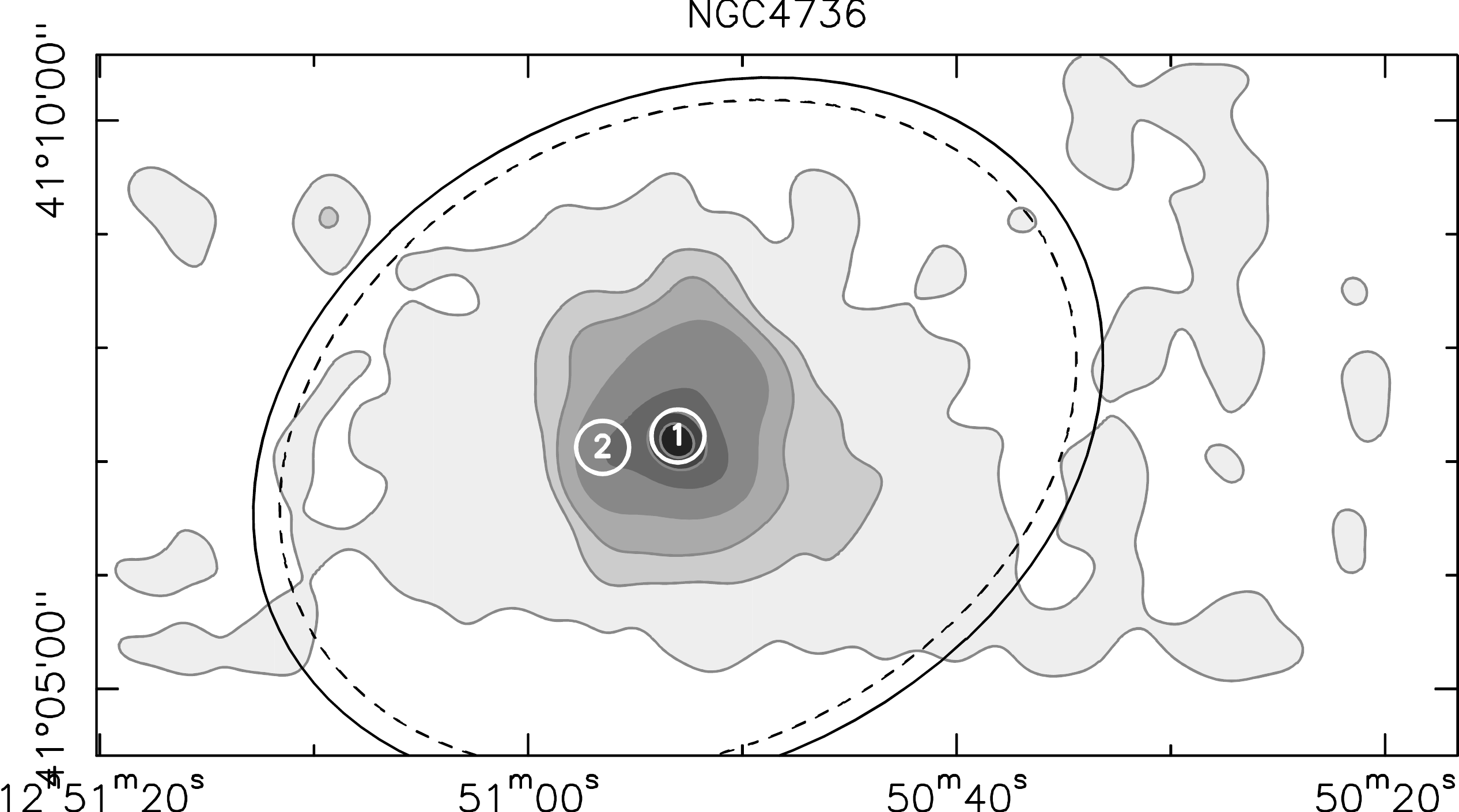}
\end{center}
\caption{Same as Fig.~\ref{f-map-1} for NGC~4569, NGC~4579, NGC~4631, NGC~4725, and NGC~4736.} 
\label{f-map-4}
\end{figure}

We summarize here the results of our observations. Figs.~\ref{f-map-1}--\ref{f-map-4} show the location of the observed positions on the HERACLES CO(2--1) maps of the target galaxies. The HCN(1--0) and CO(1--0) final spectra are shown in Figs.~\ref{f-spec-1}--\ref{f-spec-14}. The velocity-integrated intensities of the two lines are listed in Table~\ref{t-pos}.

\begin{table}
\begin{center}
\begin{threeparttable}
\caption{\label{t-pos} HCN(1--0) and CO(1--0) velocity-integrated intensities}
                                                                                                        
\begin{tabular}{rrrrrrr}                                                                                    
\hline\hline                                                                                               
\noalign{\smallskip}                                                                                       
Name & off & $r_{25}$ & \multicolumn{2}{c}{$I_\mathrm{HCN}$} & \multicolumn{2}{c}{$I_\mathrm{CO10}$} \\
NGC &  &  & \multicolumn{2}{c}{(K~km~s$^{-1}$)} & \multicolumn{2}{c}{(K~km~s$^{-1}$)} \\                 
\noalign{\smallskip}                                                                                       
\hline                                                                                                      
\noalign{\smallskip}                                                                                       
0628 & 1 & 0.00 &    0.17 & (0.03) &    7.5 & (0.32)\\
     & 2 & 0.27 &    0.09 & (0.03) &    5.8 & (0.26)\\
2146 & 1 & 0.02 &    4.44 & (0.13) &  101.2 & (1.45)\\
     & 2 & 0.42 & $<$0.24 &    ... &   12.1 & (0.95)\\
2403 & 1 & 0.06 & $<$0.12 &    ... &    7.5 & (1.58)\\
2798 & 1 & 0.20 &    1.23 & (0.11) &   31.3 & (1.54)\\
2903 & 1 & 0.01 &    3.13 & (0.10) &   54.1 & (2.01)\\
     & 2 & 0.14 &    0.77 & (0.06) &   40.0 & (1.67)\\
     & 3 & 0.14 &    0.52 & (0.04) &   32.0 & (1.79)\\
     & 4 & 0.37 & $<$0.12 &    ... &    5.5 & (1.11)\\
2976 & 1 & 0.39 &    0.15 & (0.03) &    9.6 & (0.22)\\
     & 2 & 0.40 &    0.10 & (0.03) &    7.3 & (0.24)\\
3034 & 1 & 0.02 &   29.55 & (0.18) &  507.1 & (1.51)\\
     & 2 & 1.76 & $<$0.16 &    ... &    9.0 & (0.98)\\
3049 & 1 & 0.04 &    0.11 & (0.03) & $<$4.3 &    ...\\
3077 & 1 & 0.01 & $<$0.12 &    ... &    4.9 & (0.78)\\
3184 & 1 & 0.32 & $<$0.08 &    ... &    6.3 & (0.12)\\
     & 2 & 0.25 & $<$0.16 &    ... &    5.2 & (0.53)\\
     & 3 & 0.01 &    0.19 & (0.03) &    9.2 & (0.12)\\
3198 & 1 & 0.01 &    0.19 & (0.04) &   12.3 & (0.69)\\
3351 & 1 & 0.01 &    2.43 & (0.09) &   42.8 & (1.13)\\
3521 & 1 & 0.12 &    0.59 & (0.08) &   35.2 & (0.81)\\
     & 2 & 0.11 &    0.76 & (0.11) &   39.0 & (0.79)\\
     & 3 & 0.41 &    0.15 & (0.03) &   12.3 & (0.41)\\
     & 4 & 0.38 &    0.28 & (0.05) &   13.7 & (0.42)\\
3627 & 1 & 0.01 &    2.42 & (0.11) &   58.2 & (0.78)\\
     & 2 & 0.19 &    1.59 & (0.08) &   53.8 & (0.64)\\
     & 3 & 0.51 &    0.30 & (0.04) &   14.3 & (0.46)\\
3938 & 1 & 0.04 &    0.28 & (0.05) &    7.6 & (0.45)\\
4254 & 1 & 0.44 &    0.21 & (0.03) &    9.4 & (0.28)\\
     & 2 & 0.16 &    0.77 & (0.06) &   24.0 & (0.40)\\
     & 3 & 0.86 & $<$0.12 &    ... &    4.0 & (0.20)\\
4321 & 1 & 0.45 &    0.11 & (0.03) &    9.6 & (0.33)\\
     & 2 & 0.59 &    0.11 & (0.03) &    5.2 & (0.23)\\
     & 4 & 0.01 &    2.95 & (0.11) &   55.9 & (0.73)\\
     & 5 & 0.29 &    0.41 & (0.04) &   12.3 & (0.36)\\
4536 & 1 & 0.02 &    1.45 & (0.10) &   57.4 & (0.85)\\
4569 & 1 & 0.01 &    2.57 & (0.13) &   73.9 & (0.79)\\
4579 & 1 & 0.01 &    1.20 & (0.07) &   18.6 & (0.74)\\
     & 2 & 0.48 & $<$0.16 &    ... &    6.9 & (0.49)\\
4631 & 1 & 0.24 &    0.84 & (0.10) &   47.3 & (0.64)\\
     & 2 & 0.67 & $<$0.12 &    ... &    8.1 & (0.39)\\
4725 & 1 & 0.06 & $<$0.24 &    ... &    6.1 & (0.62)\\
4736 & 1 & 0.00 &    1.16 & (0.13) &   35.1 & (0.74)\\
     & 2 & 0.18 &    0.31 & (0.09) &   16.2 & (0.45)\\
5055 & 1 & 0.00 &    2.21 & (0.11) &   67.4 & (0.72)\\
     & 2 & 0.18 &    0.33 & (0.06) &   16.9 & (0.44)\\
     & 3 & 0.32 &    0.22 & (0.04) &   15.7 & (0.34)\\
     & 4 & 0.42 & $<$0.12 &    ... &    8.5 & (0.31)\\
     & 5 & 0.48 & $<$0.12 &    ... &   12.3 & (0.36)\\
5194 & 1 & 0.15 &    2.41 & (0.09) &    ... & ...\\
     & 2 & 0.50 &    0.63 & (0.06) &    ... & ...\\
     & 3 & 0.00 &    4.88 & (0.10) &    ... & ...\\
5457 & 1 & 0.00 &    0.64 & (0.03) &    ... & ...\\
5713 & 1 & 0.11 &    1.22 & (0.06) &   39.2 & (0.64)\\
6946 & 1 & 0.00 &    9.57 & (0.22) &  224.3 & (0.87)\\
     & 2 & 0.42 &    0.47 & (0.05) &   23.5 & (0.22)\\
     & 3 & 0.73 & $<$0.15 &    ... &    2.5 & (0.20)\\
     & 4 & 0.47 &    0.58 & (0.05) &   23.7 & (0.24)\\
     & 5 & 0.75 &    0.17 & (0.02) &    9.6 & (0.23)\\
7331 & 1 & 0.15 &    1.16 & (0.11) &   40.5 & (0.35)\\
     & 2 & 0.13 &    0.93 & (0.10) &   39.9 & (0.50)\\
\noalign{\smallskip}
\hline               
\end{tabular}

\begin{tablenotes}
\item{\sc Note. ---}  Errors at $1\sigma$--level are written within brackets.  We give $4\sigma$ values for non-detections. We omit position 3 in NGC~4321, which was only observed in CO(1--0).
\end{tablenotes}
\end{threeparttable}
\end{center}
\end{table}

\begin{figure}
\begin{center}
\includegraphics[height=0.2\textheight]{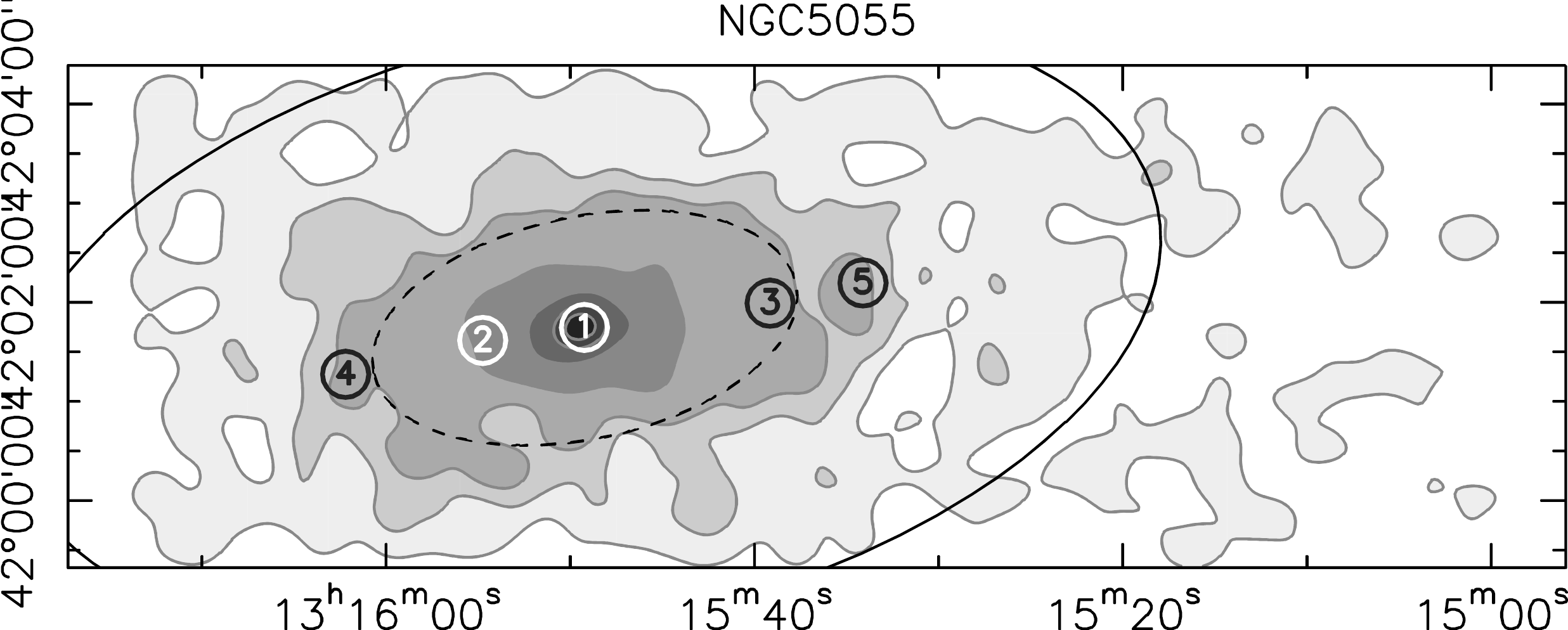}
\includegraphics[height=0.3\textheight]{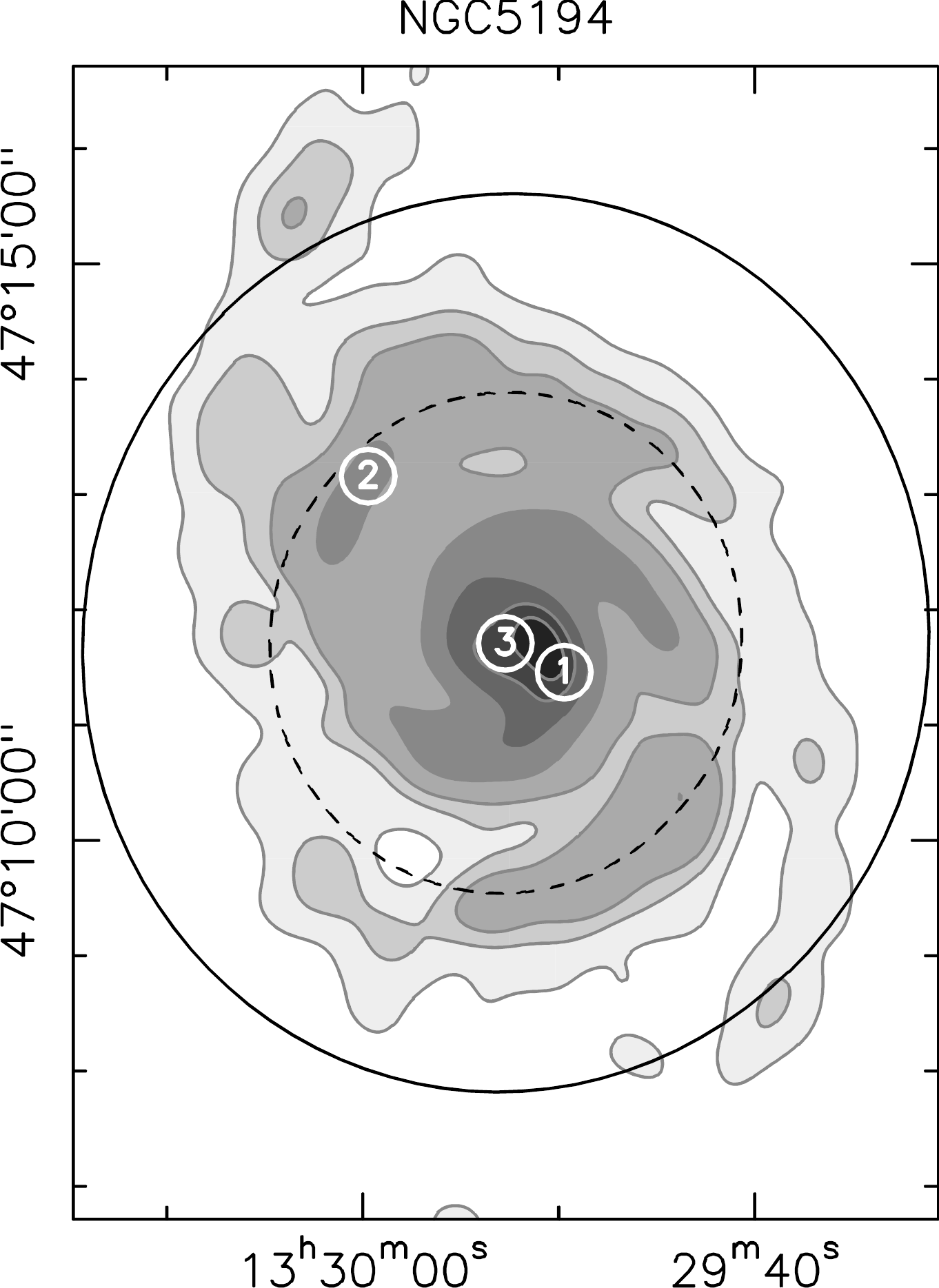}\\
\includegraphics[height=0.3\textheight]{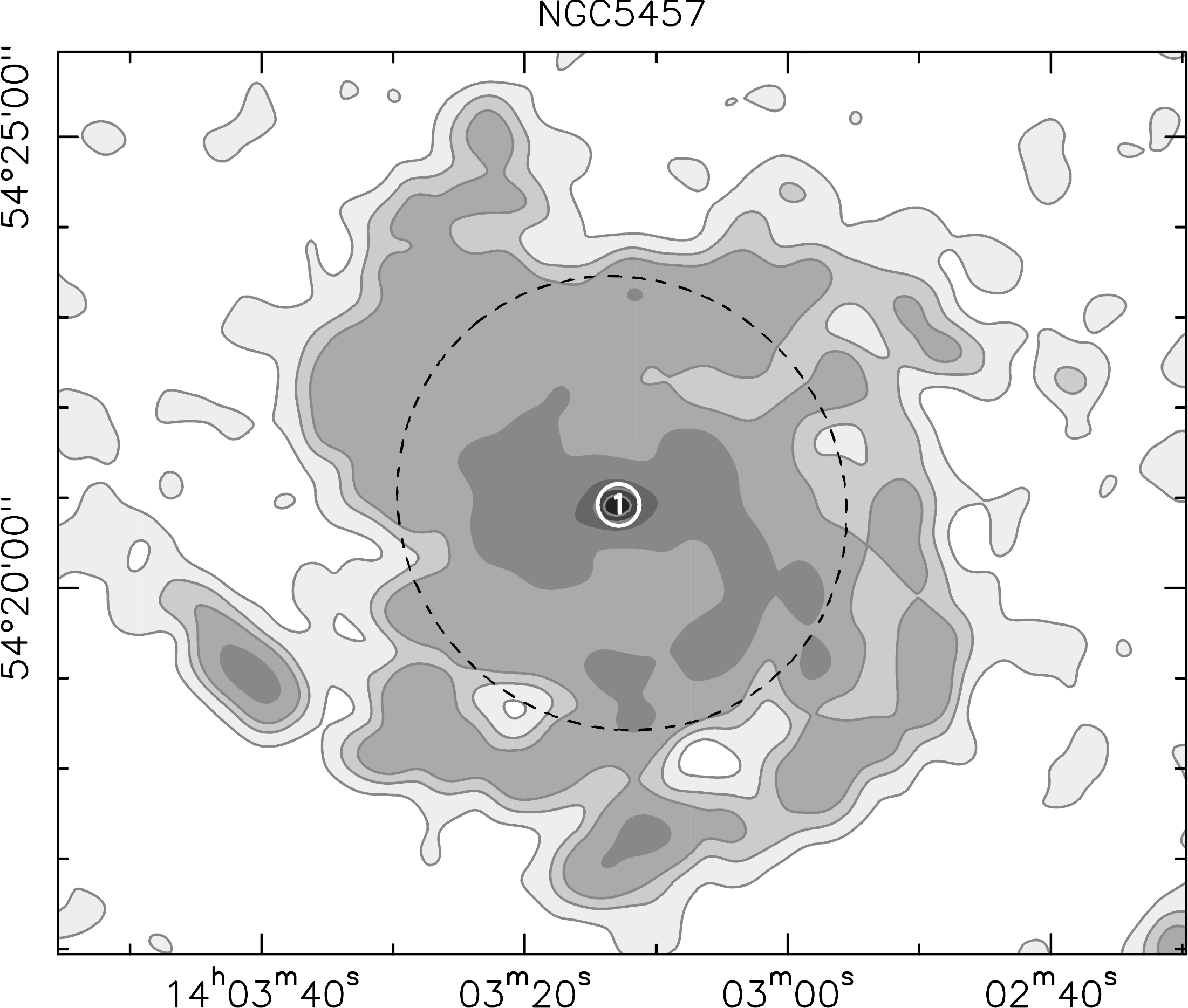}
\includegraphics[height=0.3\textheight]{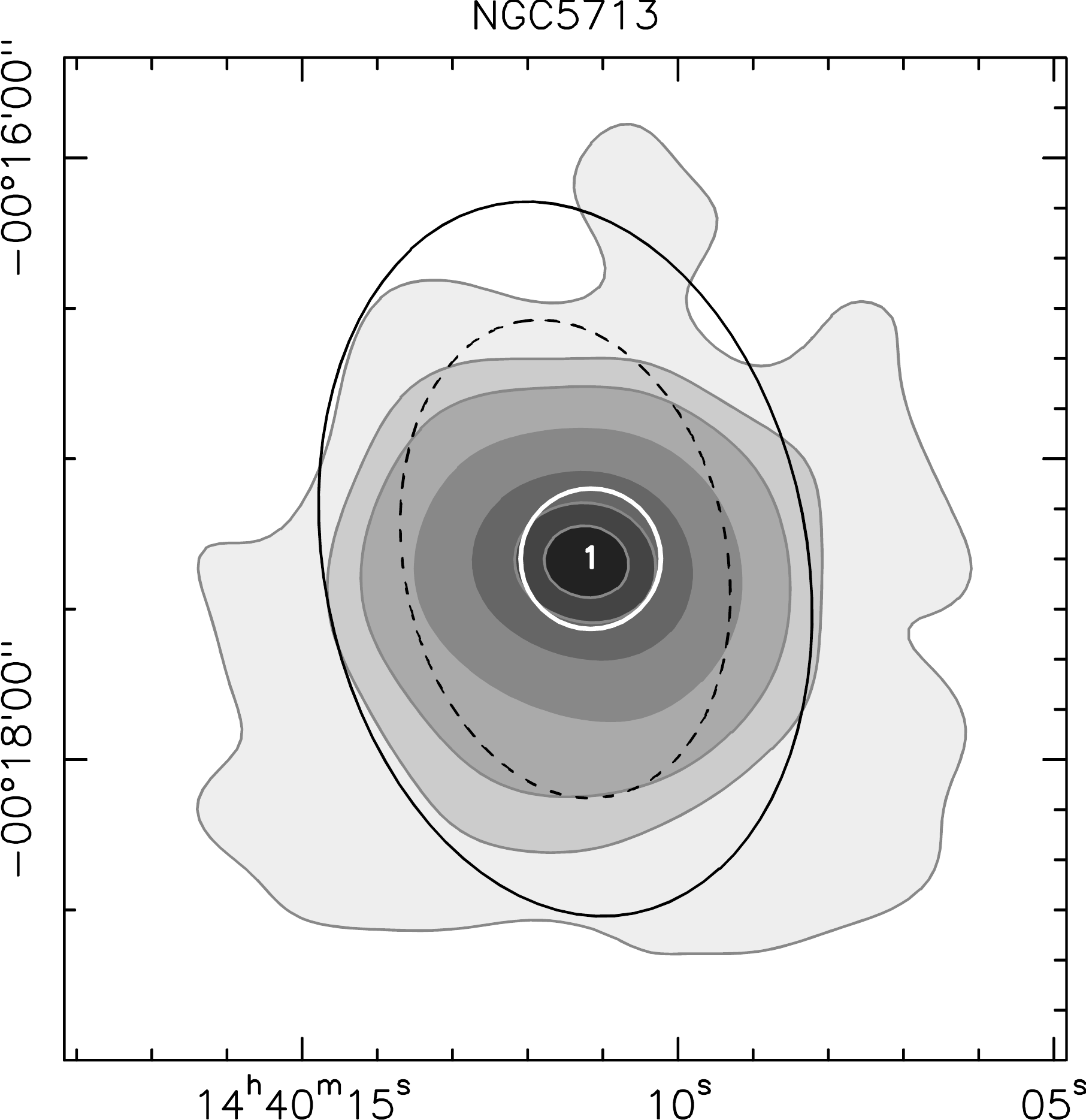}\\
\includegraphics[height=0.3\textheight]{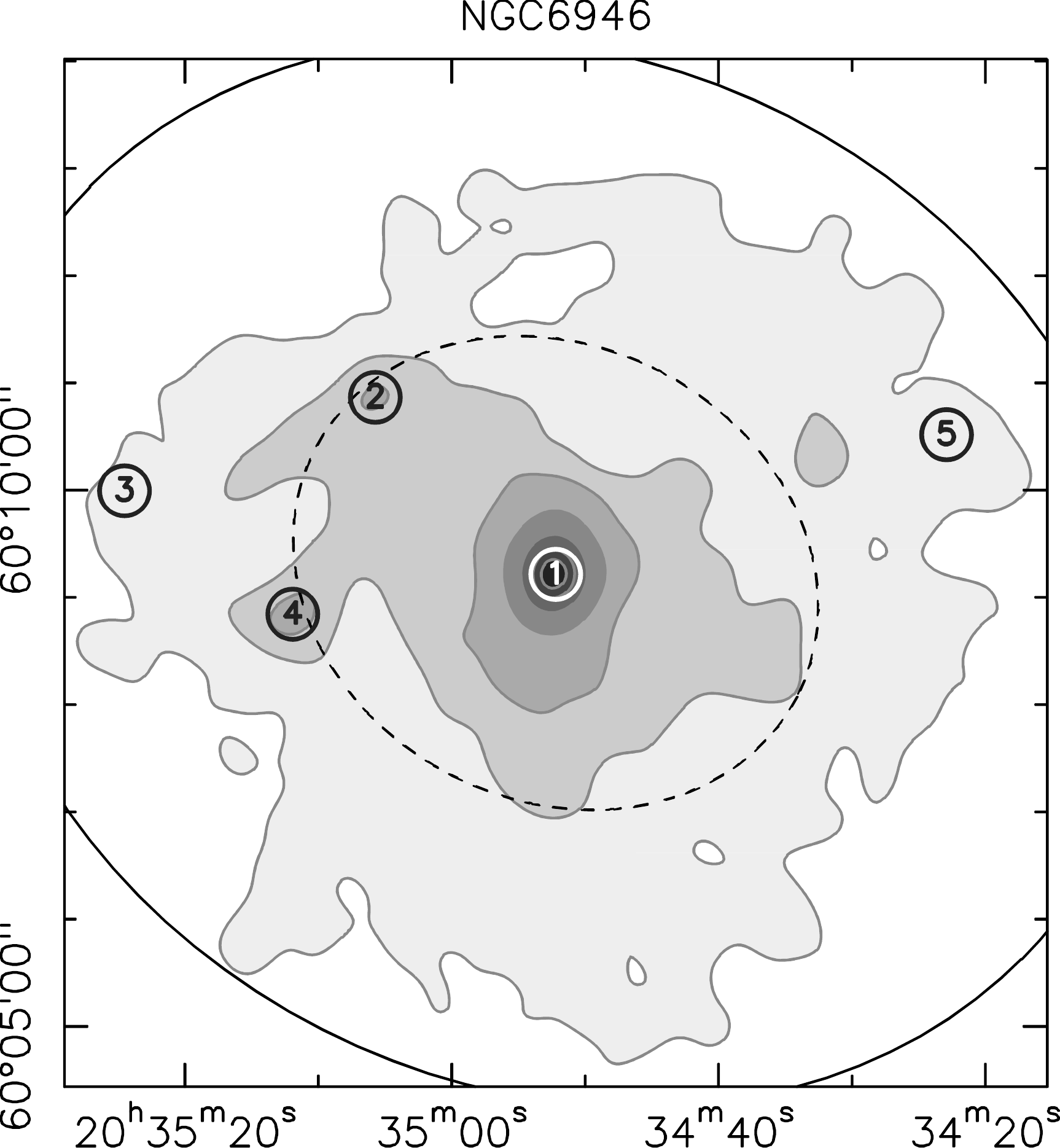}
\includegraphics[height=0.3\textheight]{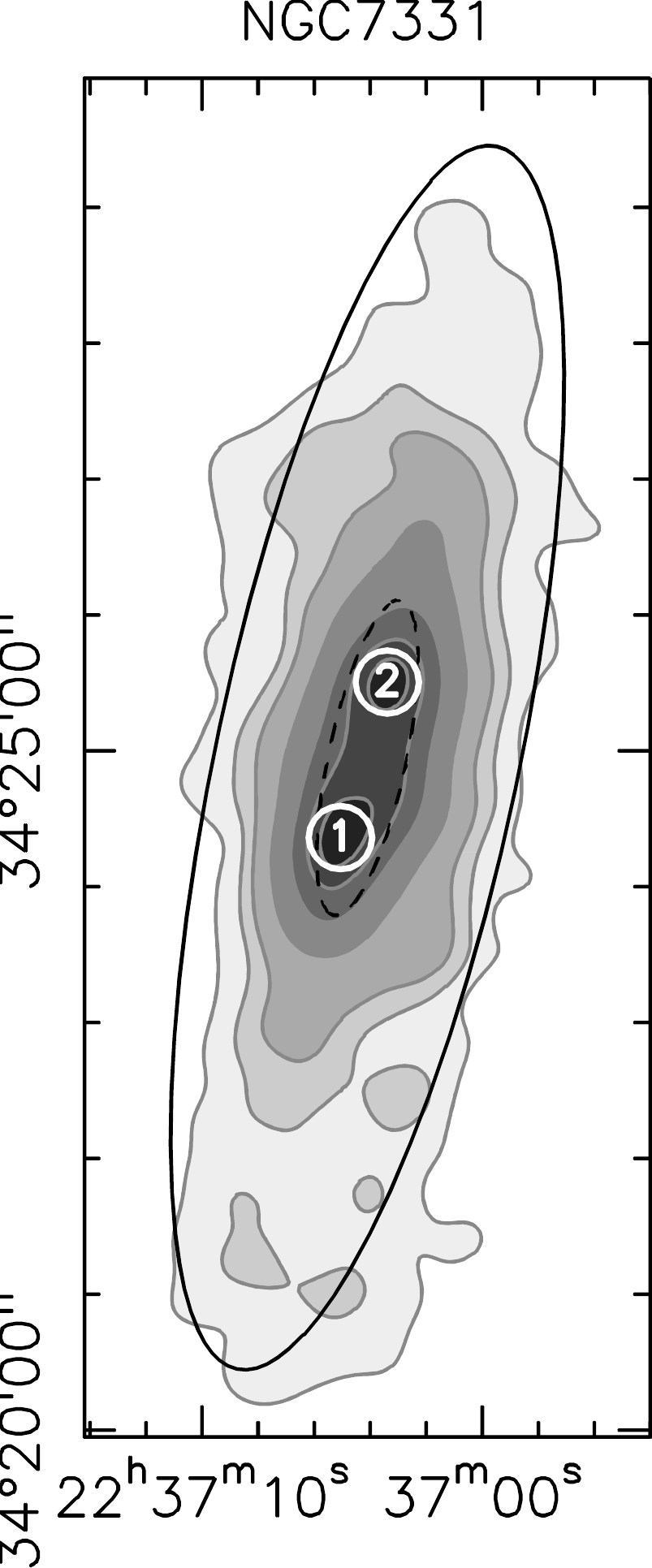}
\end{center}
\caption{Same as Fig.~\ref{f-map-1} for NGC~5055, NGC~5194, NGC~5457, NGC~5713, NGC~6946, and NGC~7331.} 
\label{f-map-5}
\end{figure}

\newpage
\begin{center}
\newpage 
\begin{figure}[h!]
\begin{center} 
\begin{tabular}{cc}
\includegraphics[width=0.42\textwidth]{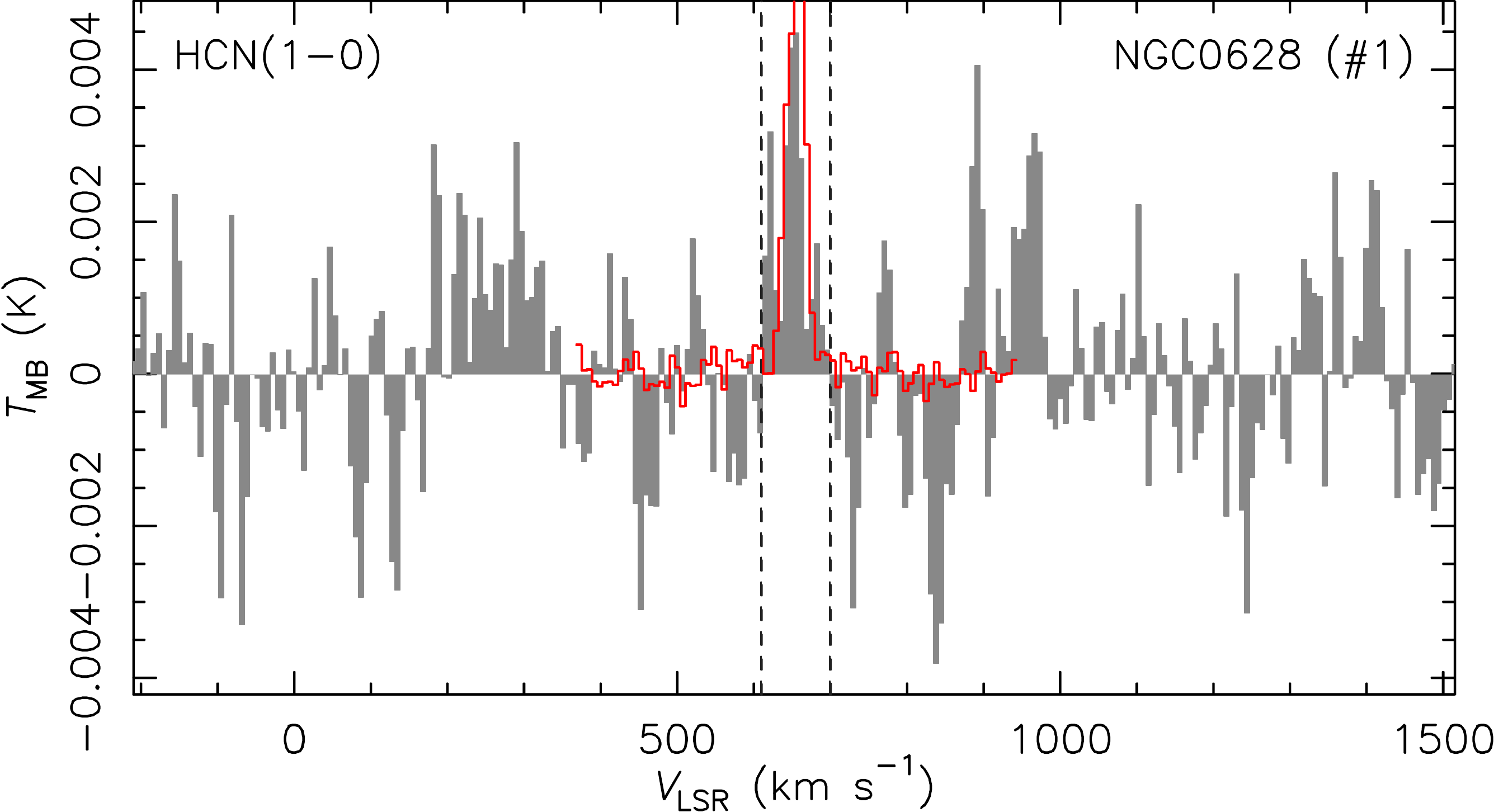} & 
\includegraphics[width=0.42\textwidth]{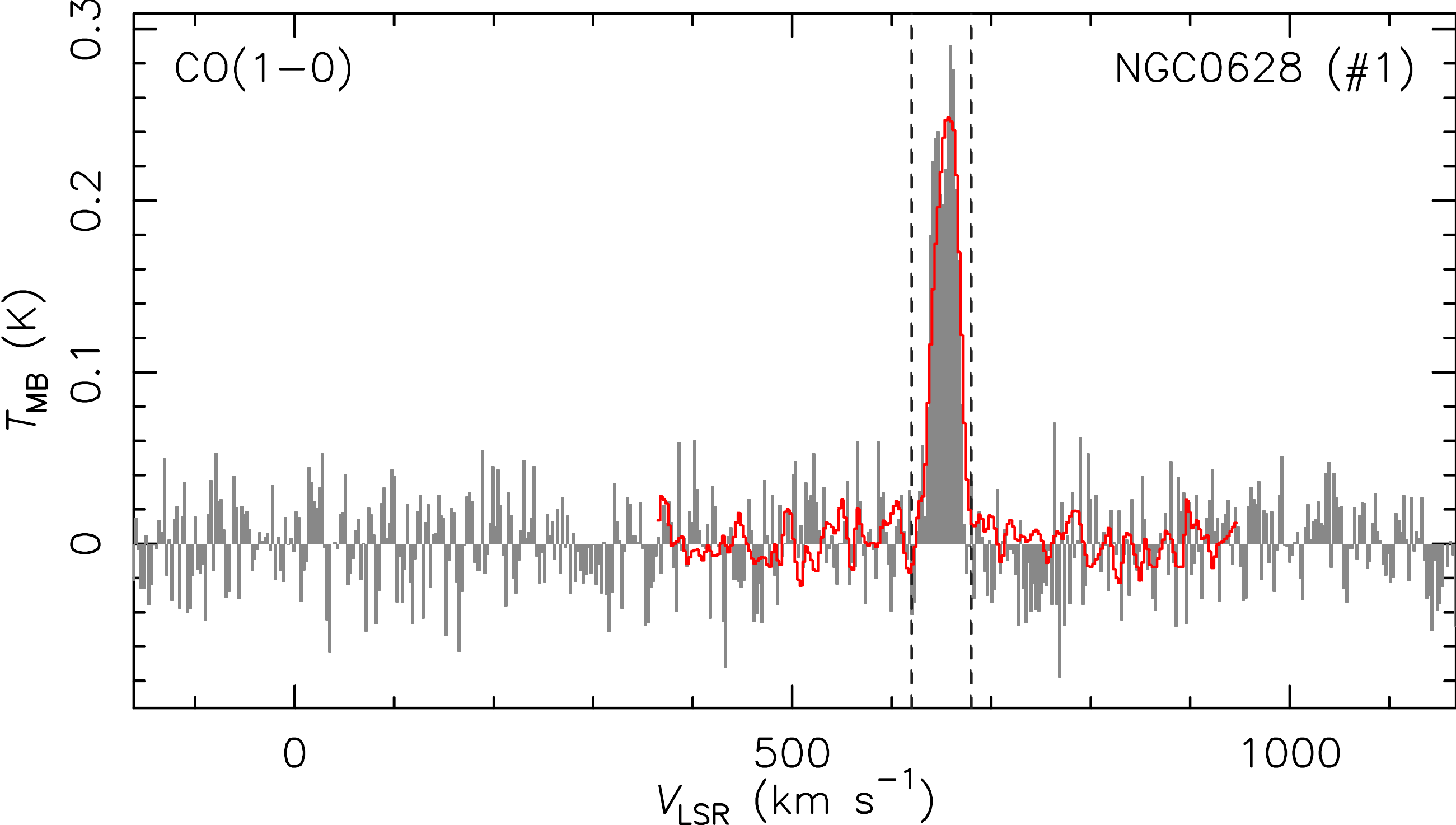} \\ 
\includegraphics[width=0.42\textwidth]{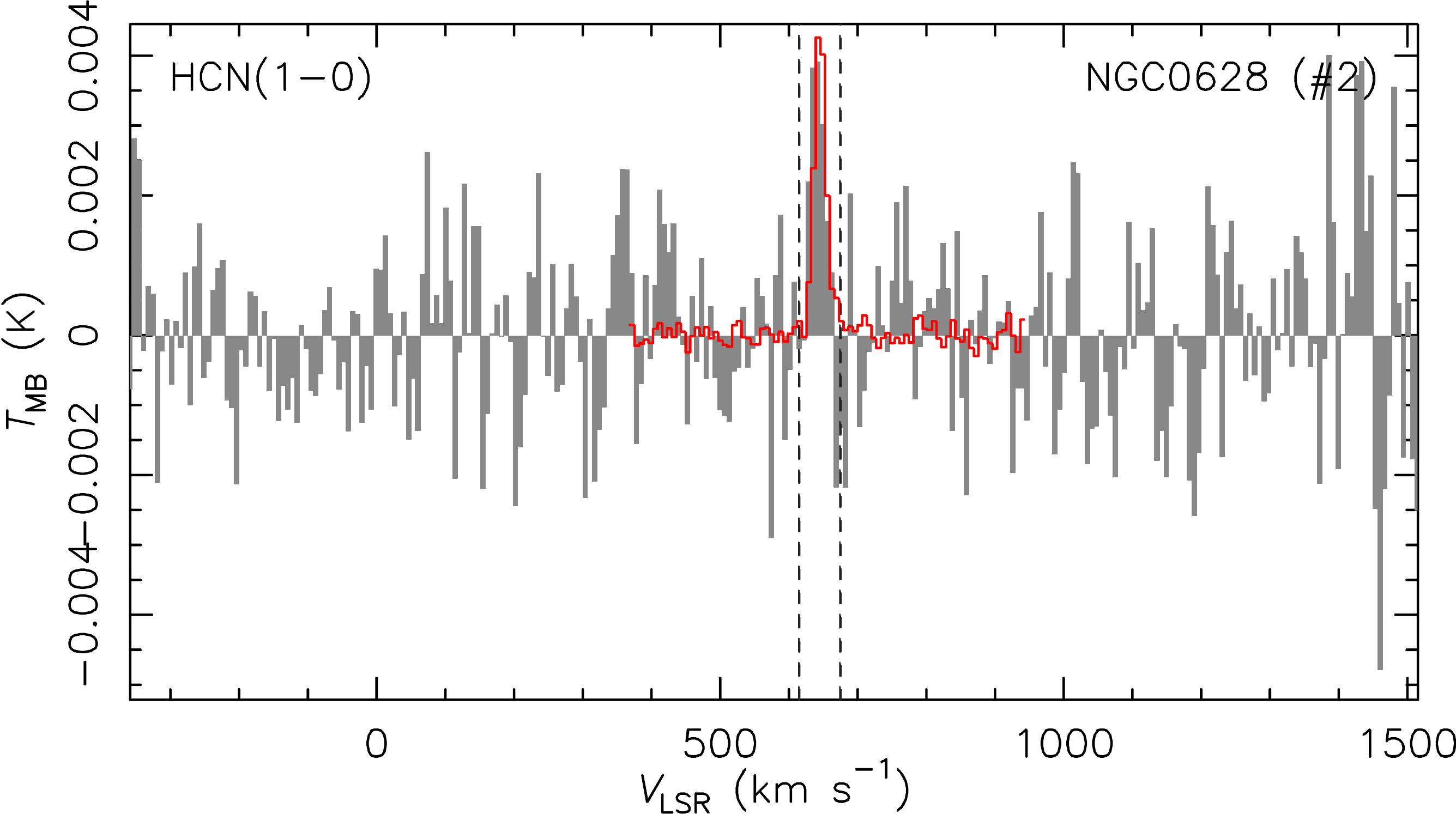} & 
\includegraphics[width=0.42\textwidth]{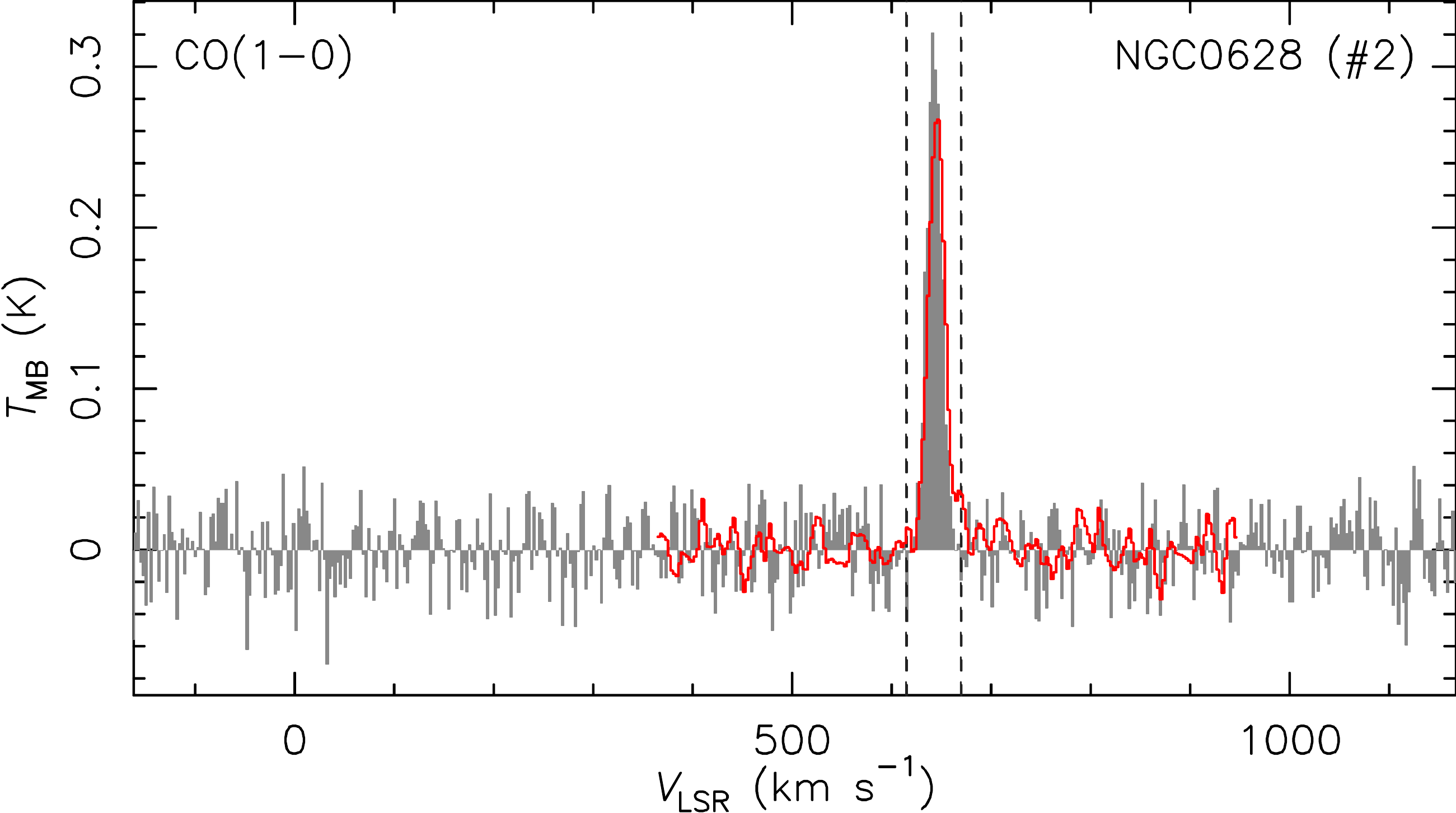} \\ 
\includegraphics[width=0.42\textwidth]{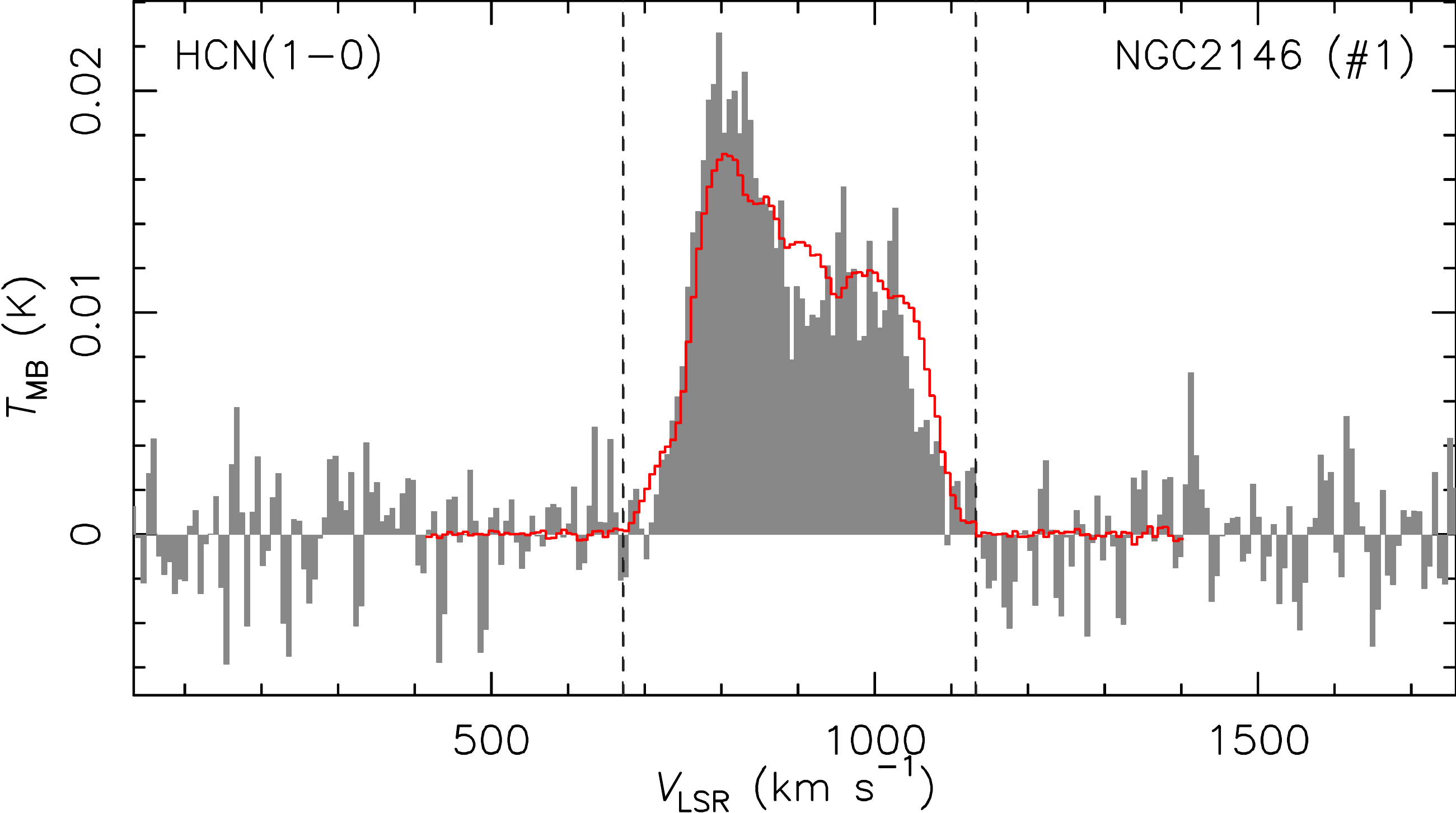} & 
\includegraphics[width=0.42\textwidth]{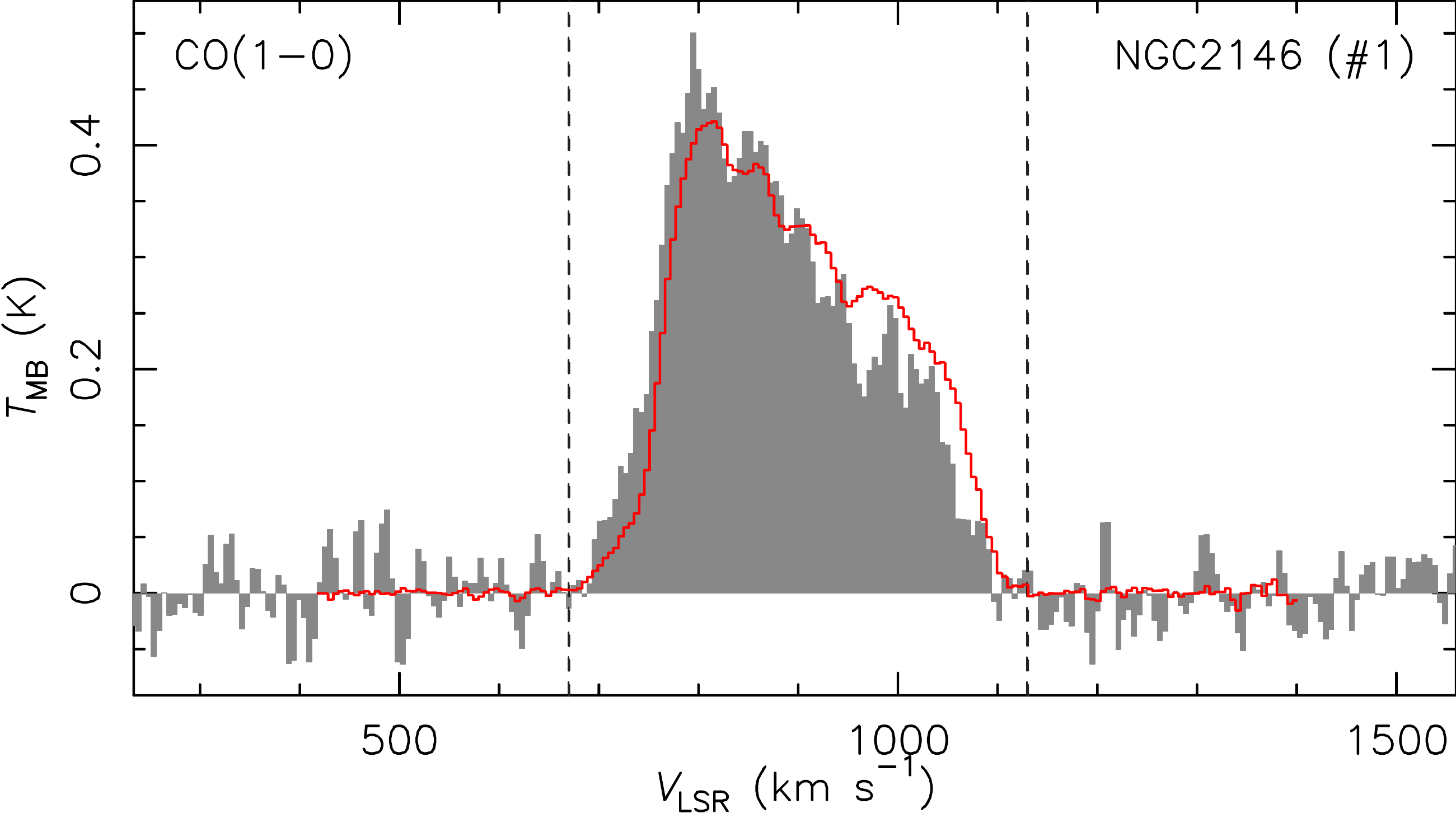} \\ 
\includegraphics[width=0.42\textwidth]{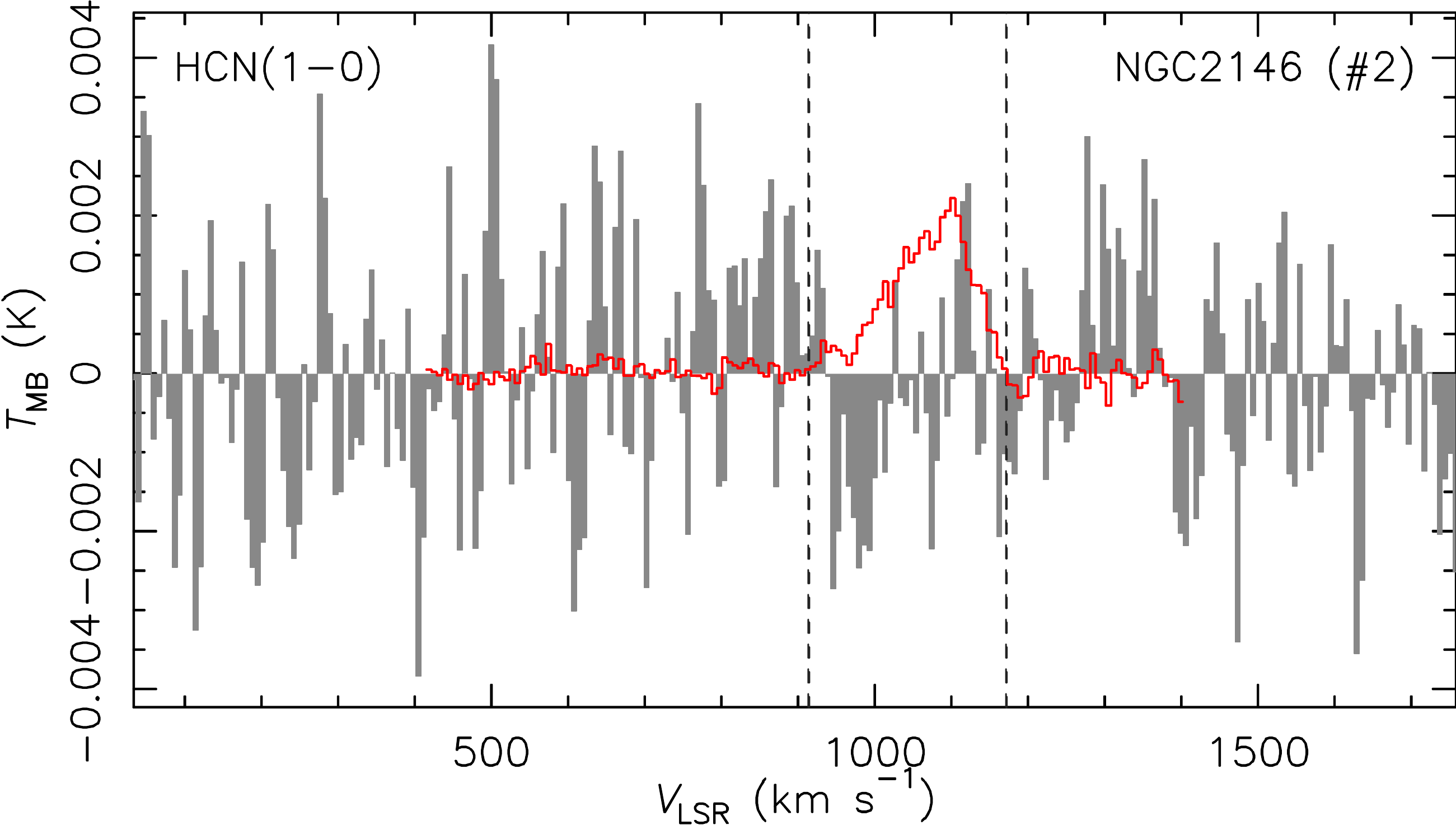} & 
\includegraphics[width=0.42\textwidth]{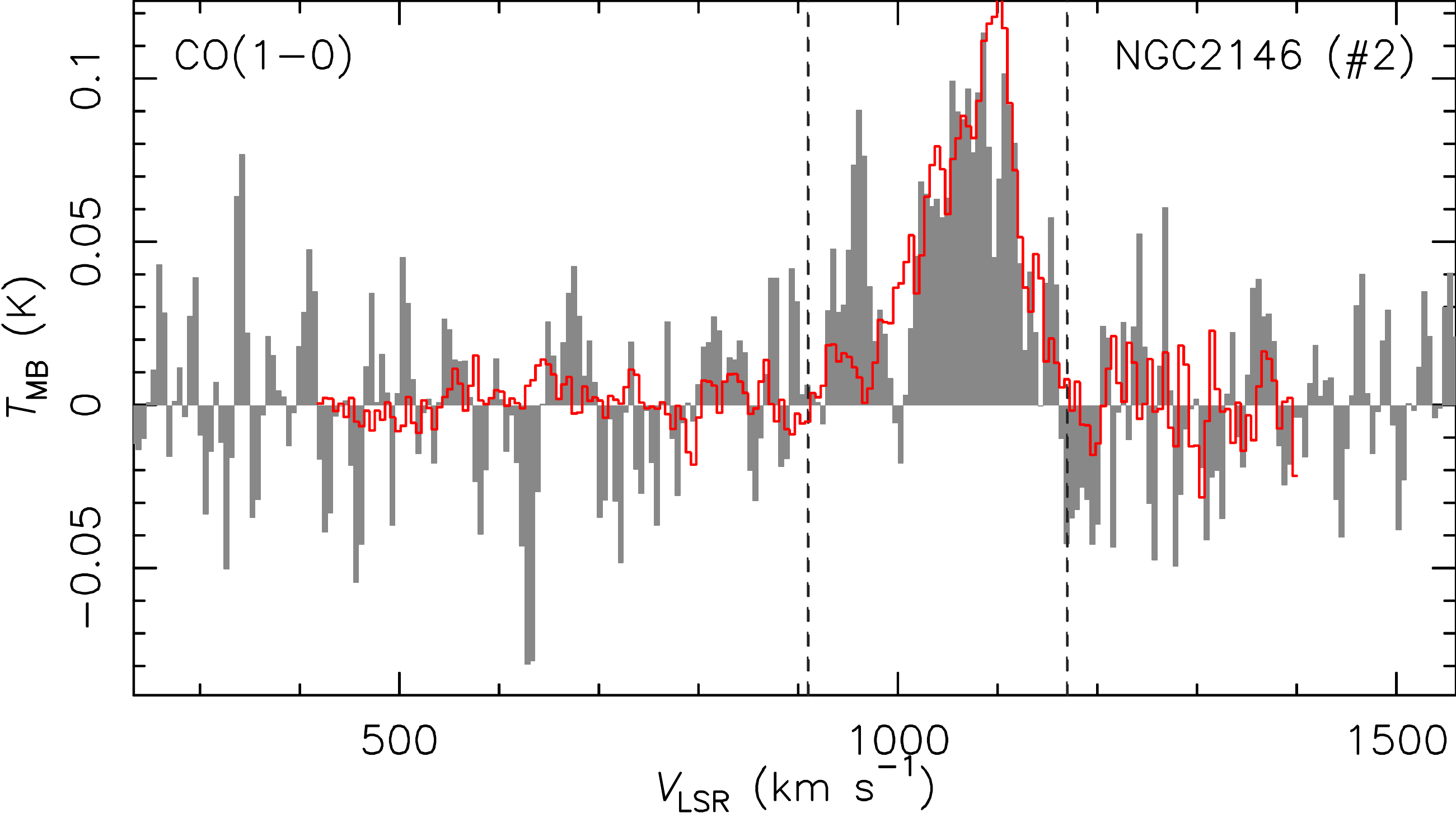} \\ 
\includegraphics[width=0.42\textwidth]{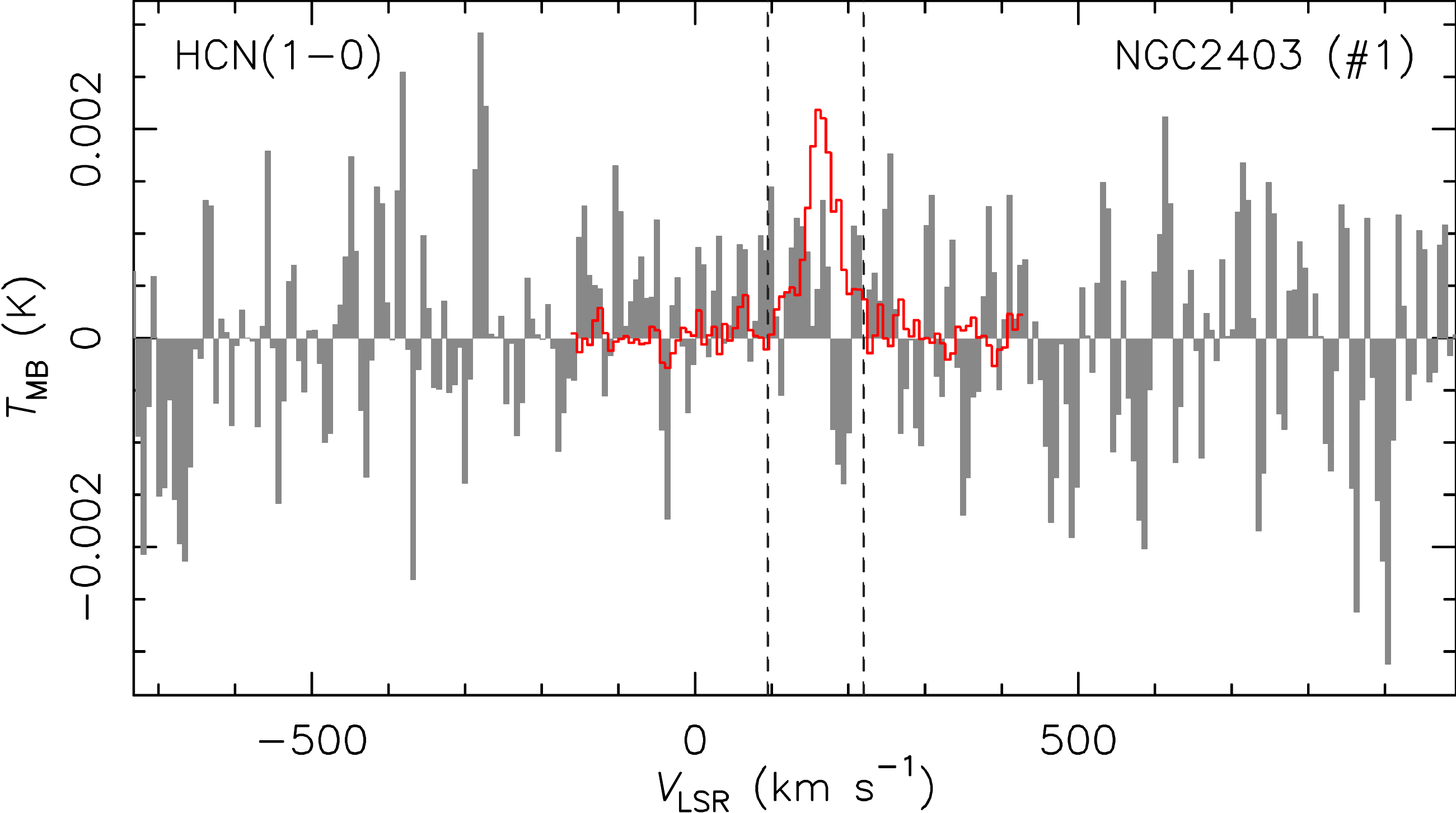} & 
\includegraphics[width=0.42\textwidth]{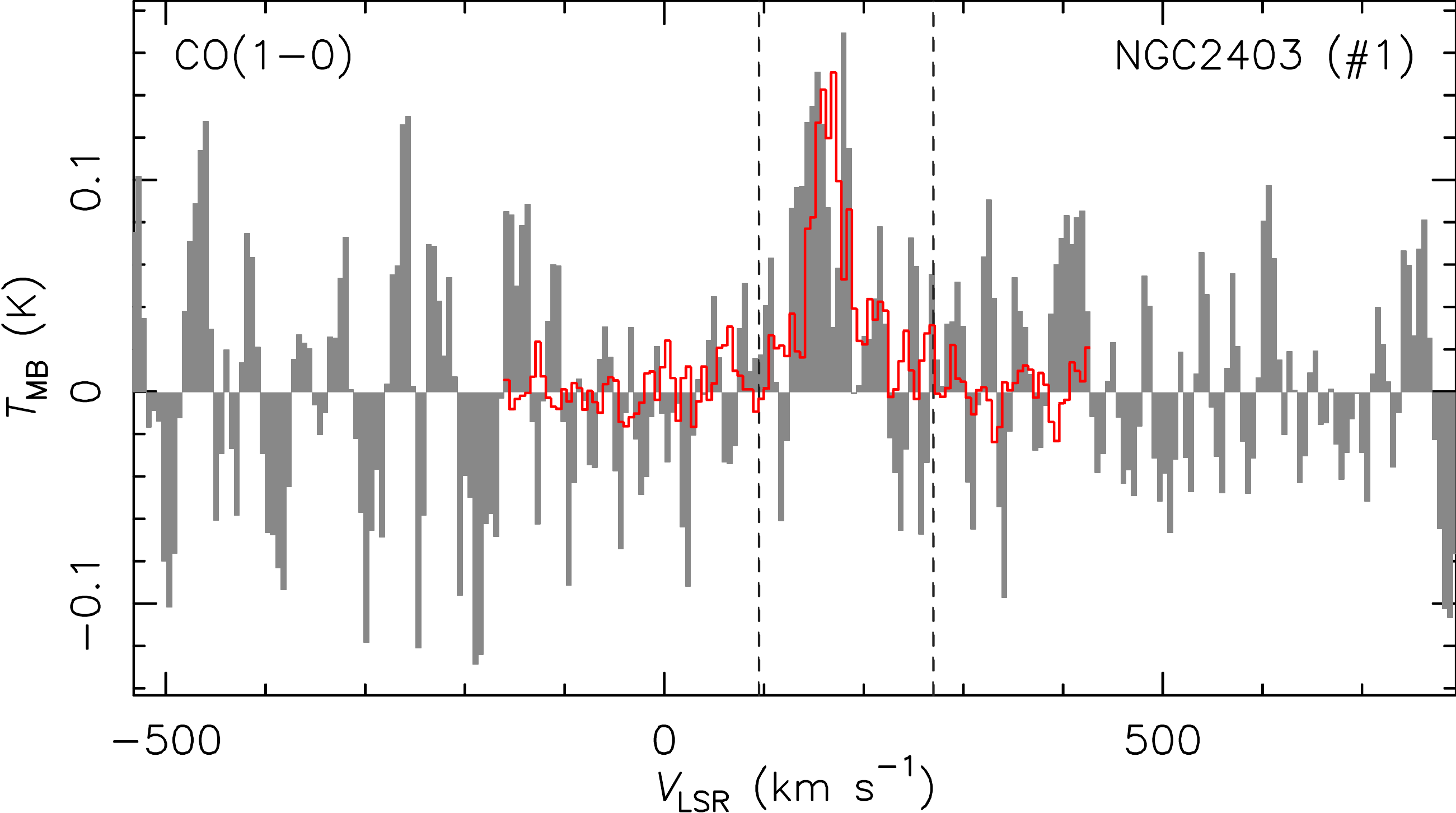} \\ 
\end{tabular}
\end{center} 
\caption{HCN(1--0) (left) and CO(1--0) (right) spectra at the positions observed in NGC~0628, NGC~2146, and NGC2403.  
The vertical dashed lines delimit the line windows used to reduce the data and calculate the line parameters. In each panel, the red spectrum is the HERACLES CO(2--1) line at the same position, convolved to matched spectral and spatial resolution. 
 The CO(2--1) line is rescaled so that its velocity-integrated intensity within the line window is: (1) the same as that of the other line, if the latter has $SNR_\mathrm{line}\geq4$; (2) equivalent to a $4\sigma$ value for the other line, otherwise. 
}
\label{f-spec-1} 
\end{figure} 
\newpage 
\begin{figure}[h!]
\begin{center} 
\begin{tabular}{cc} 
\includegraphics[width=0.42\textwidth]{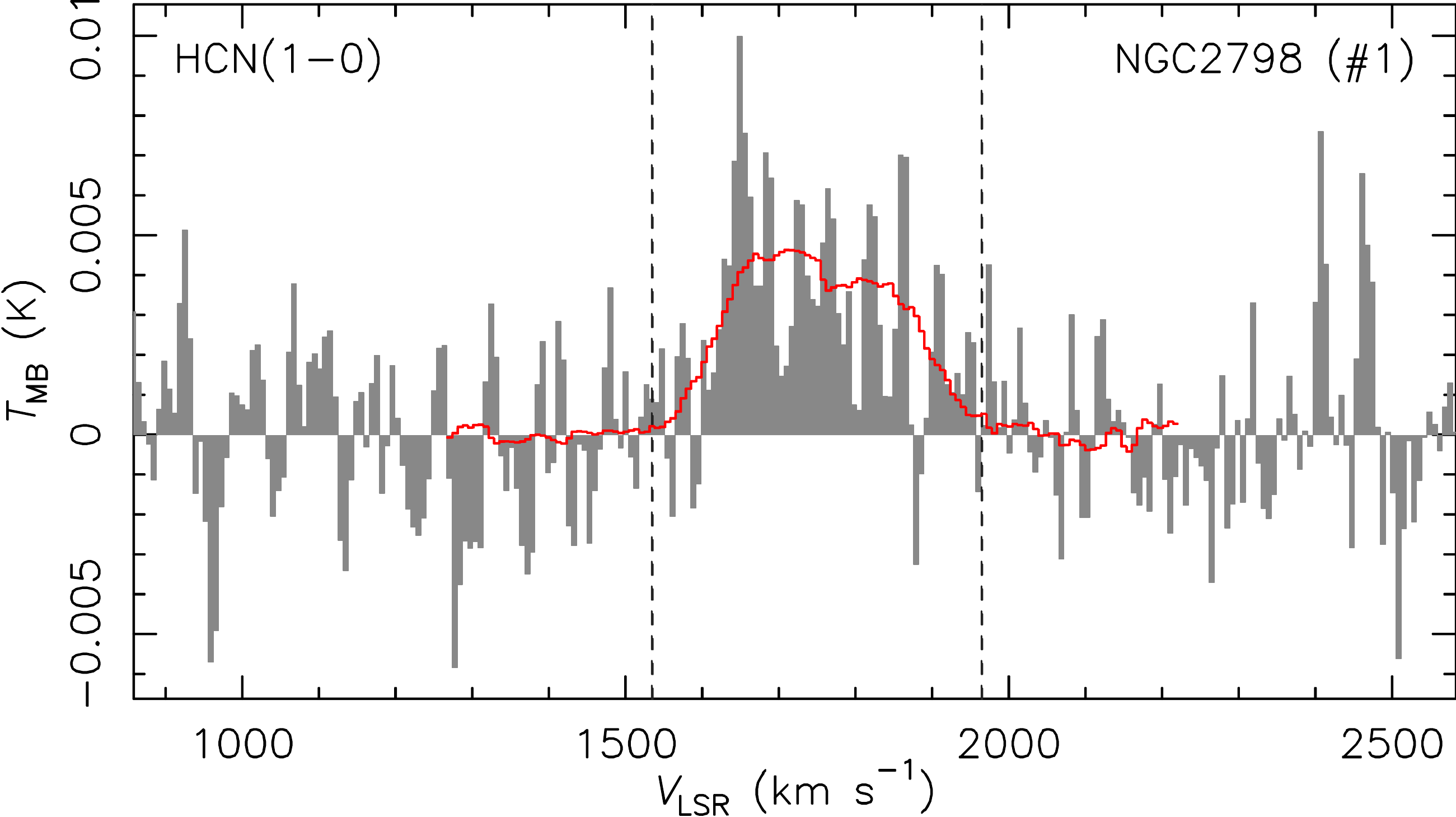} & 
\includegraphics[width=0.42\textwidth]{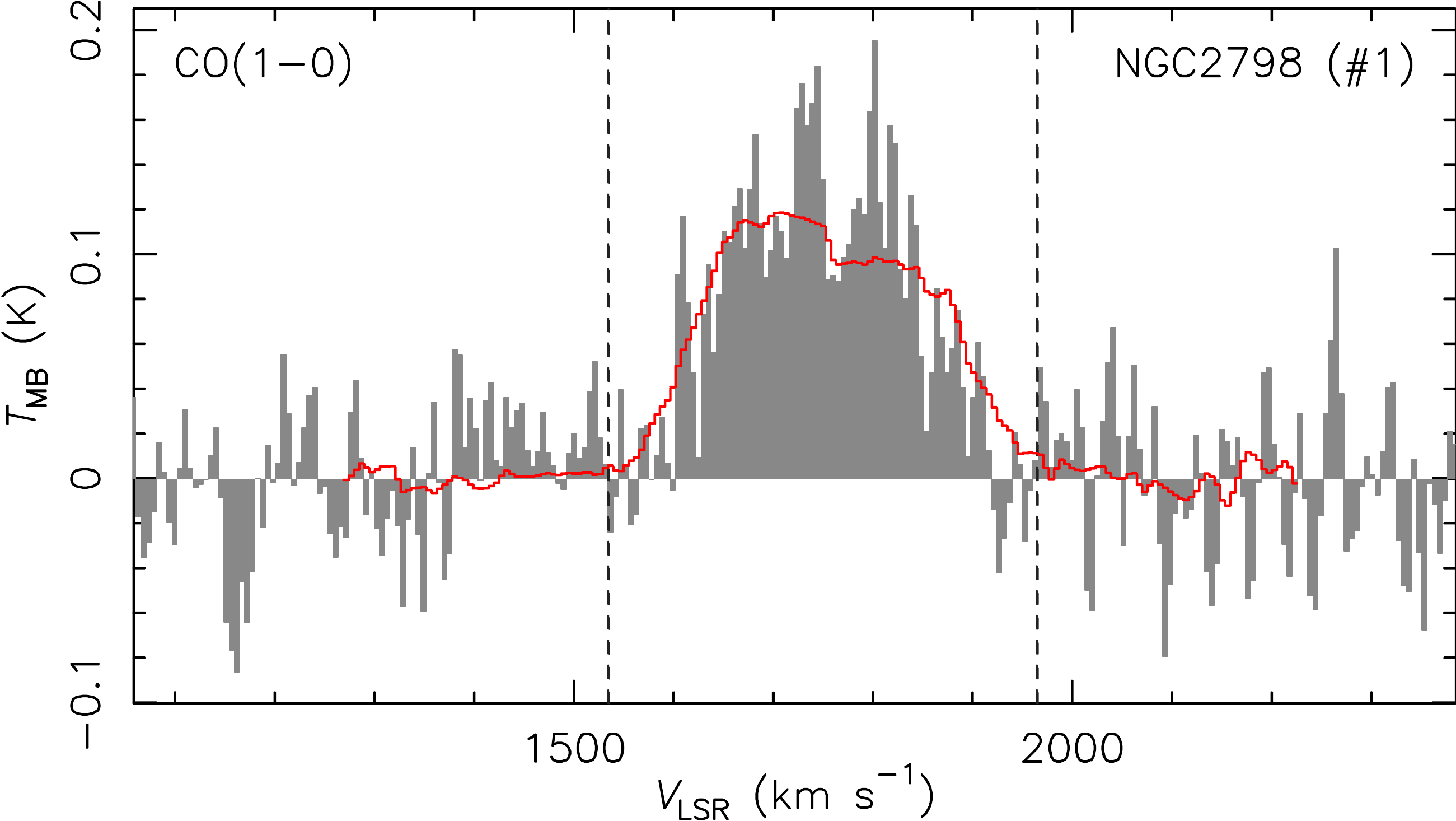} \\ 
\includegraphics[width=0.42\textwidth]{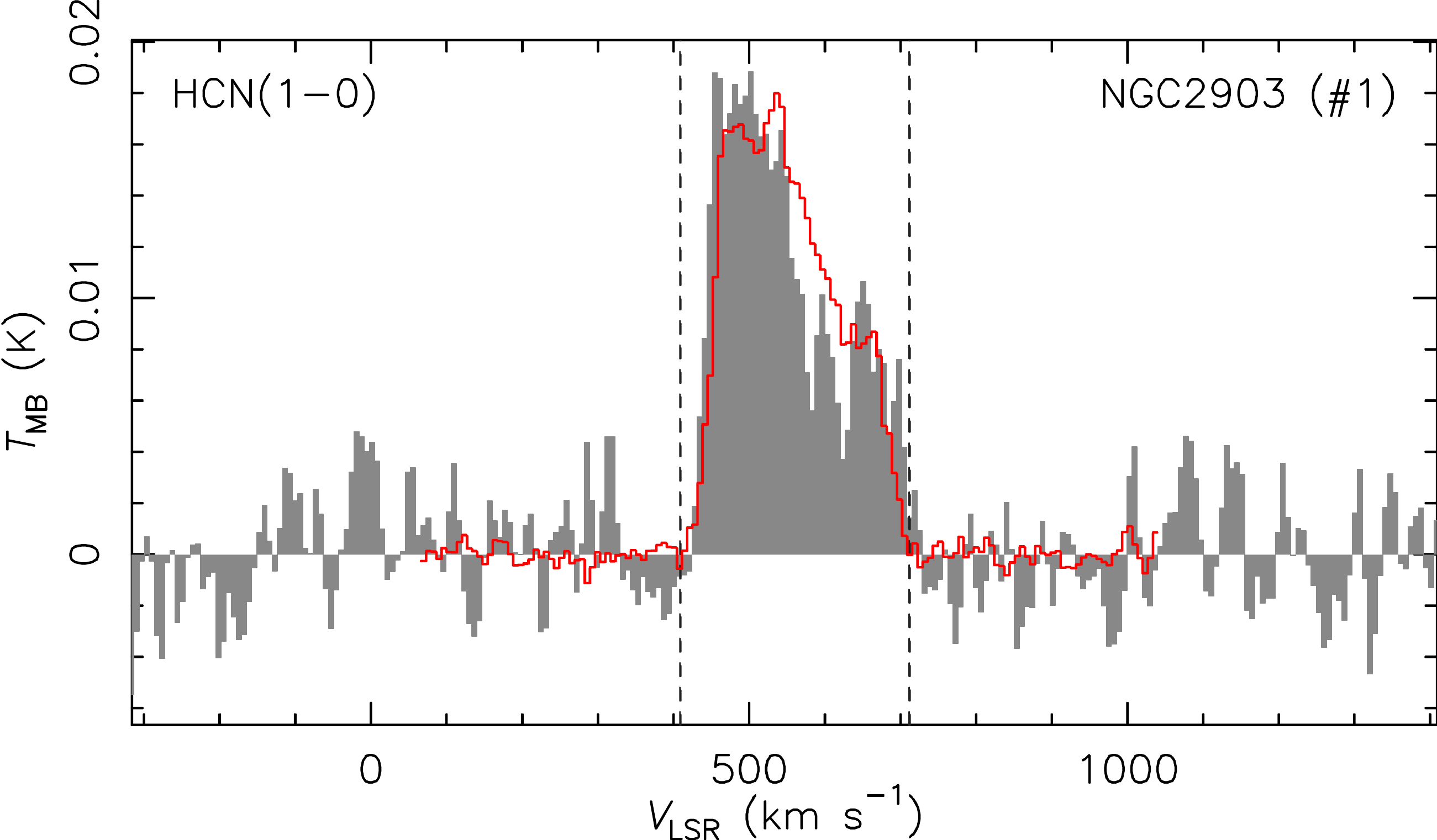} & 
\includegraphics[width=0.42\textwidth]{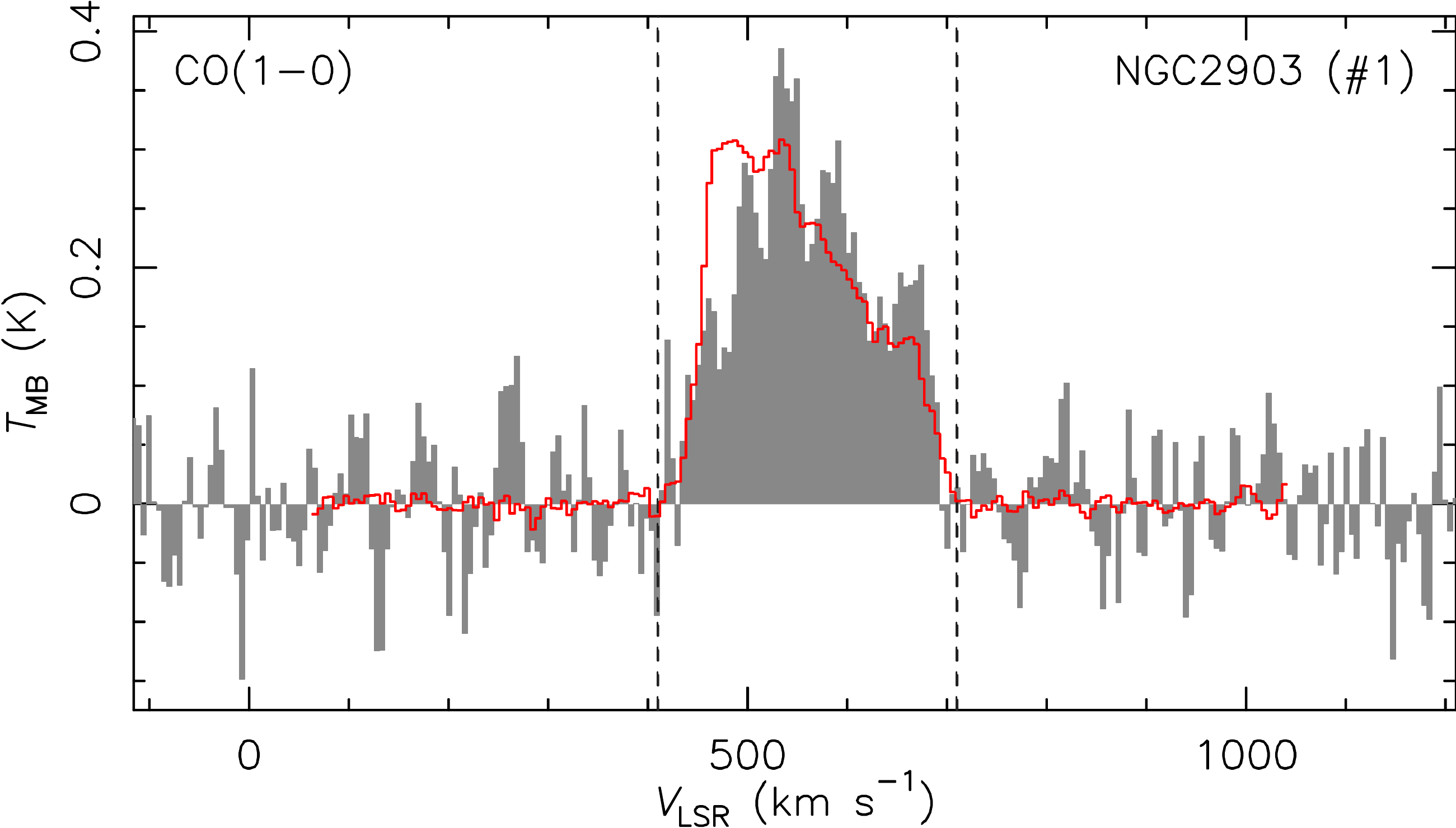} \\ 
\includegraphics[width=0.42\textwidth]{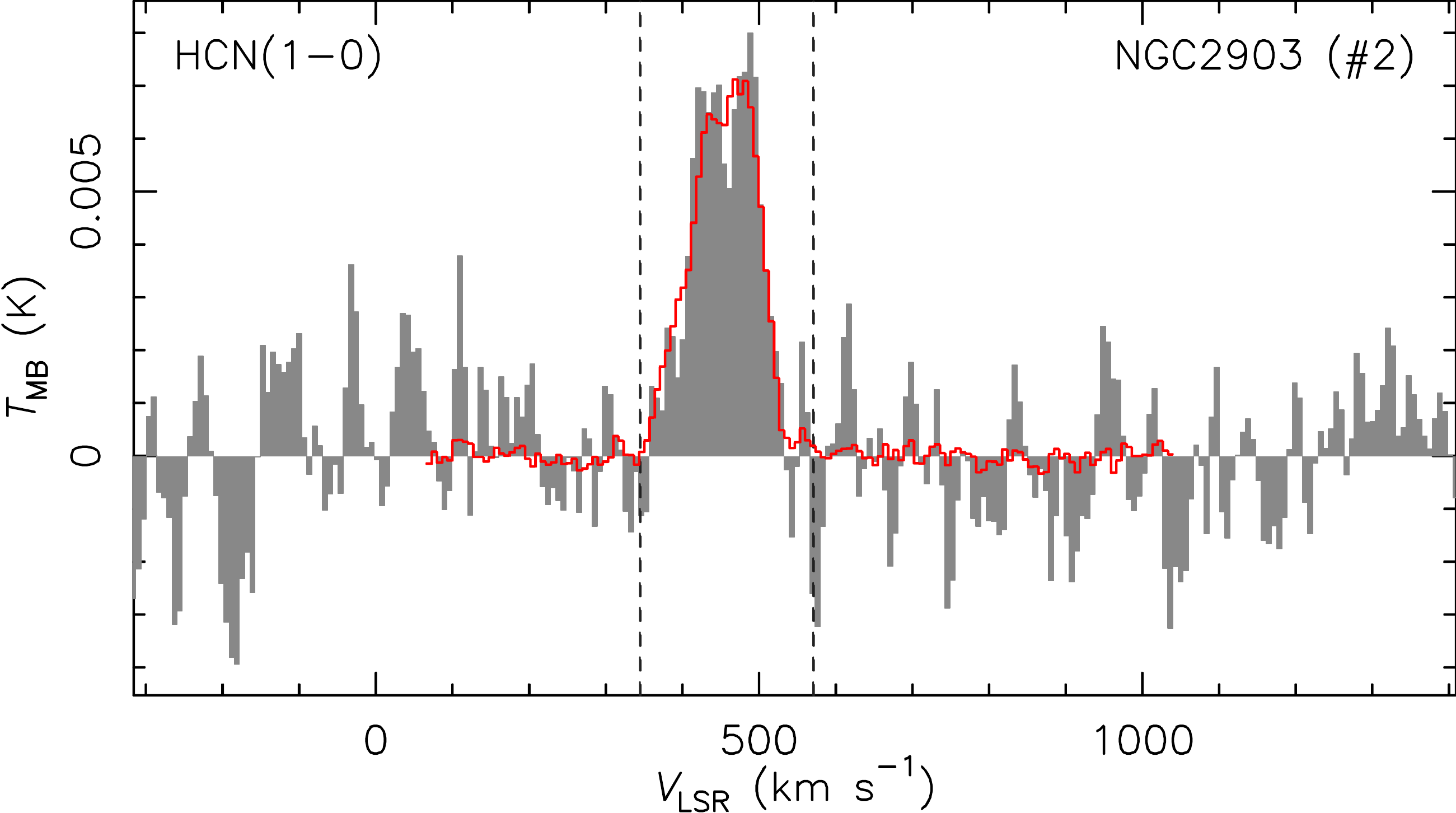} & 
\includegraphics[width=0.42\textwidth]{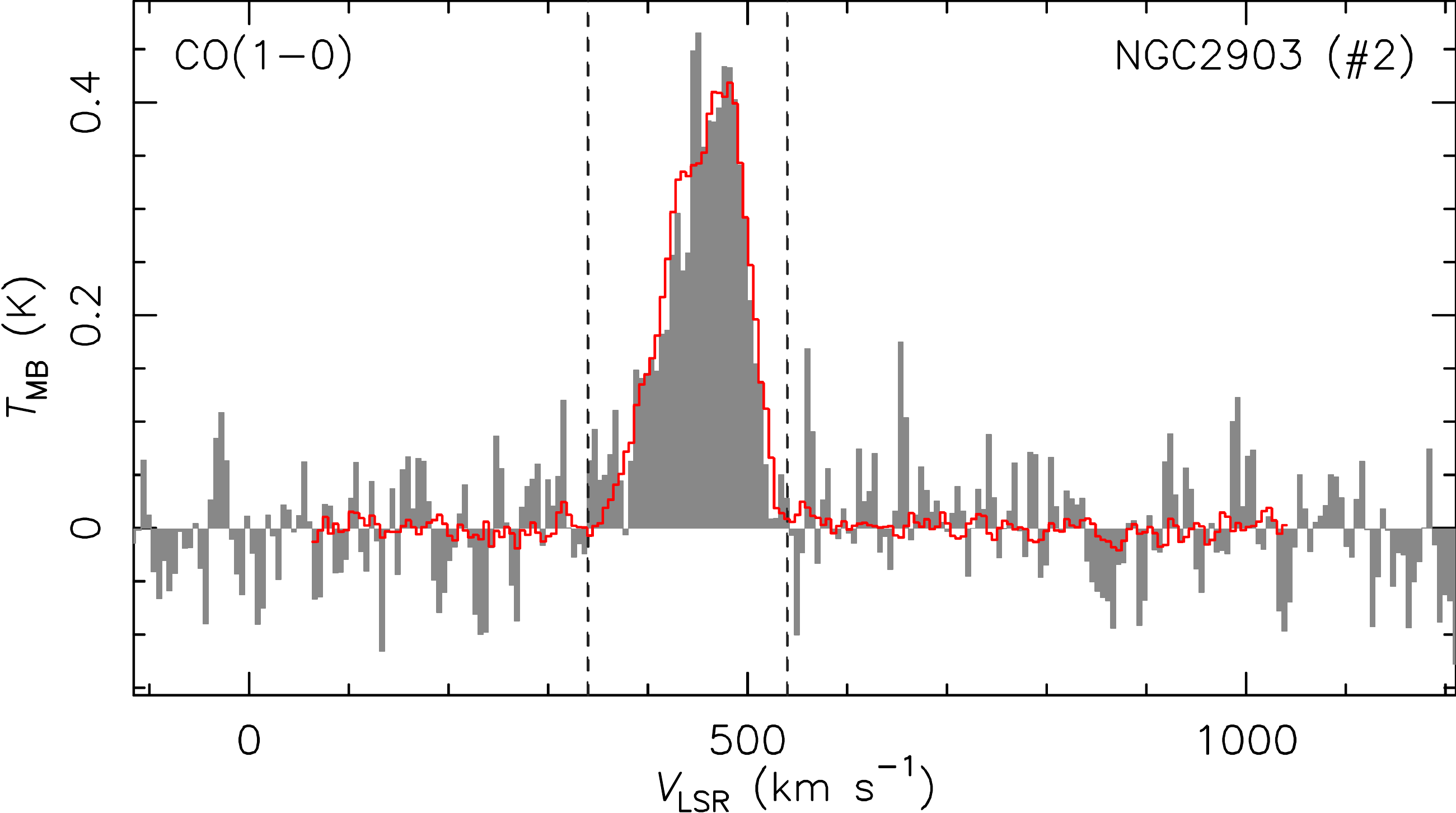} \\ 
\includegraphics[width=0.42\textwidth]{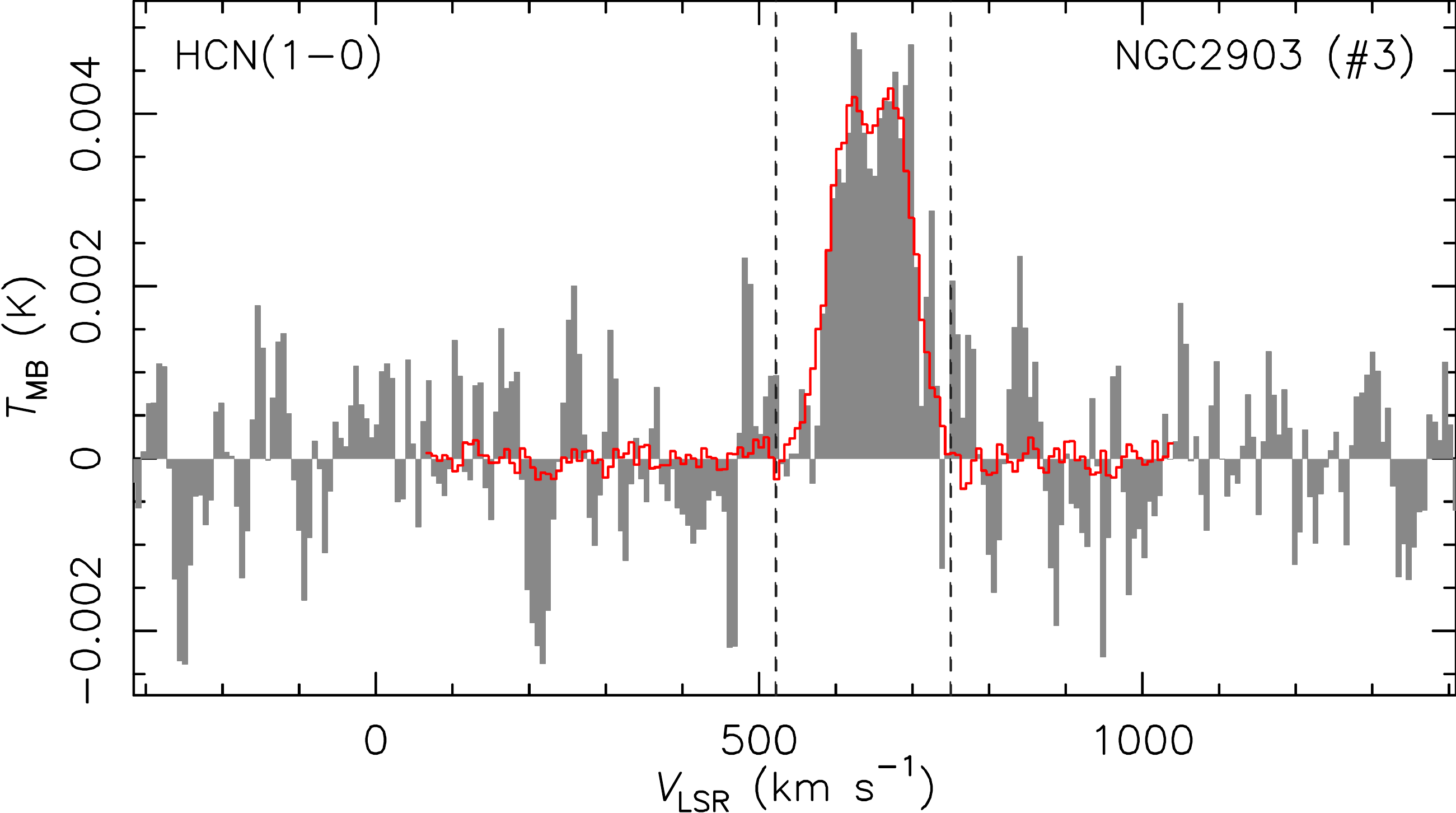} & 
\includegraphics[width=0.42\textwidth]{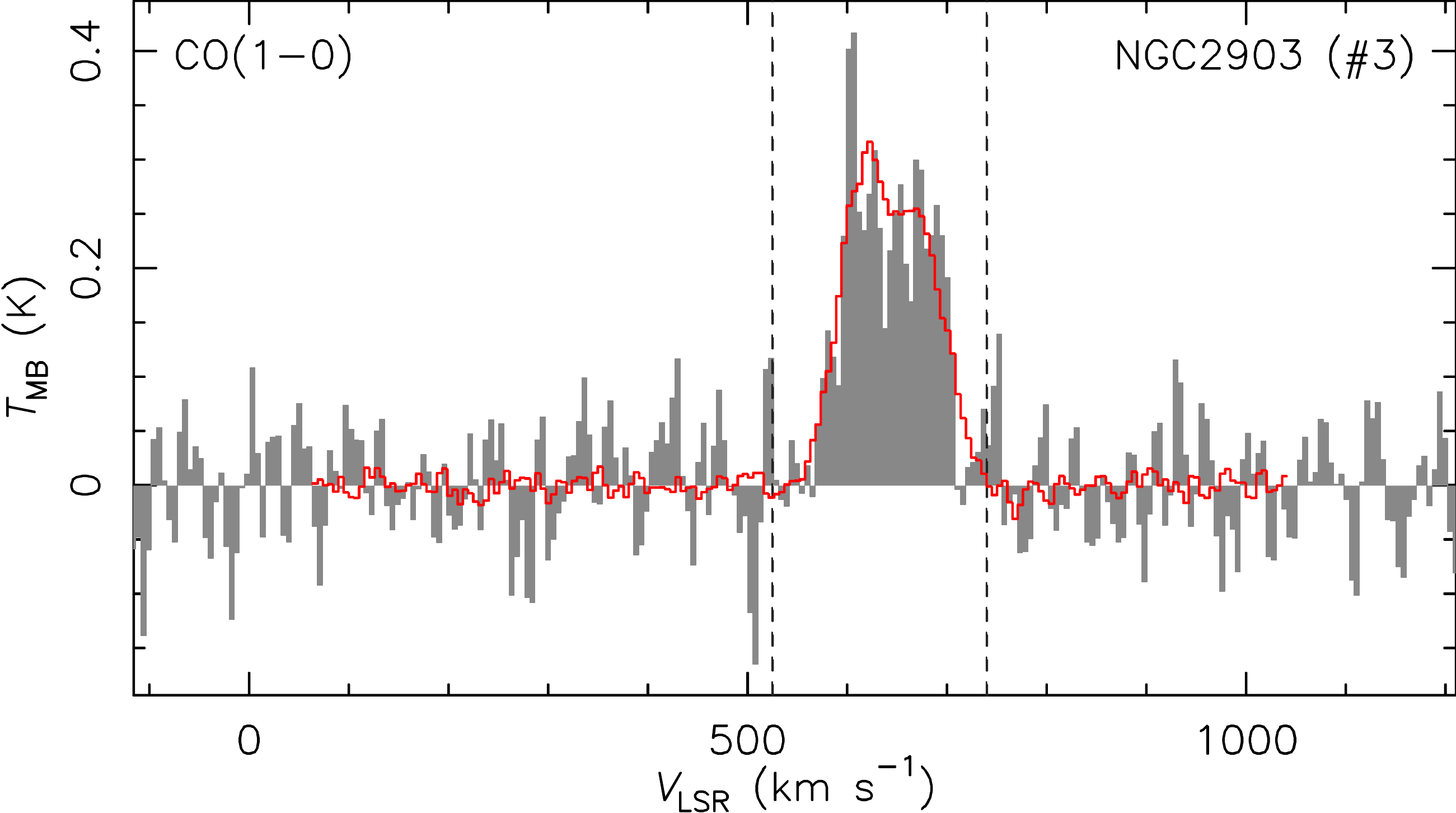} \\ 
\includegraphics[width=0.42\textwidth]{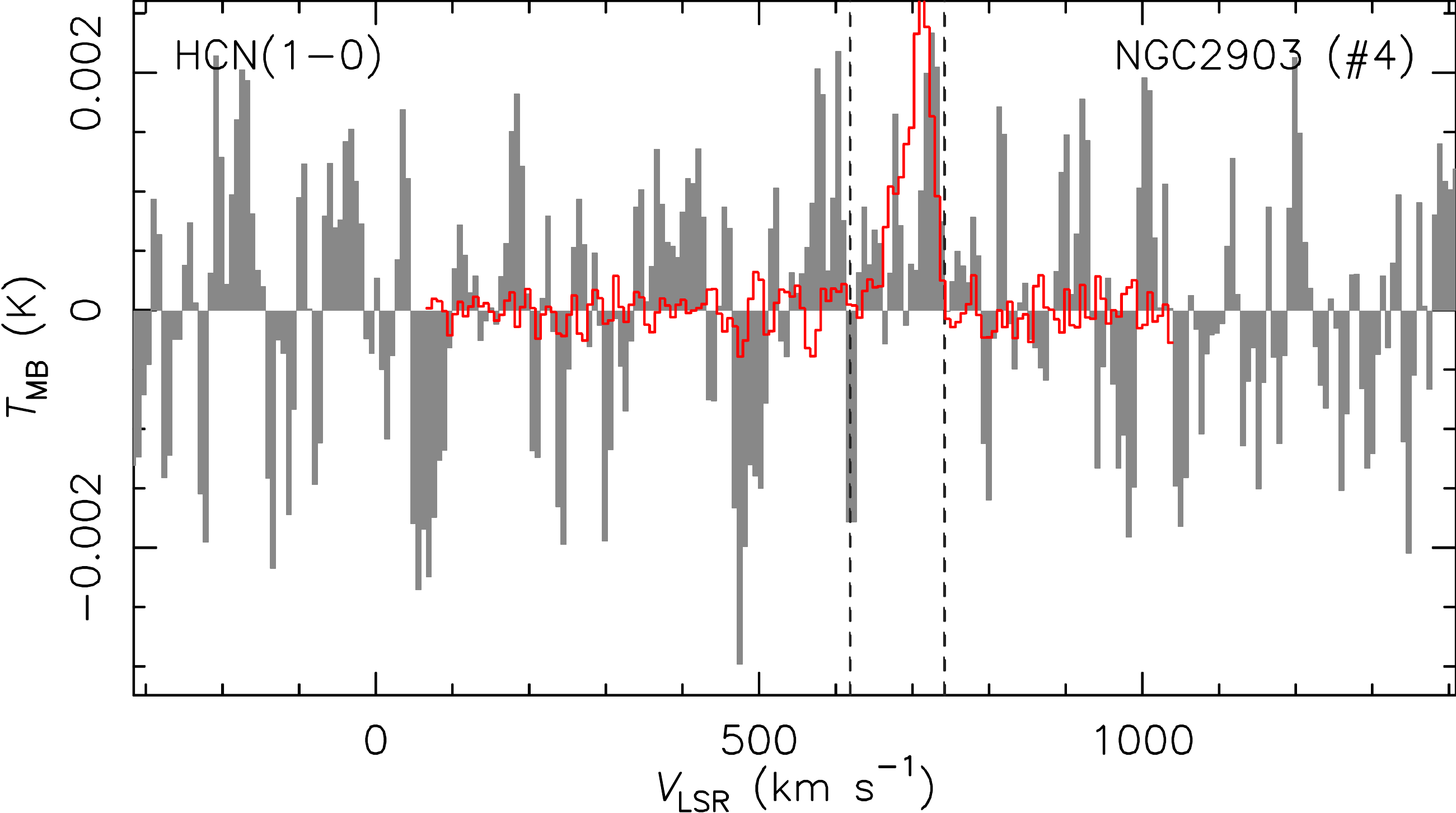} & 
\includegraphics[width=0.42\textwidth]{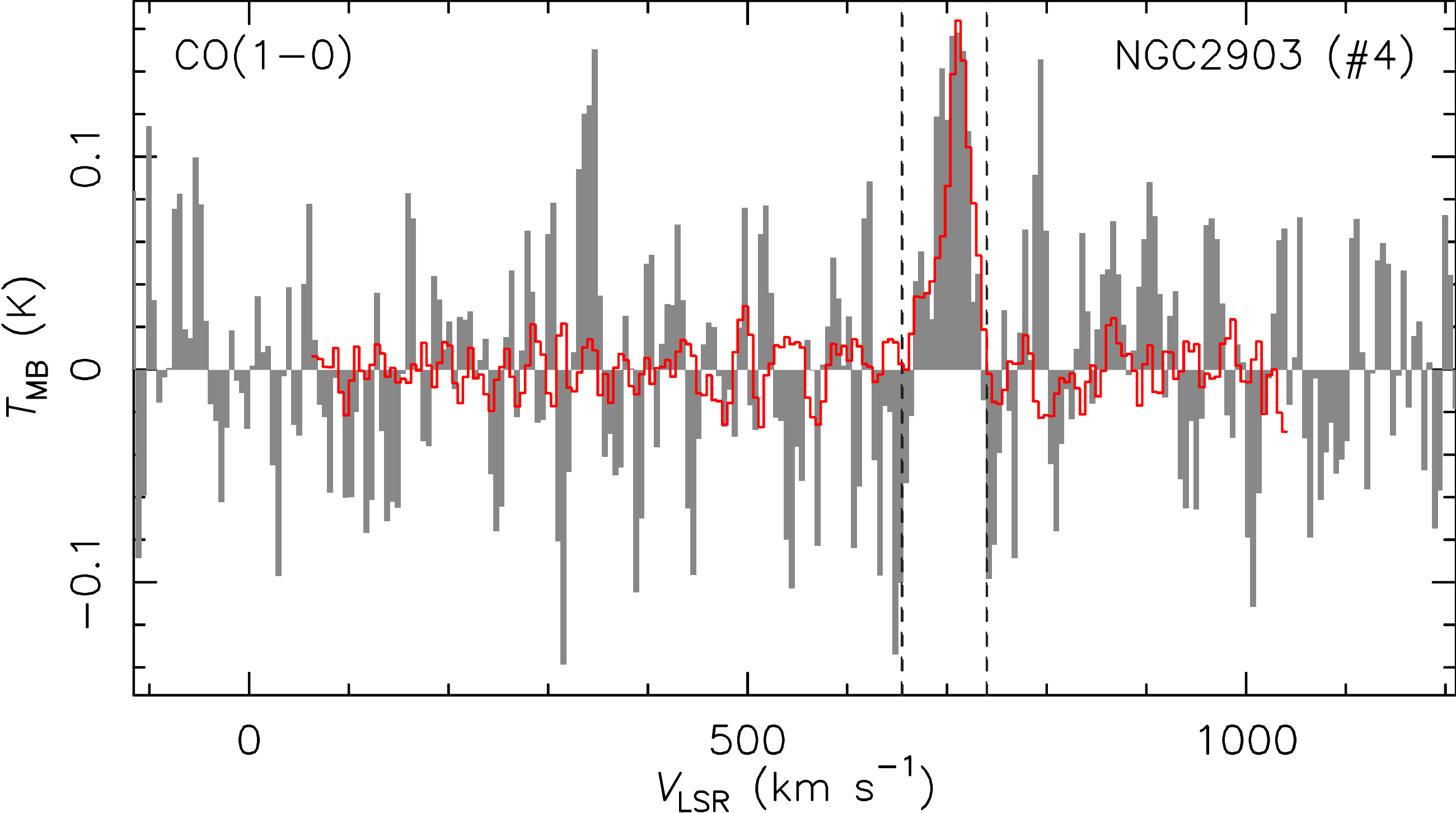} \\ 
\end{tabular}
\end{center}  
\caption{Same as Fig.~\ref{f-spec-1} for NGC~2798 and NGC~2903.} 
\end{figure} 
\newpage 
\begin{figure}[h!]
\begin{center}  
\begin{tabular}{cc} 
\includegraphics[width=0.42\textwidth]{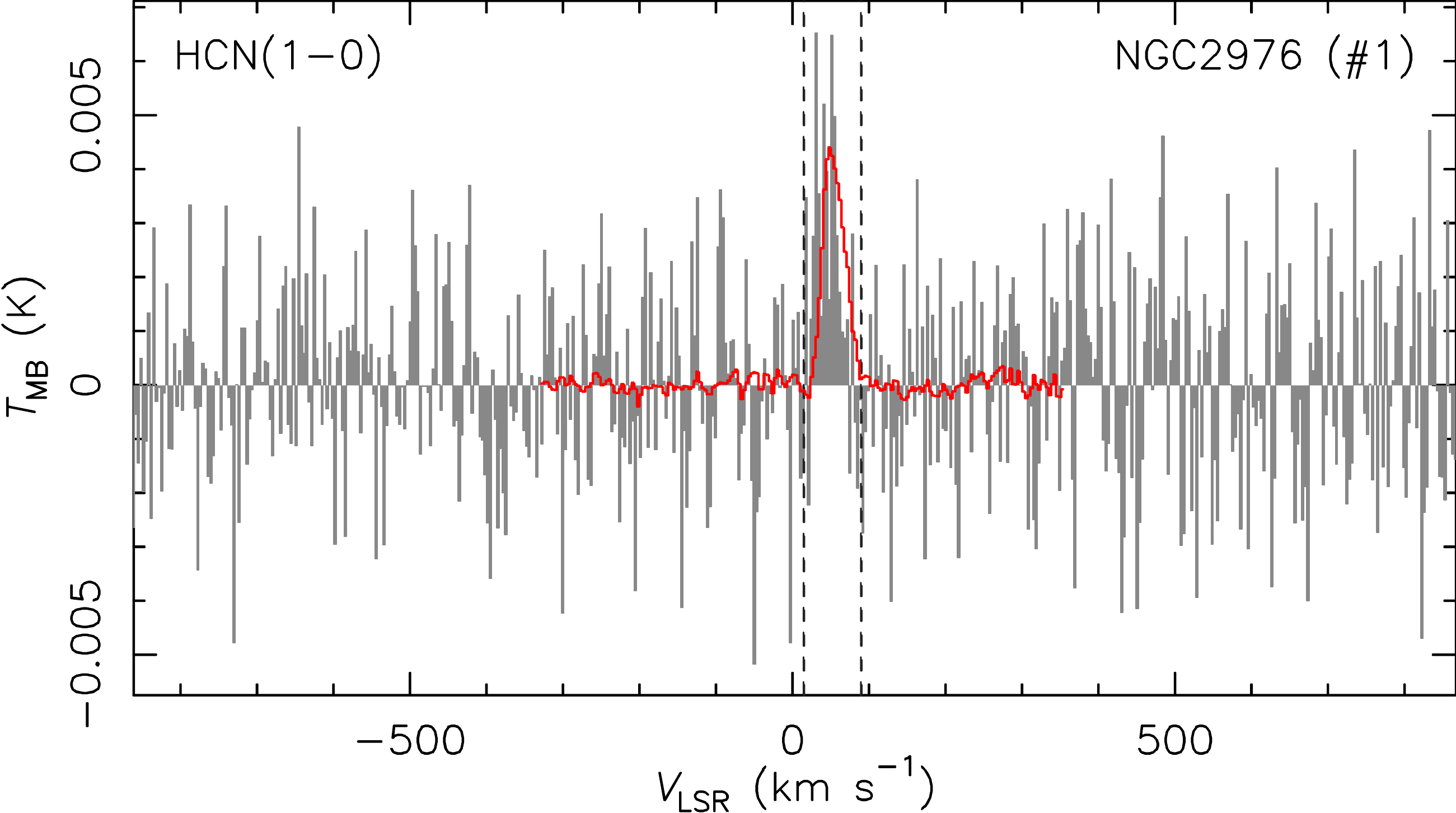} & 
\includegraphics[width=0.42\textwidth]{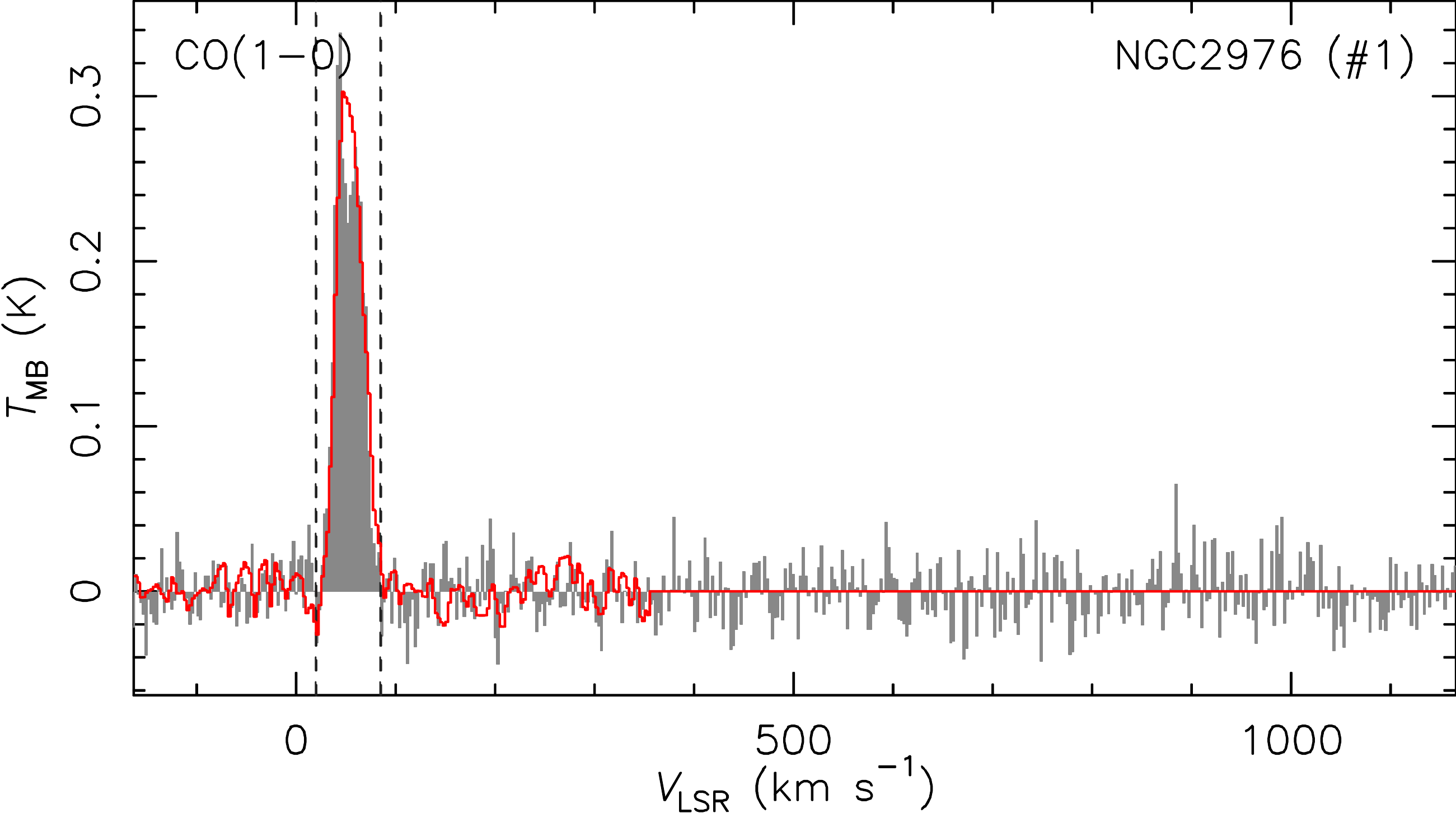} \\ 
\includegraphics[width=0.42\textwidth]{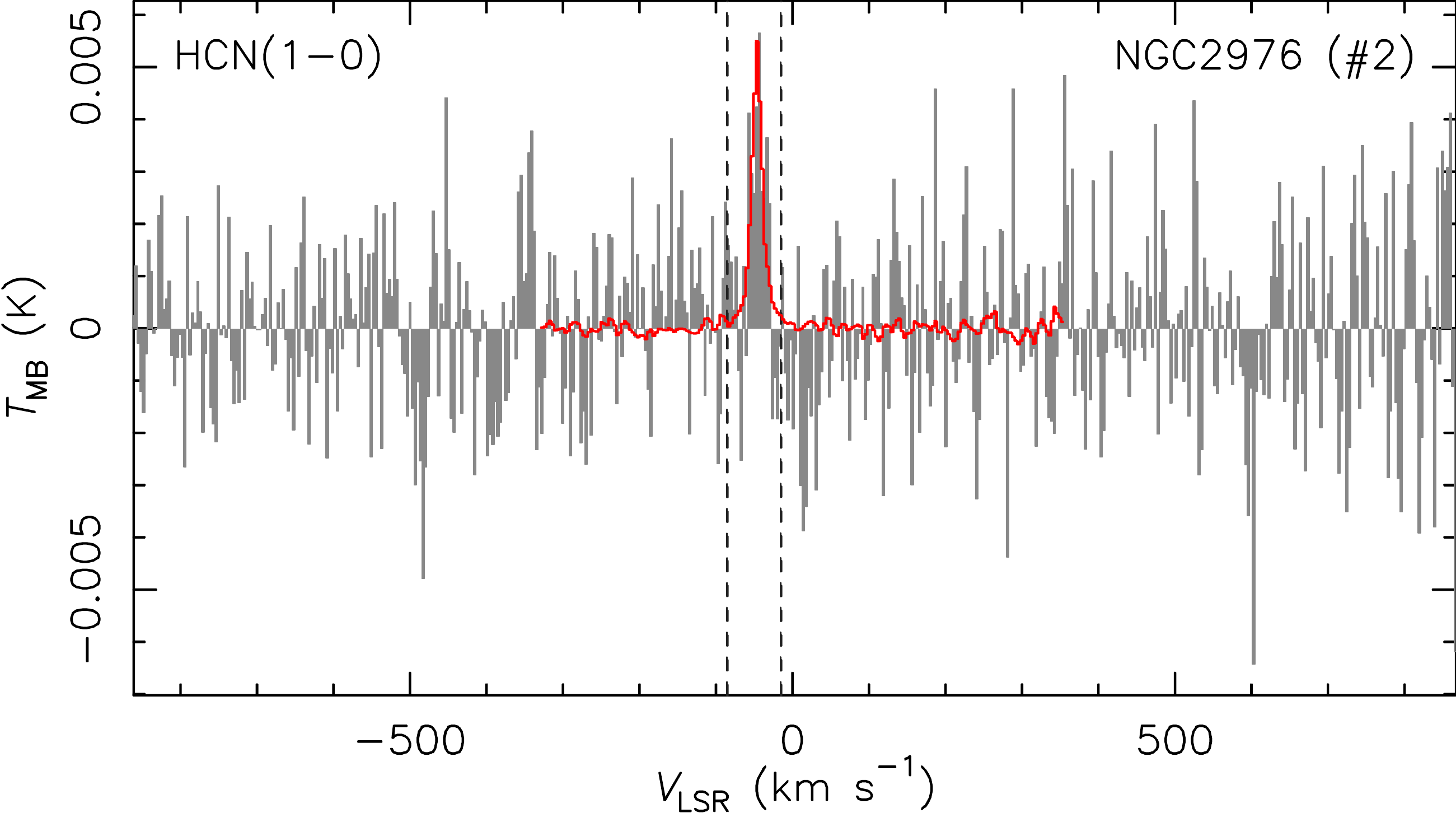} & 
\includegraphics[width=0.42\textwidth]{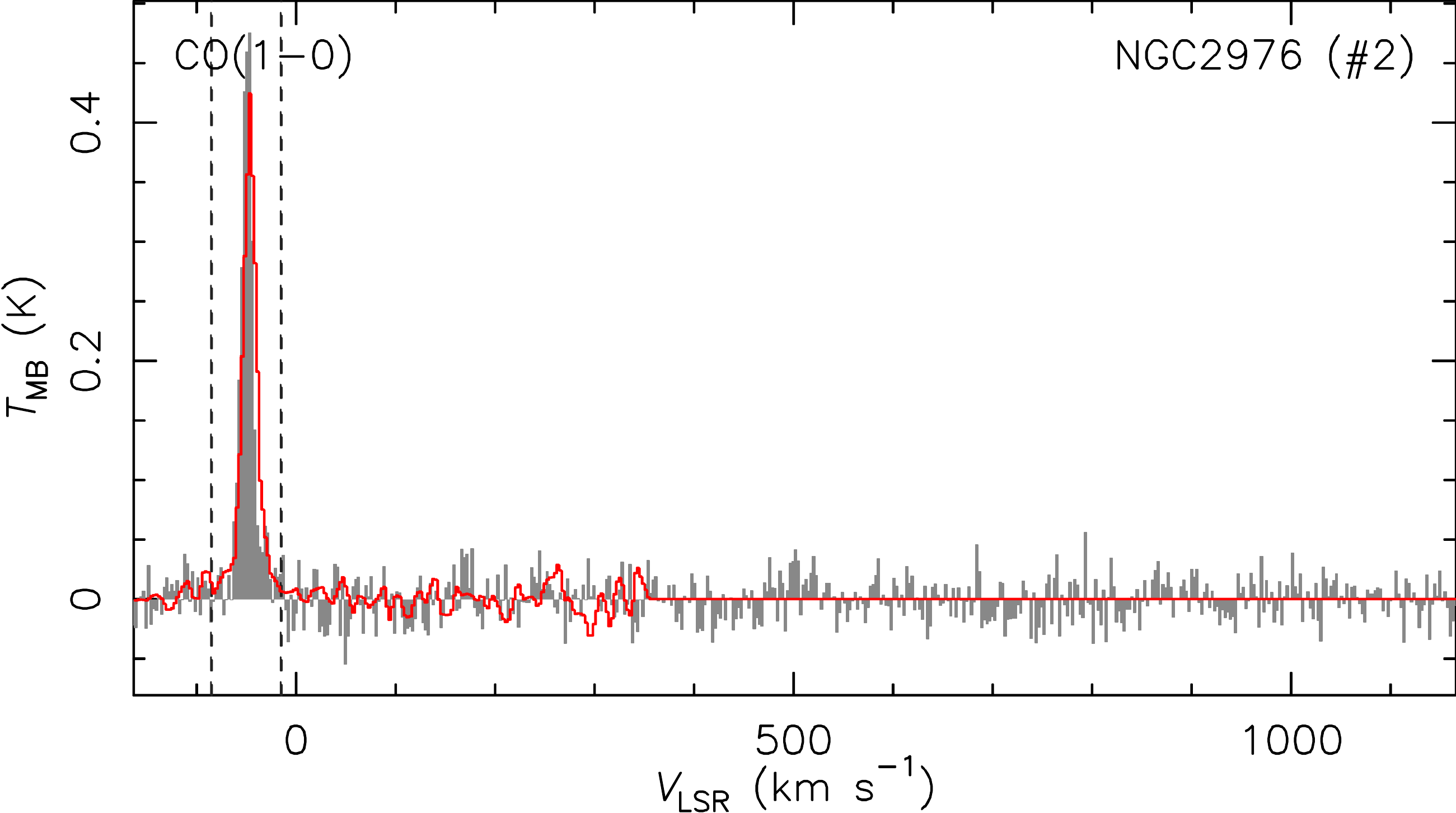} \\ 
\includegraphics[width=0.42\textwidth]{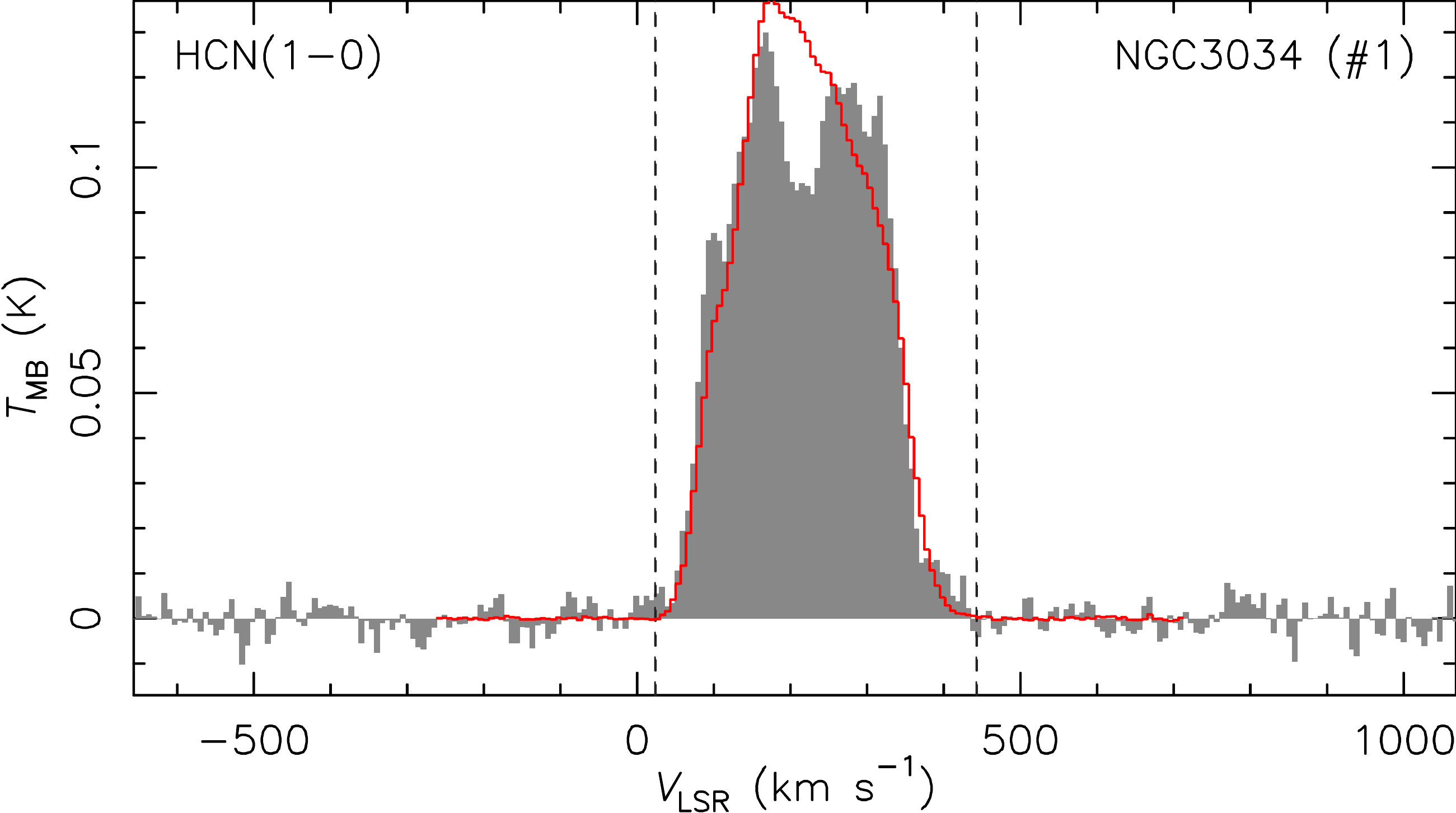} & 
\includegraphics[width=0.42\textwidth]{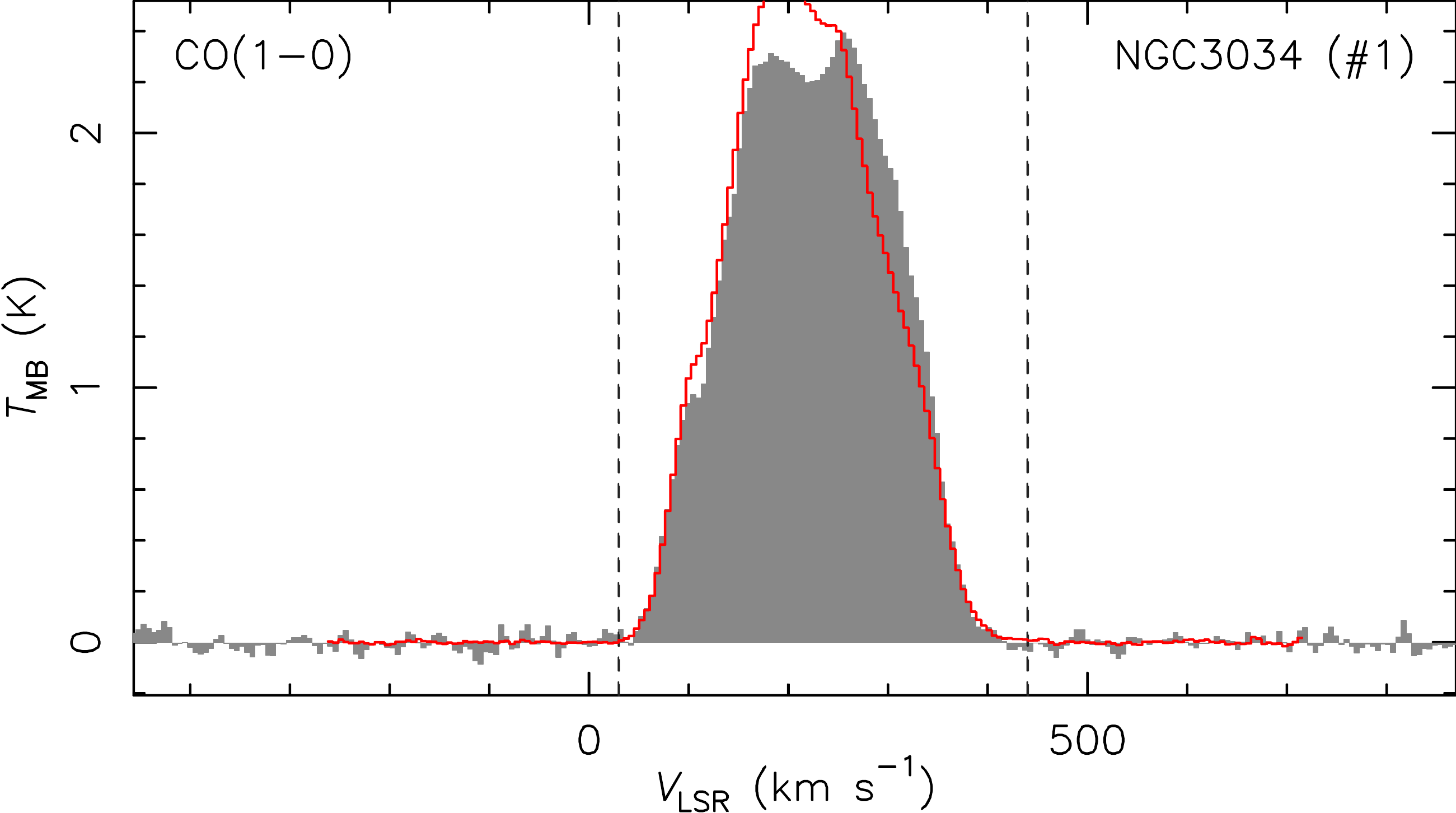} \\ 
\includegraphics[width=0.42\textwidth]{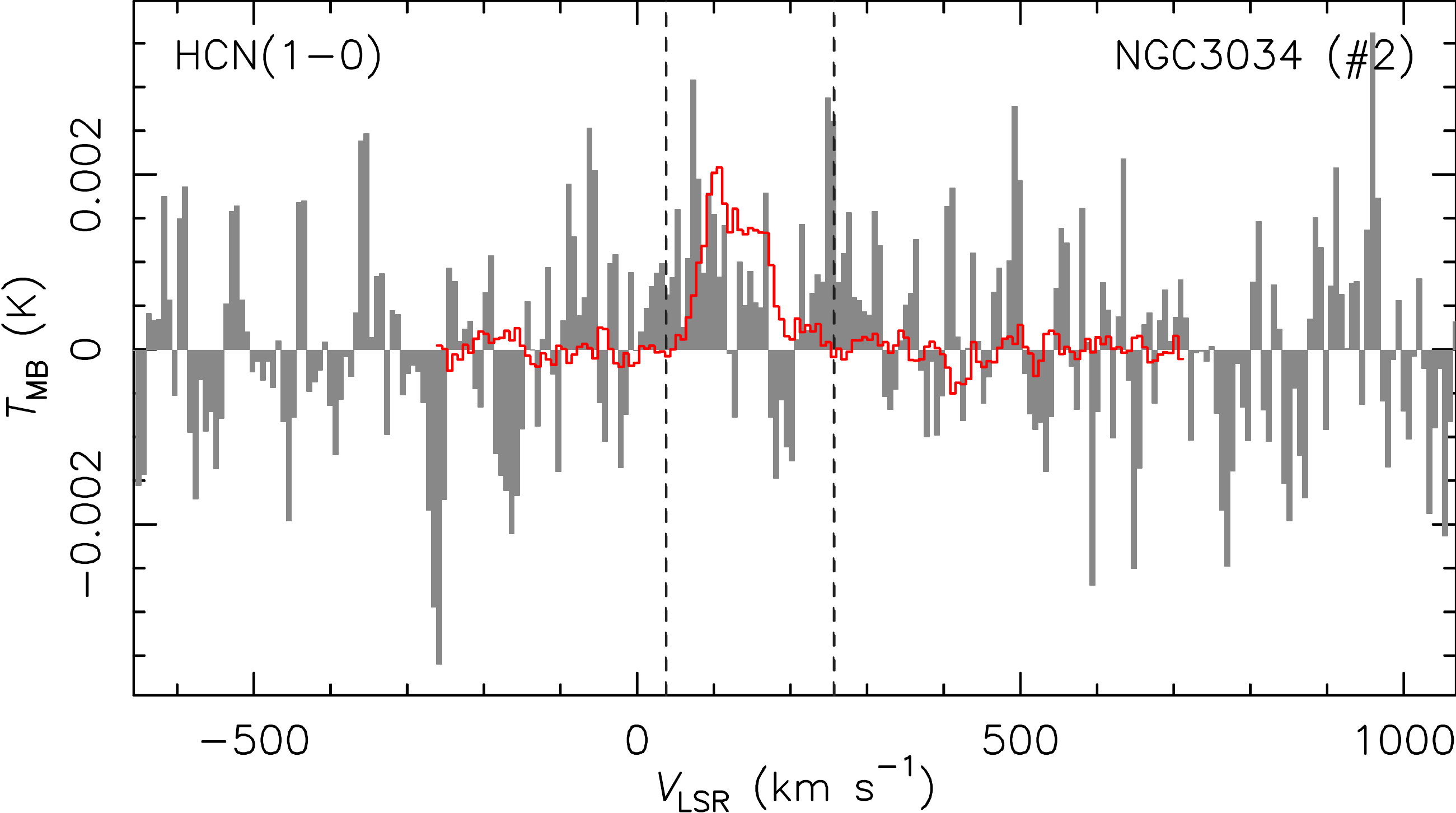} & 
\includegraphics[width=0.42\textwidth]{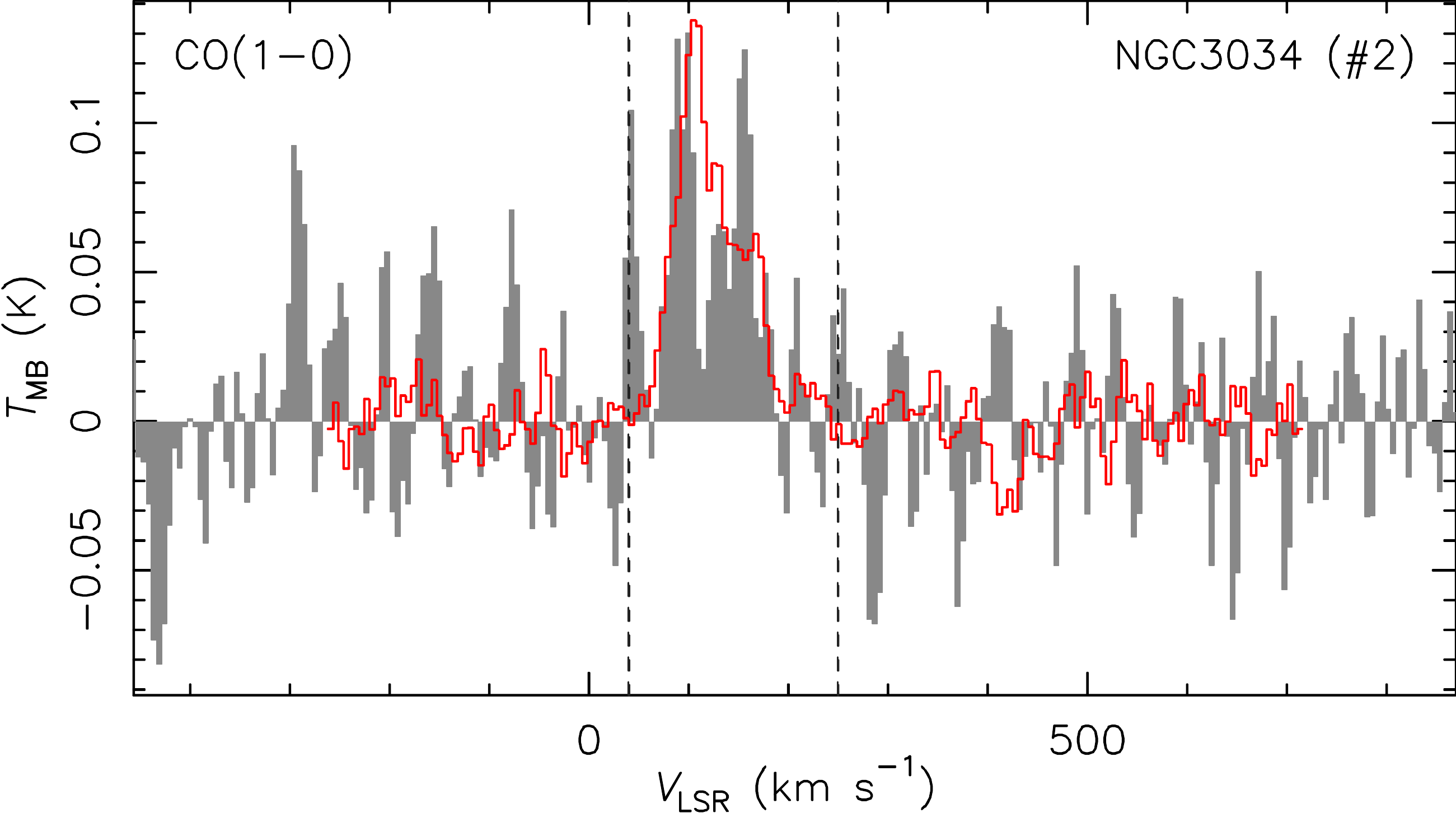} \\ 
\includegraphics[width=0.42\textwidth]{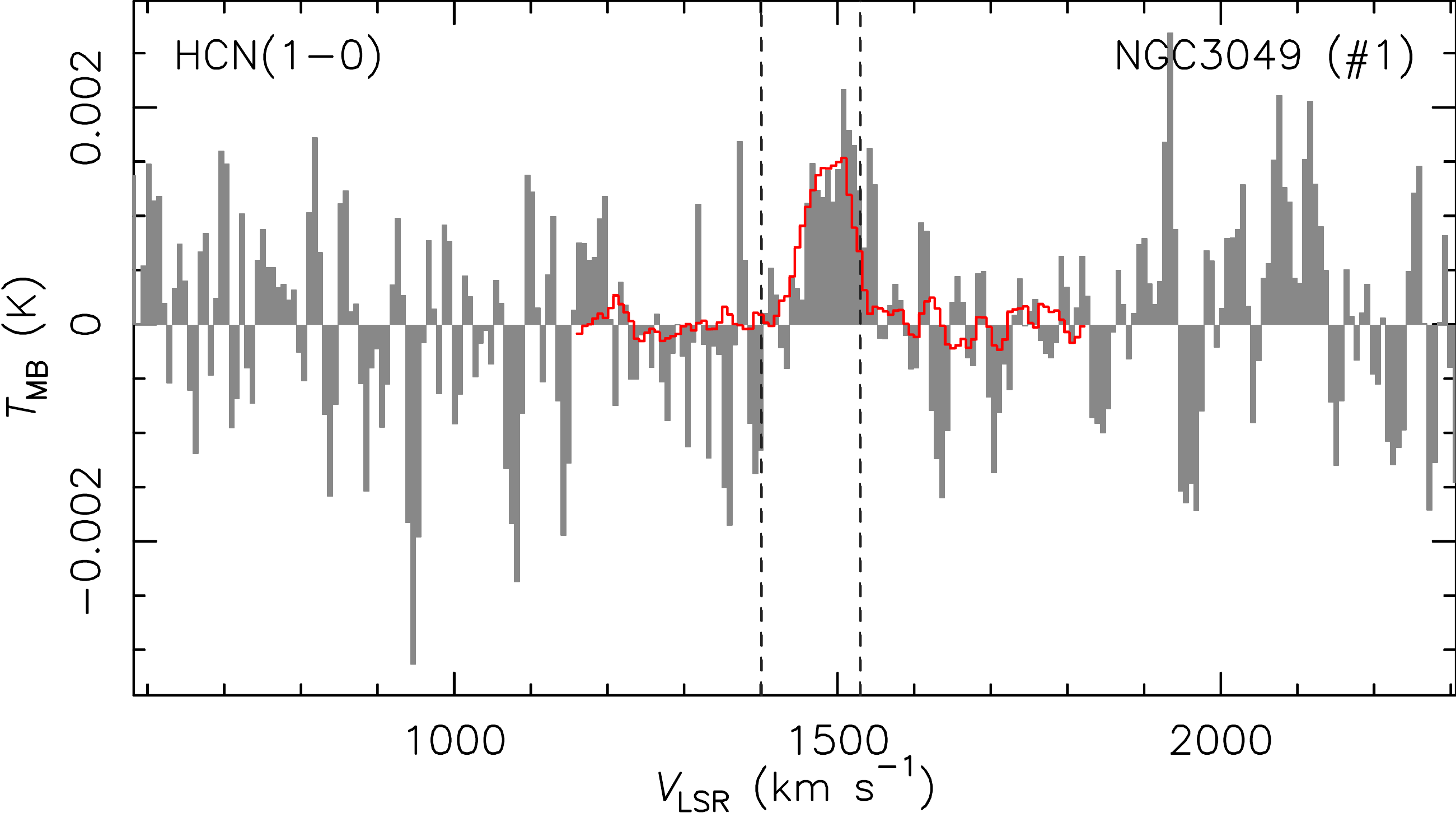} & 
\includegraphics[width=0.42\textwidth]{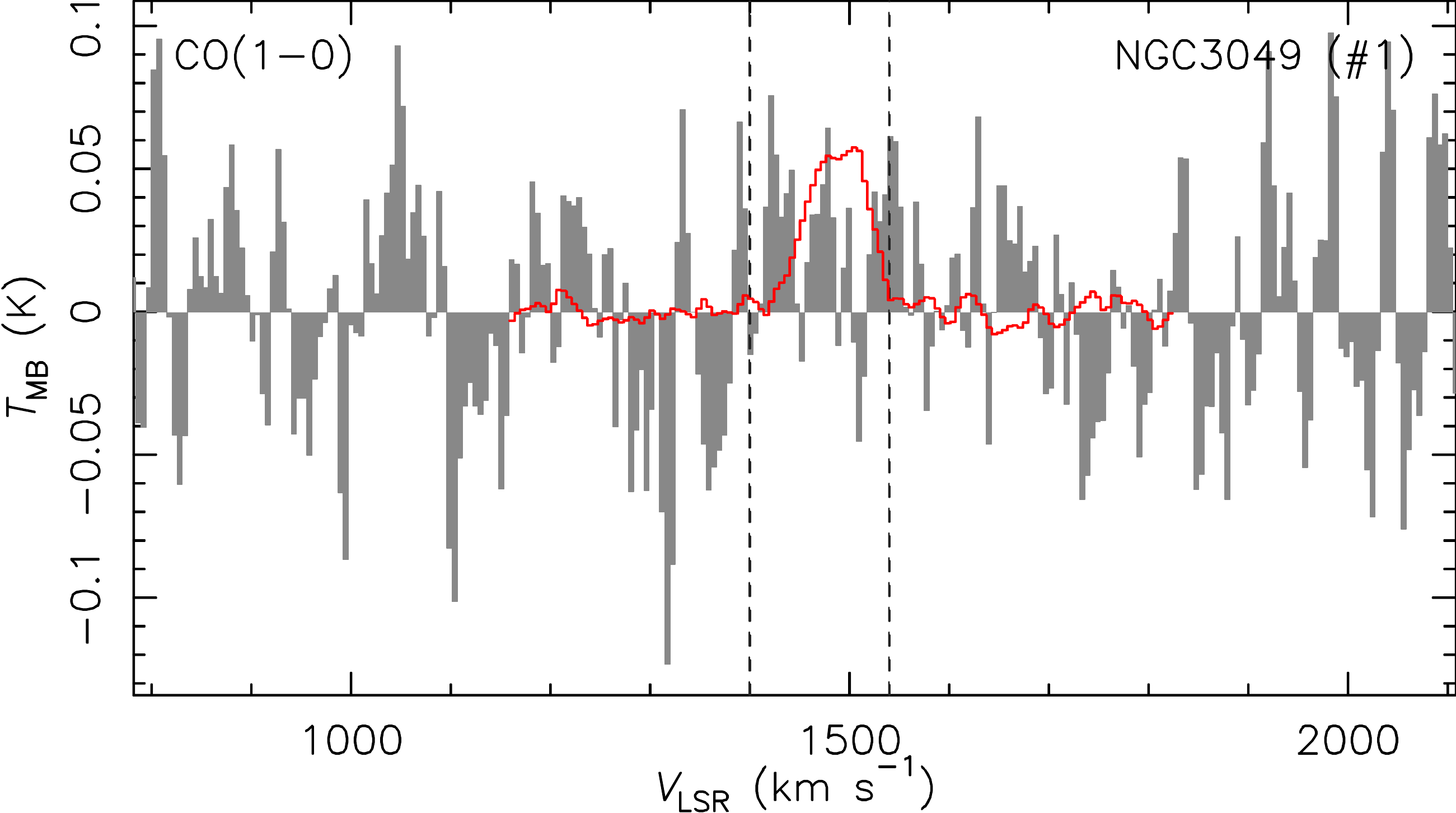} \\ 
\end{tabular}
\end{center}  
\caption{Same as Fig.~\ref{f-spec-1} for NGC~2976, NGC~3034, and NGC3049. } 
\end{figure} 
\newpage 
\begin{figure}[h!]
\begin{center}  
\begin{tabular}{cc} 
\includegraphics[width=0.42\textwidth]{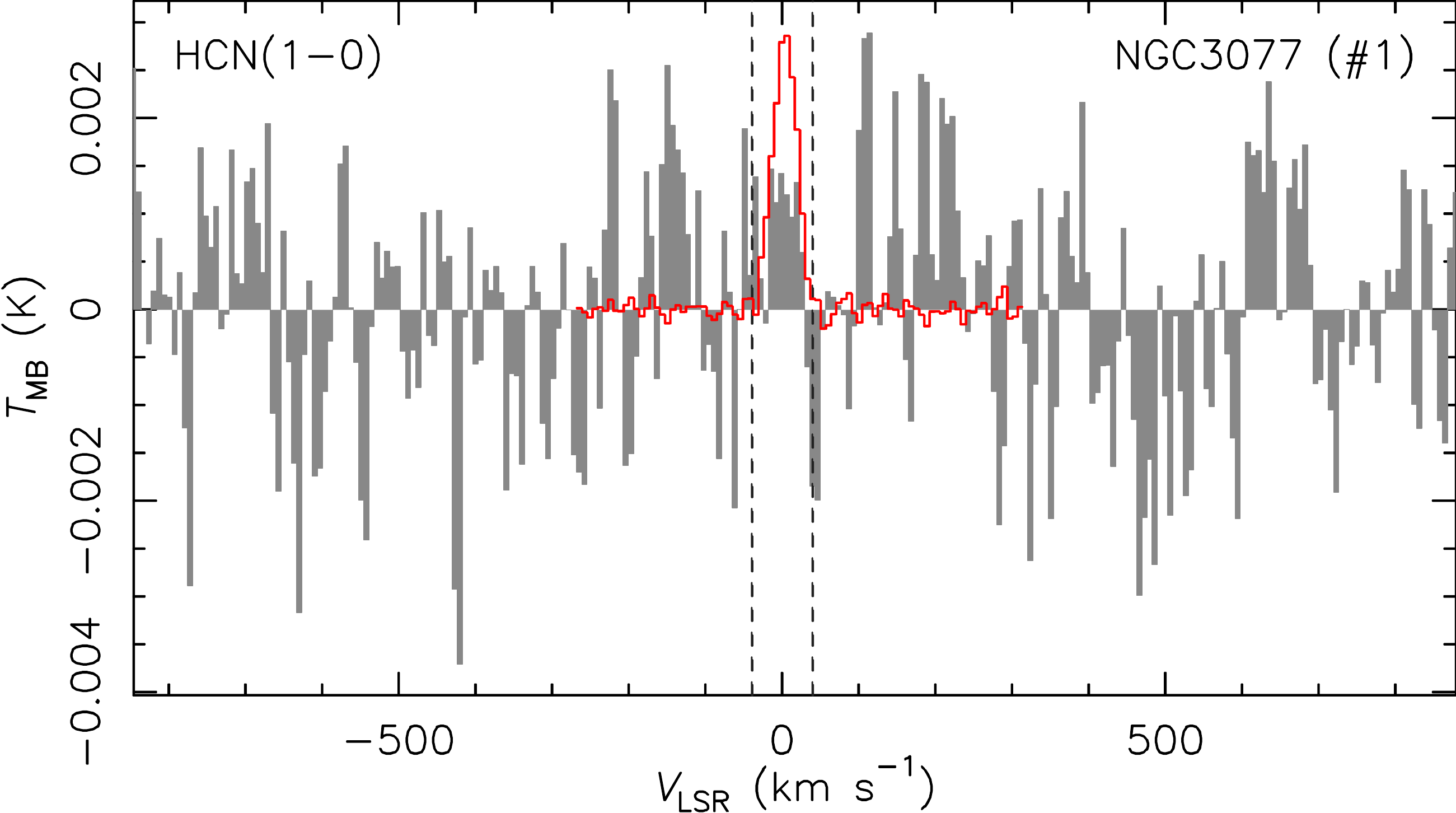} & 
\includegraphics[width=0.42\textwidth]{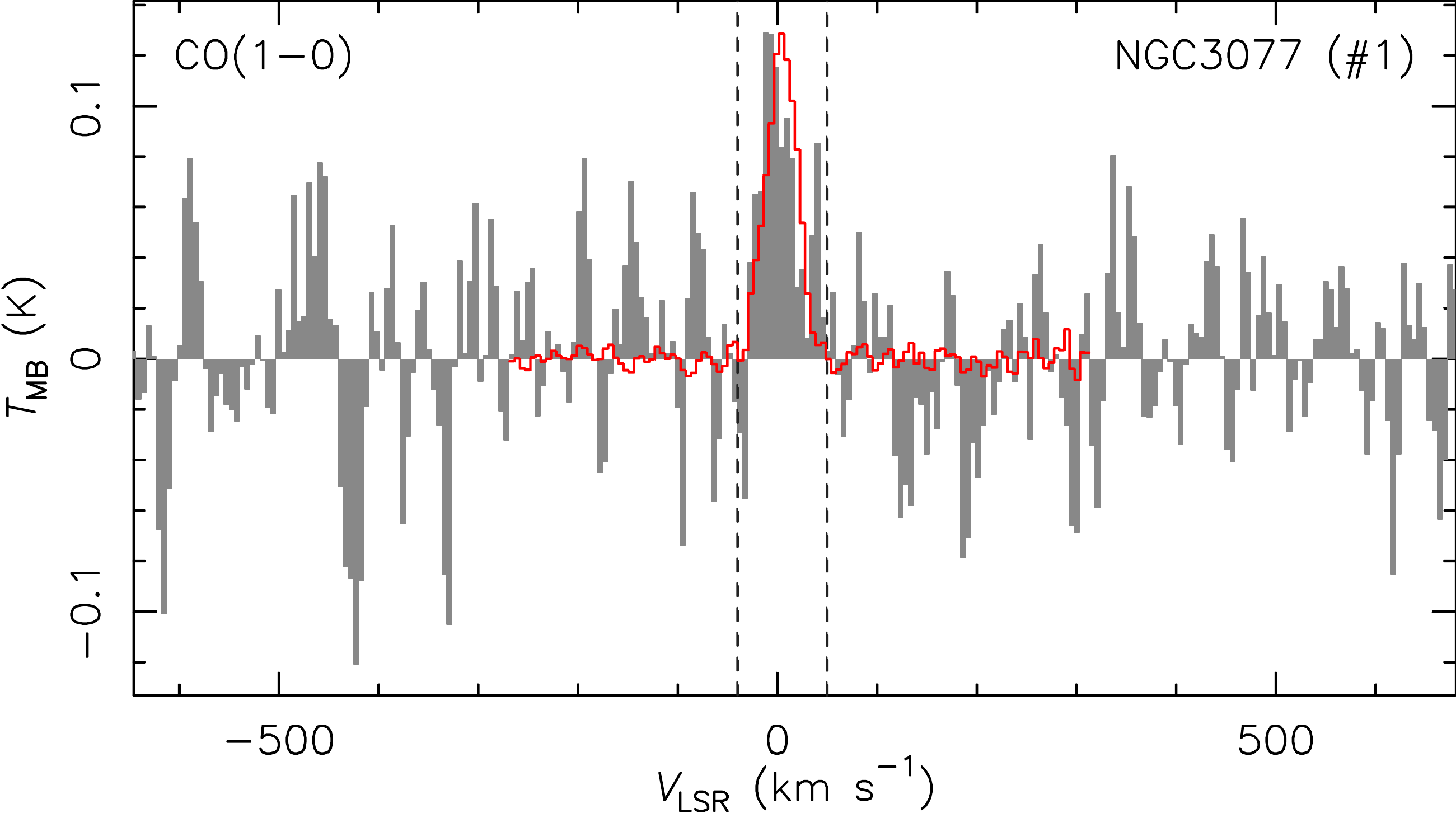} \\ 
\includegraphics[width=0.42\textwidth]{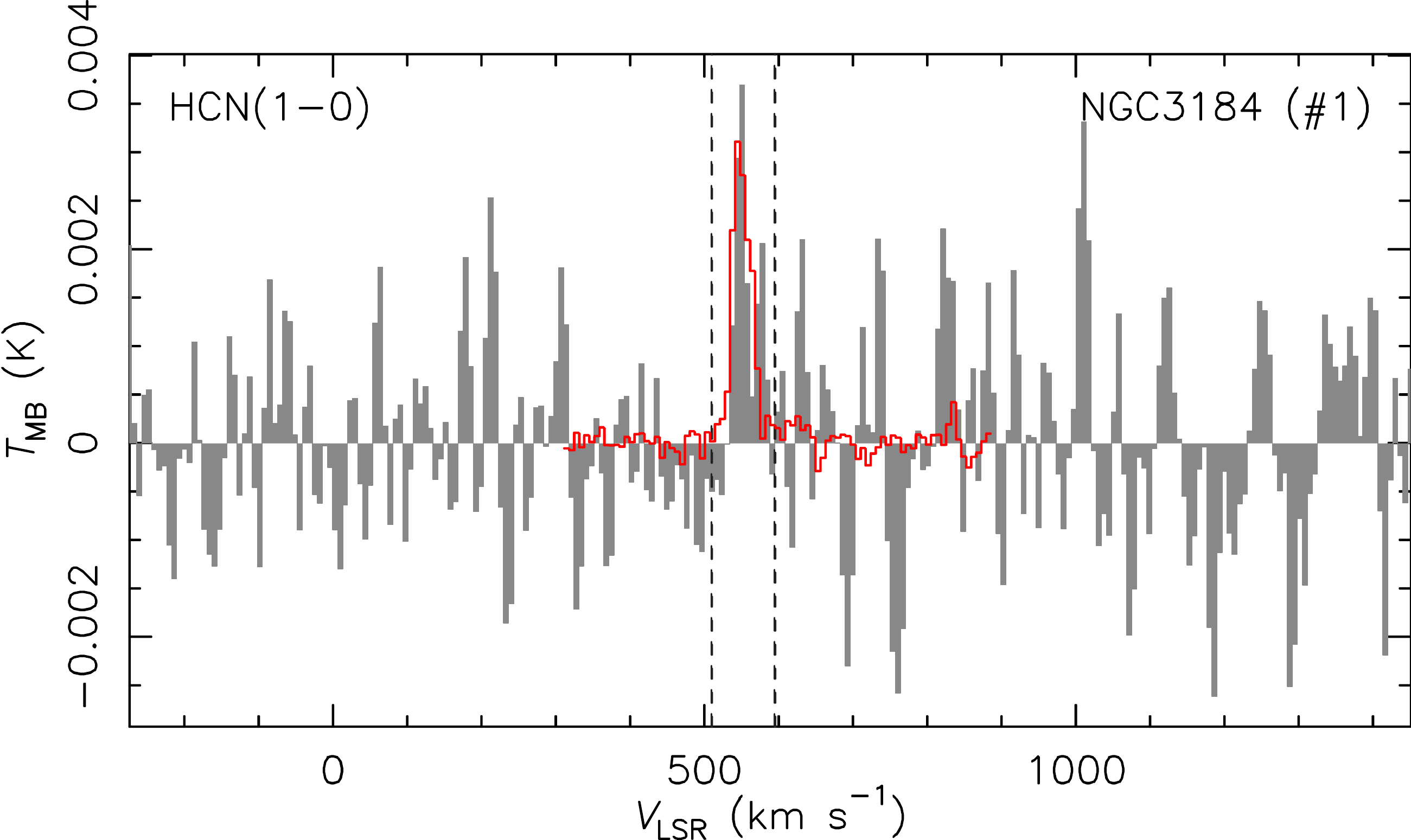} & 
\includegraphics[width=0.42\textwidth]{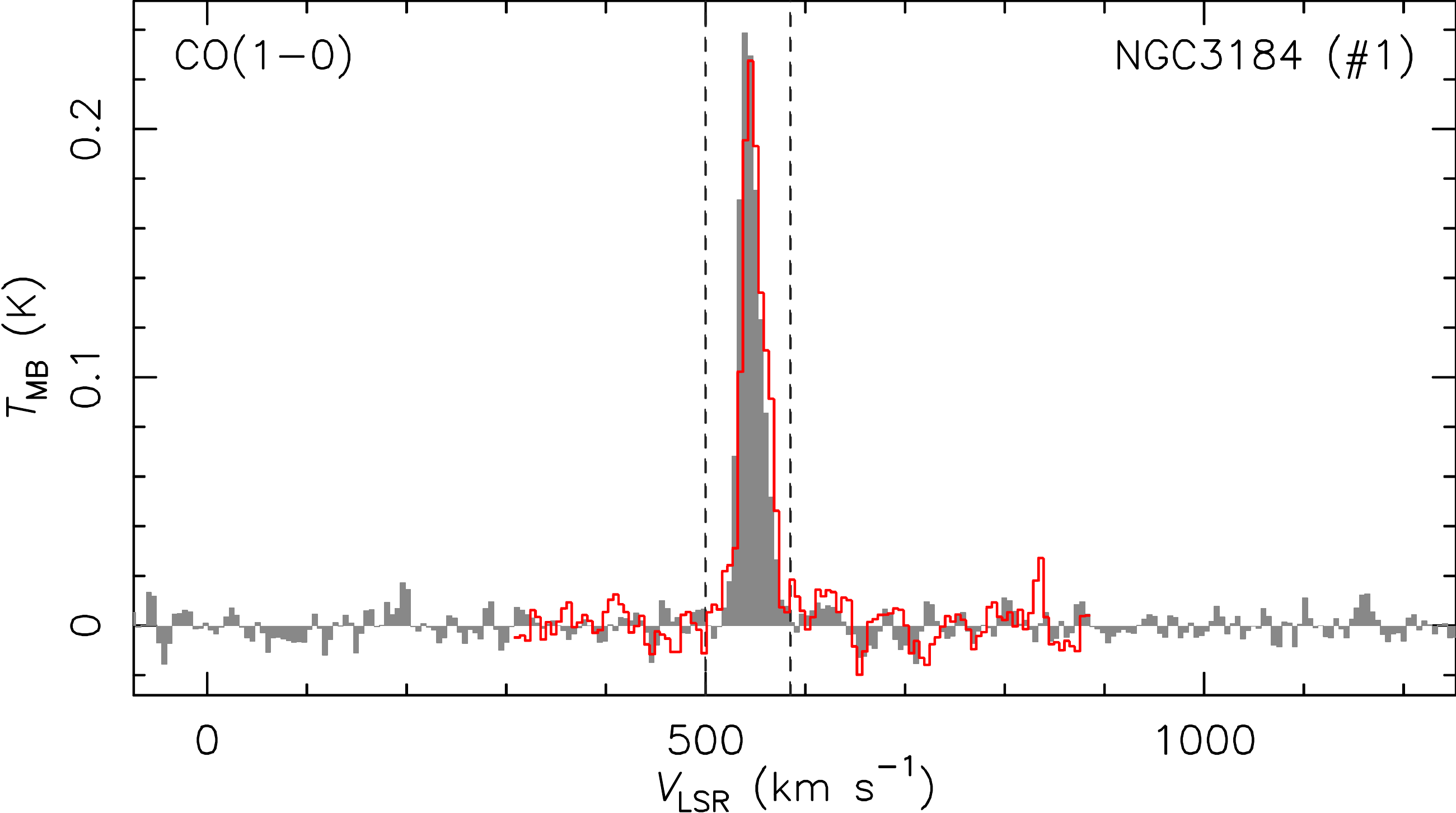} \\ 
\includegraphics[width=0.42\textwidth]{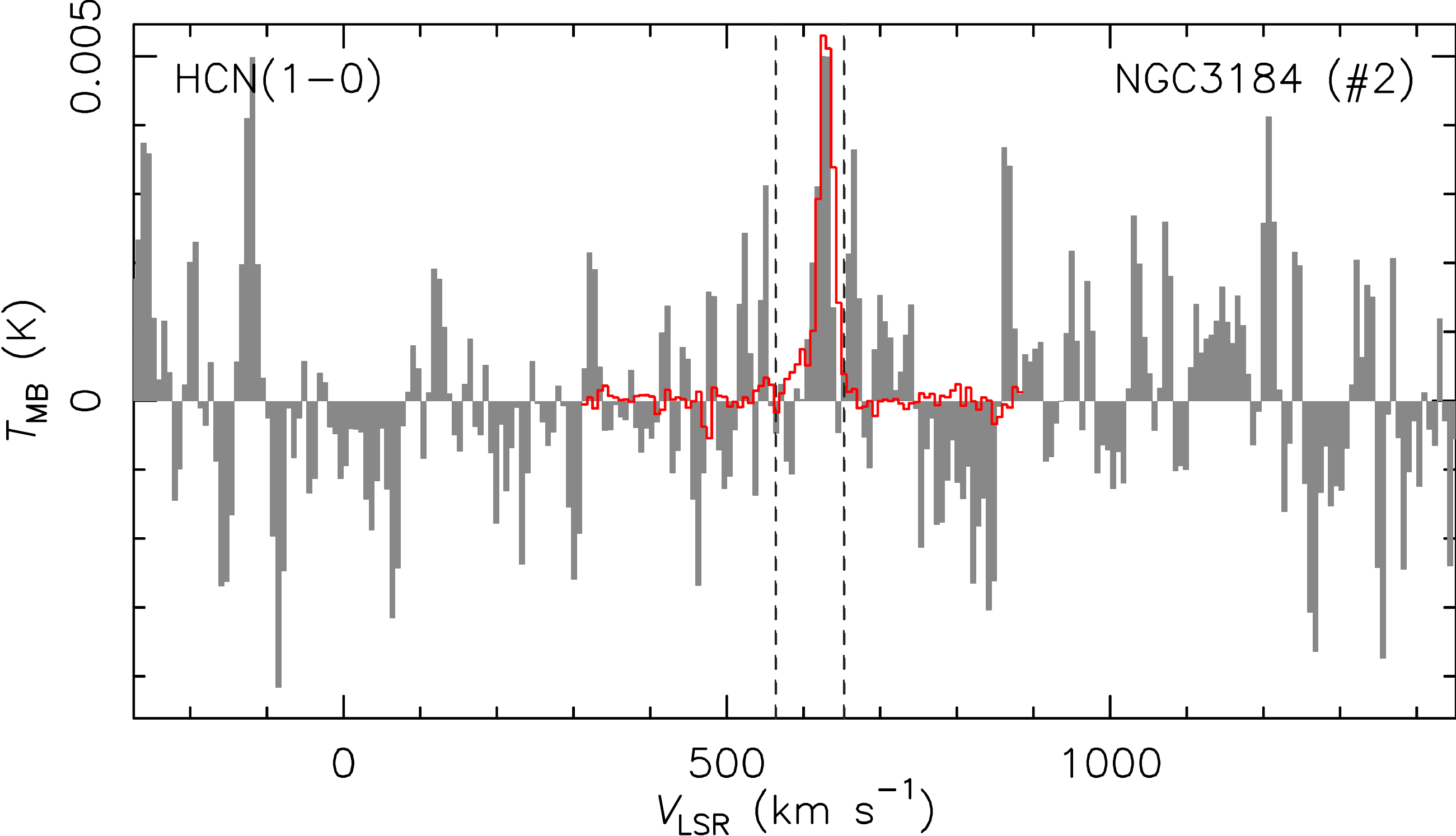} & 
\includegraphics[width=0.42\textwidth]{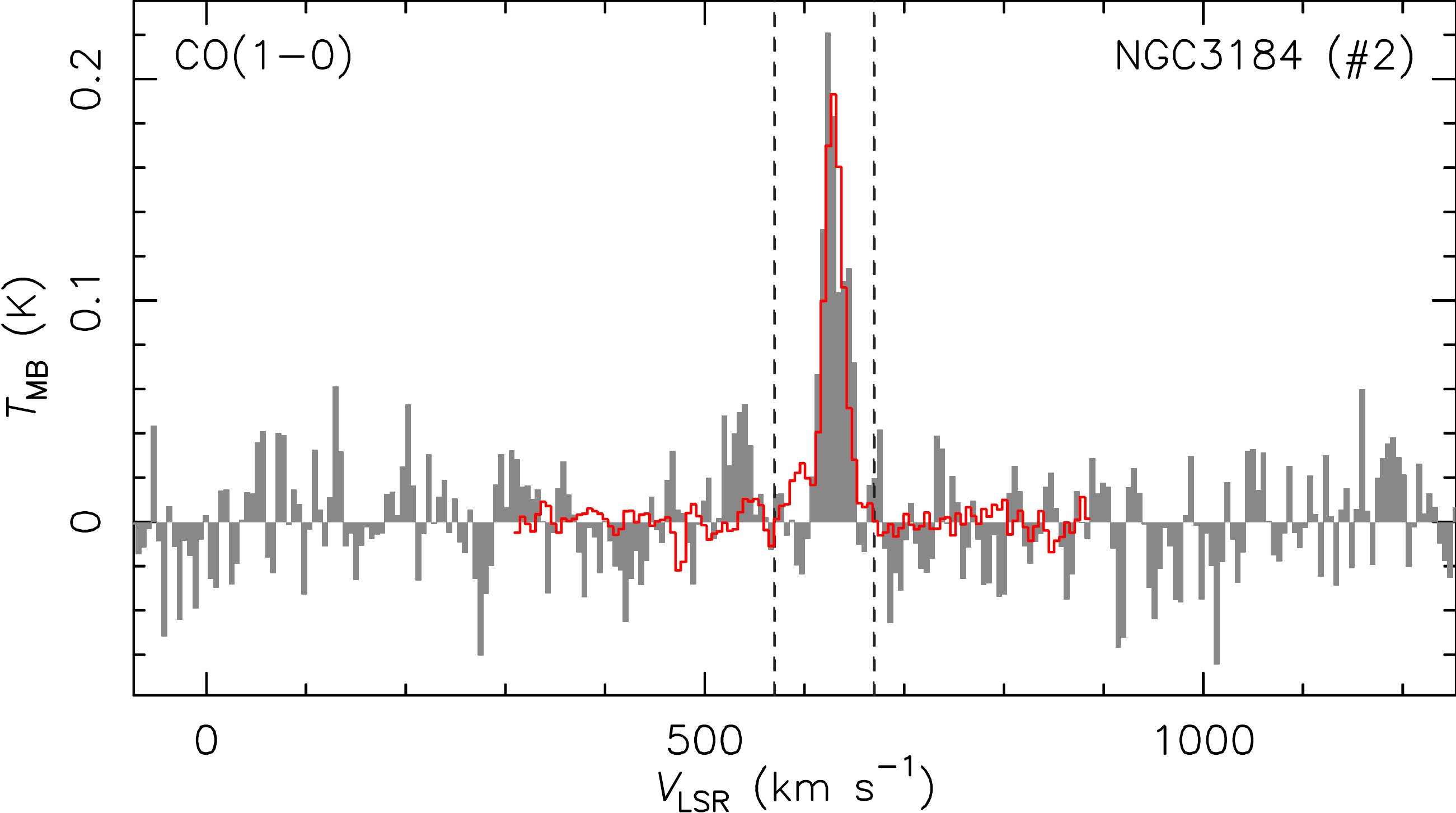} \\ 
\includegraphics[width=0.42\textwidth]{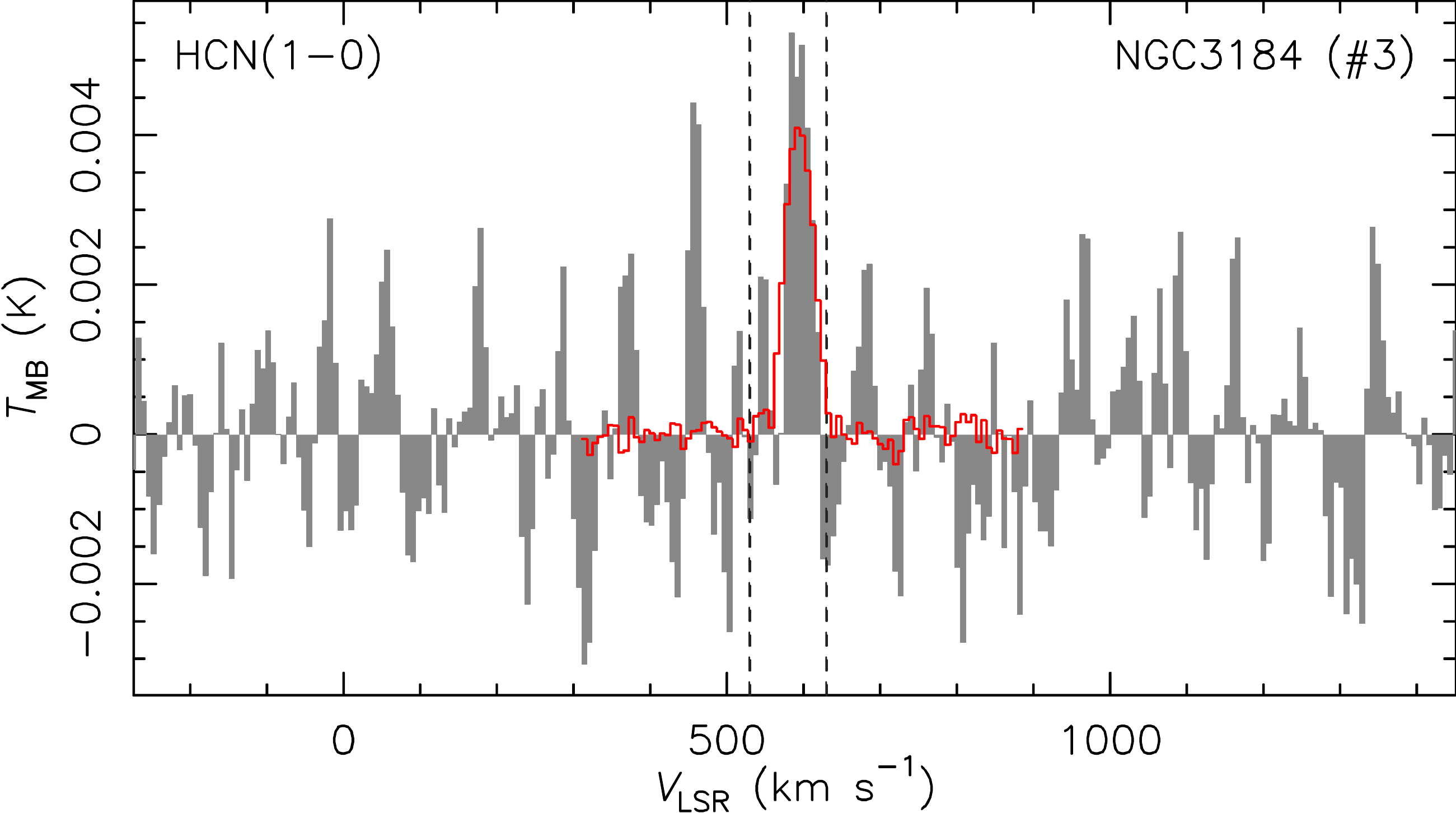} & 
\includegraphics[width=0.42\textwidth]{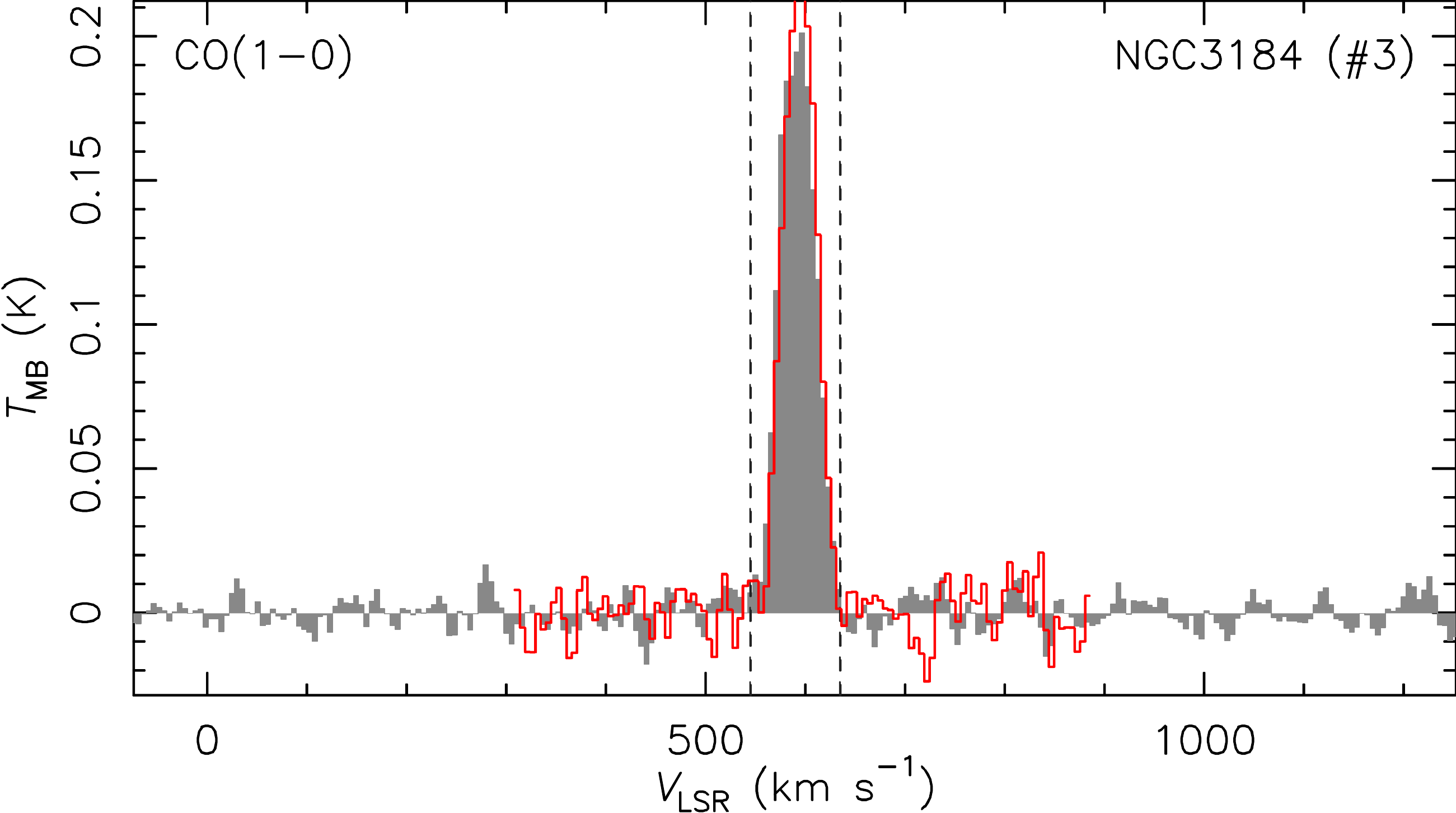} \\ 
\includegraphics[width=0.42\textwidth]{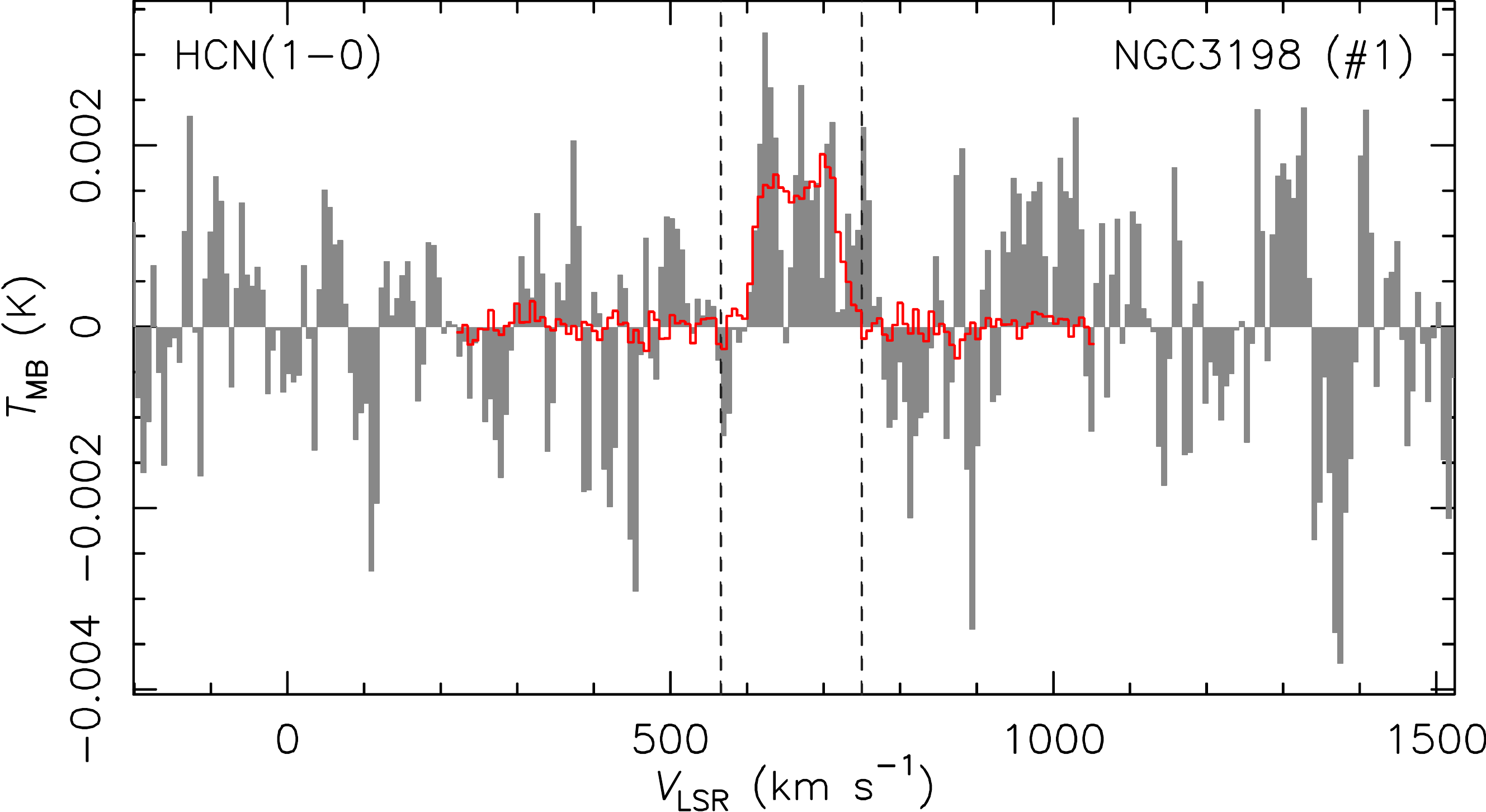} & 
\includegraphics[width=0.42\textwidth]{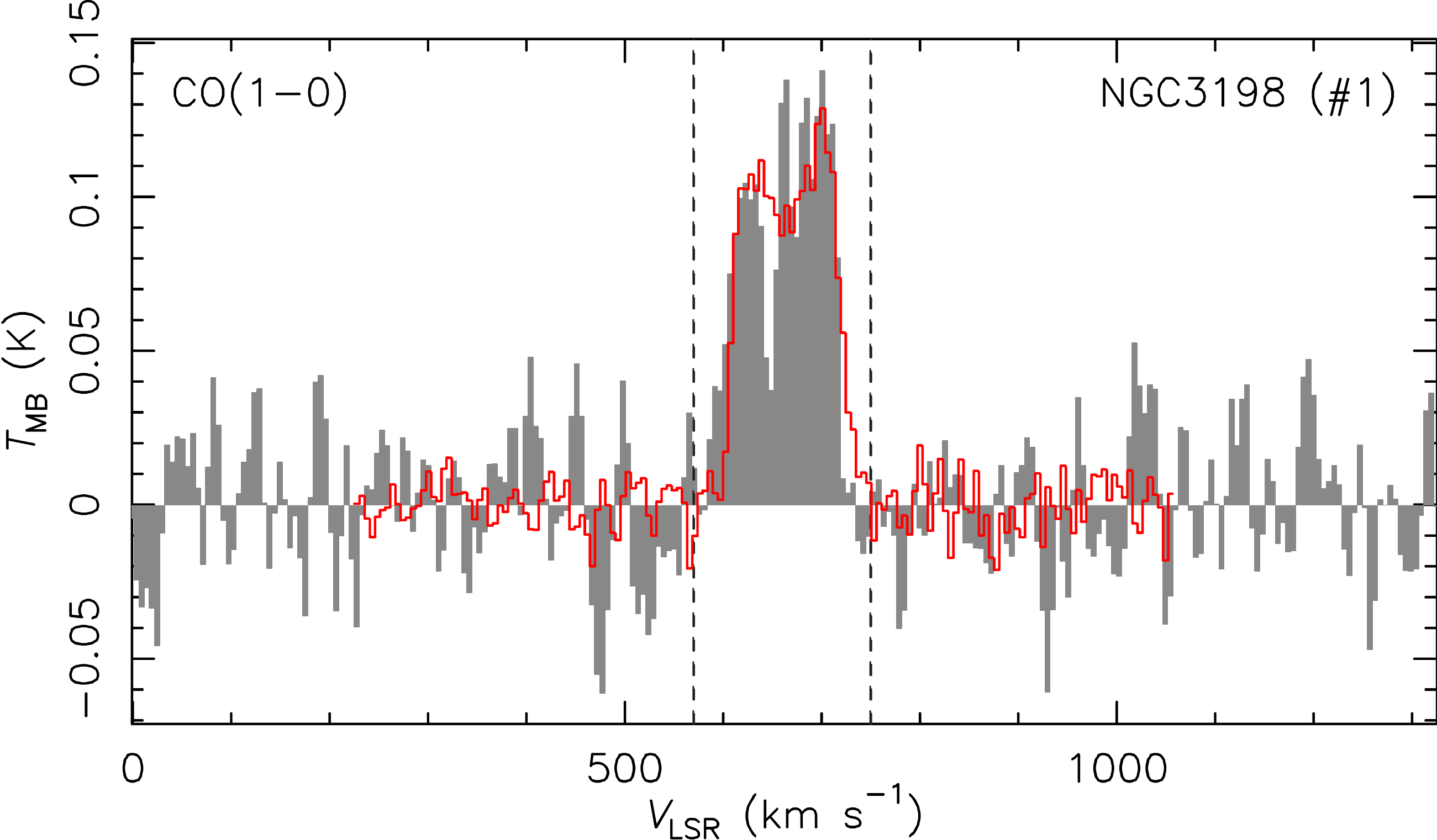} \\ 
\end{tabular}
\end{center}  
\caption{Same as Fig.~\ref{f-spec-1} for NGC~3077, NGC~3184, and NGC~3198.} 
\end{figure} 
\newpage 
\begin{figure}[h!]
\begin{center}  
\begin{tabular}{cc} 
\includegraphics[width=0.42\textwidth]{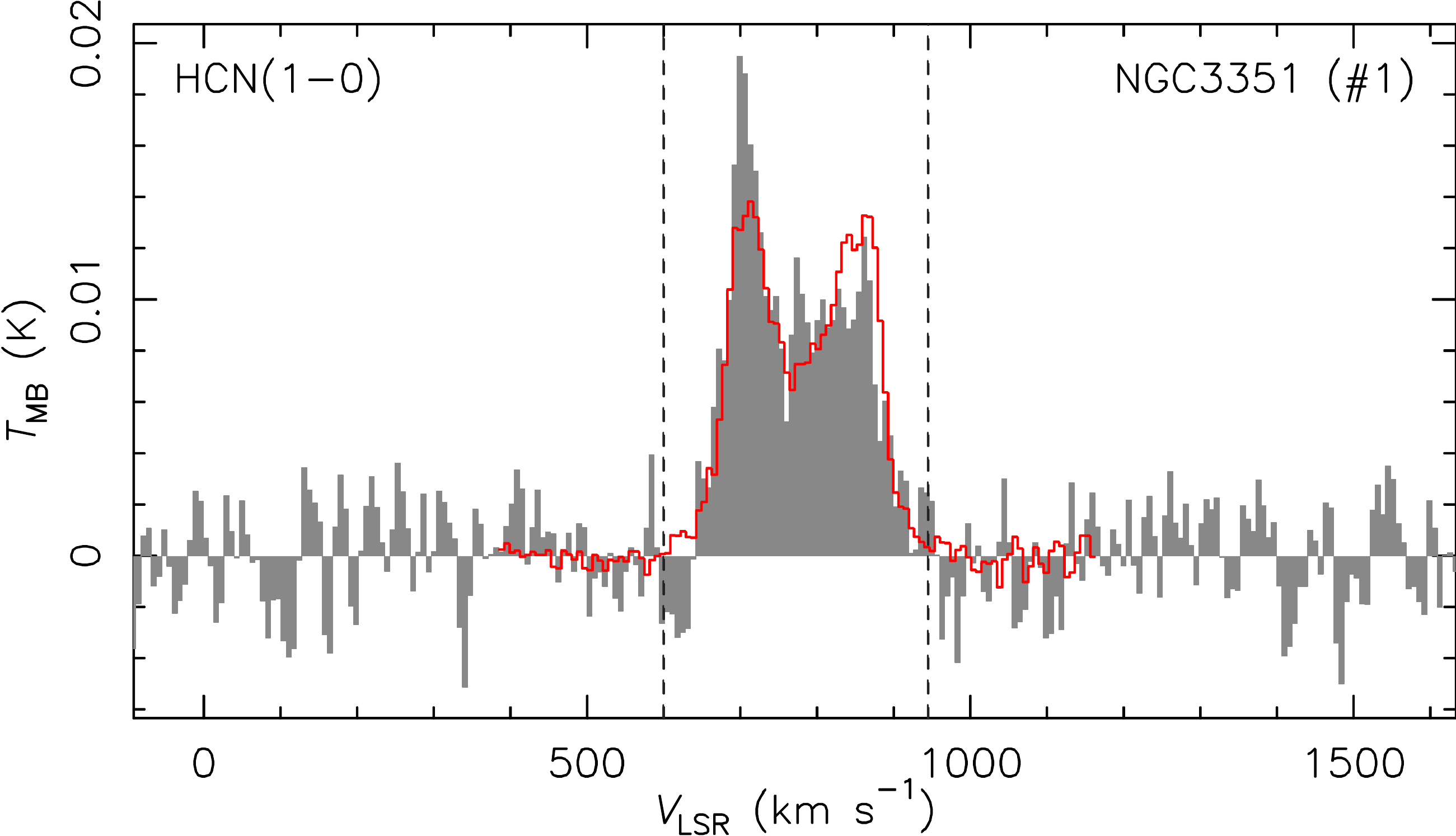} & 
\includegraphics[width=0.42\textwidth]{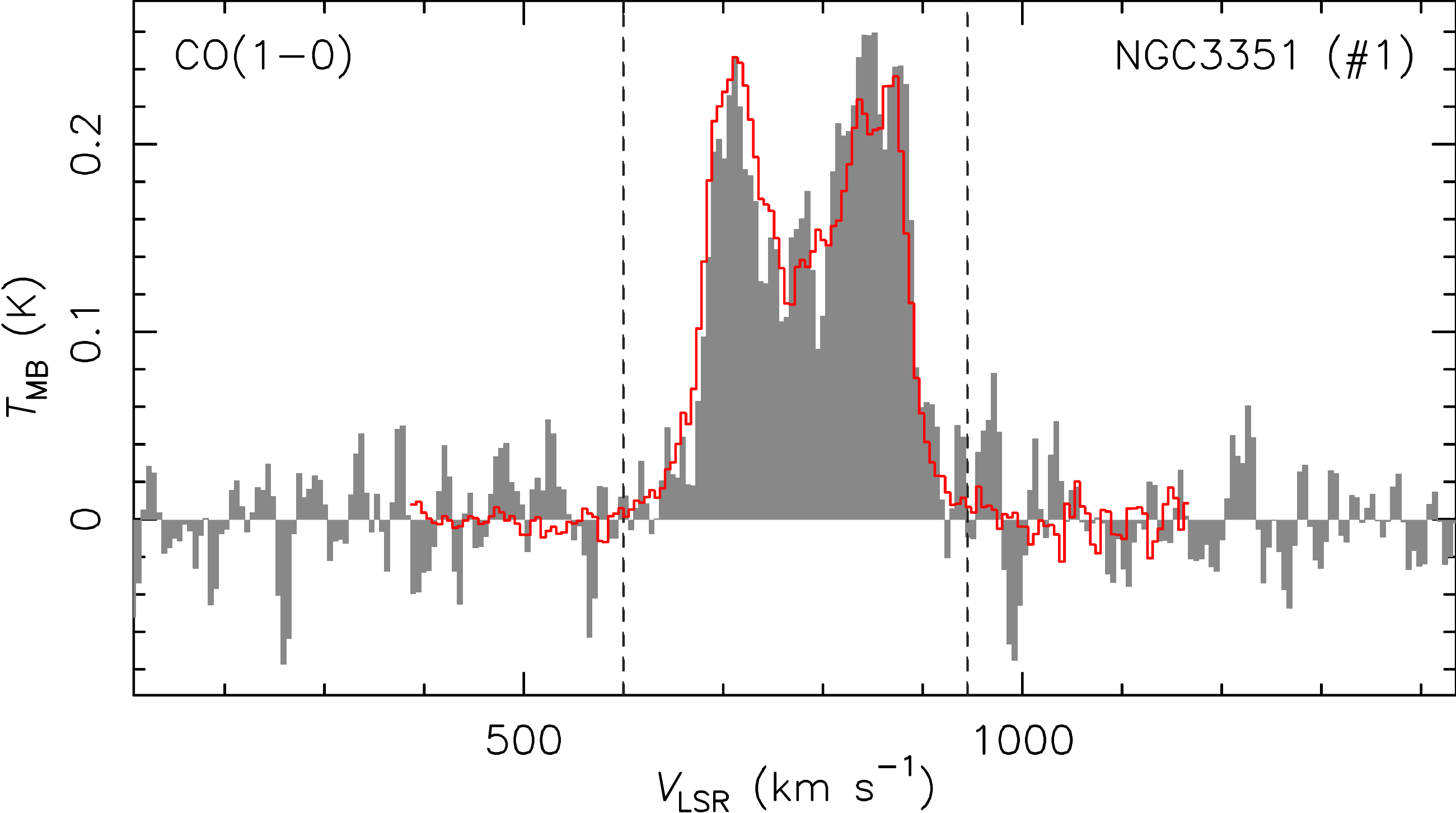} \\ 
\includegraphics[width=0.42\textwidth]{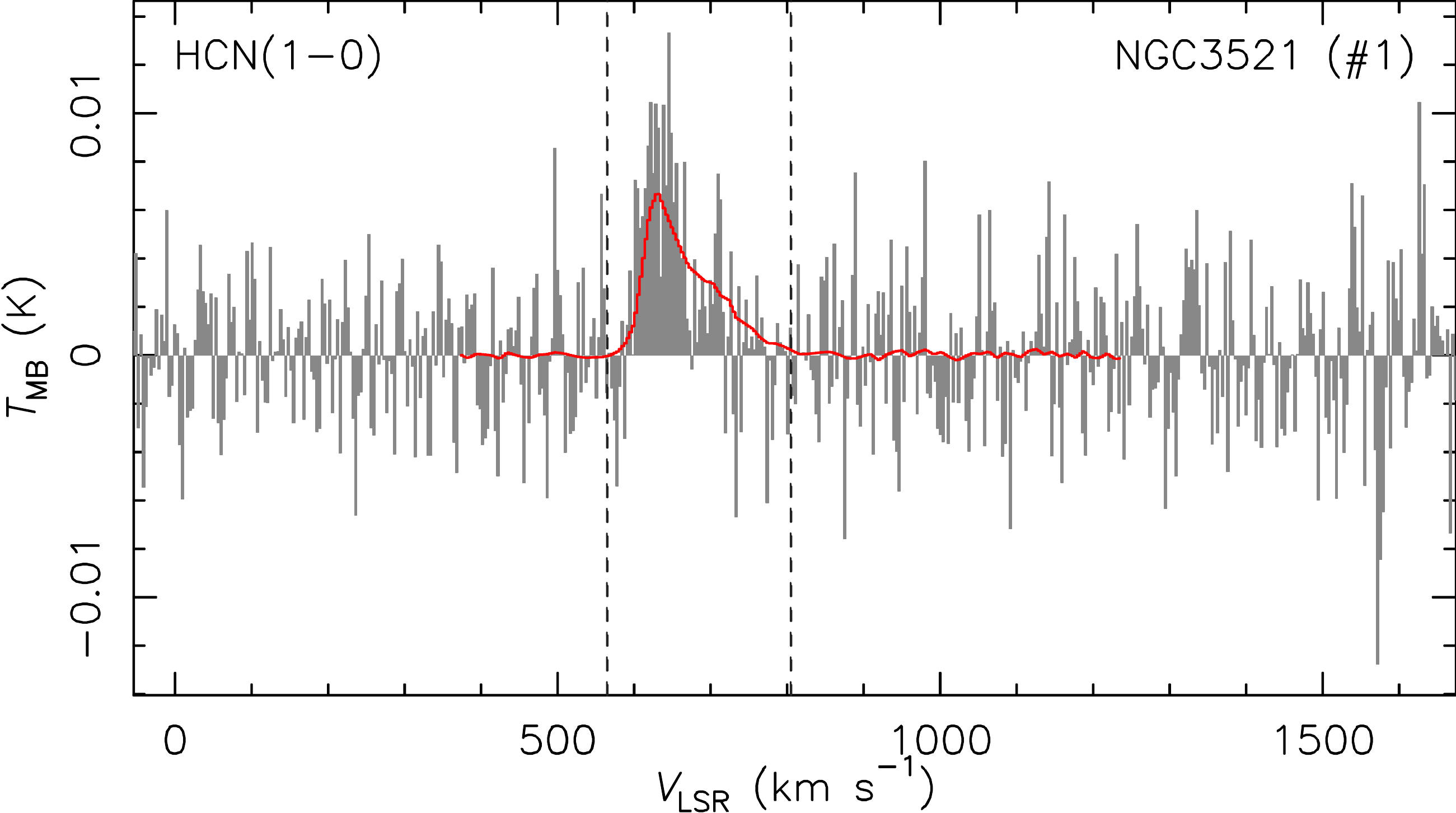} & 
\includegraphics[width=0.42\textwidth]{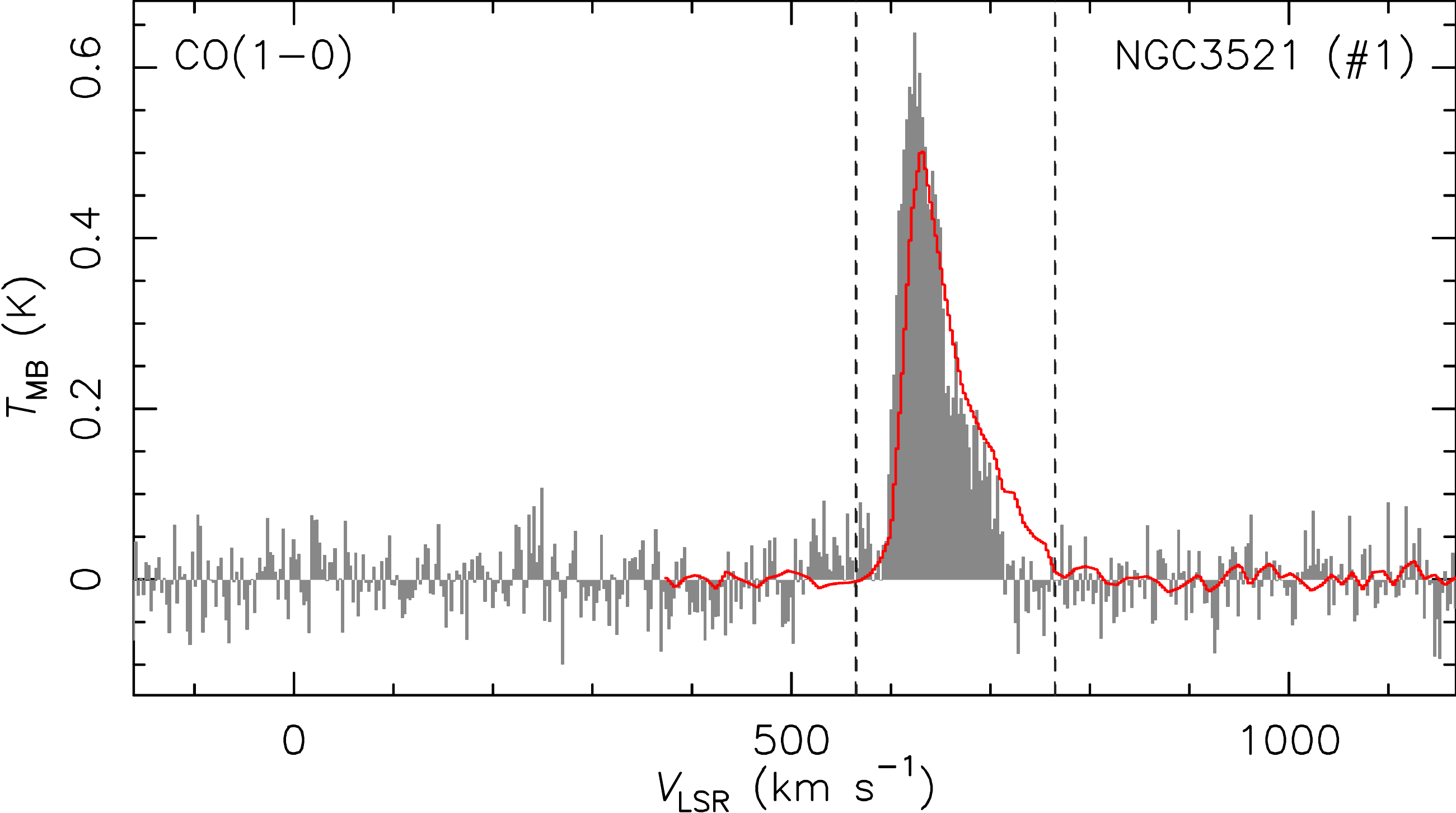} \\ 
\includegraphics[width=0.42\textwidth]{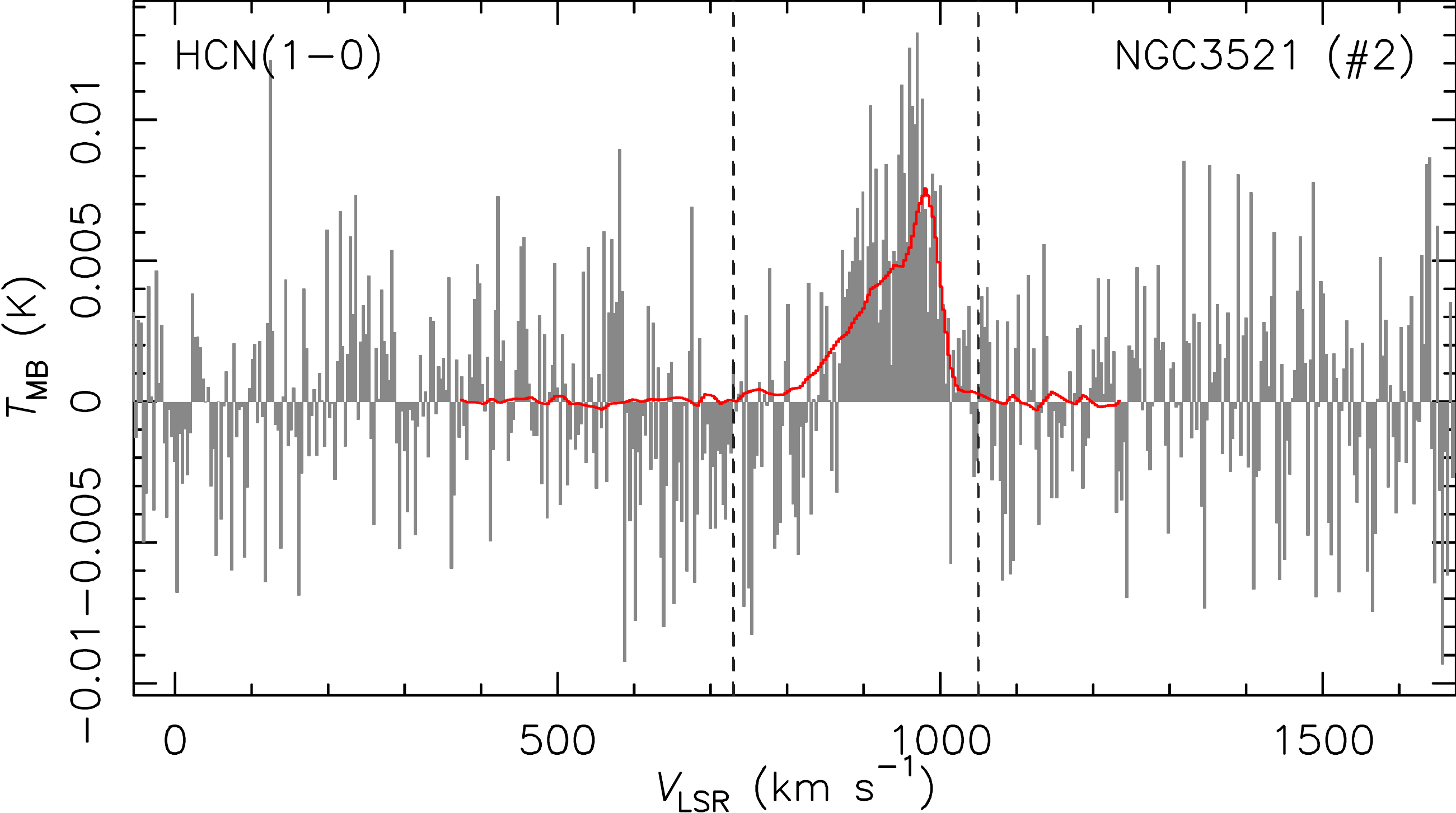} & 
\includegraphics[width=0.42\textwidth]{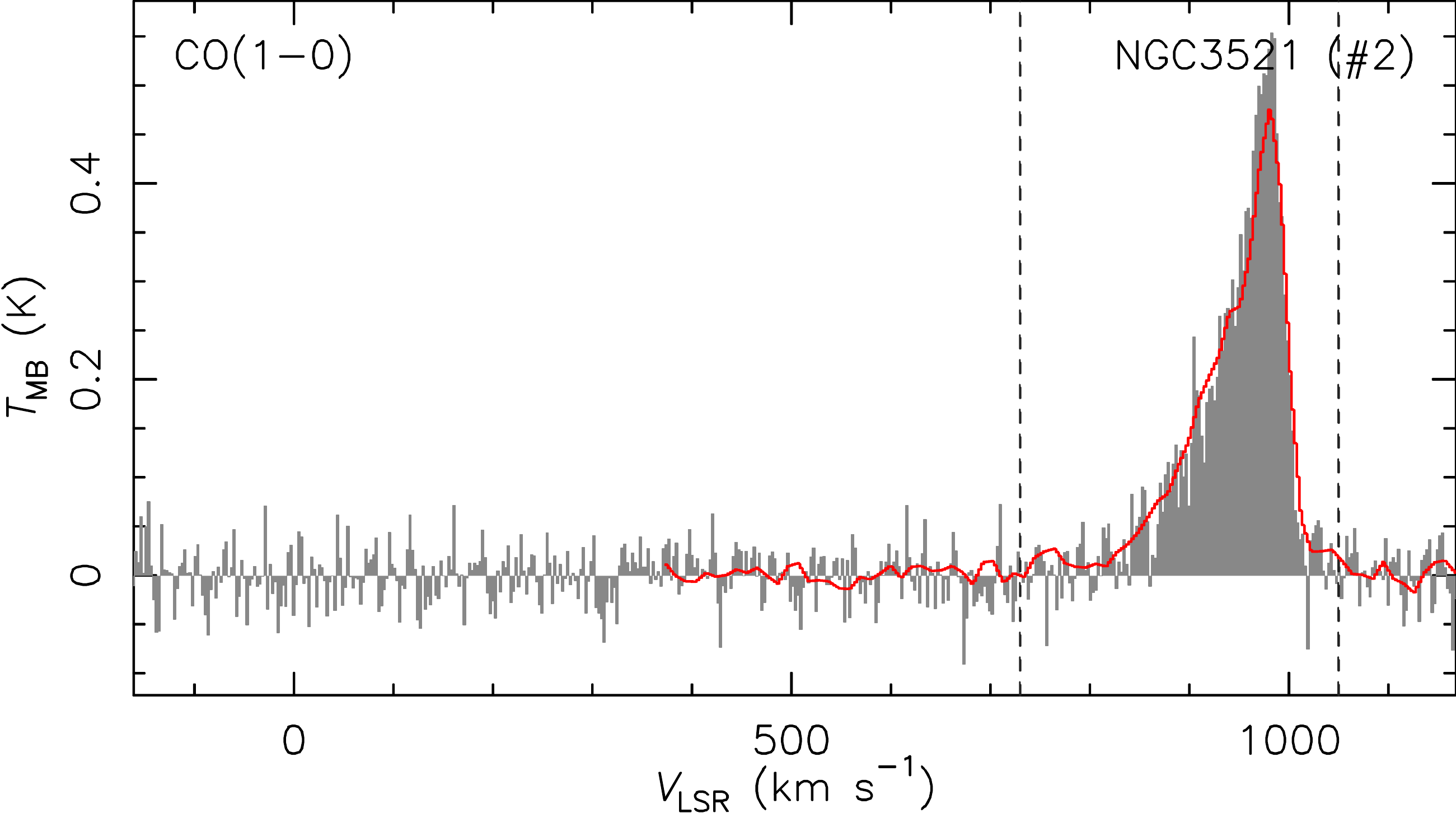} \\ 
\includegraphics[width=0.42\textwidth]{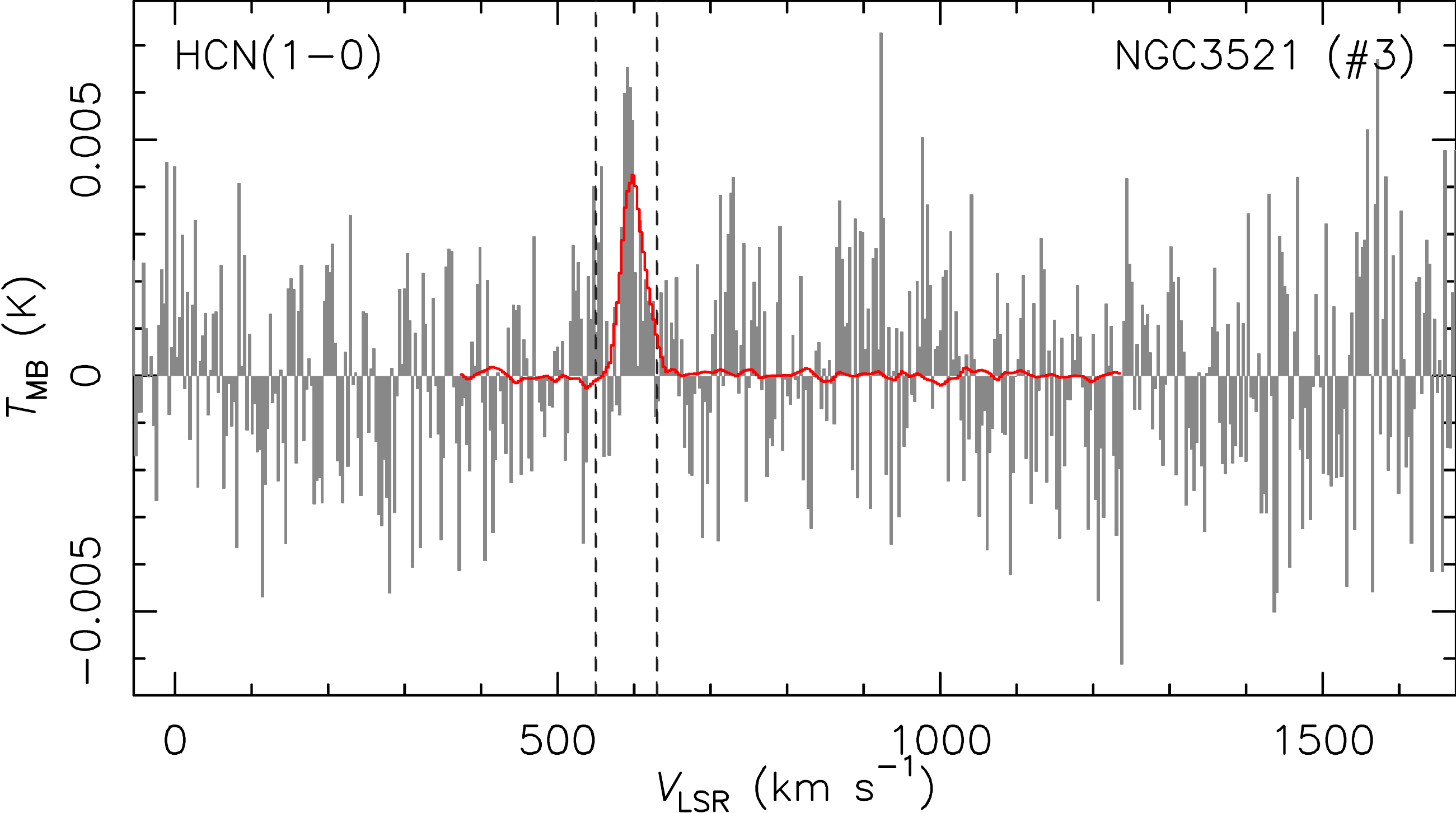} & 
\includegraphics[width=0.42\textwidth]{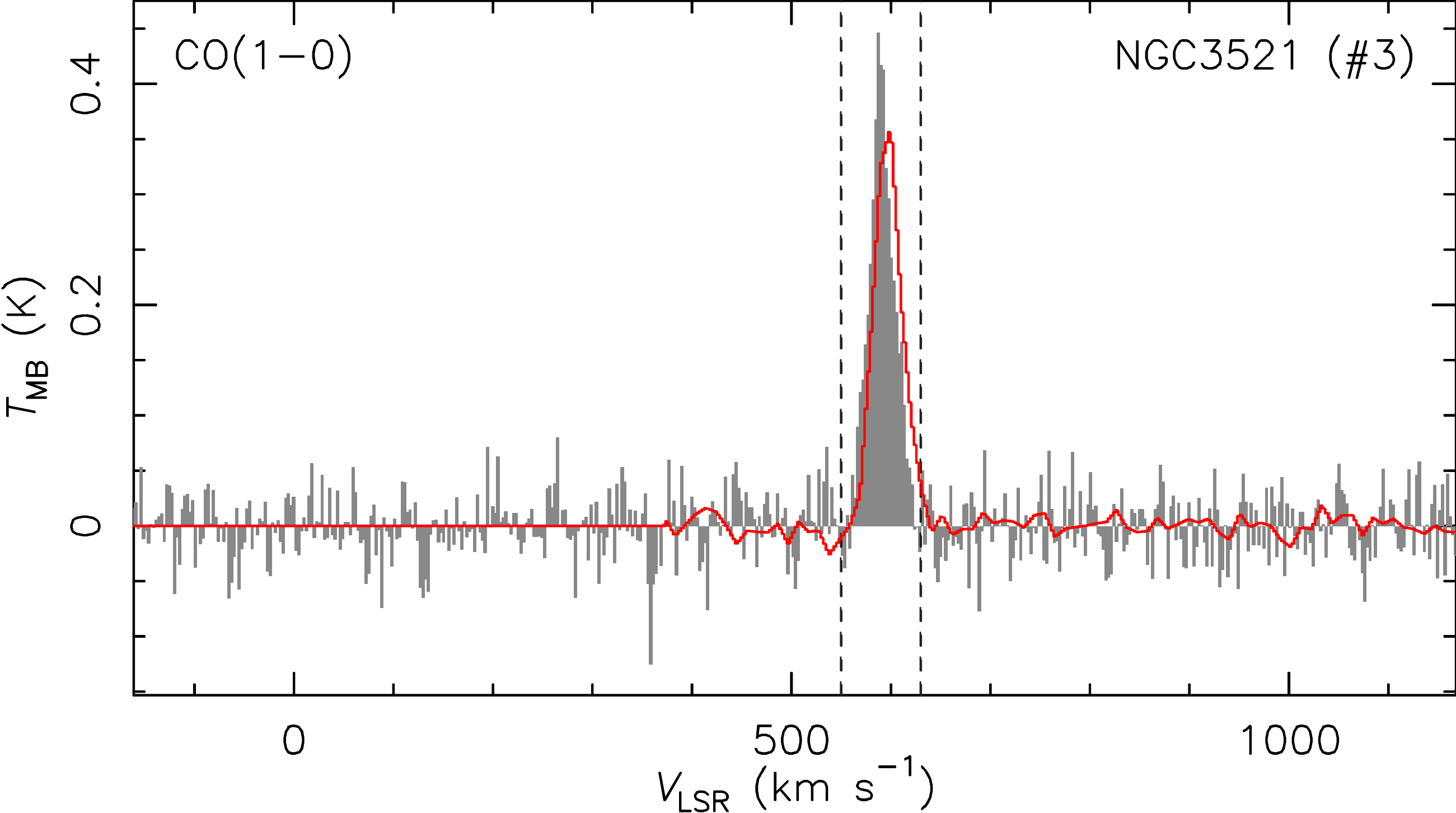} \\ 
\includegraphics[width=0.42\textwidth]{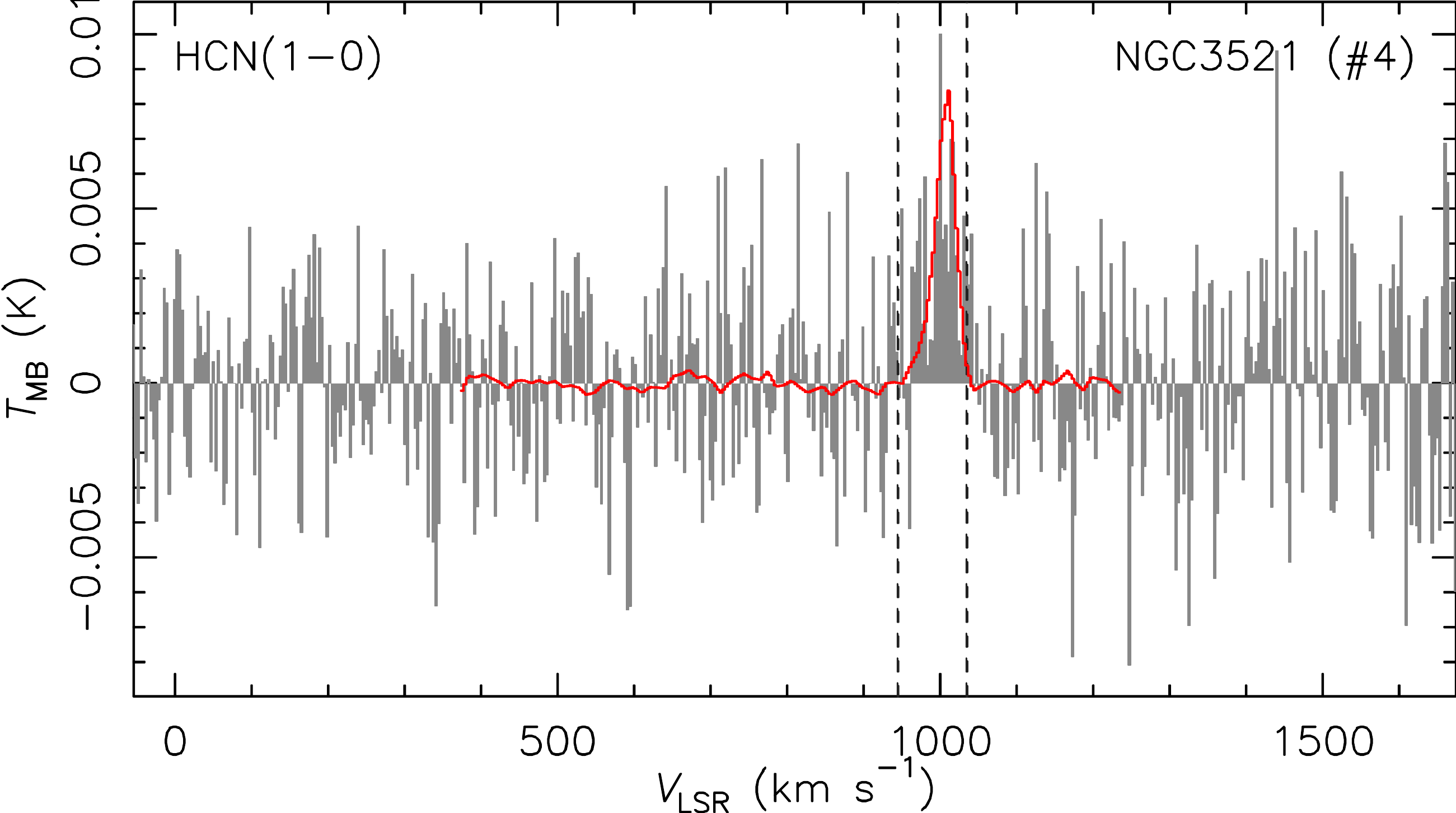} & 
\includegraphics[width=0.42\textwidth]{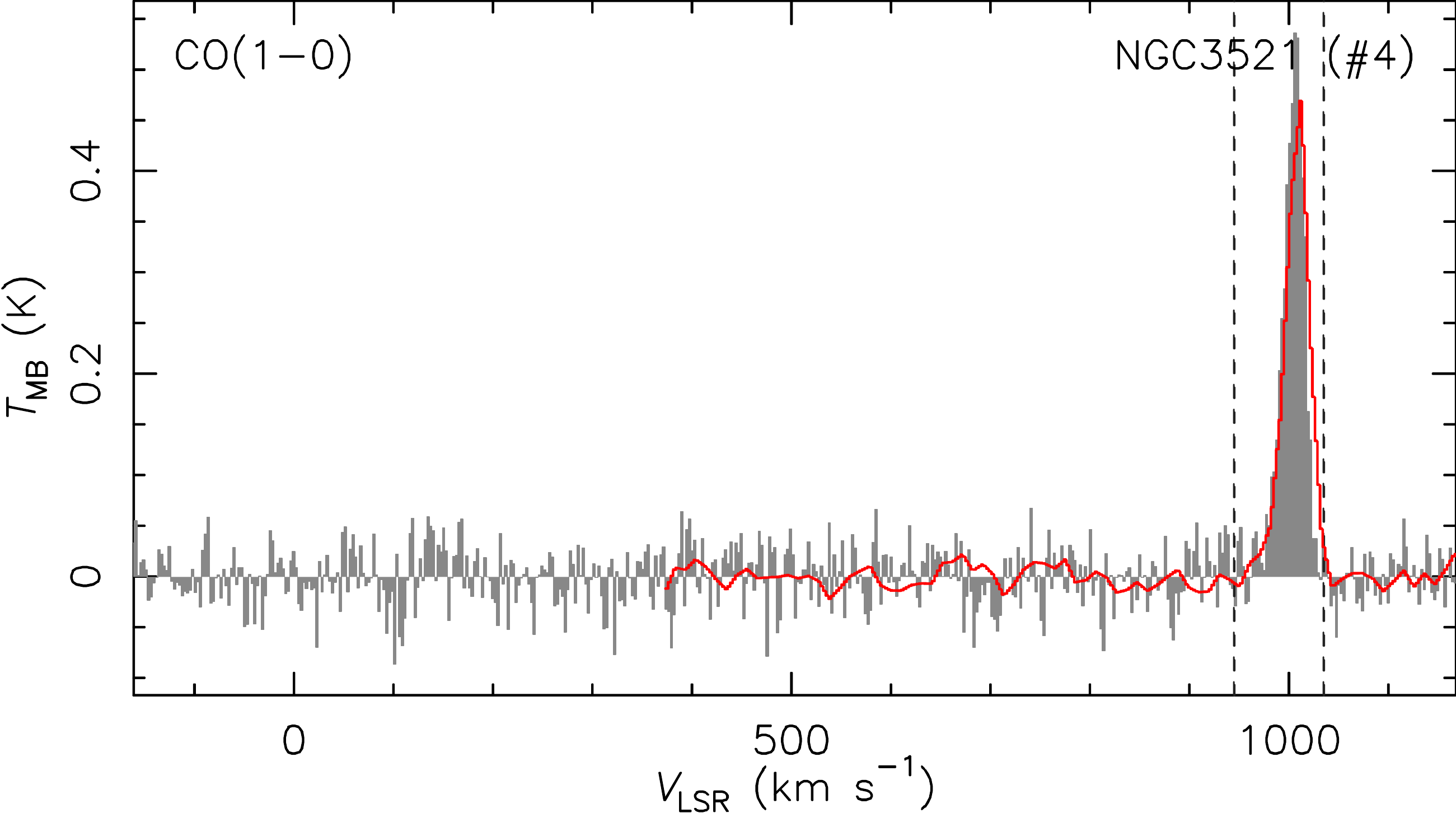} \\ 
\end{tabular}
\end{center}  
\caption{Same as Fig.~\ref{f-spec-1} for NGC~3351 and NGC3521.} 
\end{figure} 
\newpage 
\begin{figure}[h!]
\begin{center}  
\begin{tabular}{cc} 
\includegraphics[width=0.42\textwidth]{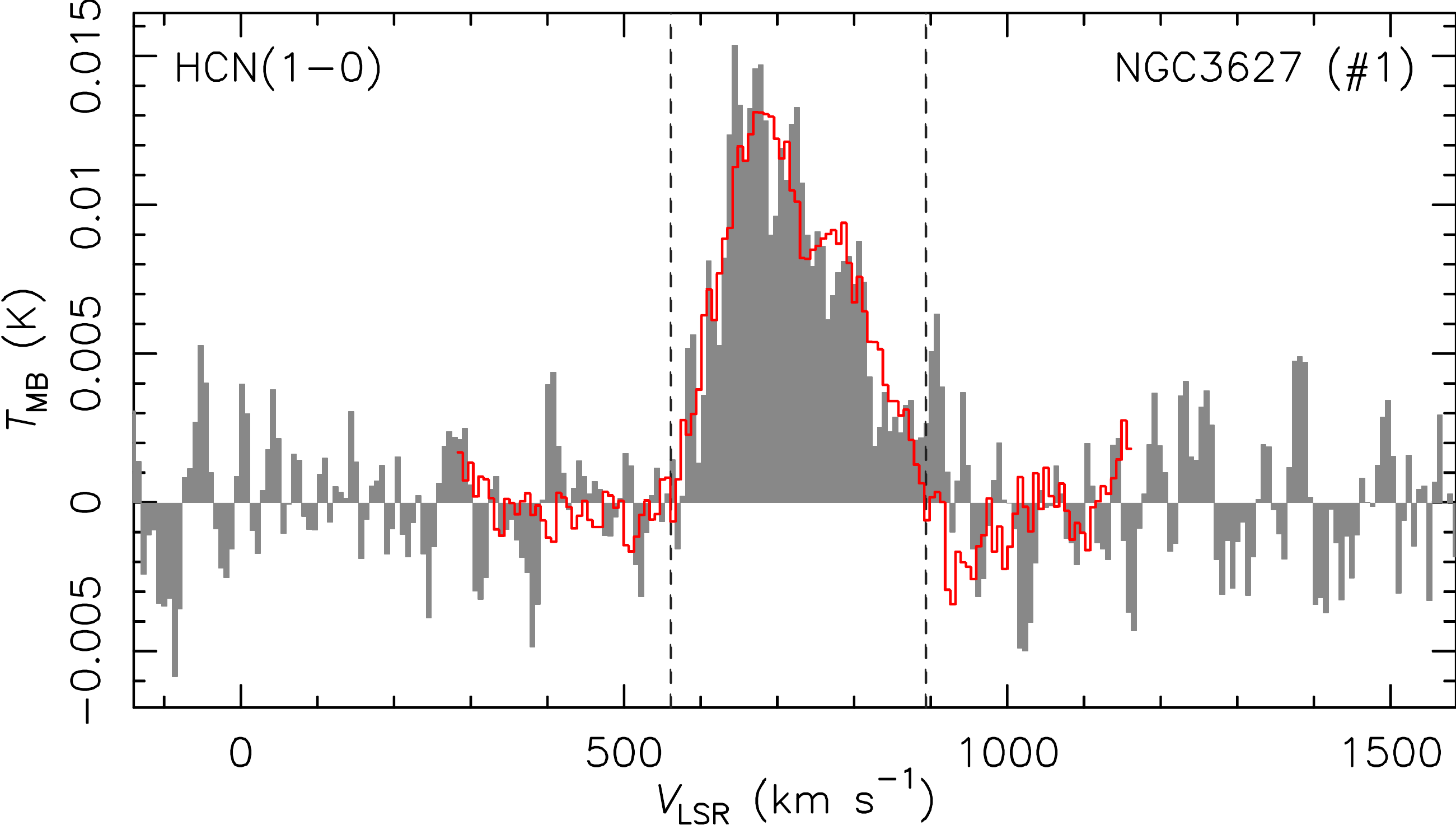} & 
\includegraphics[width=0.42\textwidth]{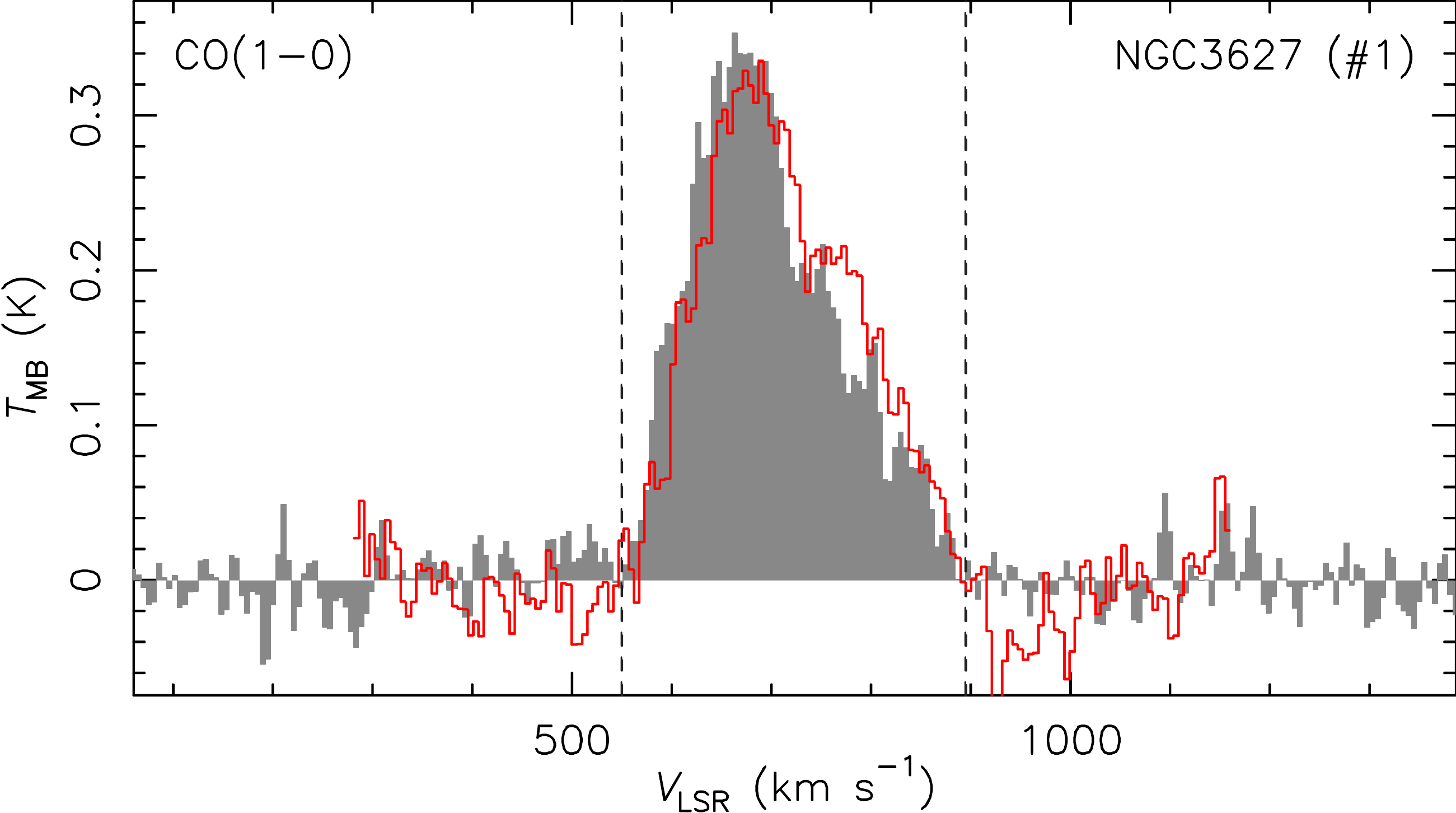} \\ 
\includegraphics[width=0.42\textwidth]{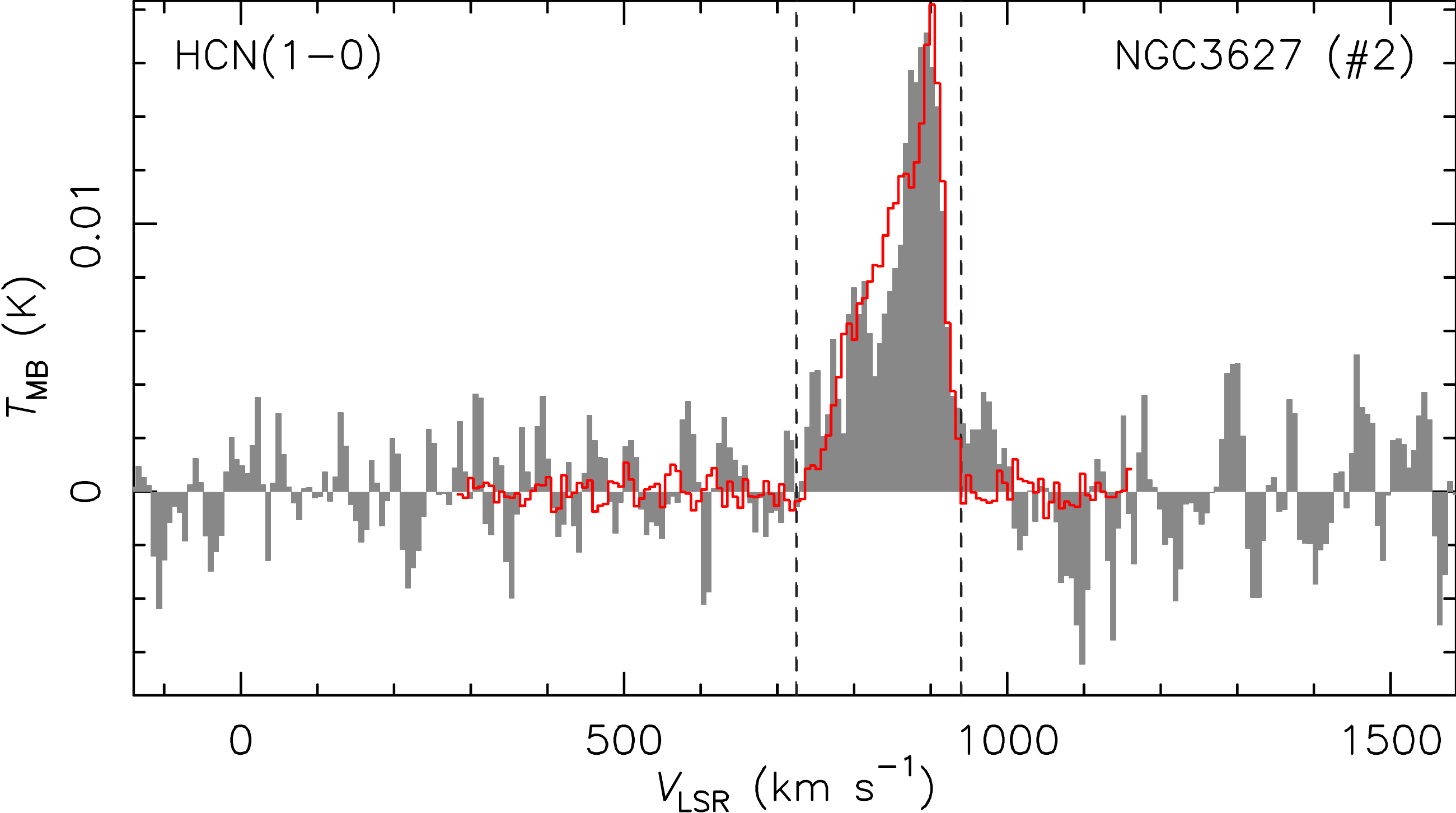} & 
\includegraphics[width=0.42\textwidth]{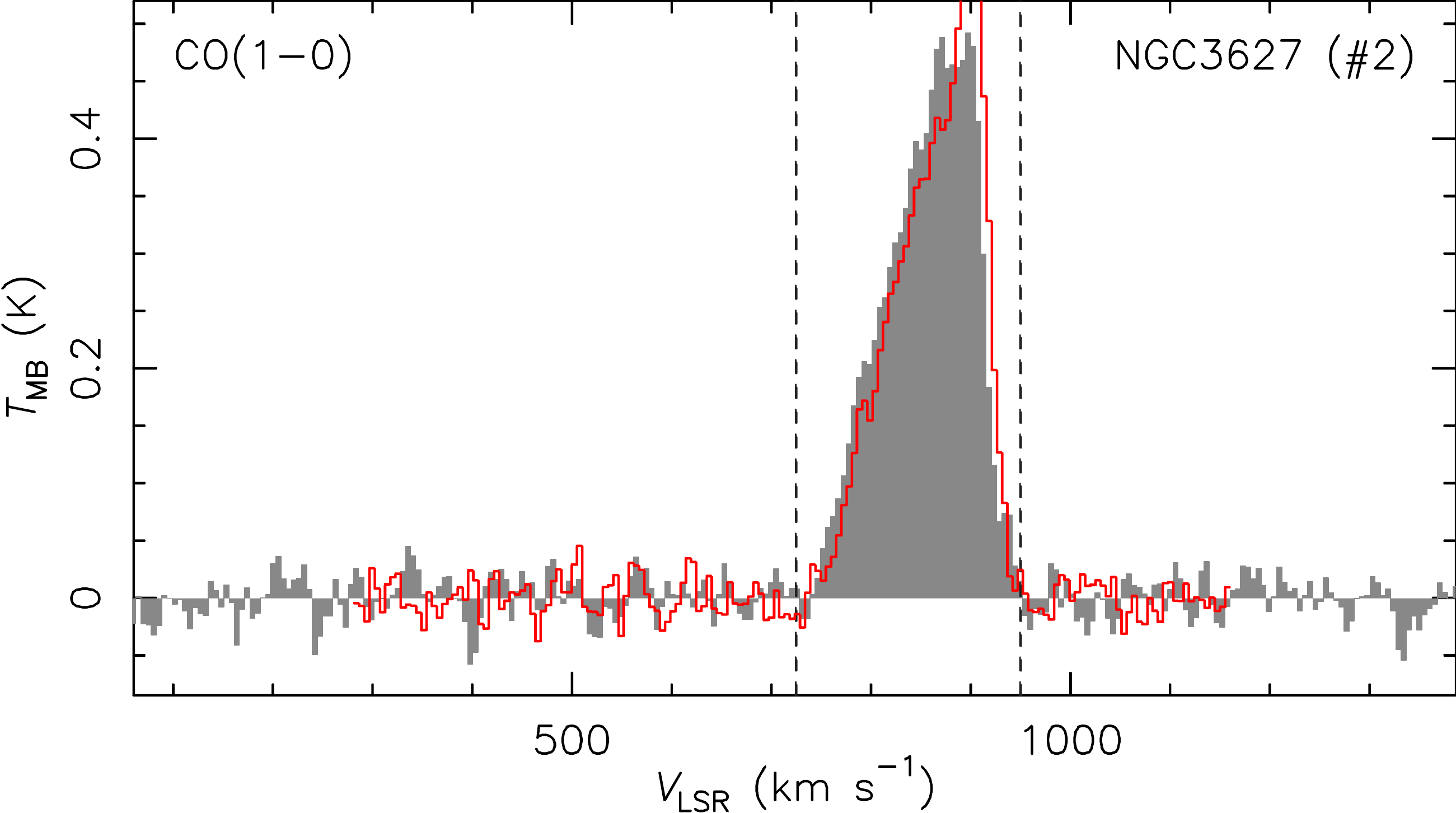} \\ 
\includegraphics[width=0.42\textwidth]{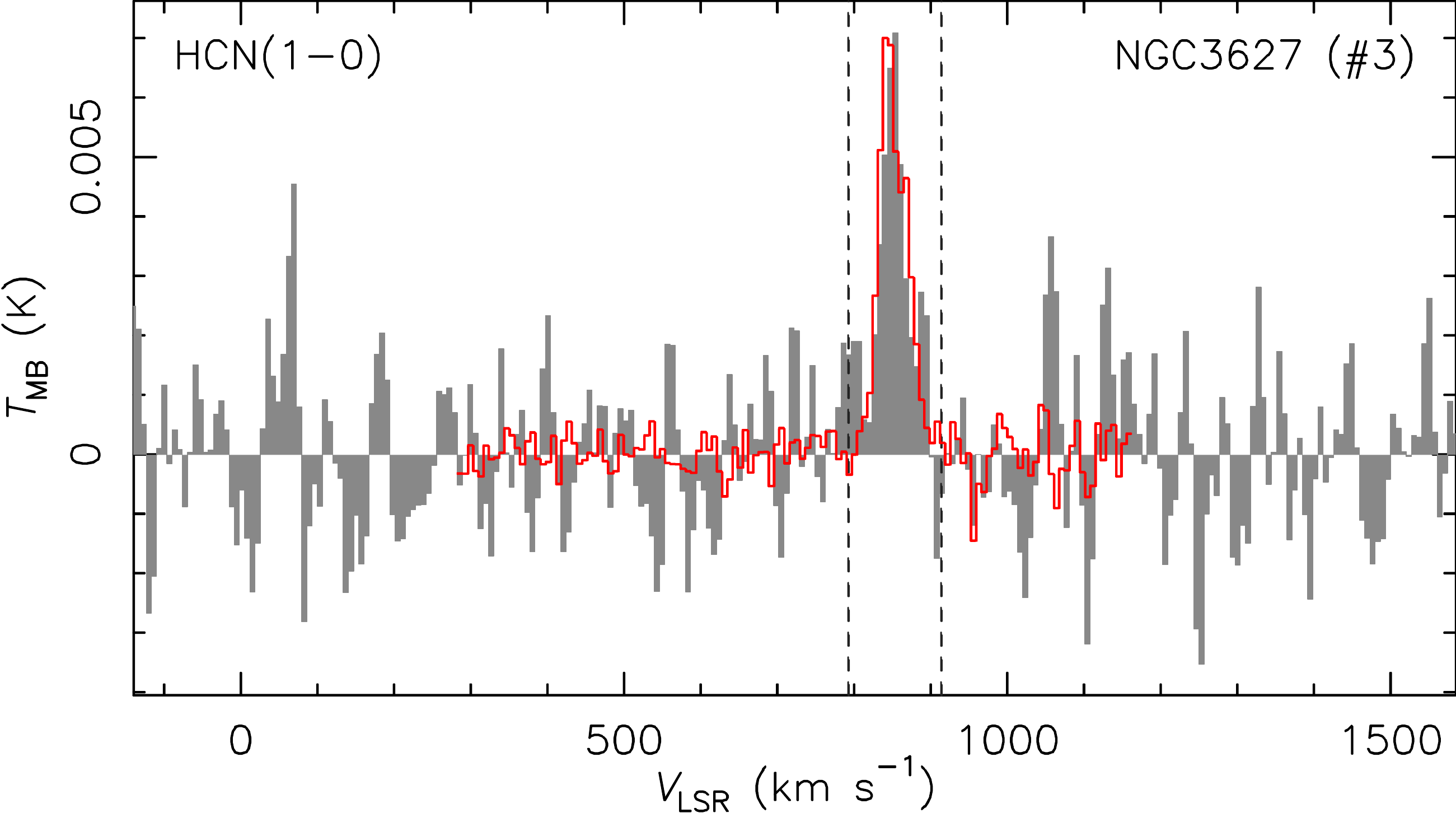} & 
\includegraphics[width=0.42\textwidth]{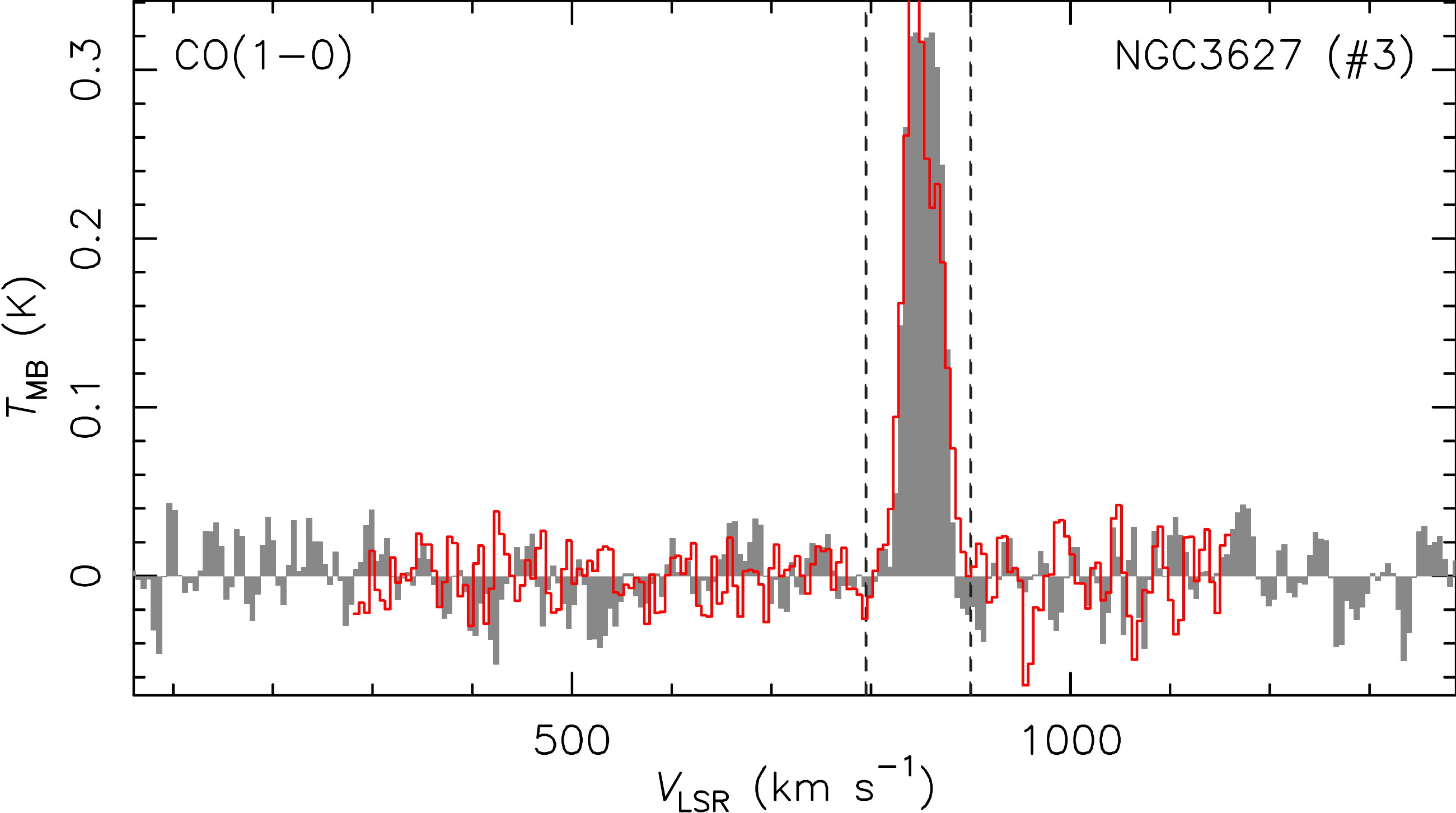} \\ 
\includegraphics[width=0.42\textwidth]{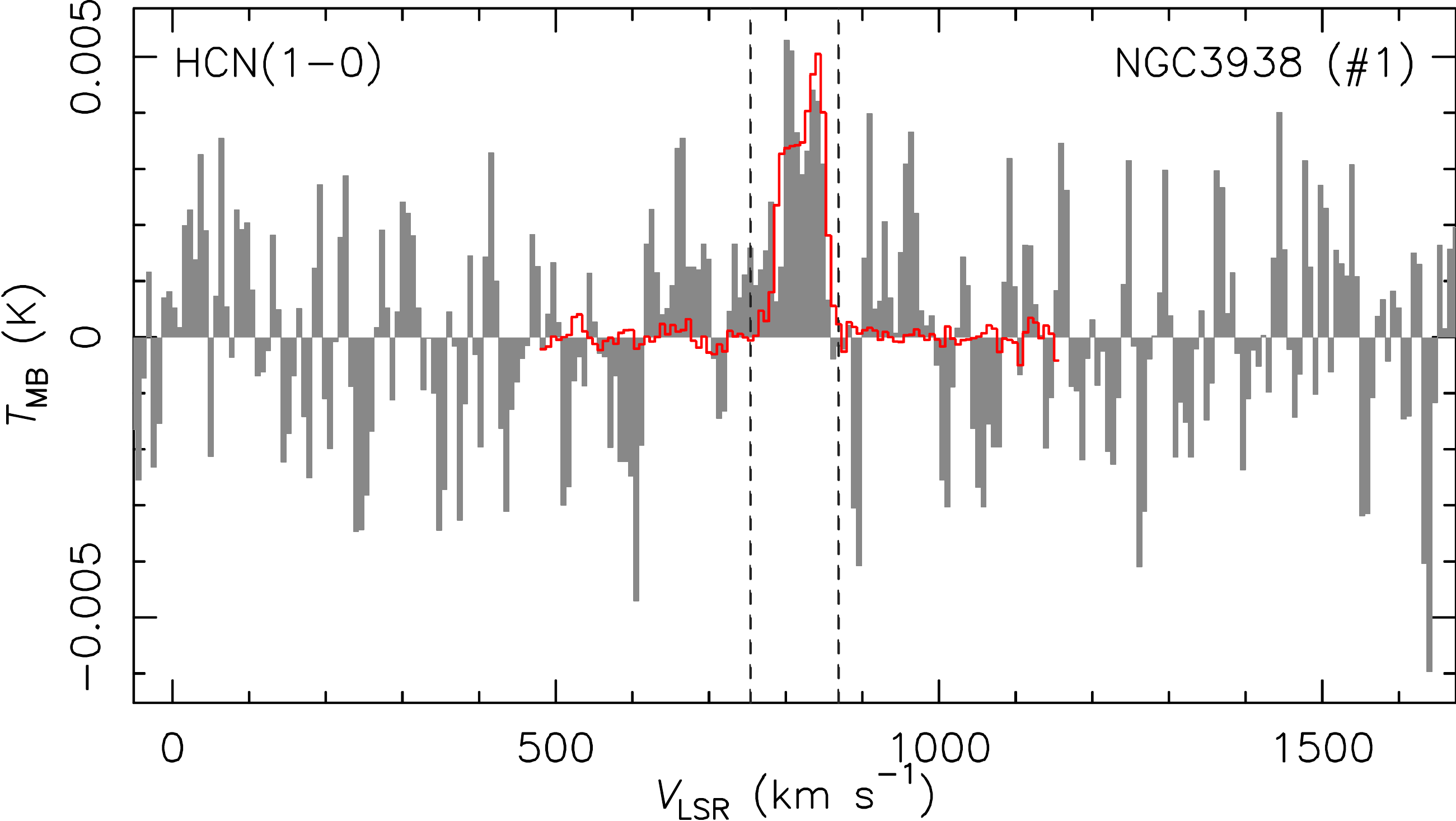} & 
\includegraphics[width=0.42\textwidth]{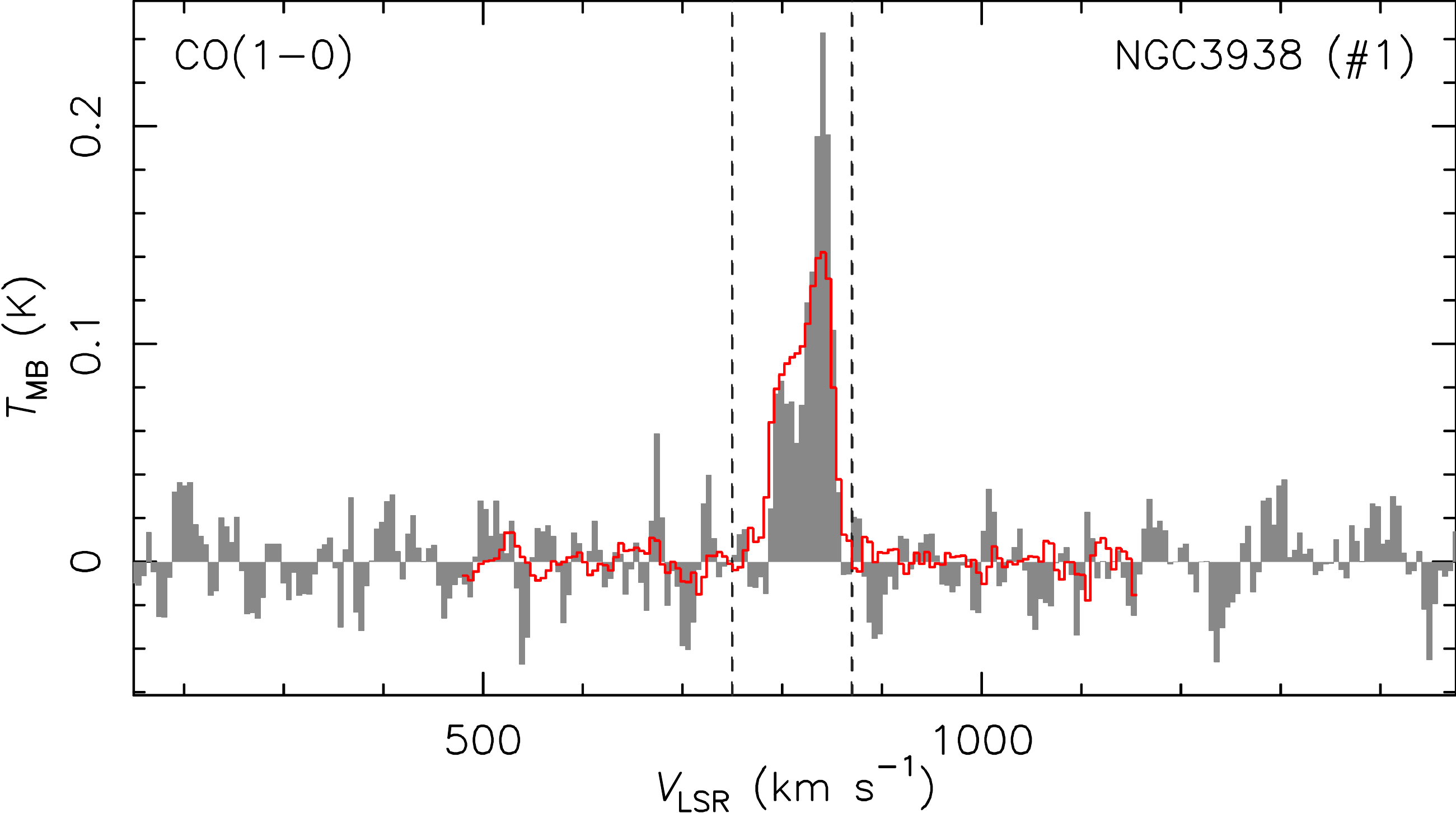} \\ 
\end{tabular}
\end{center}  
\caption{Same as Fig.~\ref{f-spec-1} for NGC~3627 and NGC~3938.} 
\end{figure} 
\newpage 
\begin{figure}[h!]
\begin{center}  
\begin{tabular}{cc} 
\includegraphics[width=0.42\textwidth]{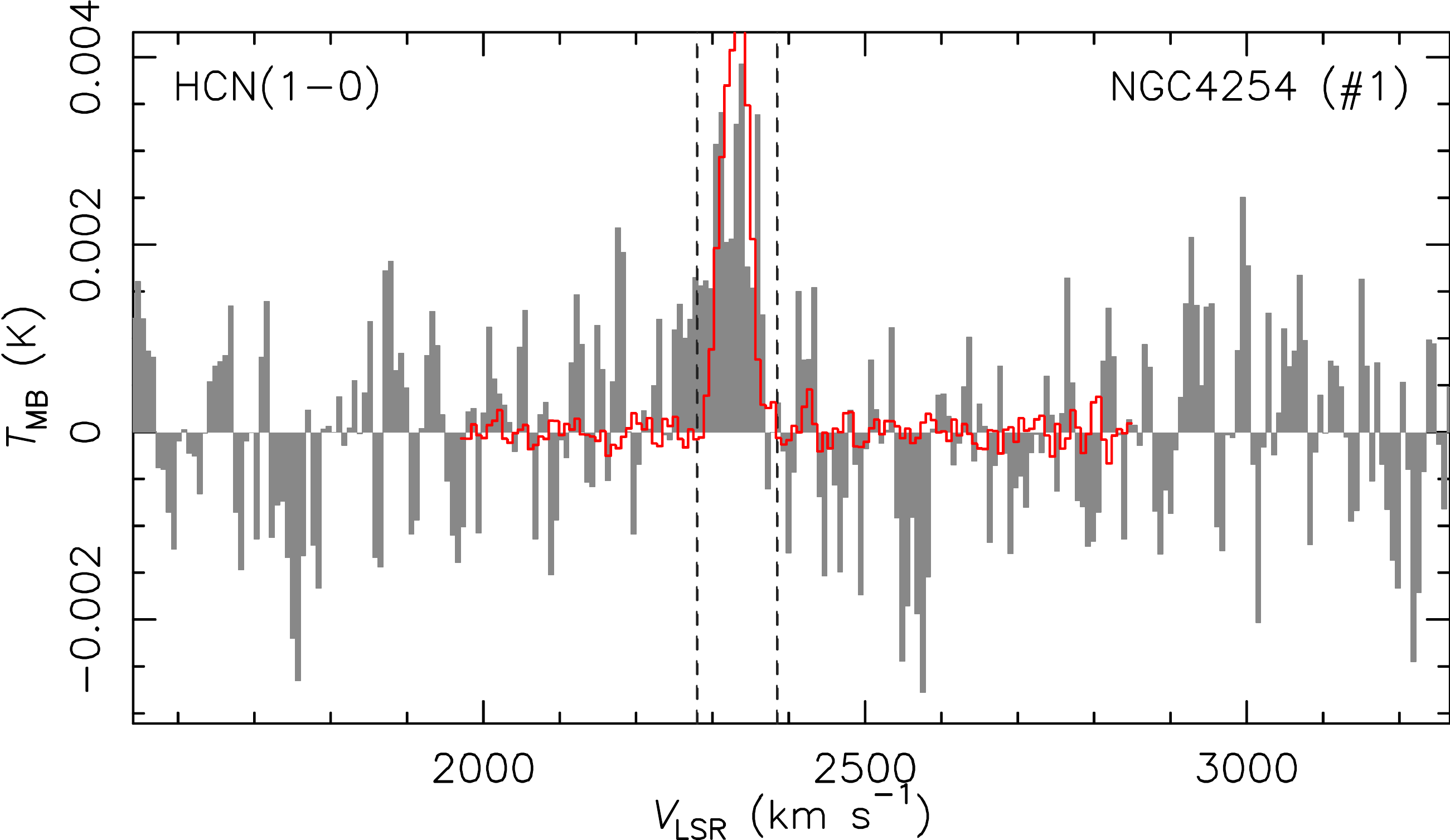} & 
\includegraphics[width=0.42\textwidth]{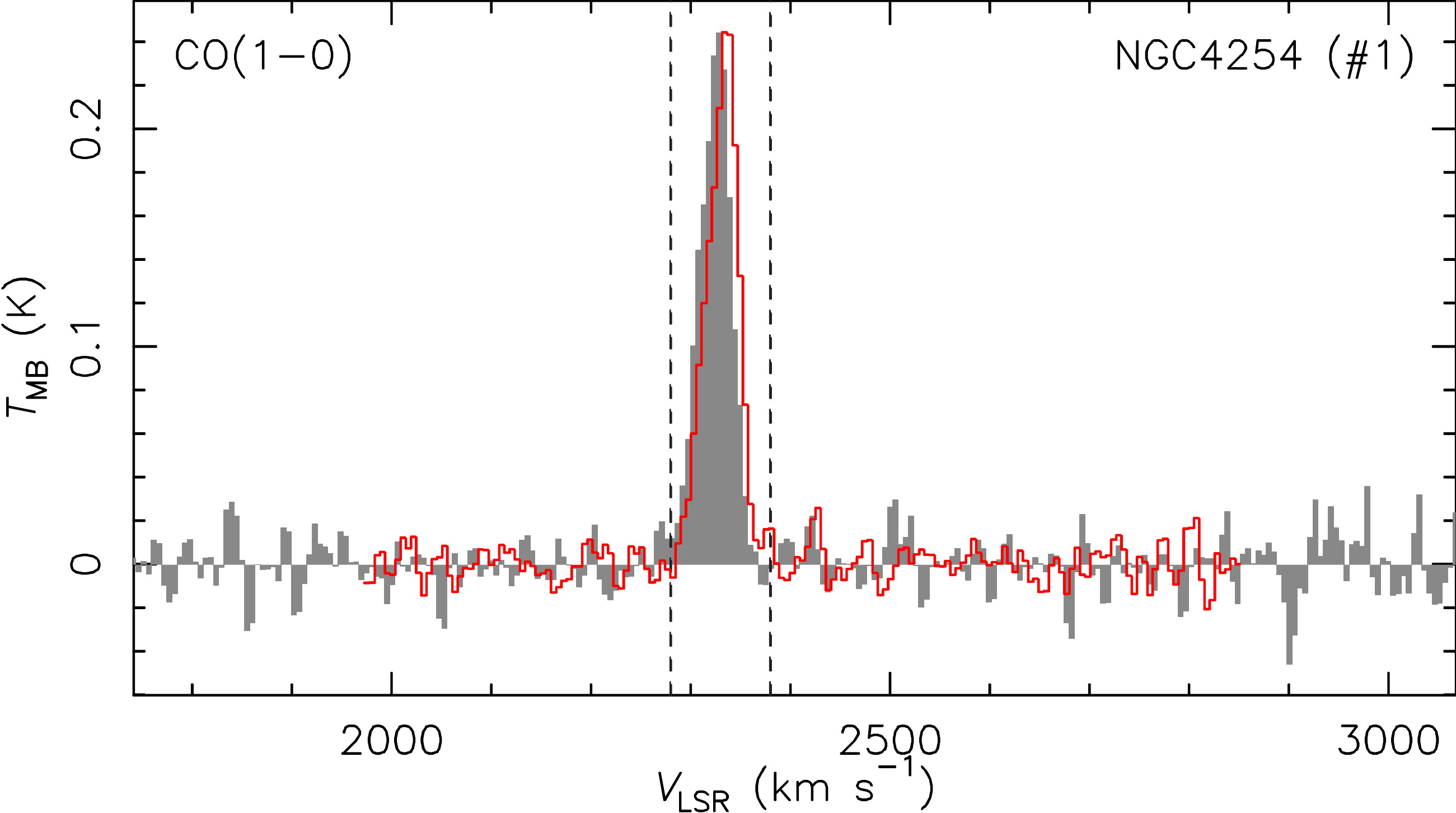} \\ 
\includegraphics[width=0.42\textwidth]{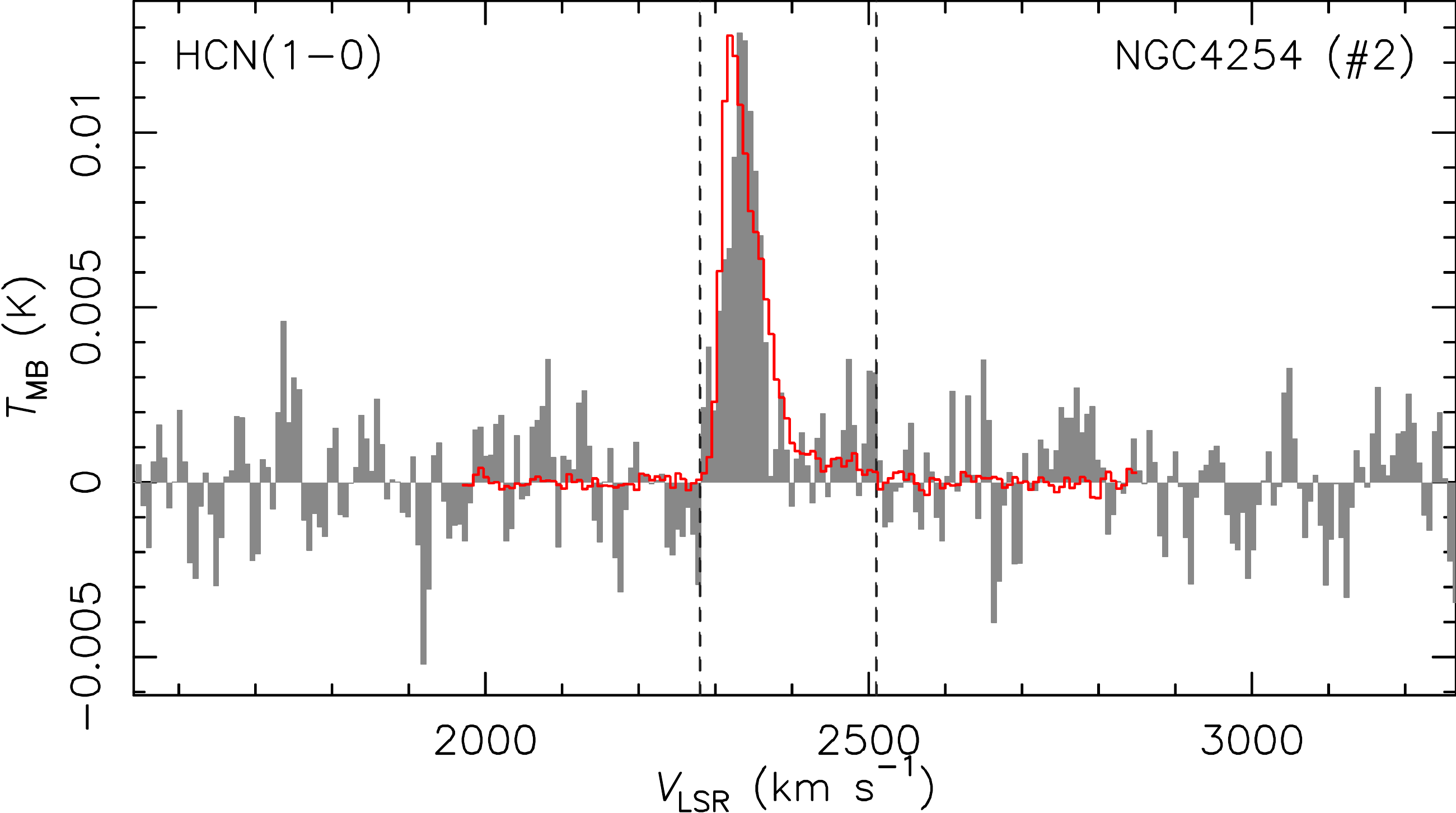} & 
\includegraphics[width=0.42\textwidth]{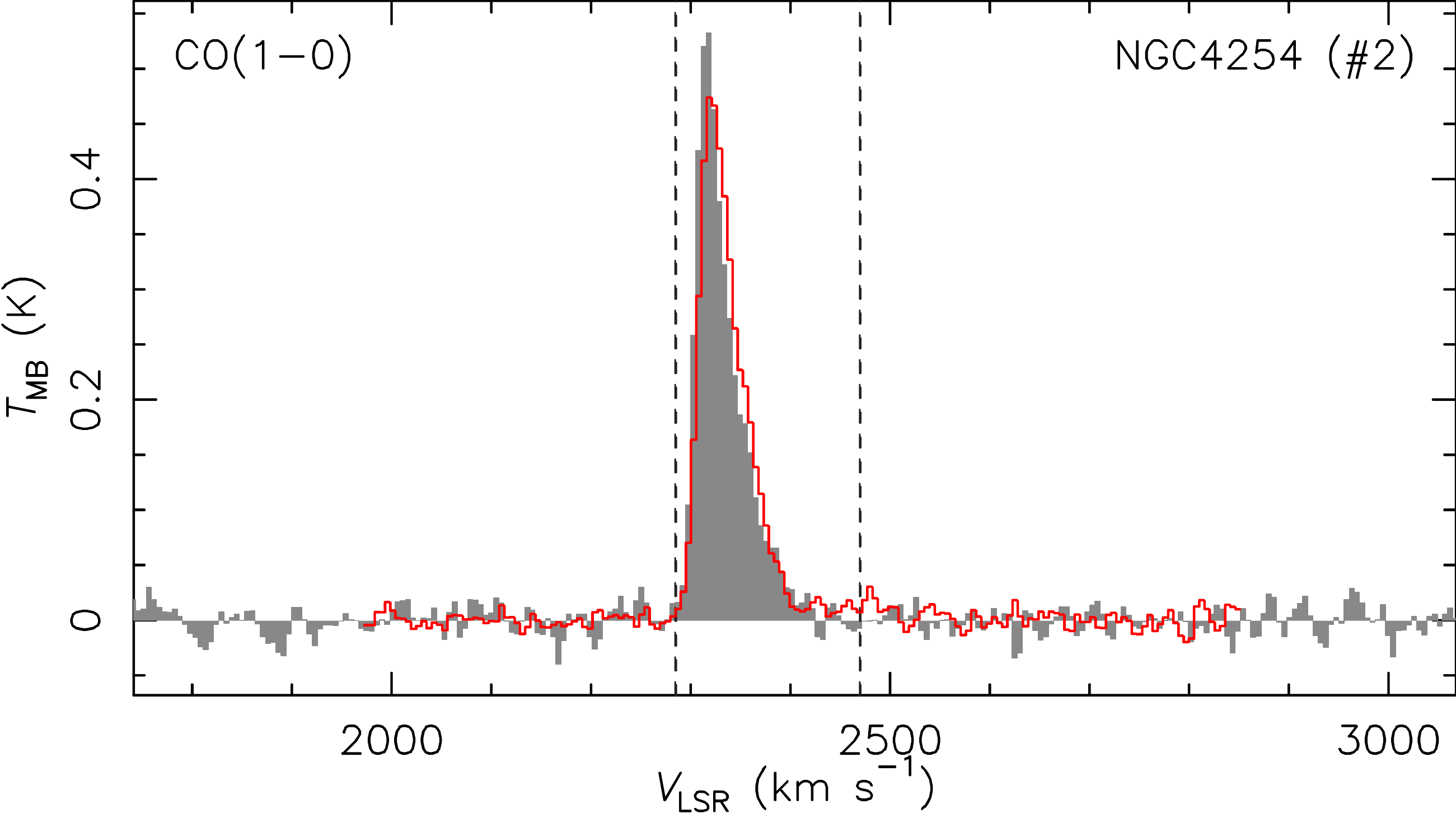} \\ 
\includegraphics[width=0.42\textwidth]{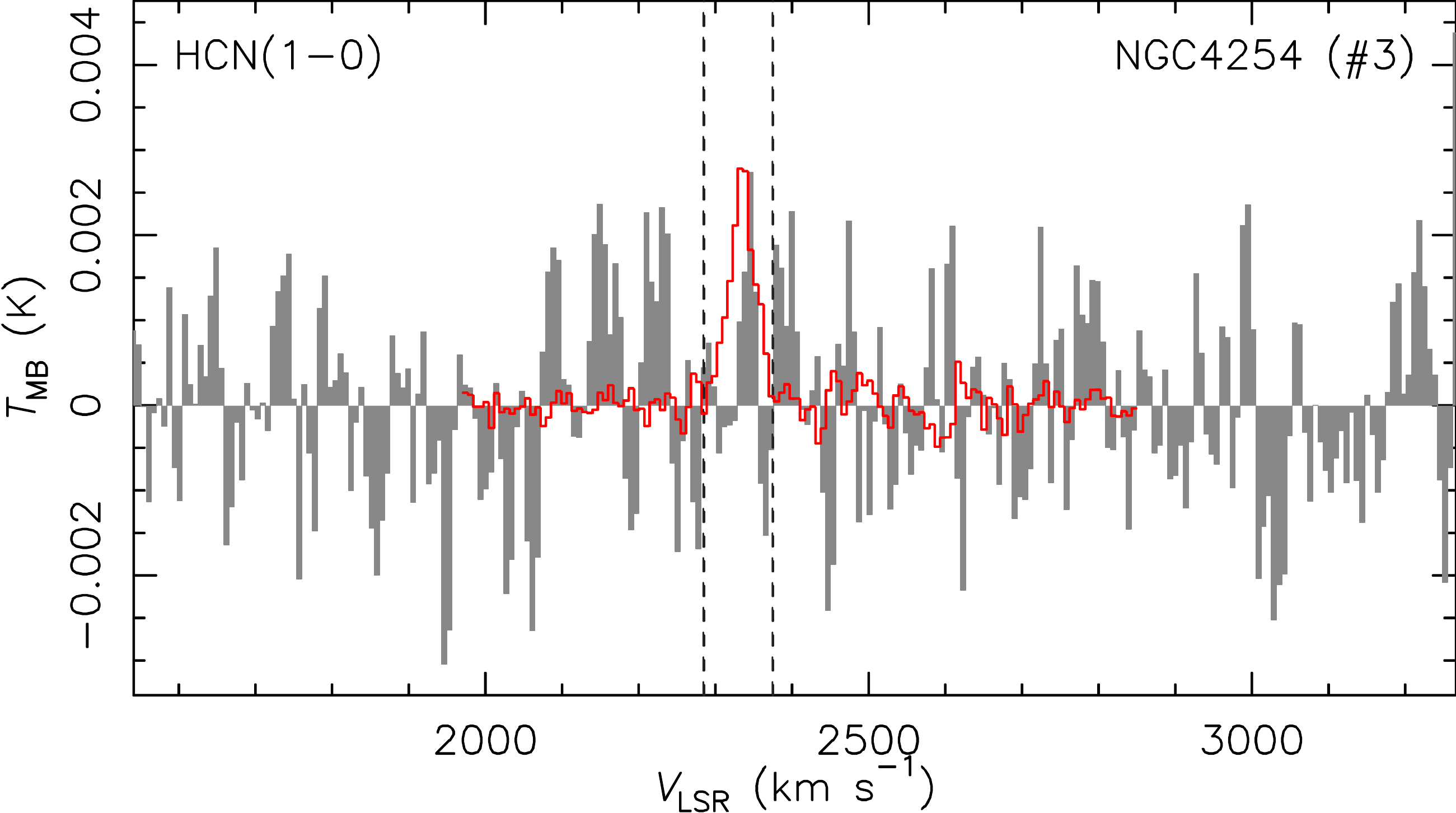} & 
\includegraphics[width=0.42\textwidth]{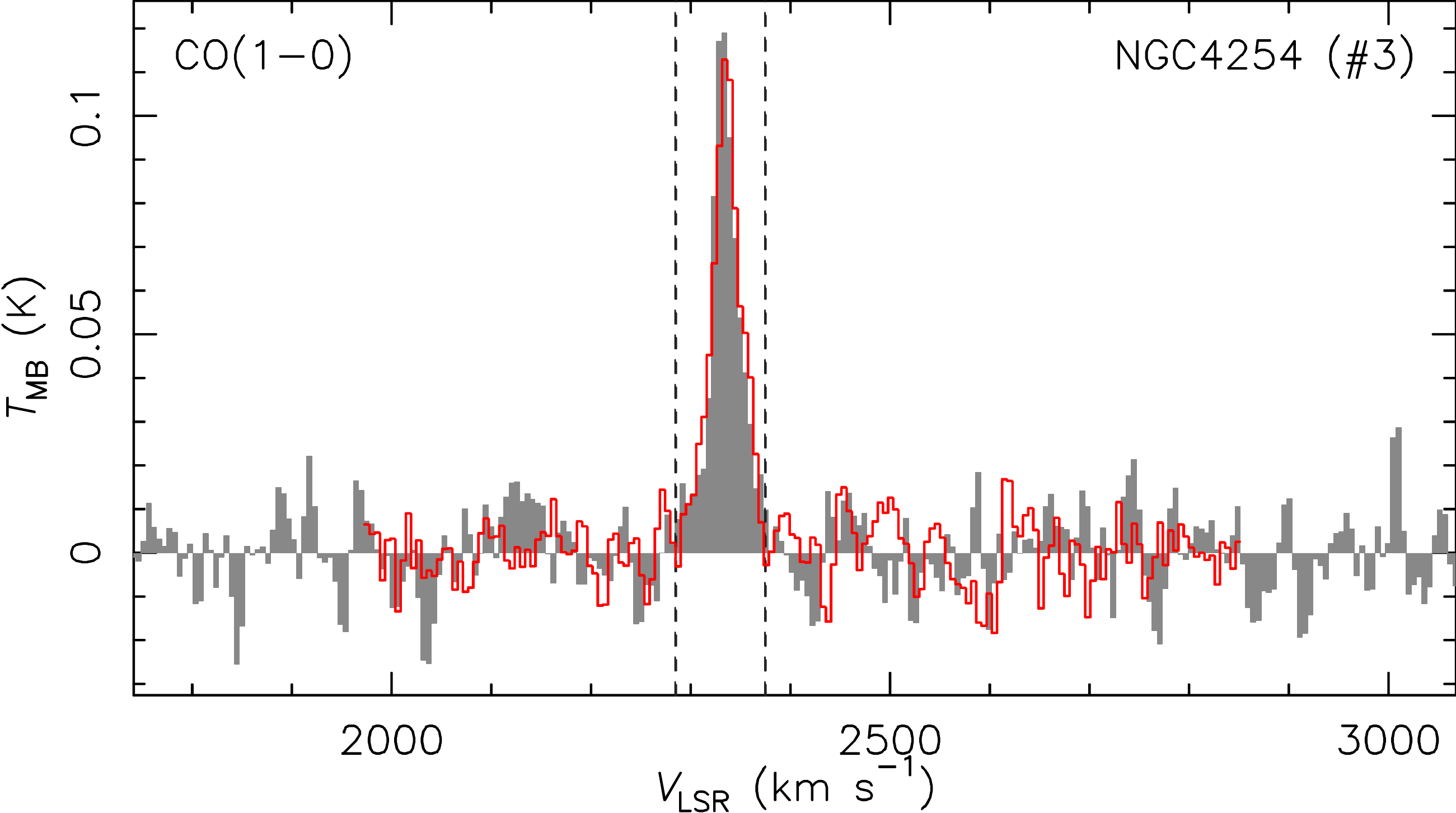} \\ 
\end{tabular}
\end{center}  
\caption{Same as Fig.~\ref{f-spec-1} for NGC~4254.} 
\end{figure} 
\newpage
\begin{figure}[h!]
\begin{center}  
\begin{tabular}{cc} 
\includegraphics[width=0.42\textwidth]{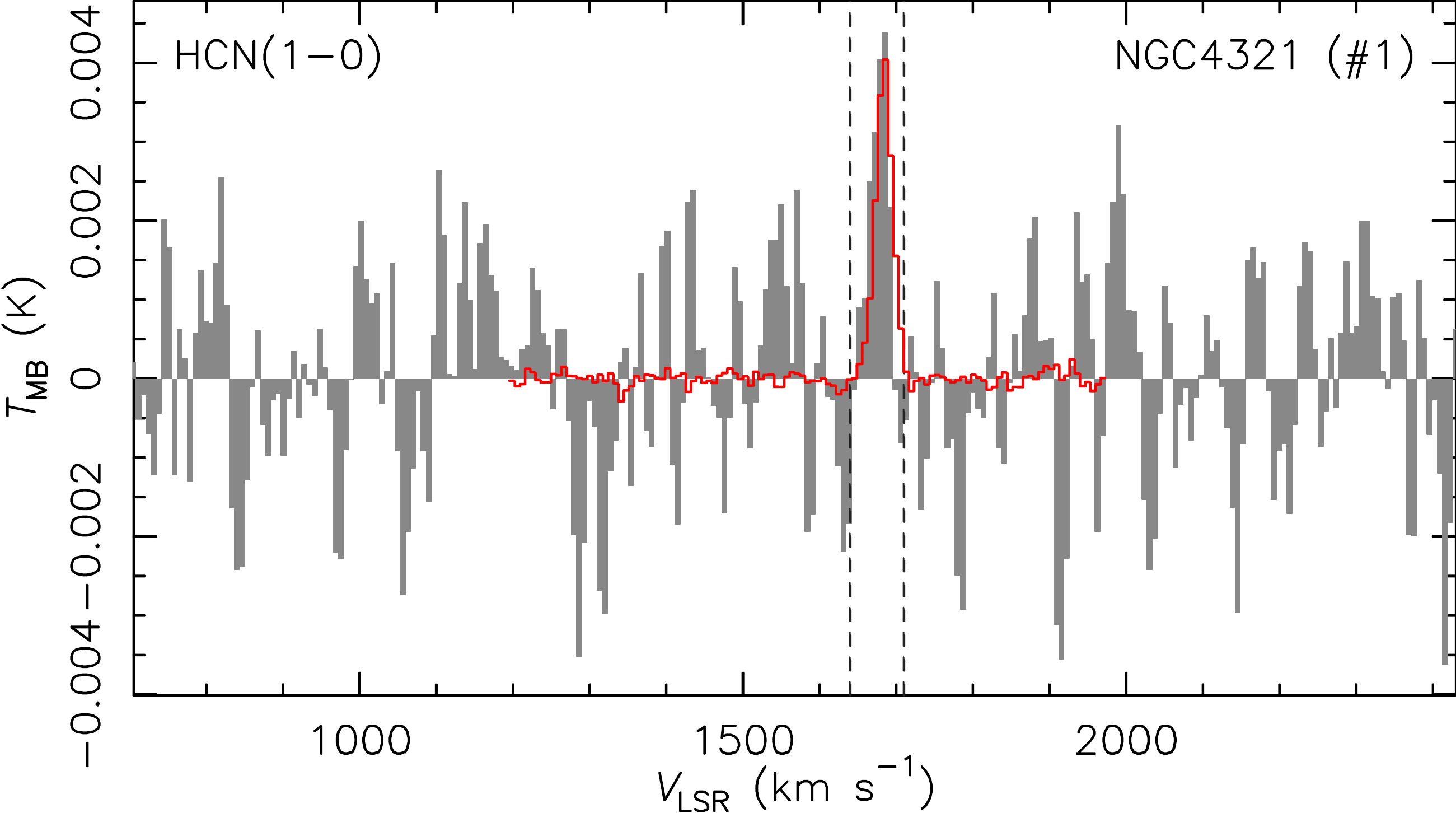} & 
\includegraphics[width=0.42\textwidth]{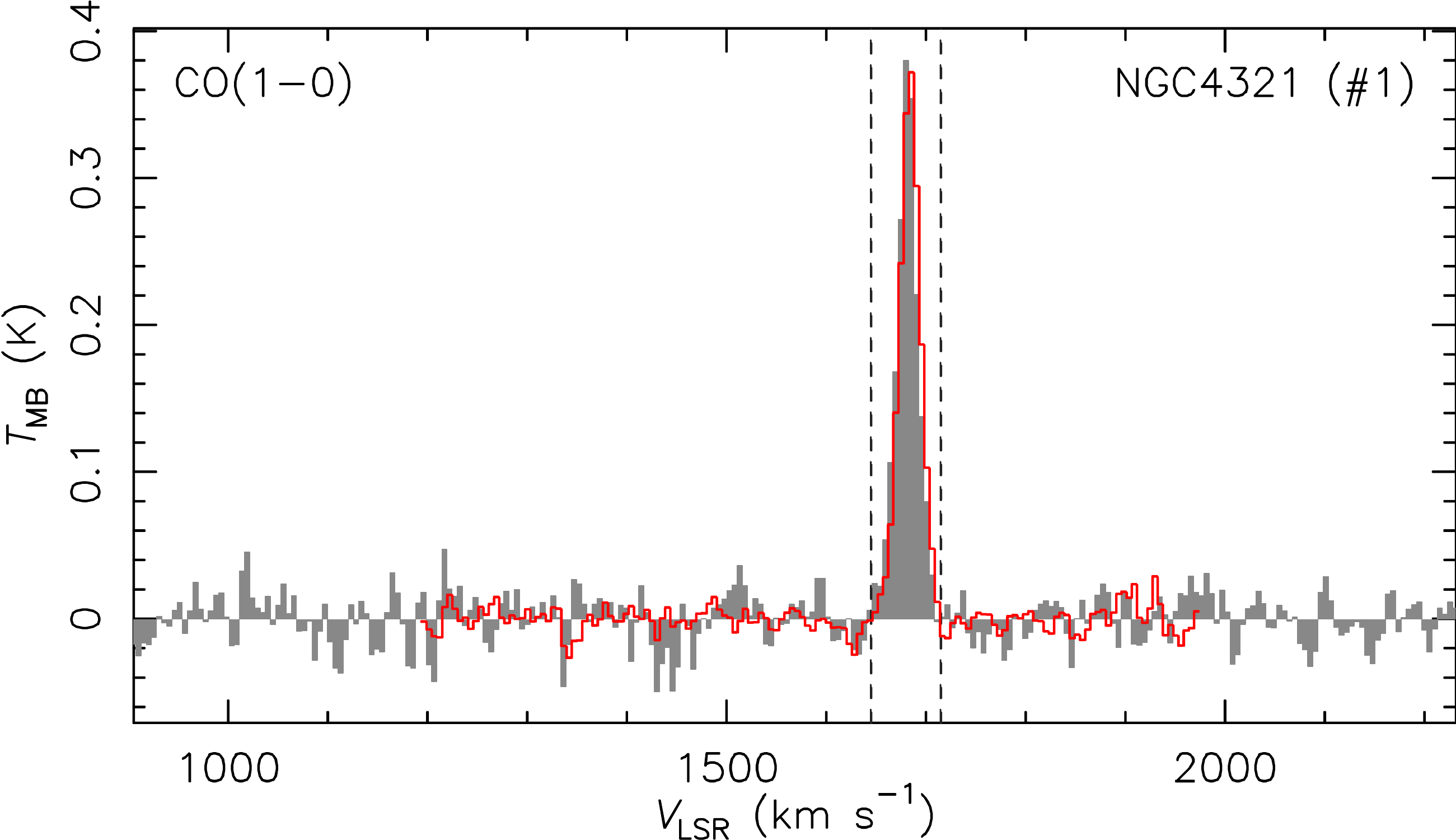} \\ 
\includegraphics[width=0.42\textwidth]{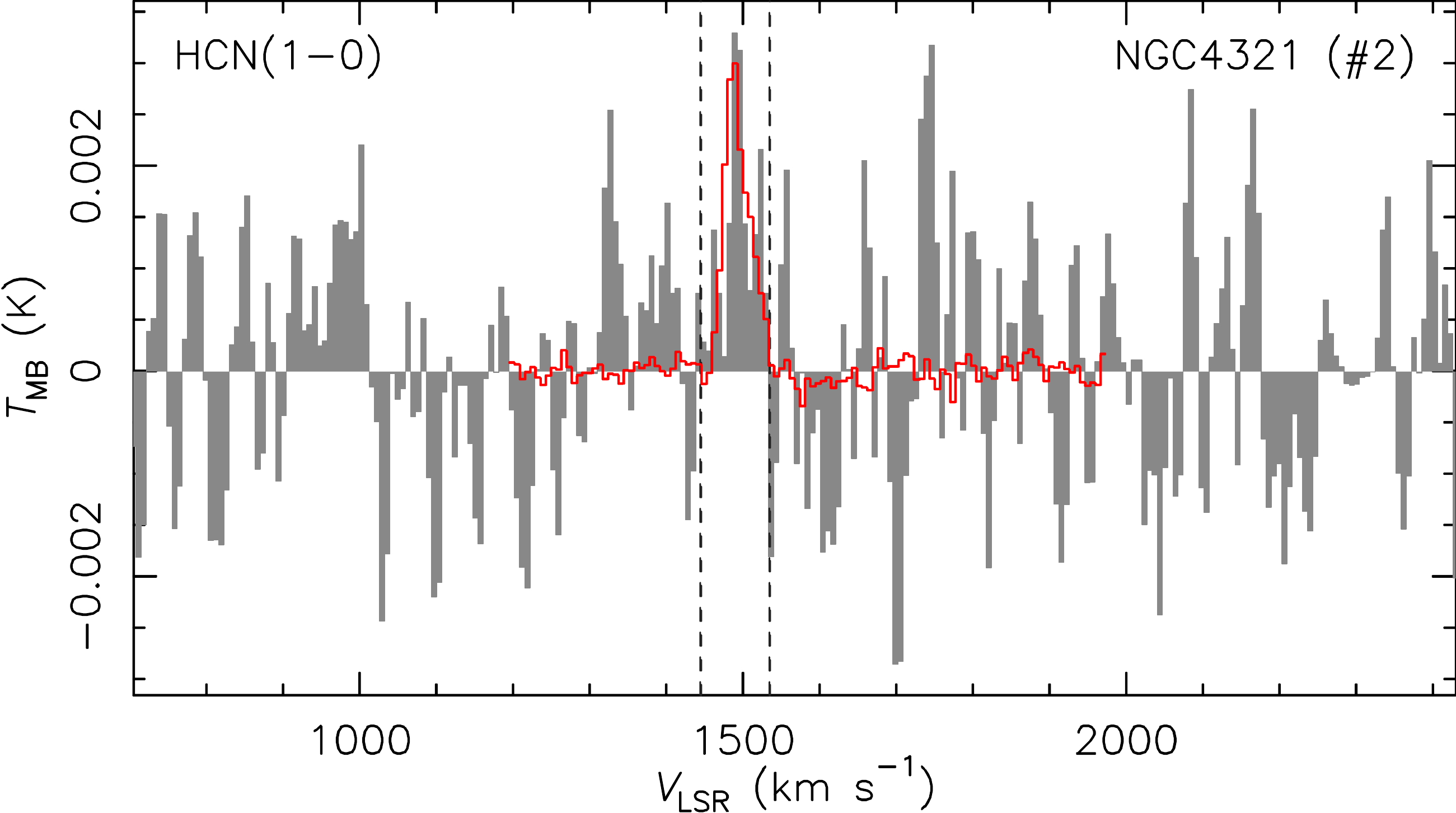} & 
\includegraphics[width=0.42\textwidth]{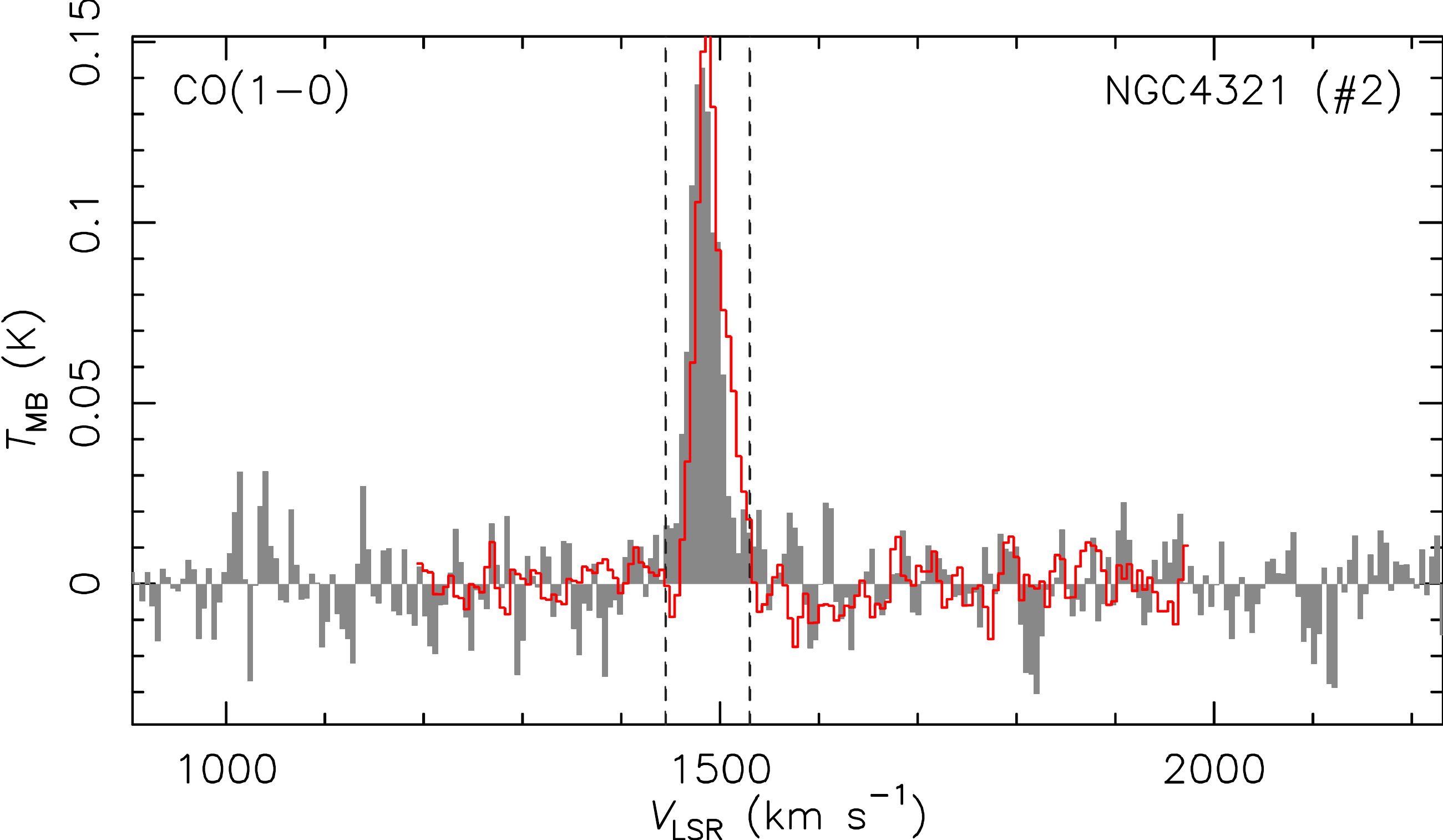} \\ 
\includegraphics[width=0.42\textwidth]{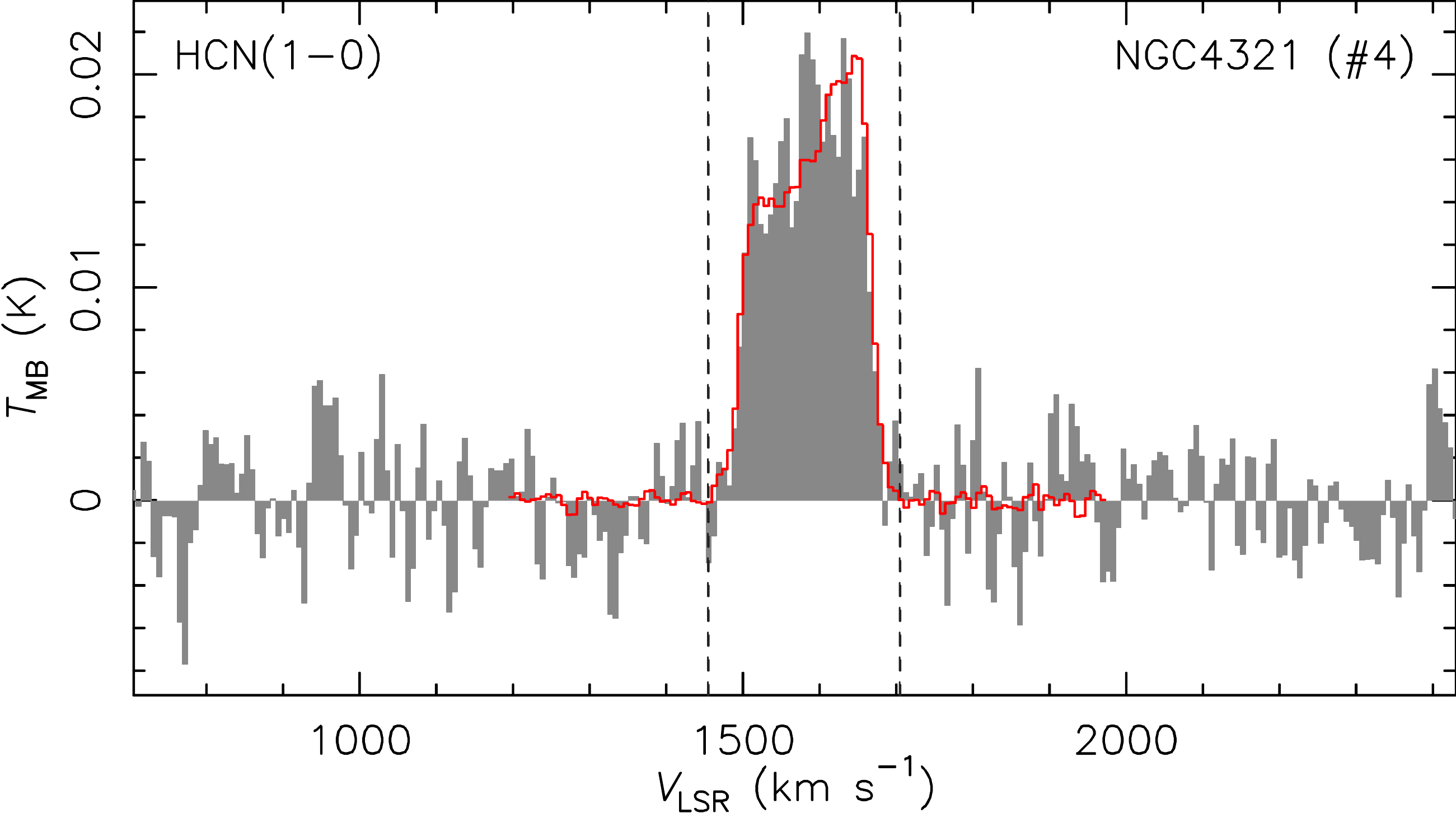} & 
\includegraphics[width=0.42\textwidth]{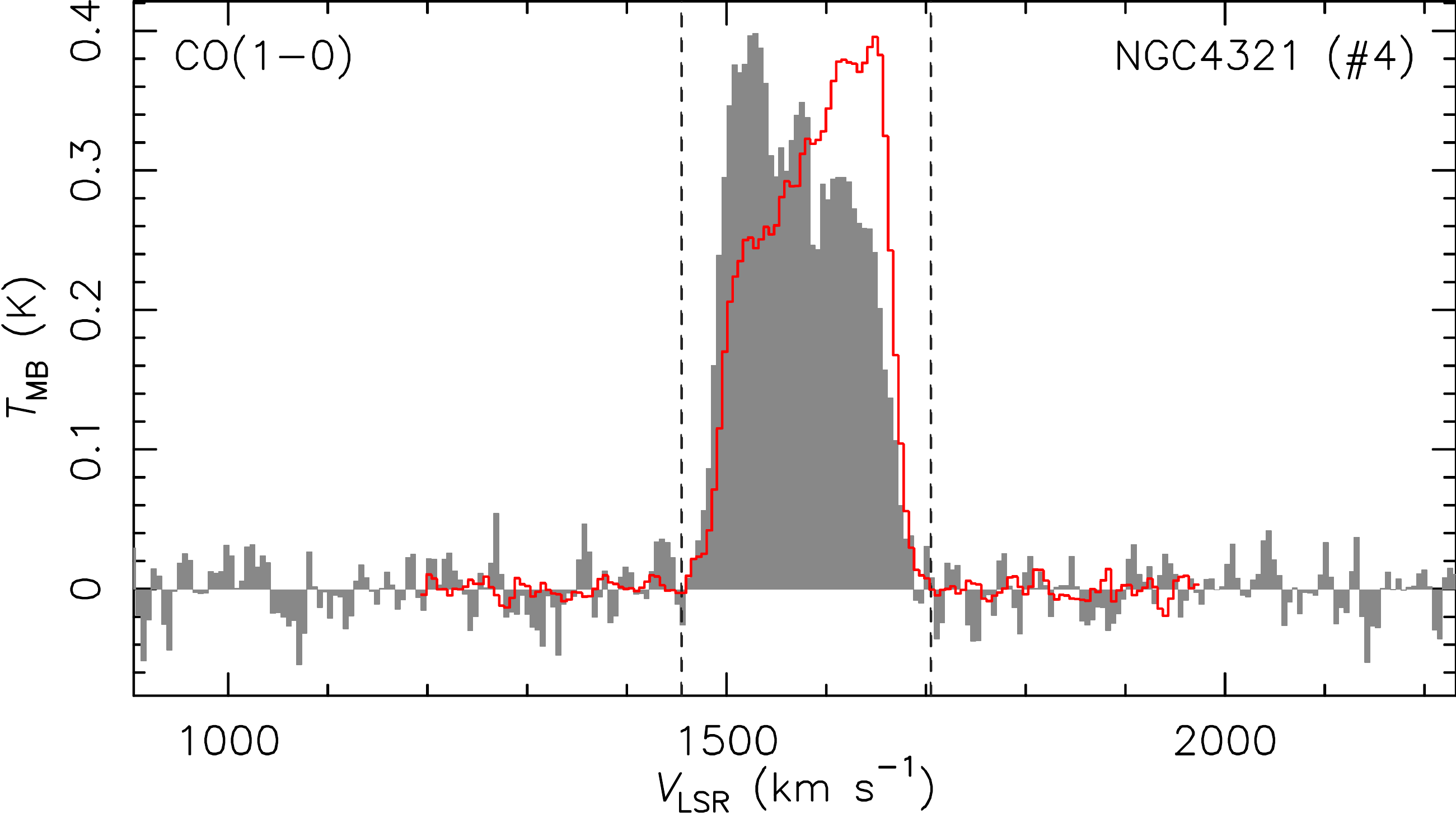} \\ 
\includegraphics[width=0.42\textwidth]{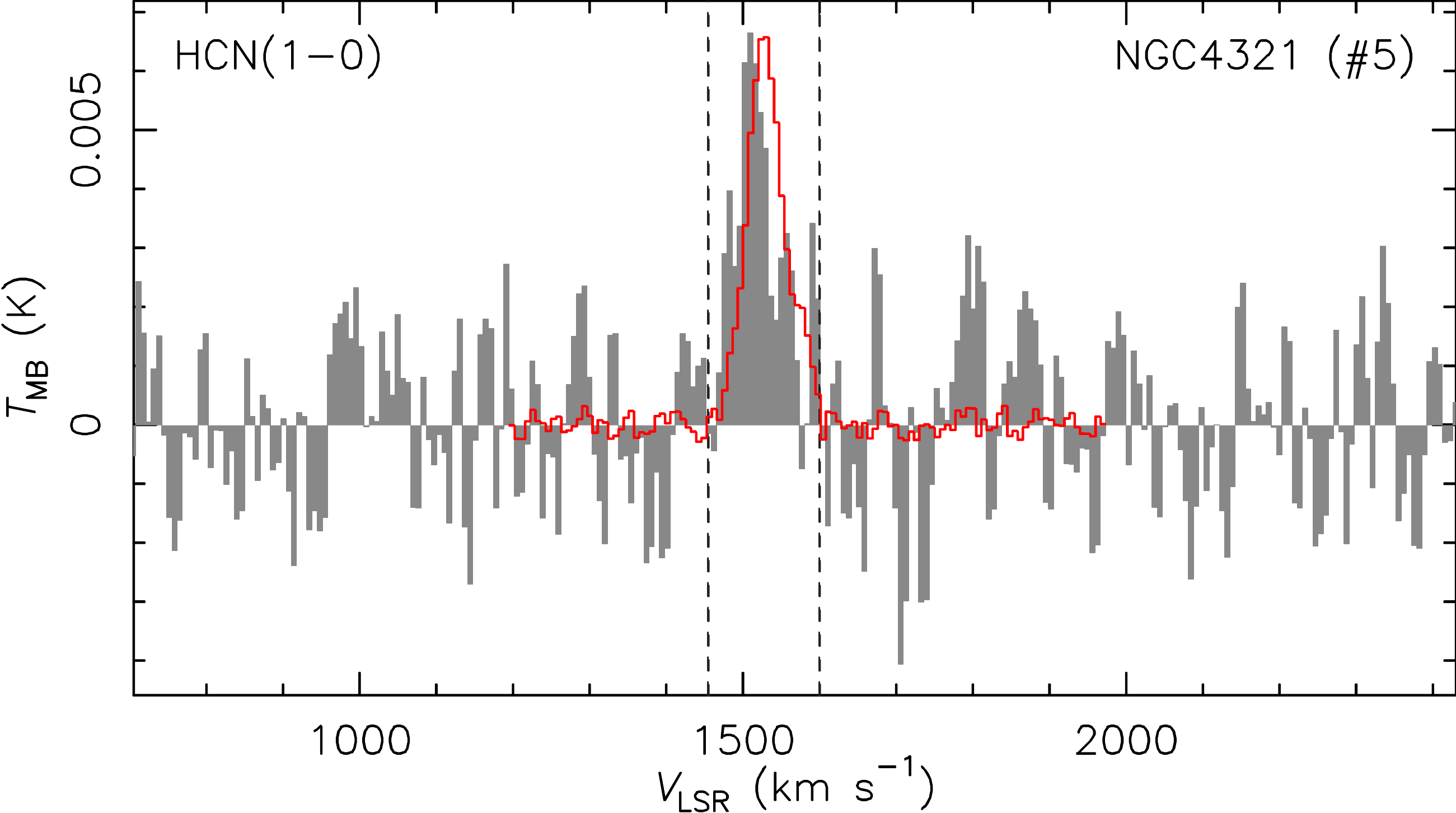} & 
\includegraphics[width=0.42\textwidth]{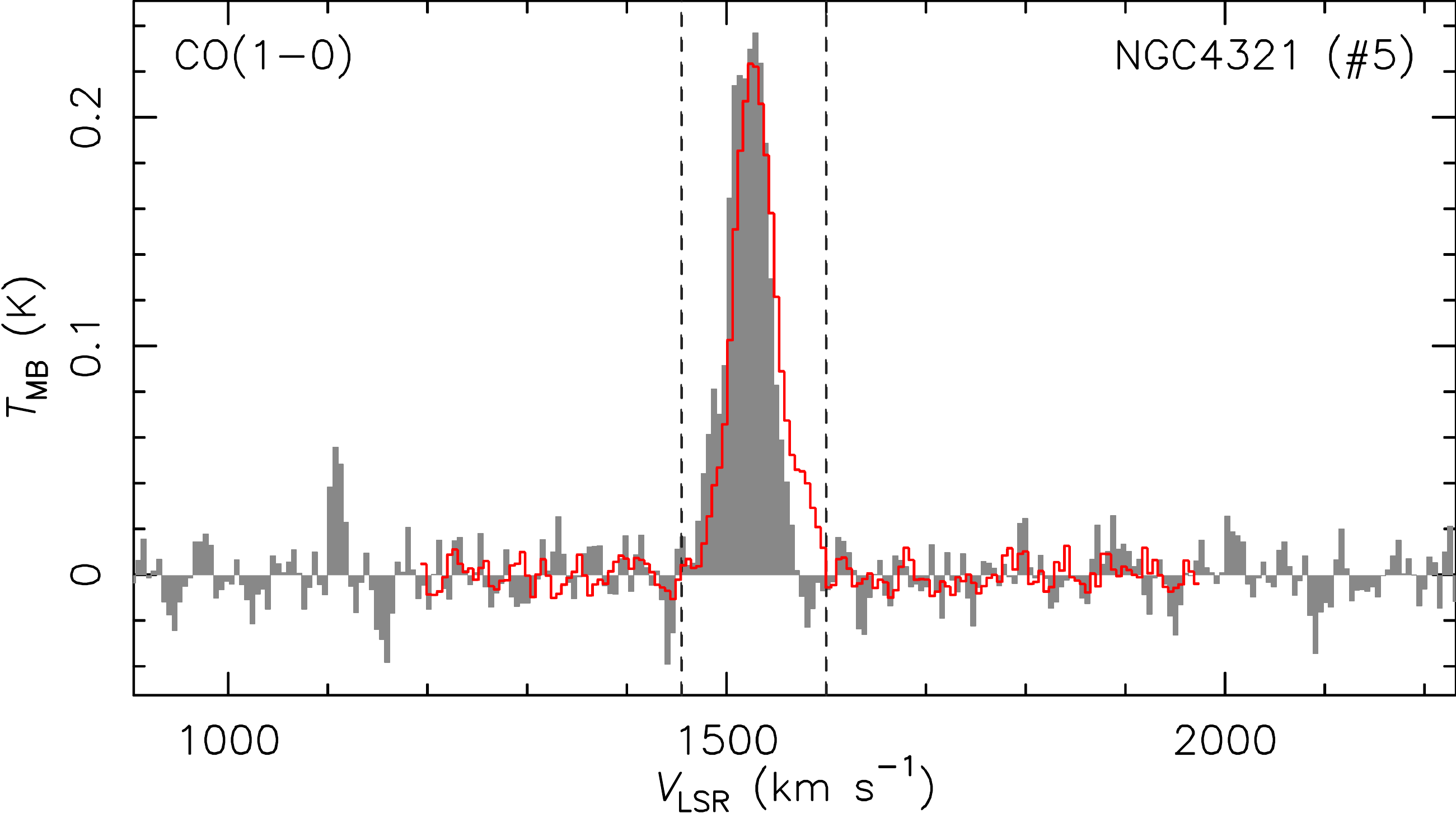} \\ 

\includegraphics[width=0.42\textwidth]{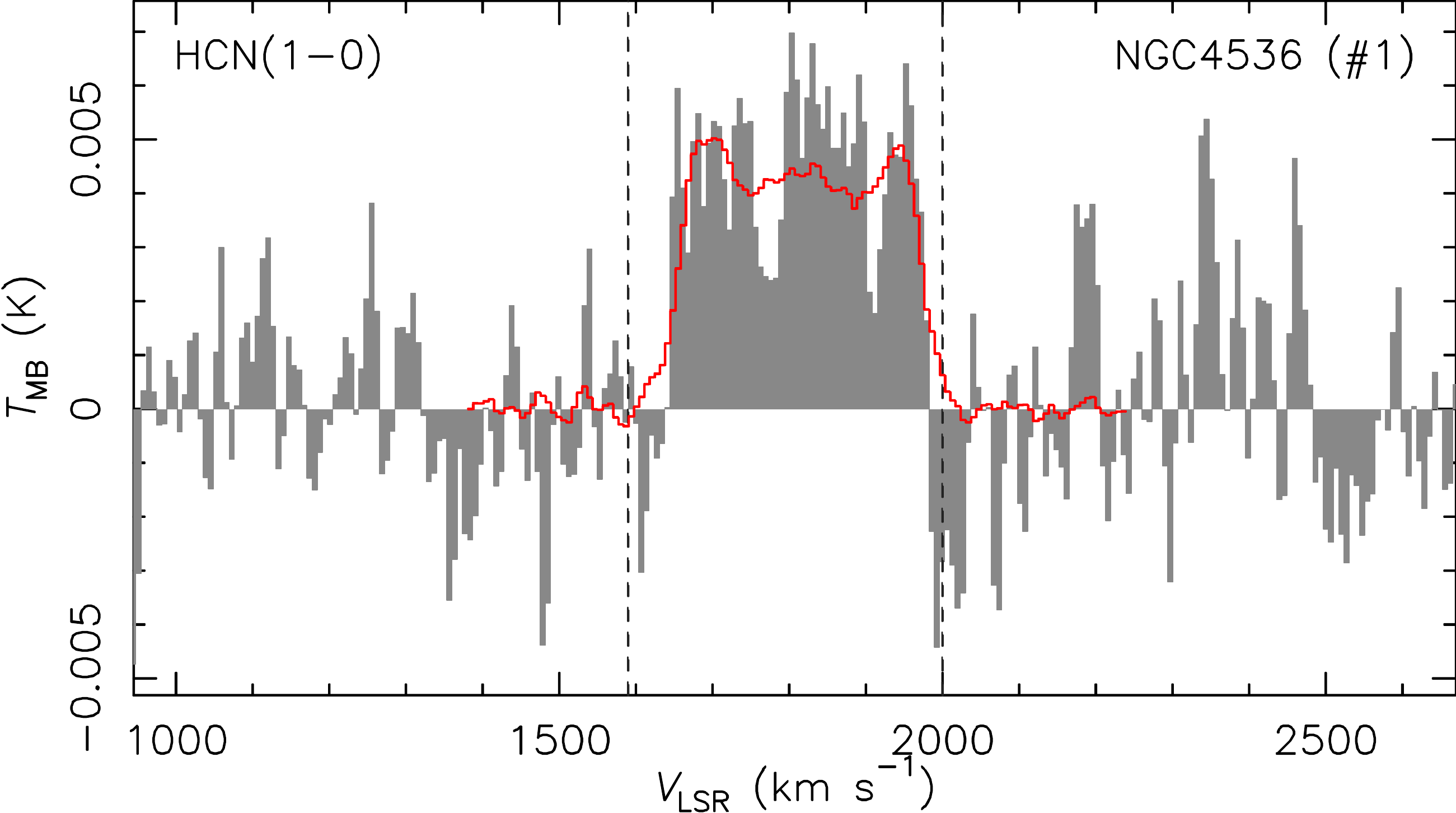} & 
\includegraphics[width=0.42\textwidth]{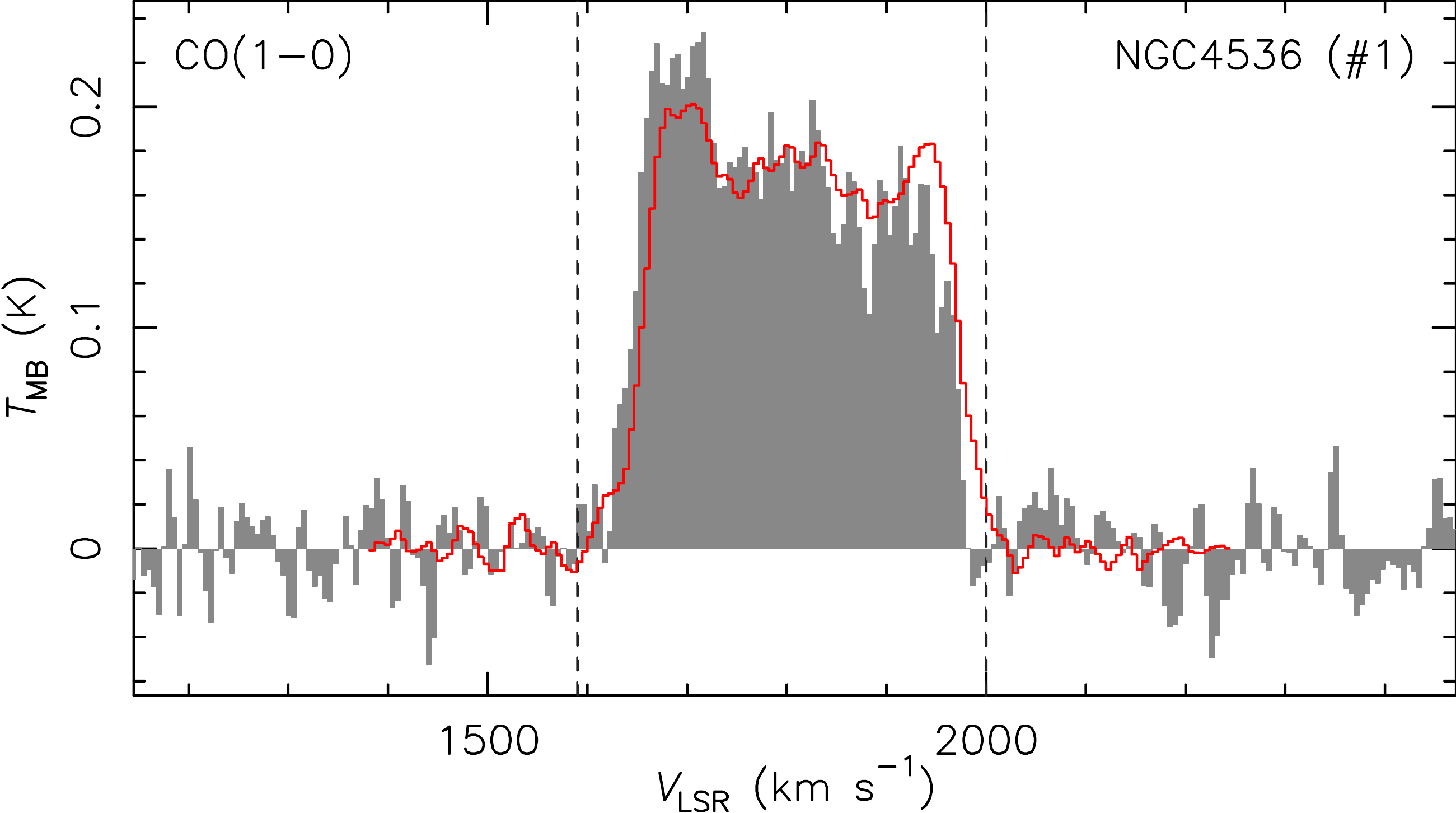} \\ 
\end{tabular}
\end{center}  
\caption{Same as Fig.~\ref{f-spec-1} for NGC~4321 and NGC~4536.} 
\end{figure} 
\newpage
\begin{figure}[h!]
\begin{center}  
\begin{tabular}{cc} 
\includegraphics[width=0.42\textwidth]{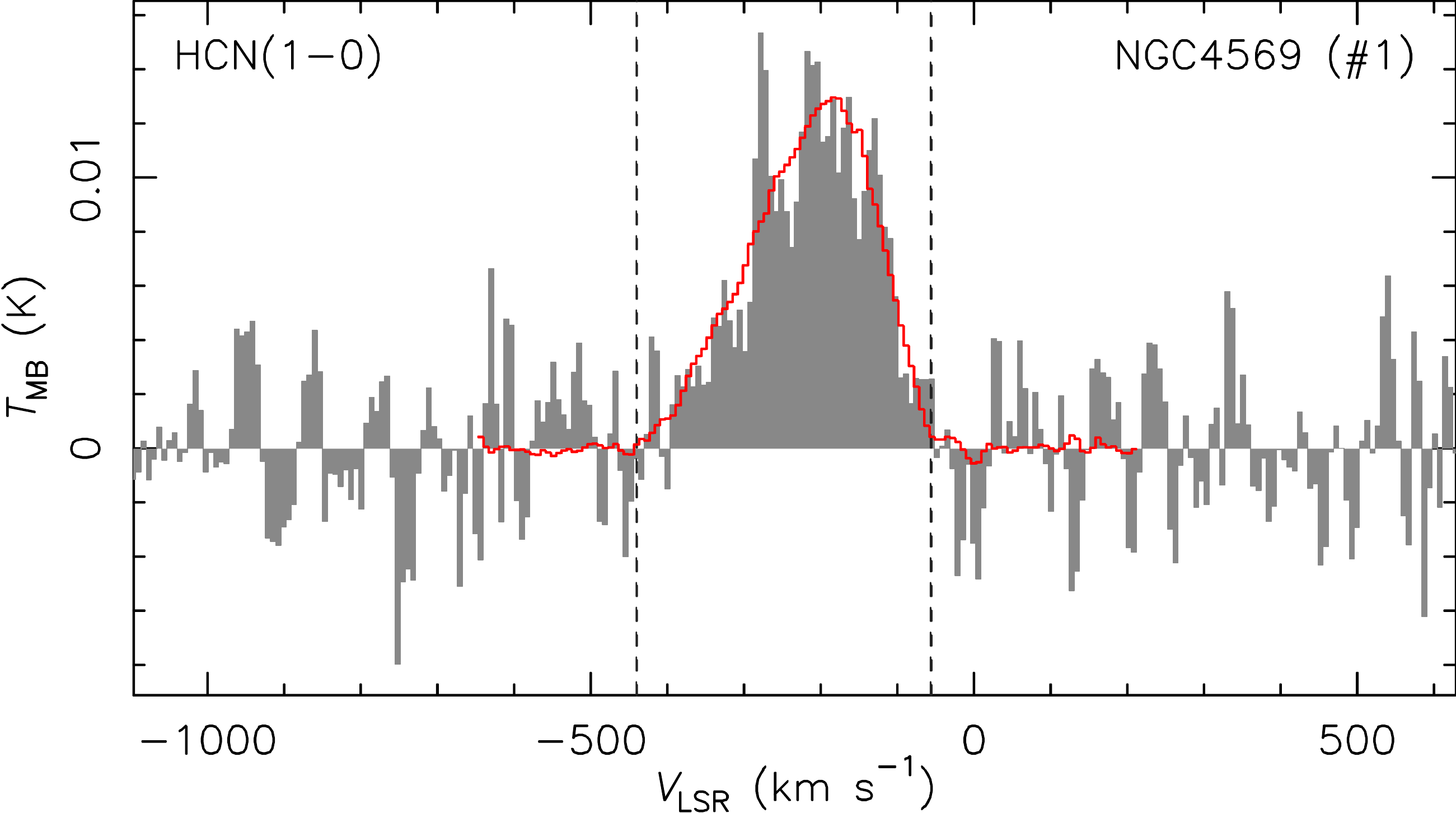} & 
\includegraphics[width=0.42\textwidth]{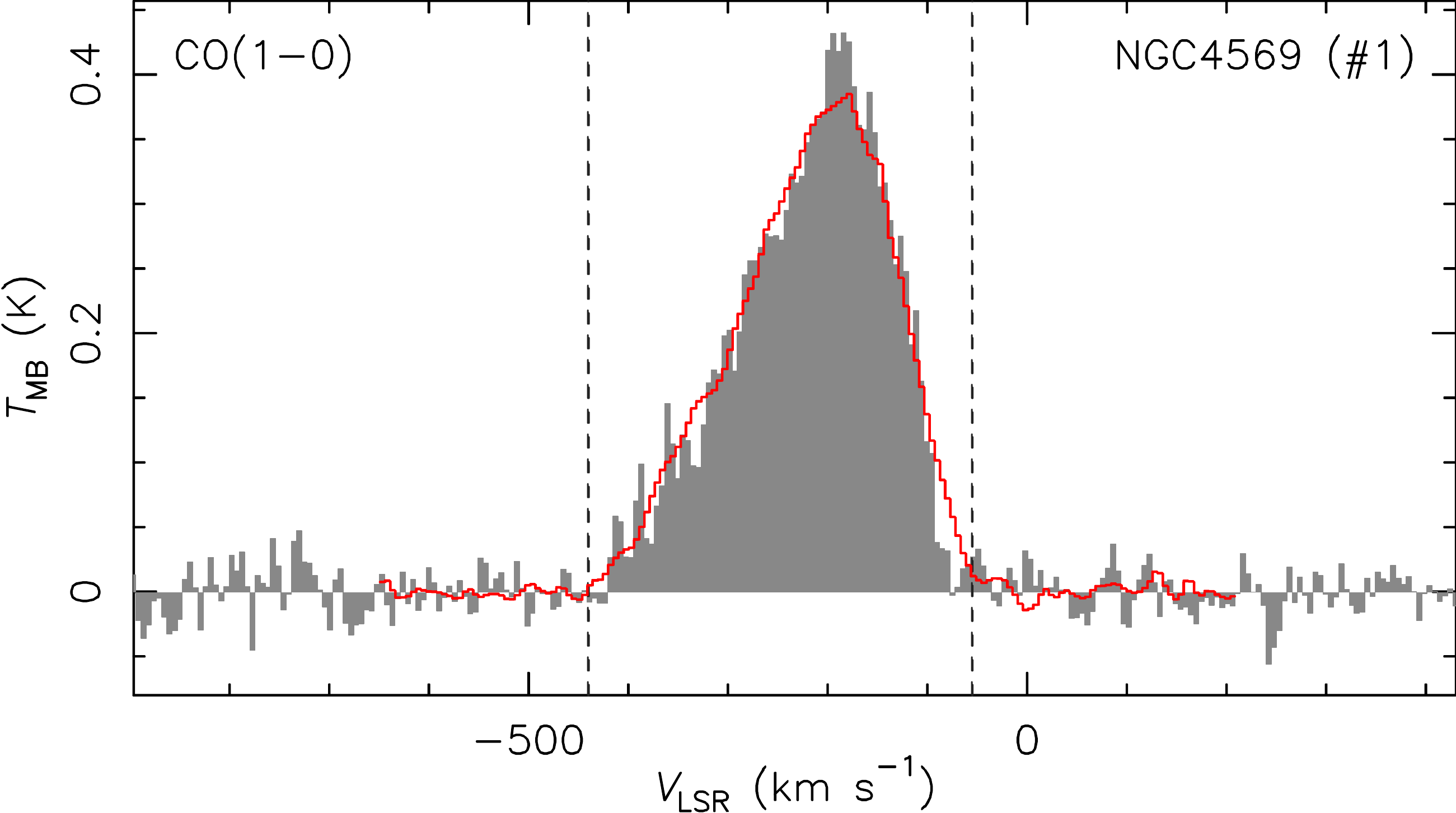} \\ 
\includegraphics[width=0.42\textwidth]{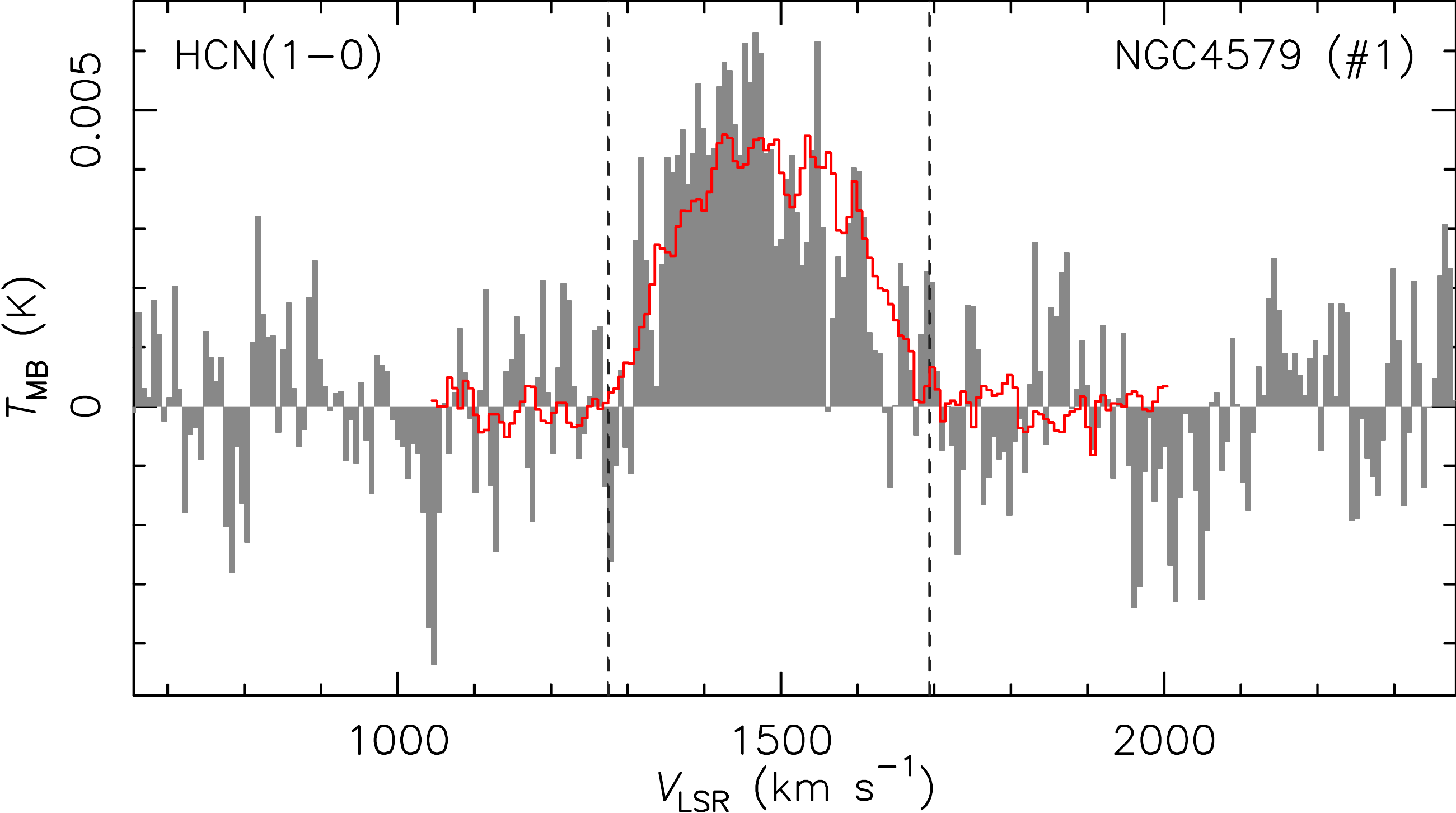} & 
\includegraphics[width=0.42\textwidth]{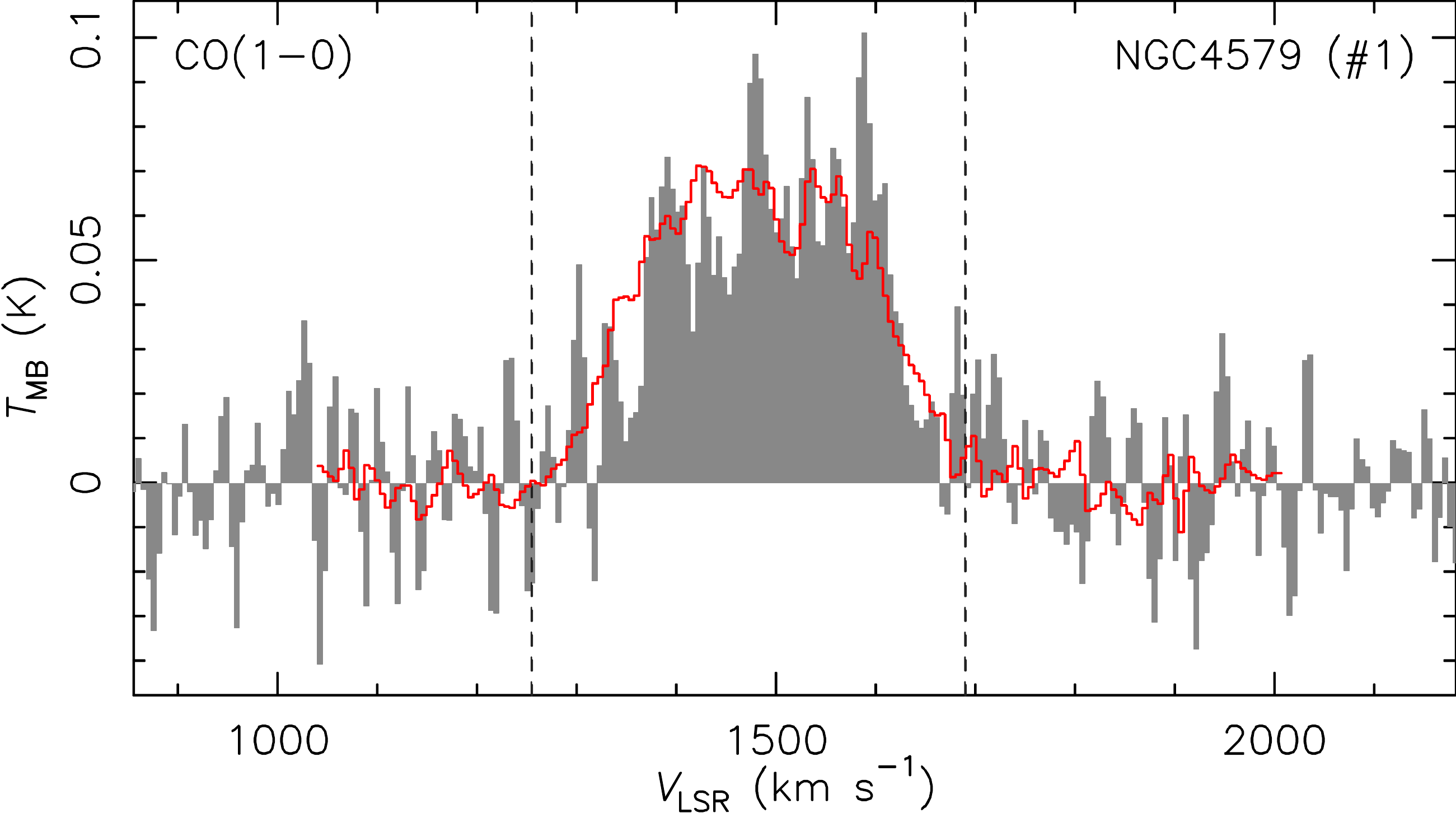} \\ 
\includegraphics[width=0.42\textwidth]{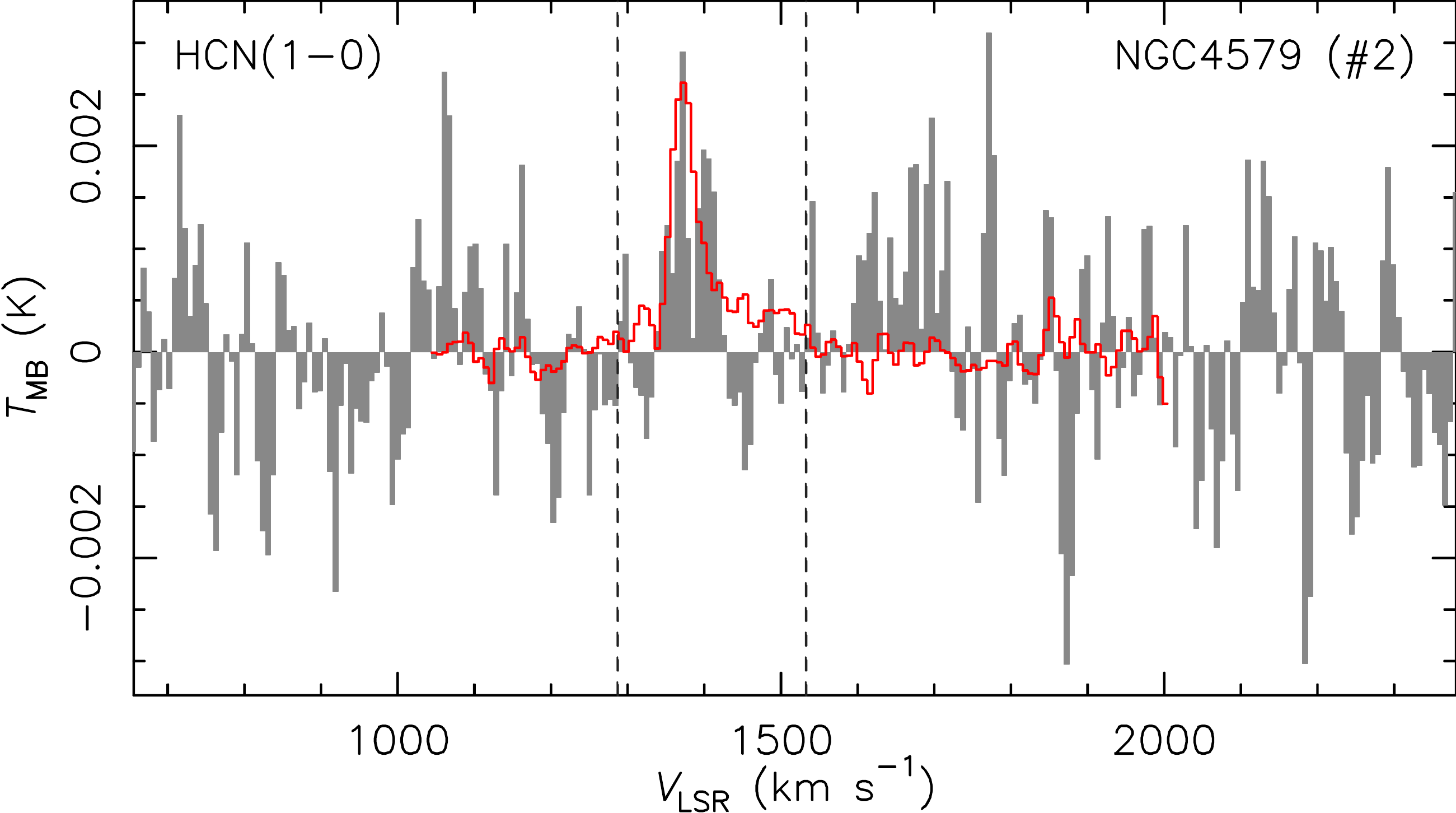} & 
\includegraphics[width=0.42\textwidth]{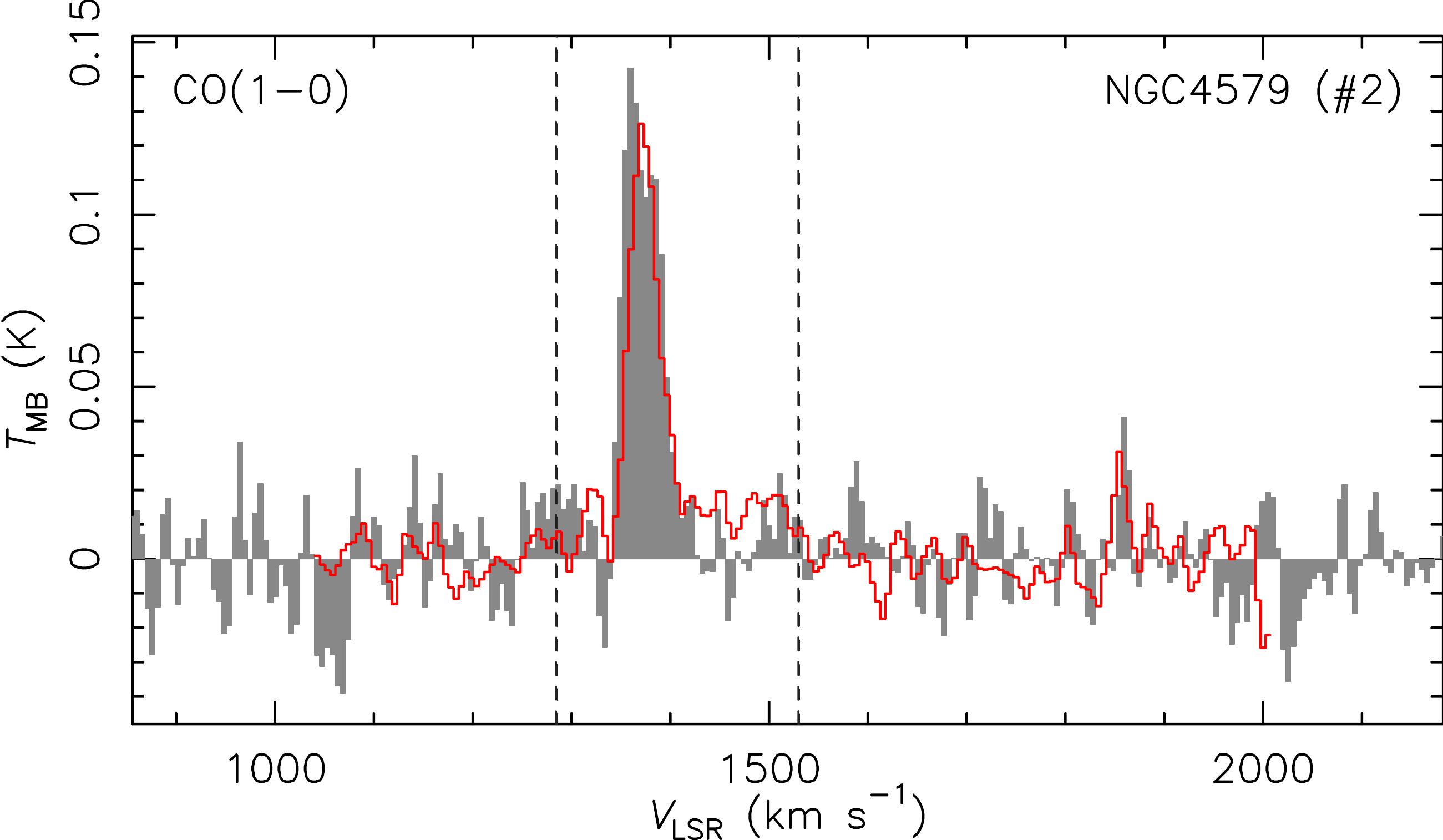} \\ 
\includegraphics[width=0.42\textwidth]{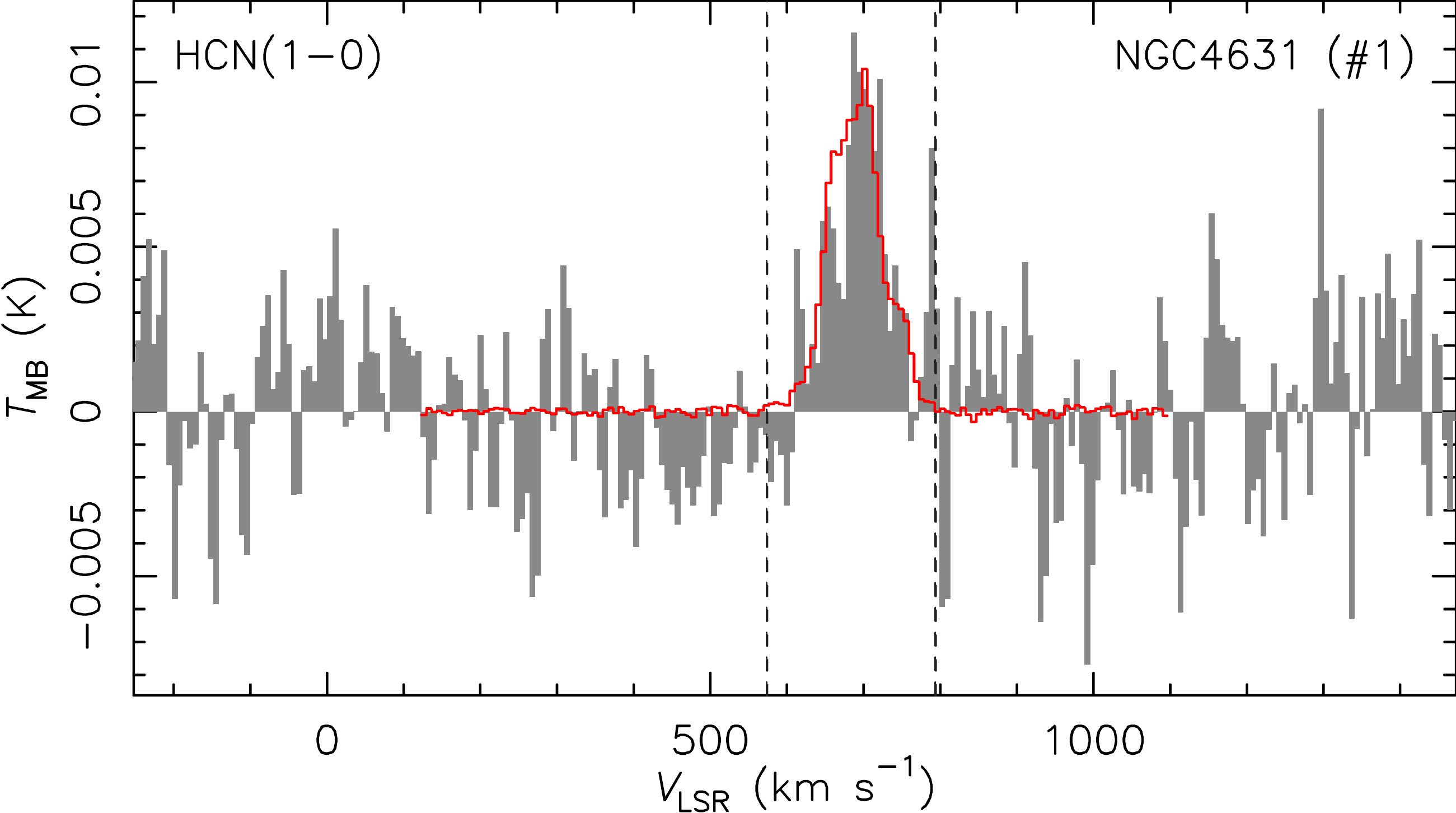} & 
\includegraphics[width=0.42\textwidth]{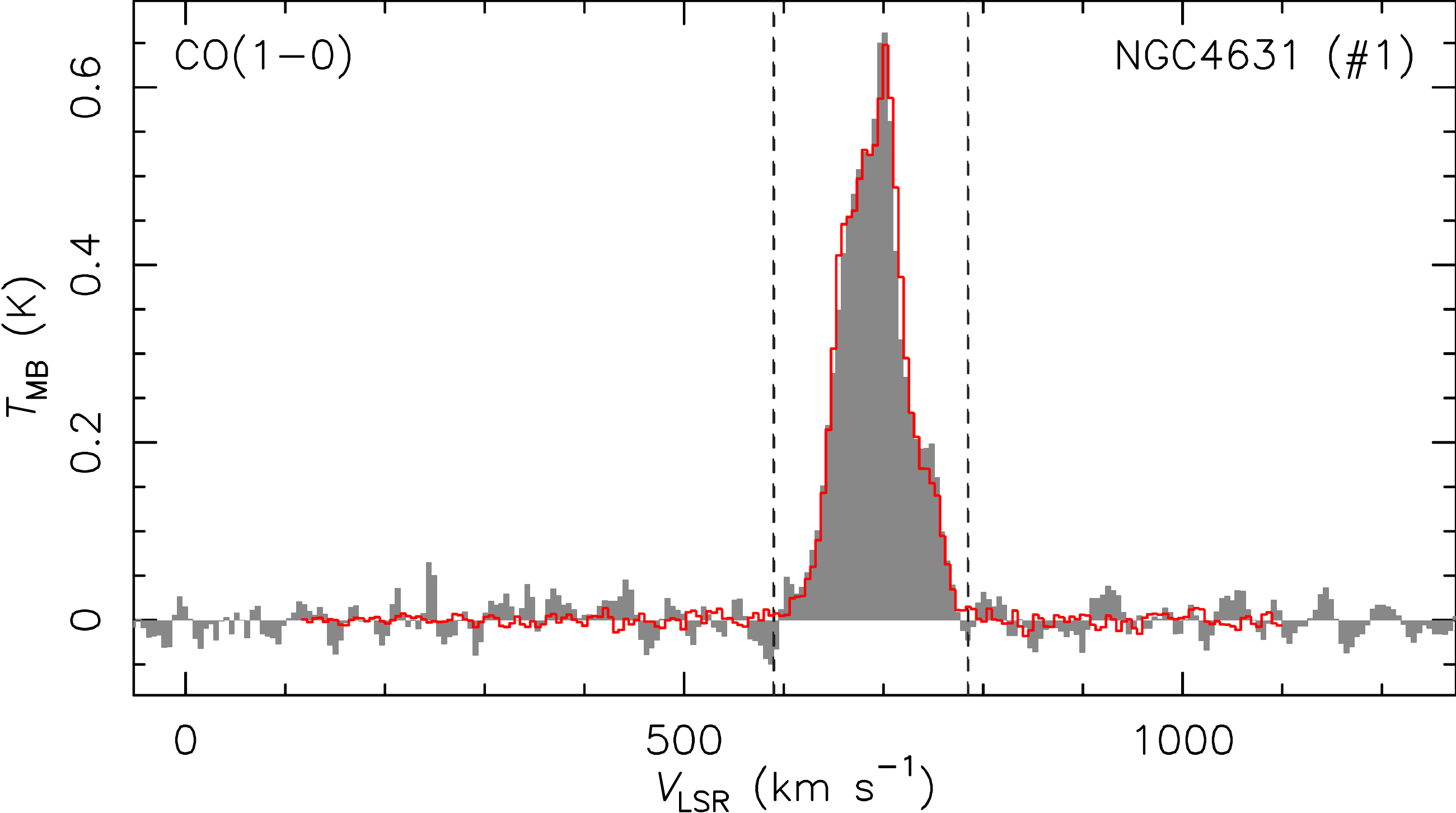} \\ 
\includegraphics[width=0.42\textwidth]{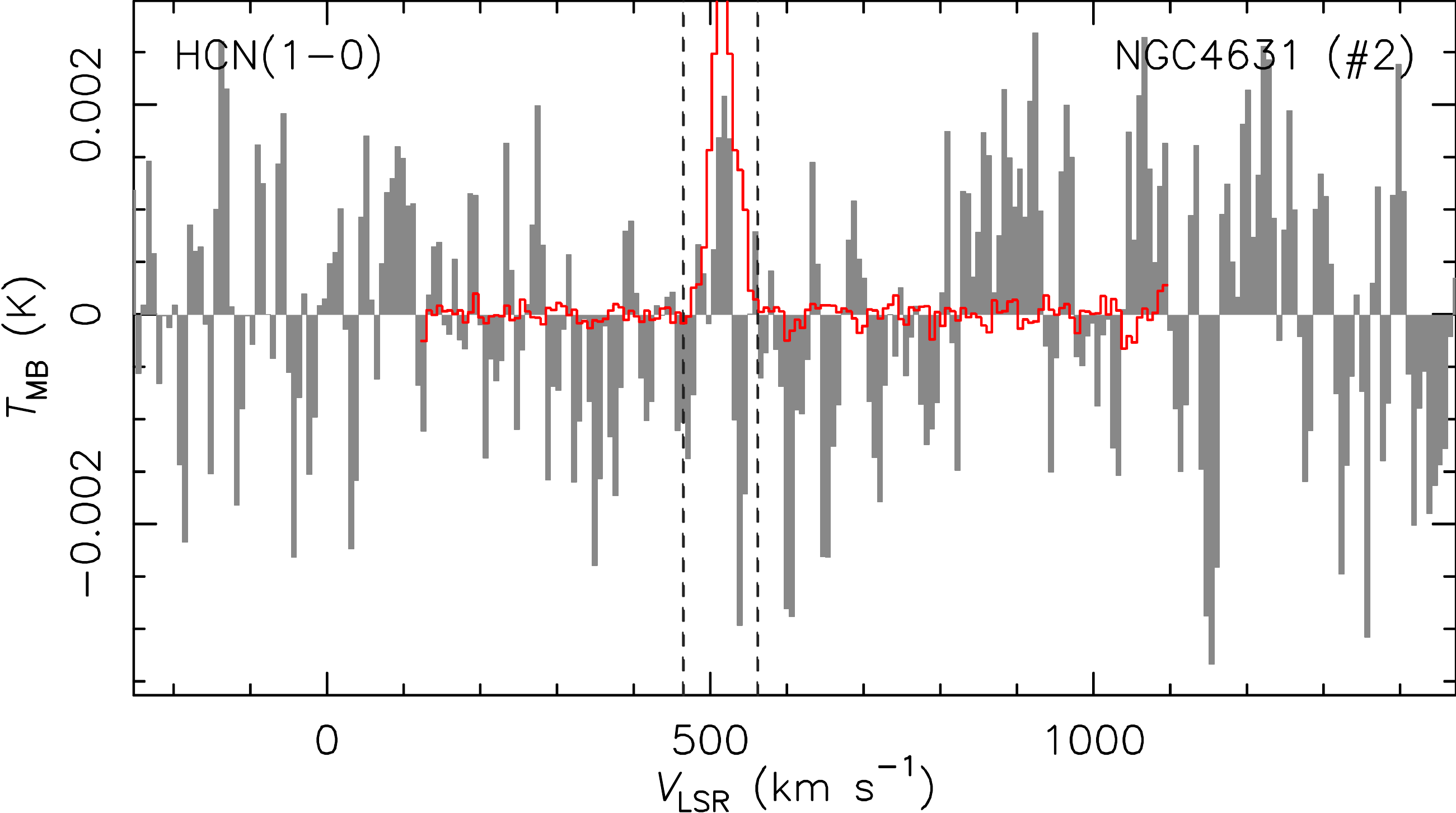} & 
\includegraphics[width=0.42\textwidth]{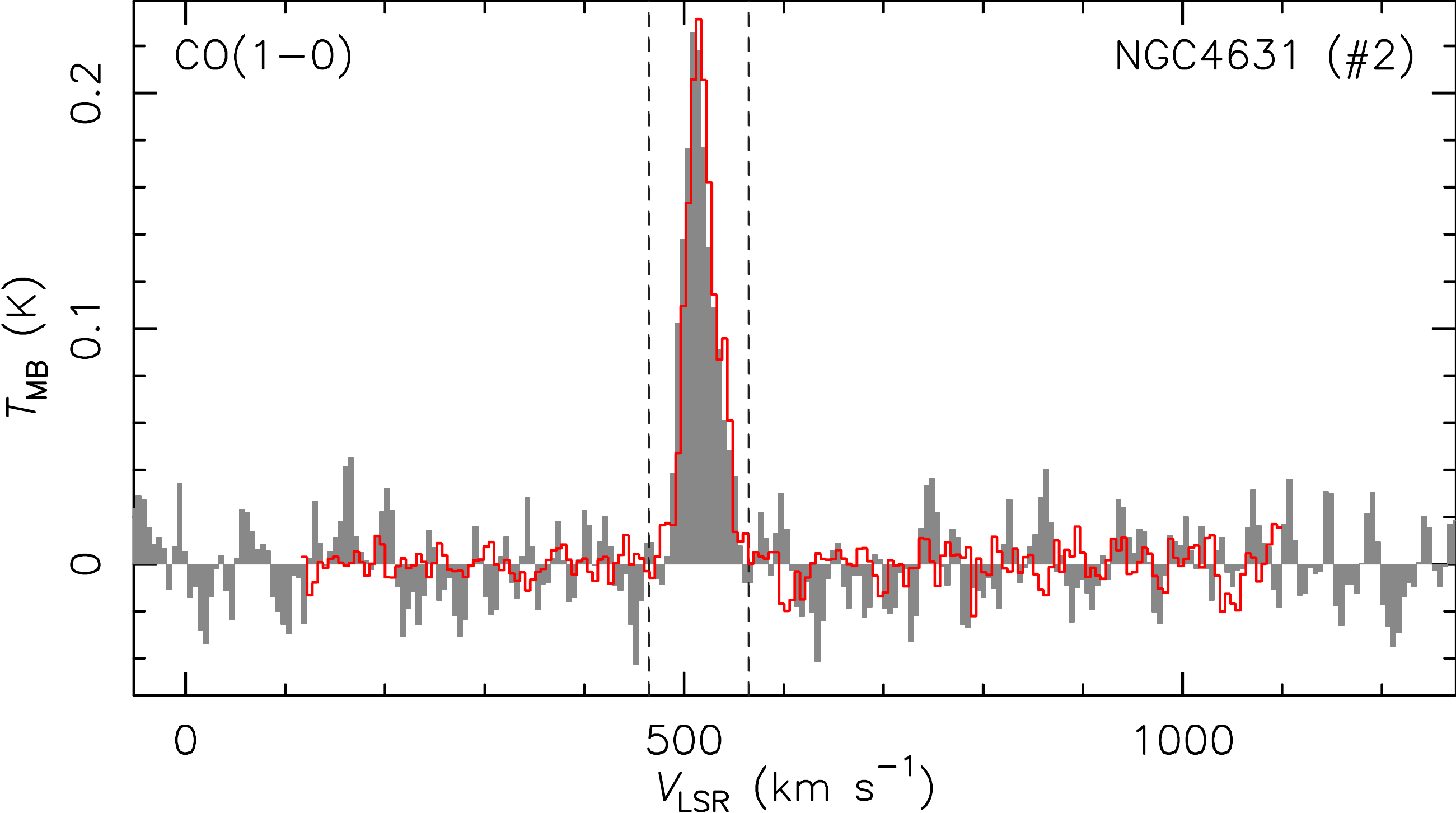} \\ 
\end{tabular}
\end{center}  
\caption{Same as Fig.~\ref{f-spec-1} for NGC~4569, NGC~4579, and NGC~4631.} 
\end{figure} 
\newpage
\begin{figure}[h!]
\begin{center}  
\begin{tabular}{cc} 
\includegraphics[width=0.42\textwidth]{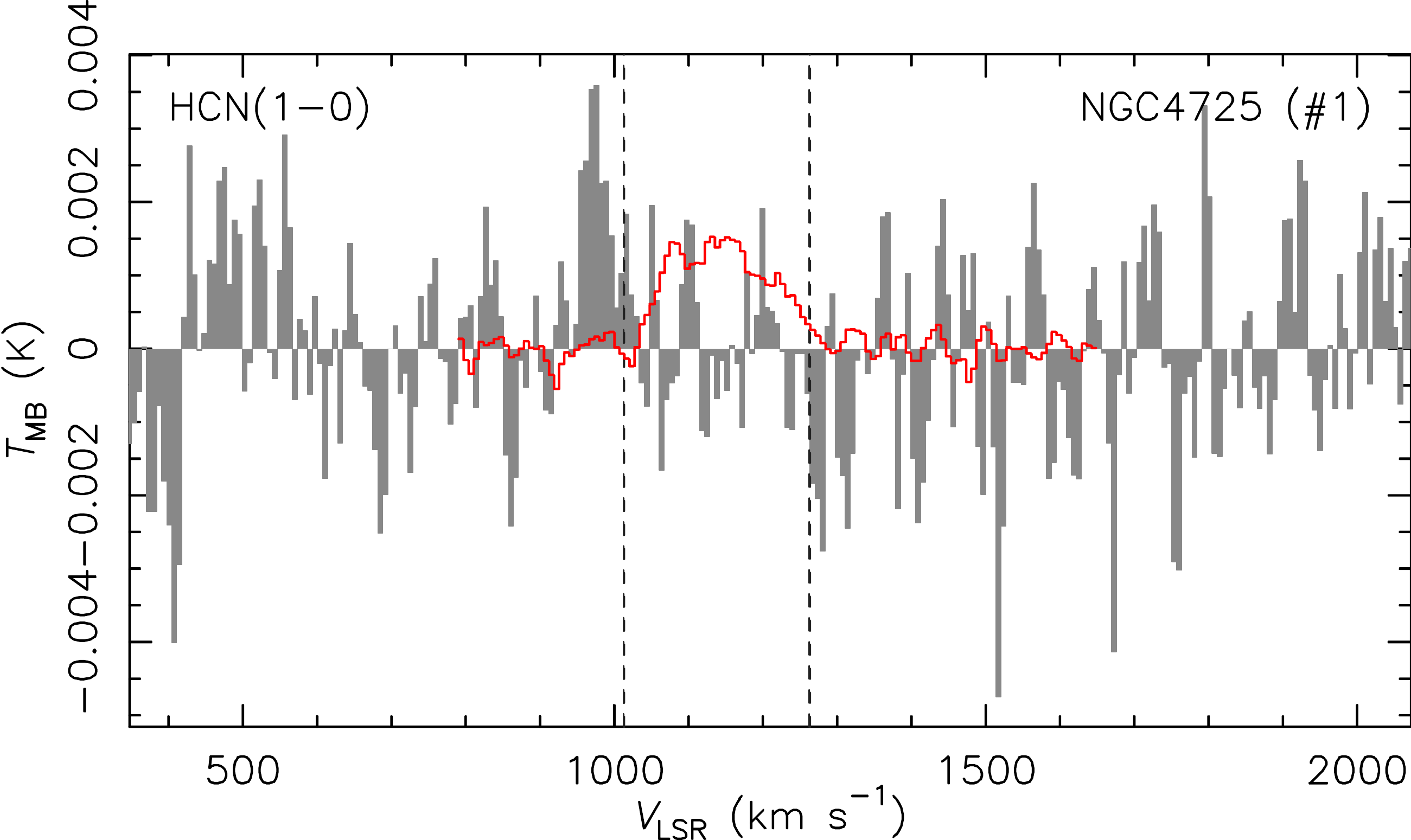} & 
\includegraphics[width=0.42\textwidth]{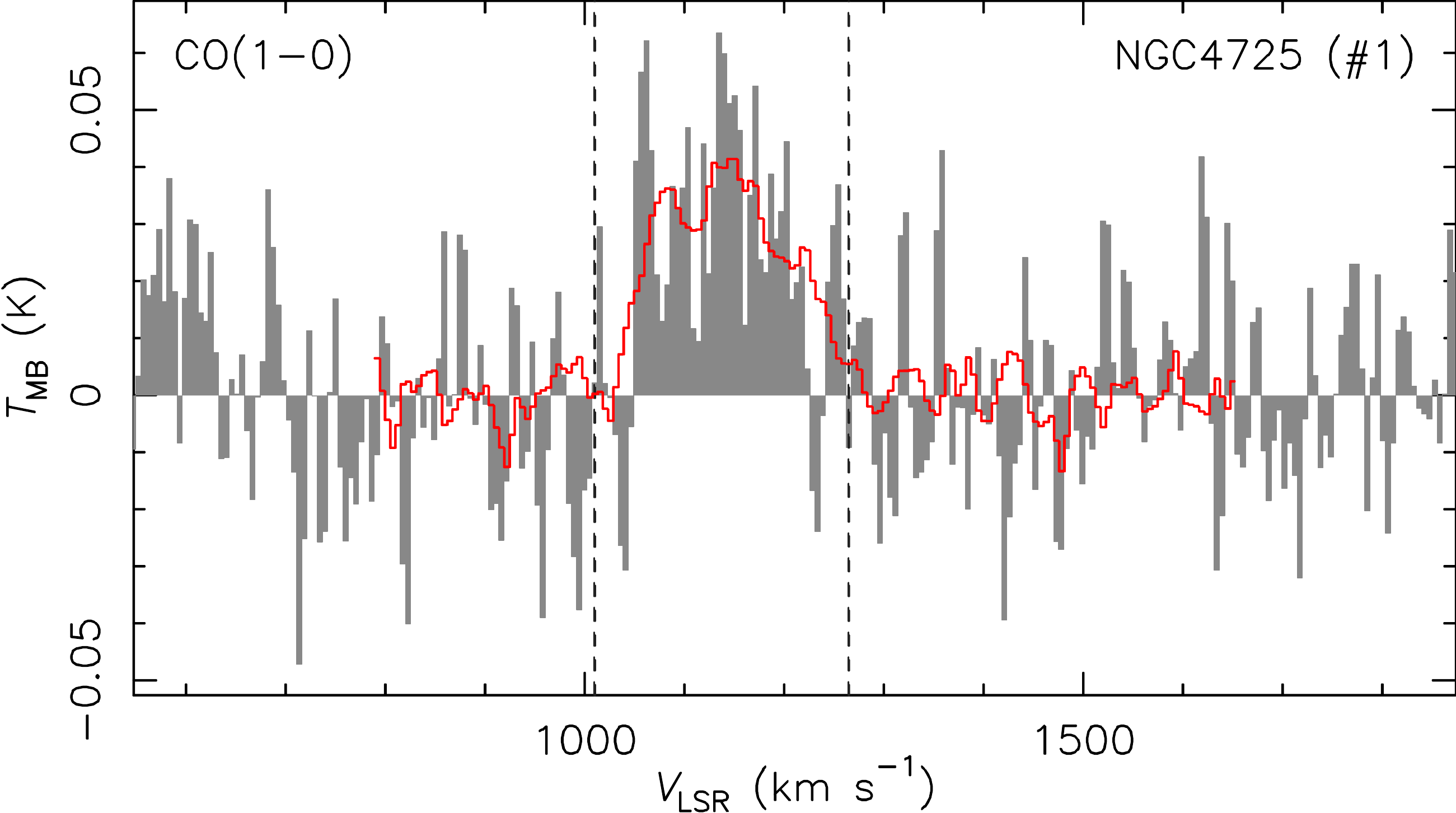} \\ 
\includegraphics[width=0.42\textwidth]{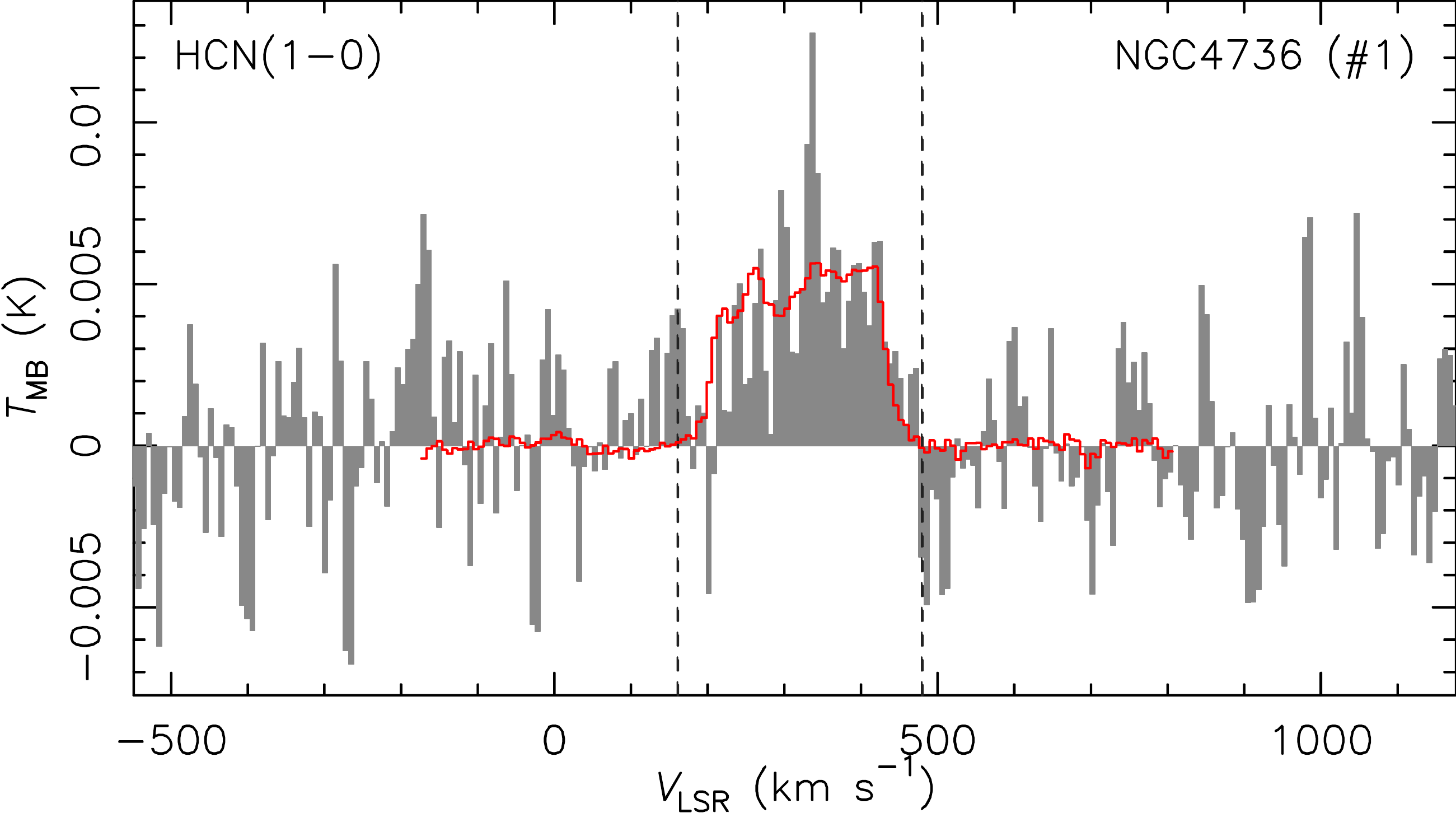} & 
\includegraphics[width=0.42\textwidth]{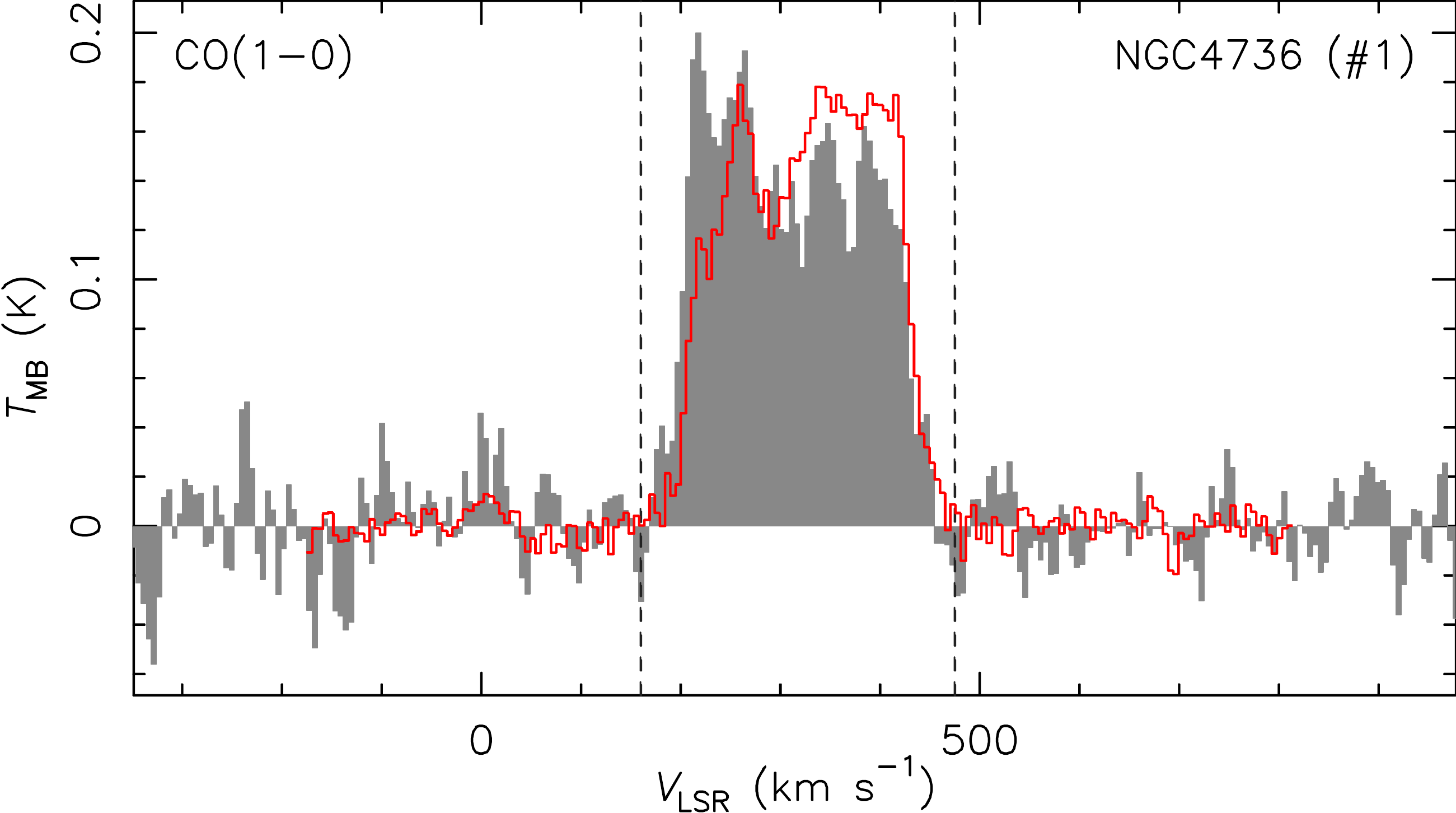} \\ 
\includegraphics[width=0.42\textwidth]{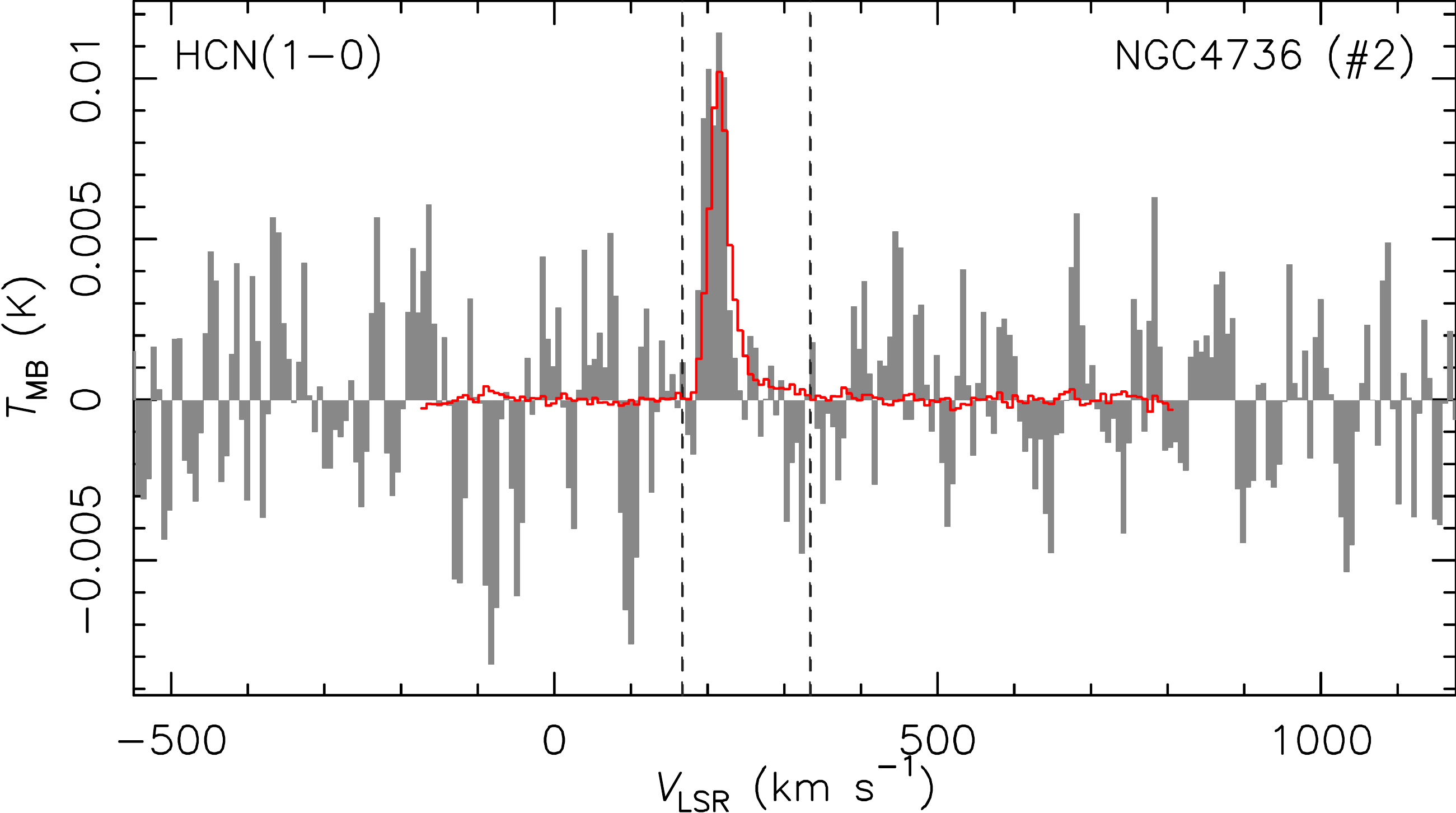} & 
\includegraphics[width=0.42\textwidth]{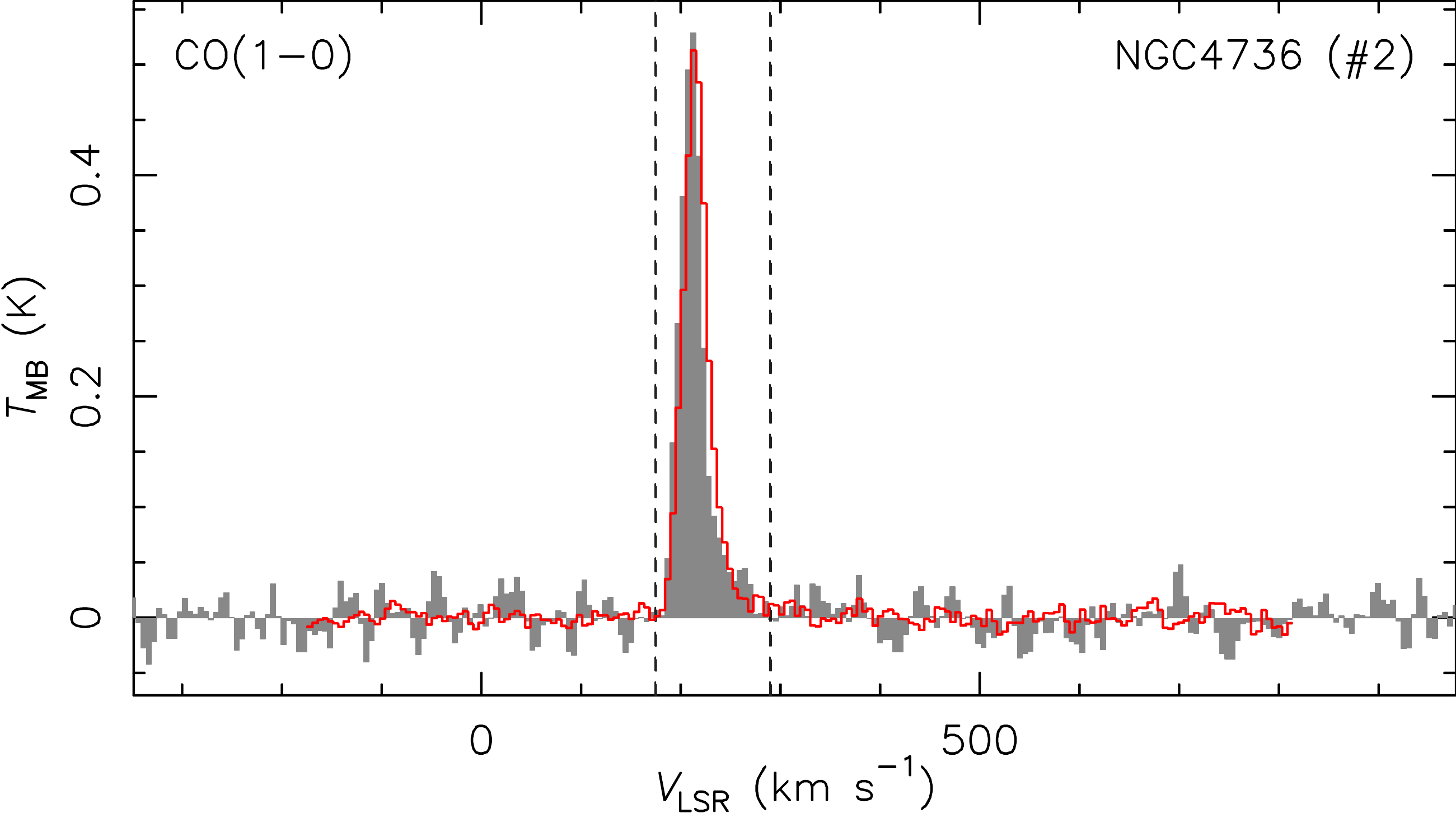} \\ 
\end{tabular}
\end{center}  
\caption{Same as Fig.~\ref{f-spec-1} for NGC~4725 and NGC~4736.} 
\end{figure} 
\newpage 
\begin{figure}[h!]
\begin{center}  
\begin{tabular}{cc} 
\includegraphics[width=0.42\textwidth]{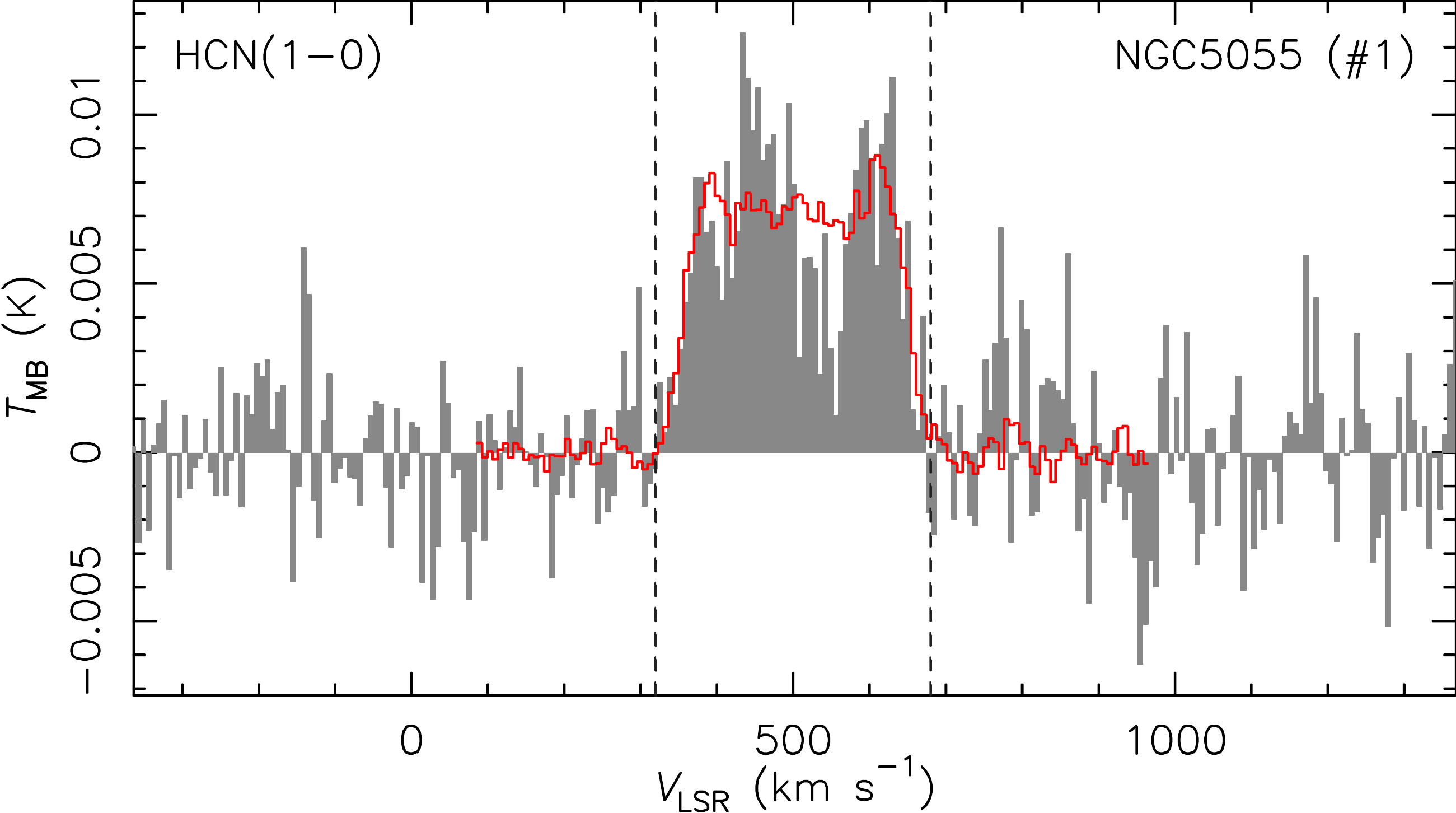} & 
\includegraphics[width=0.42\textwidth]{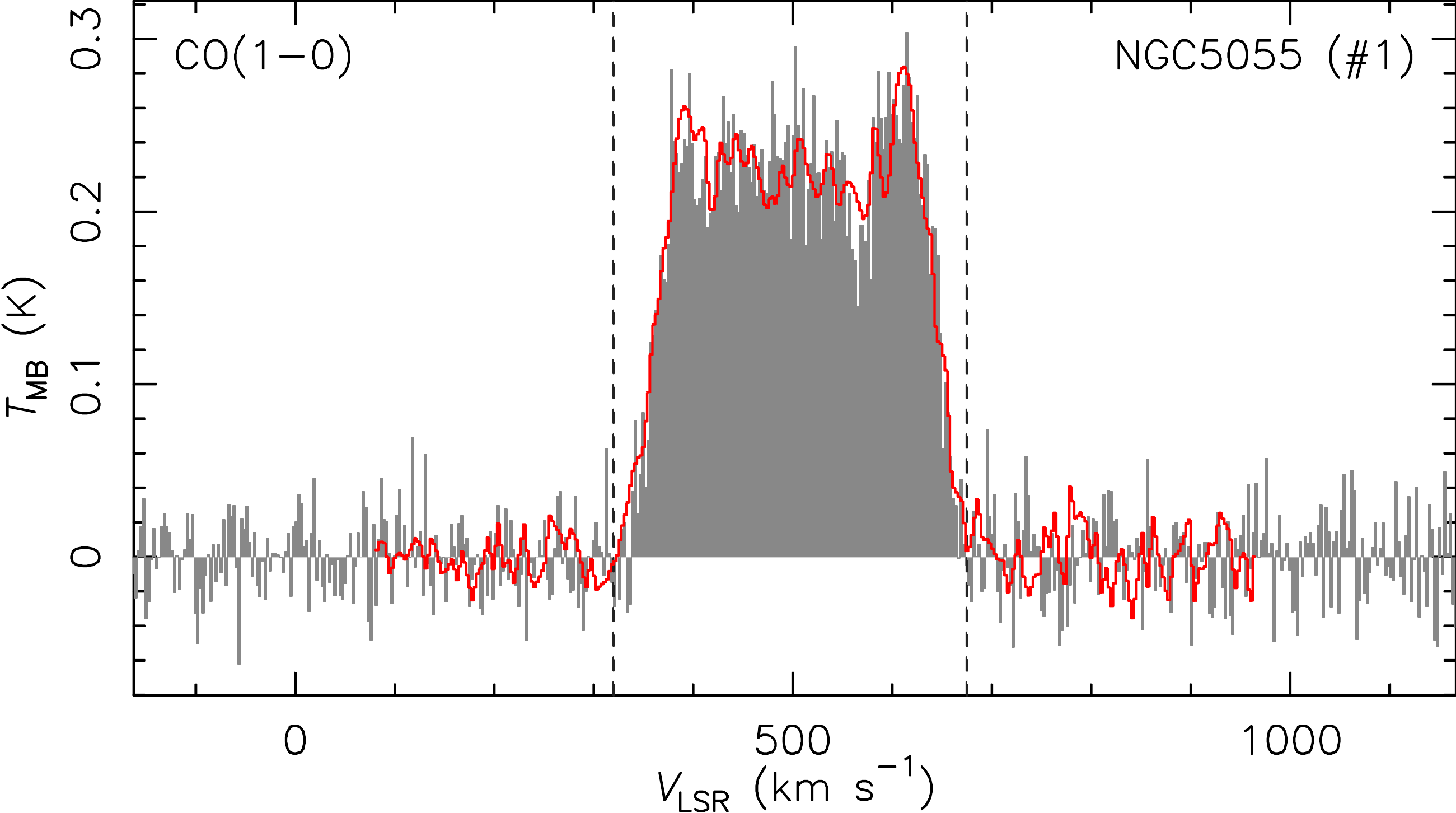} \\ 
\includegraphics[width=0.42\textwidth]{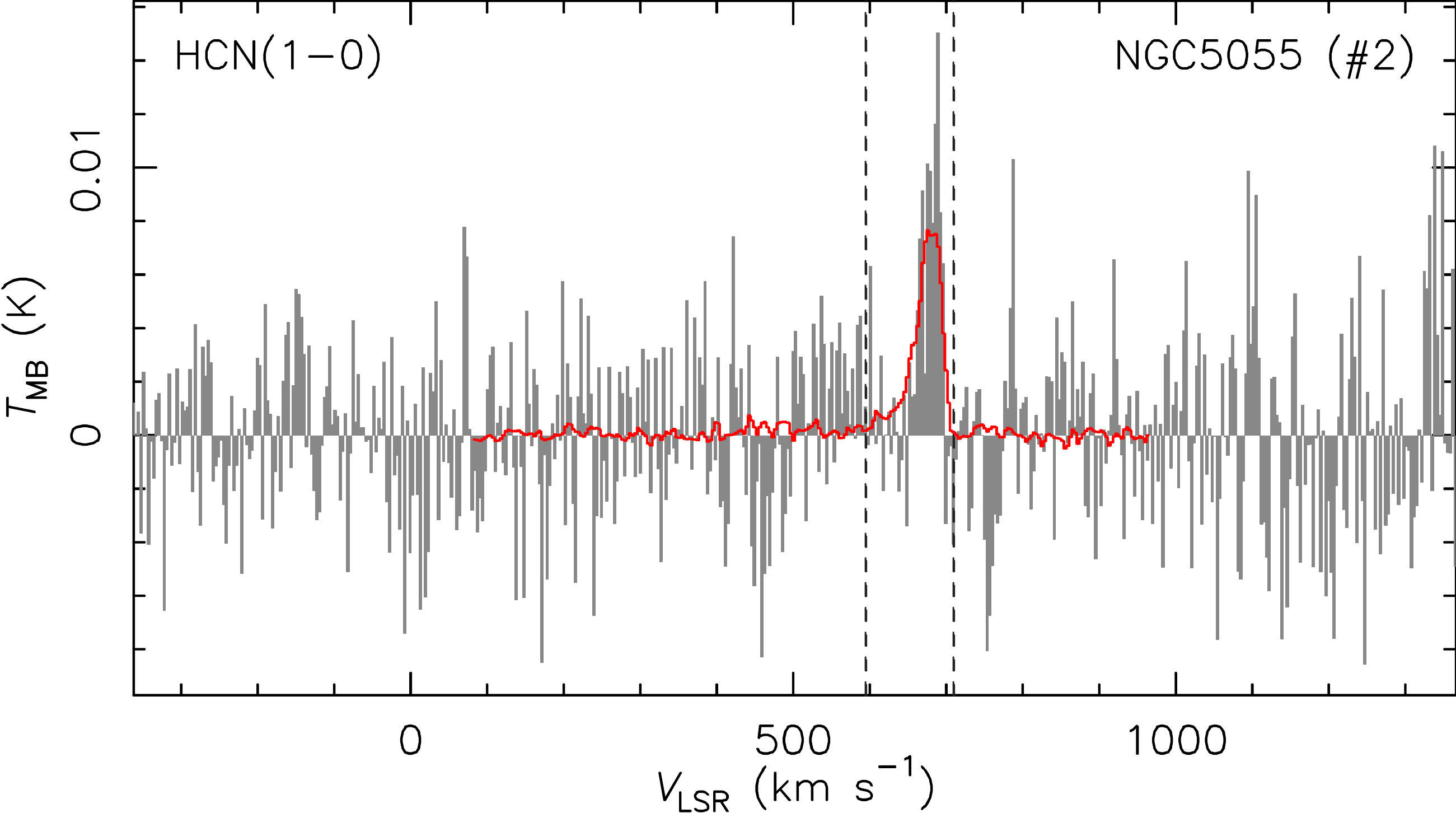} & 
\includegraphics[width=0.42\textwidth]{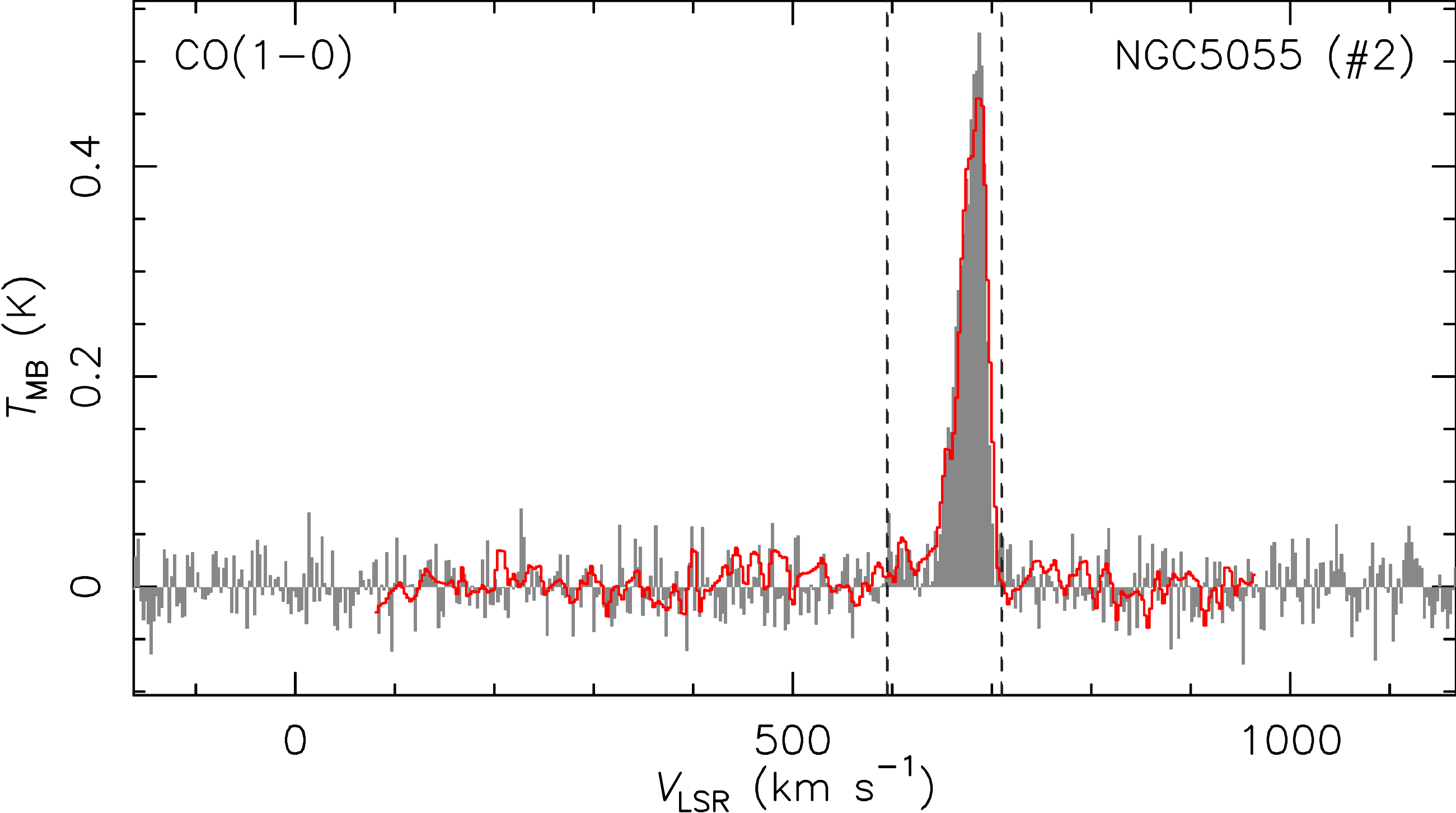} \\ 
\includegraphics[width=0.42\textwidth]{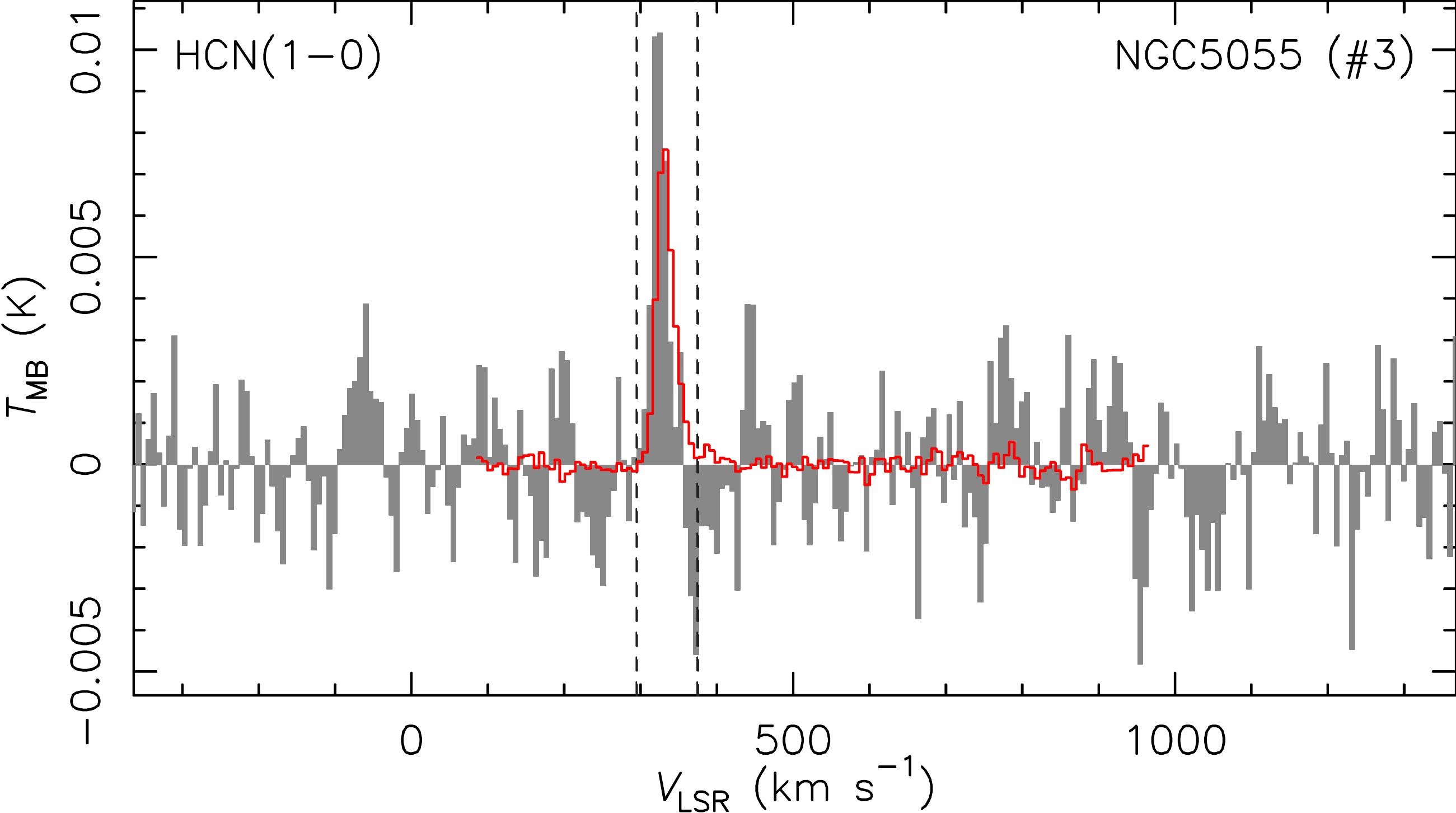} & 
\includegraphics[width=0.42\textwidth]{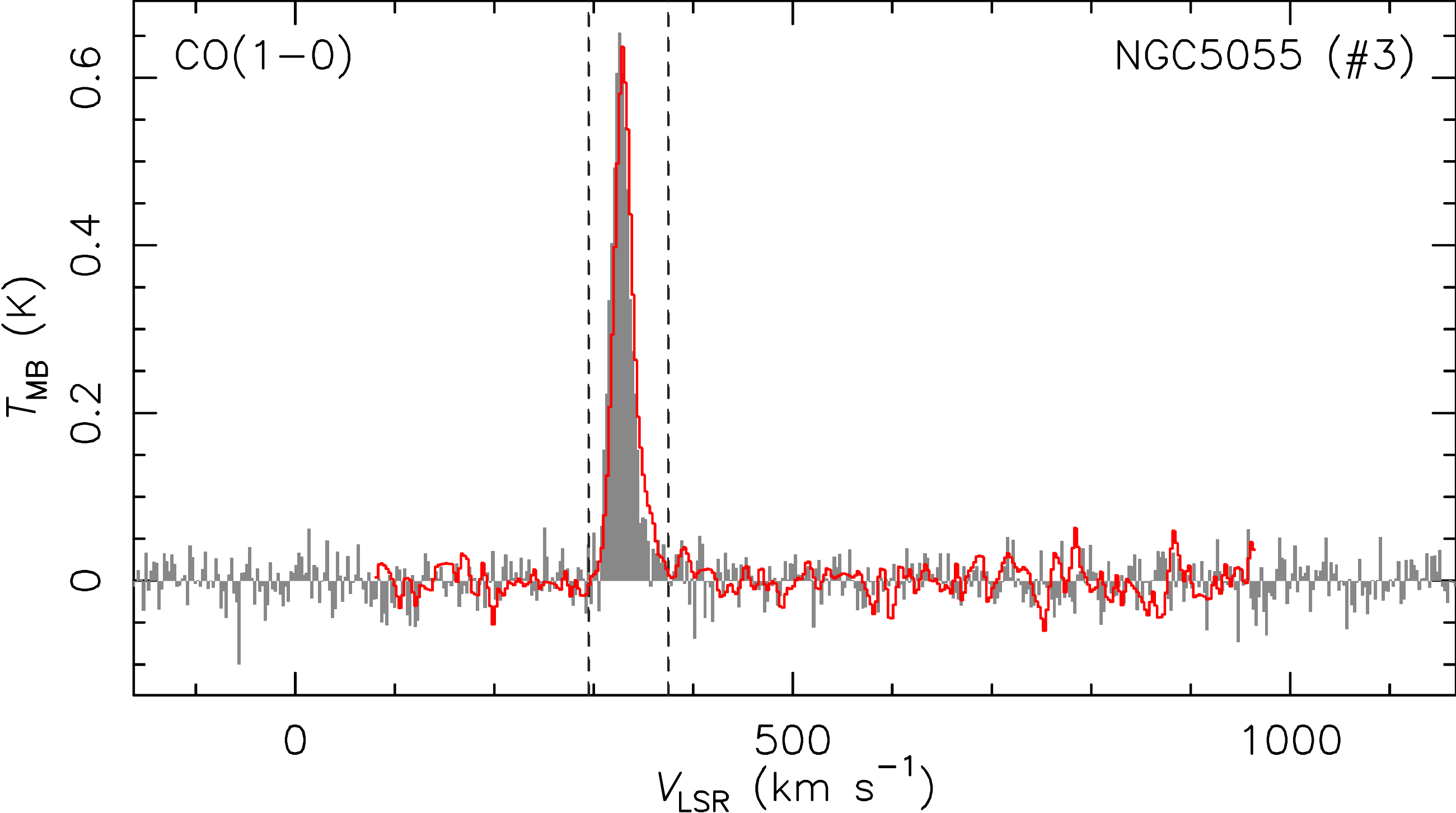} \\ 
\includegraphics[width=0.42\textwidth]{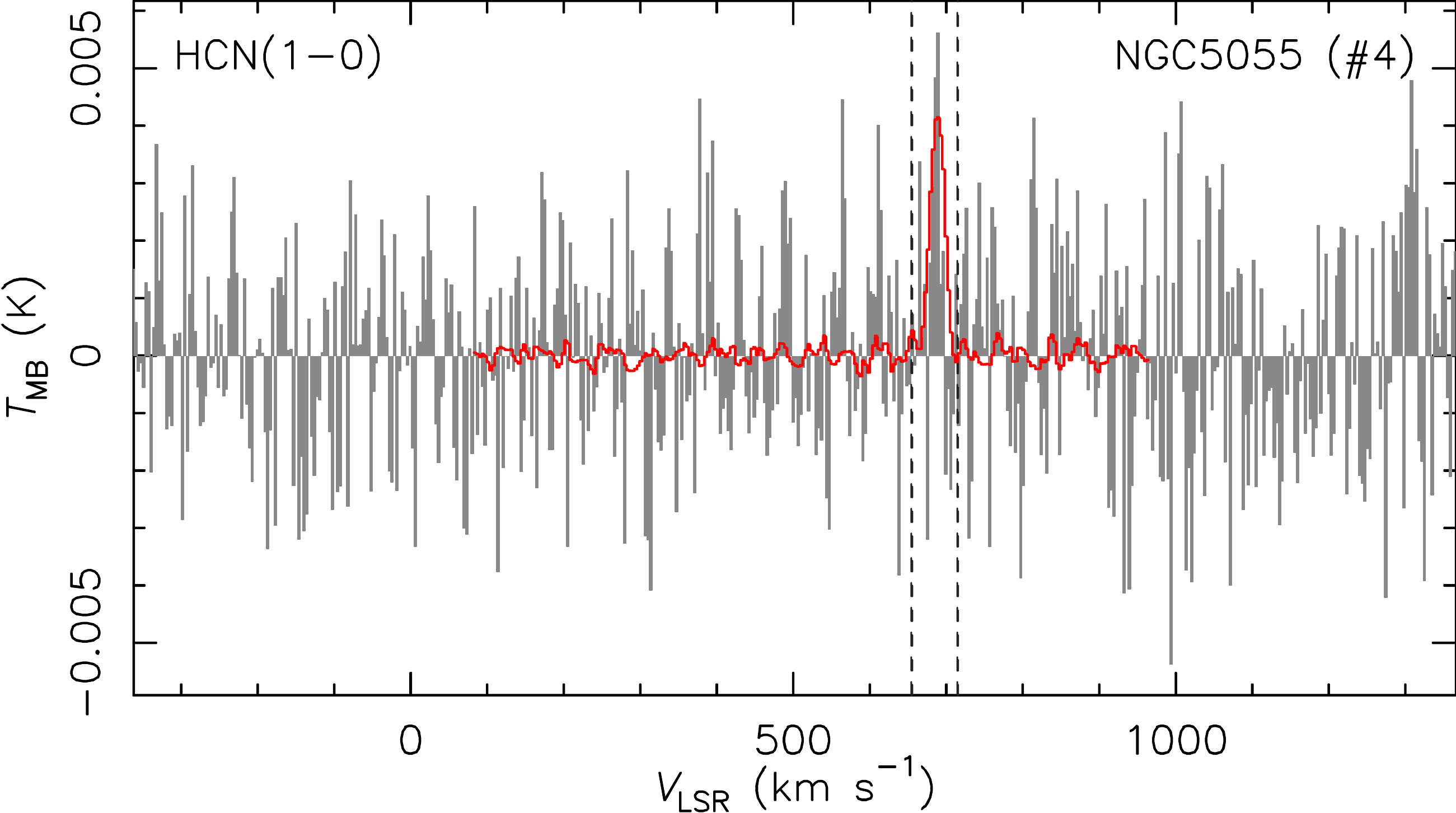} & 
\includegraphics[width=0.42\textwidth]{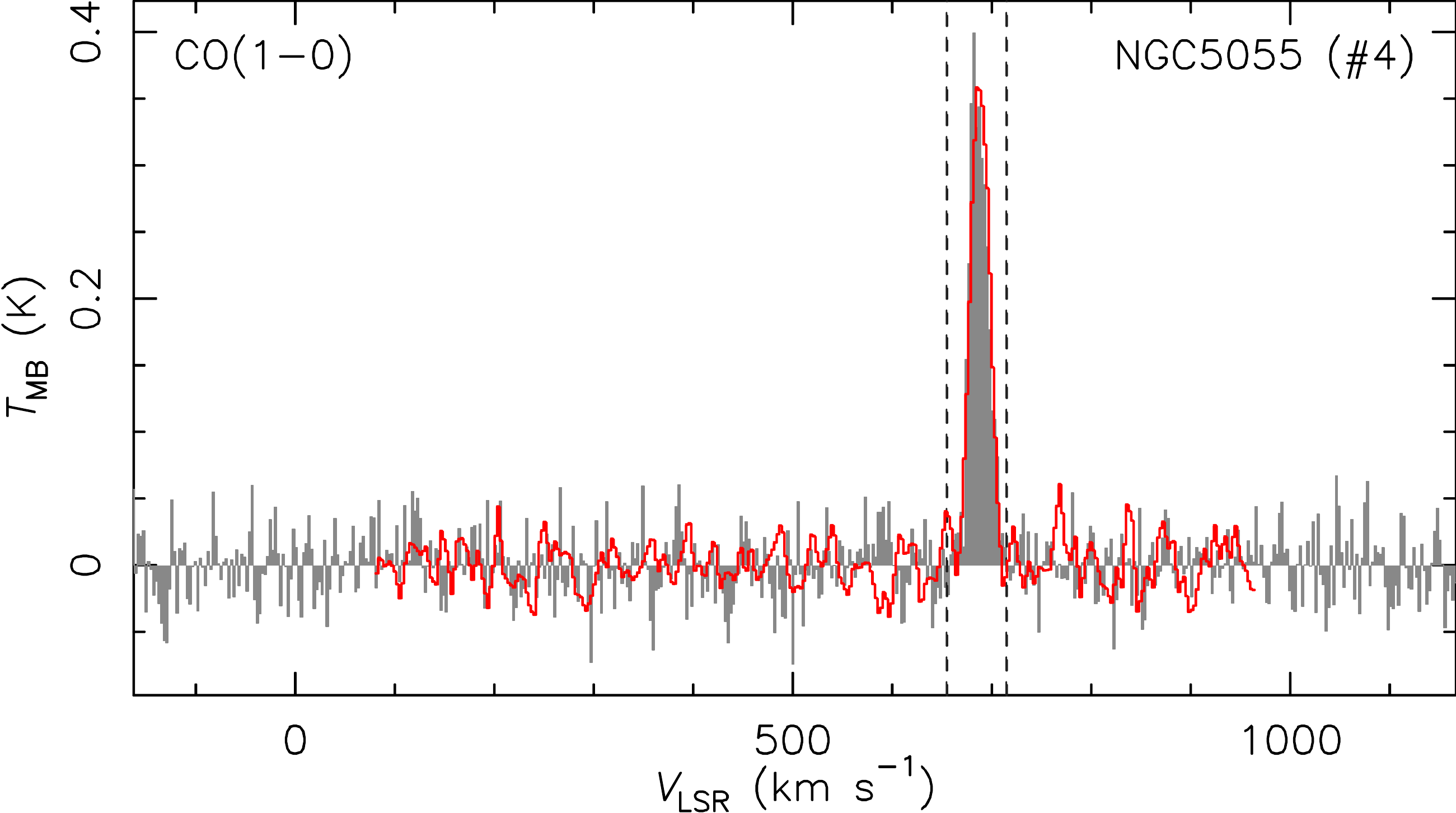} \\ 
\includegraphics[width=0.42\textwidth]{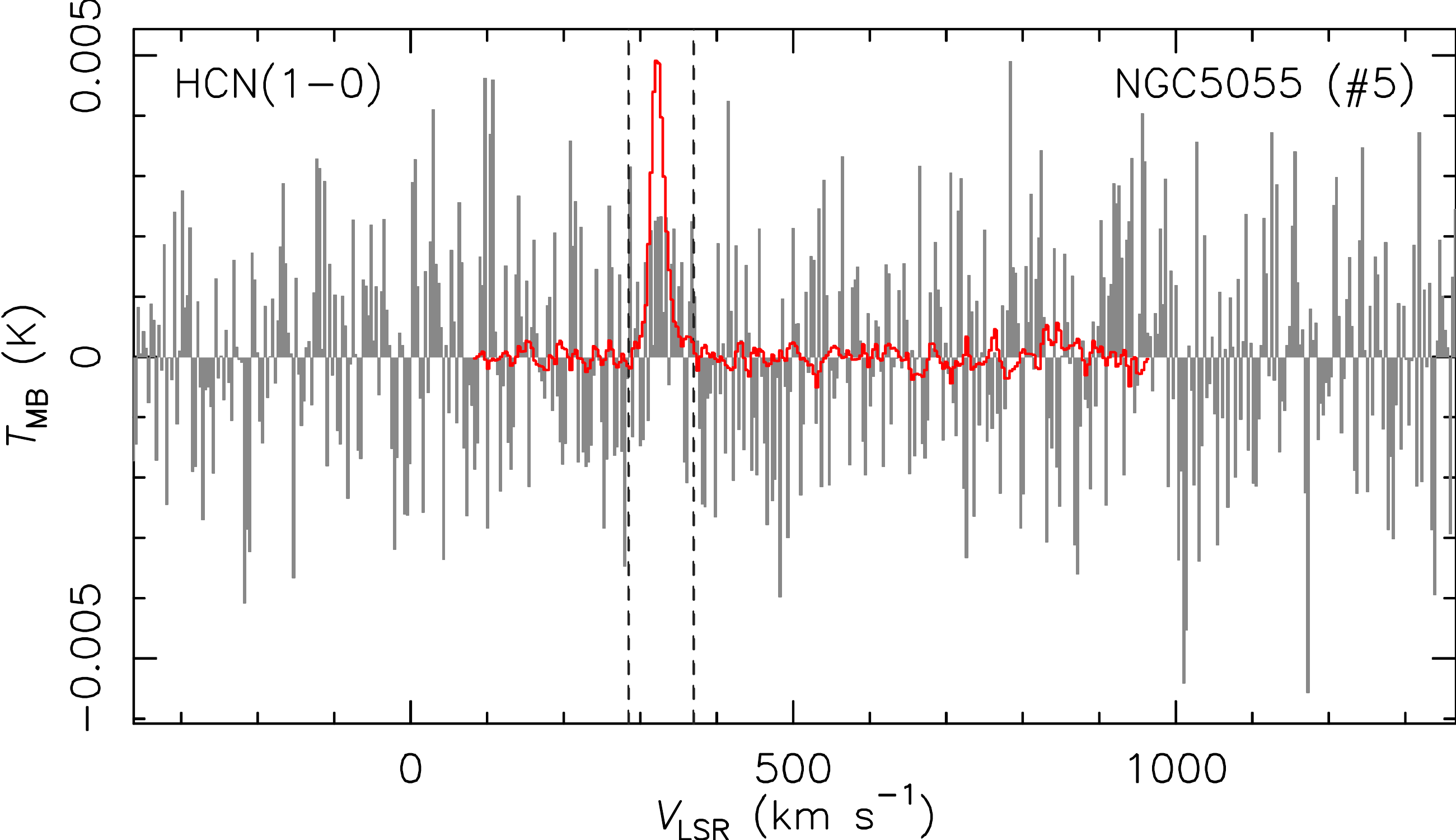} & 
\includegraphics[width=0.42\textwidth]{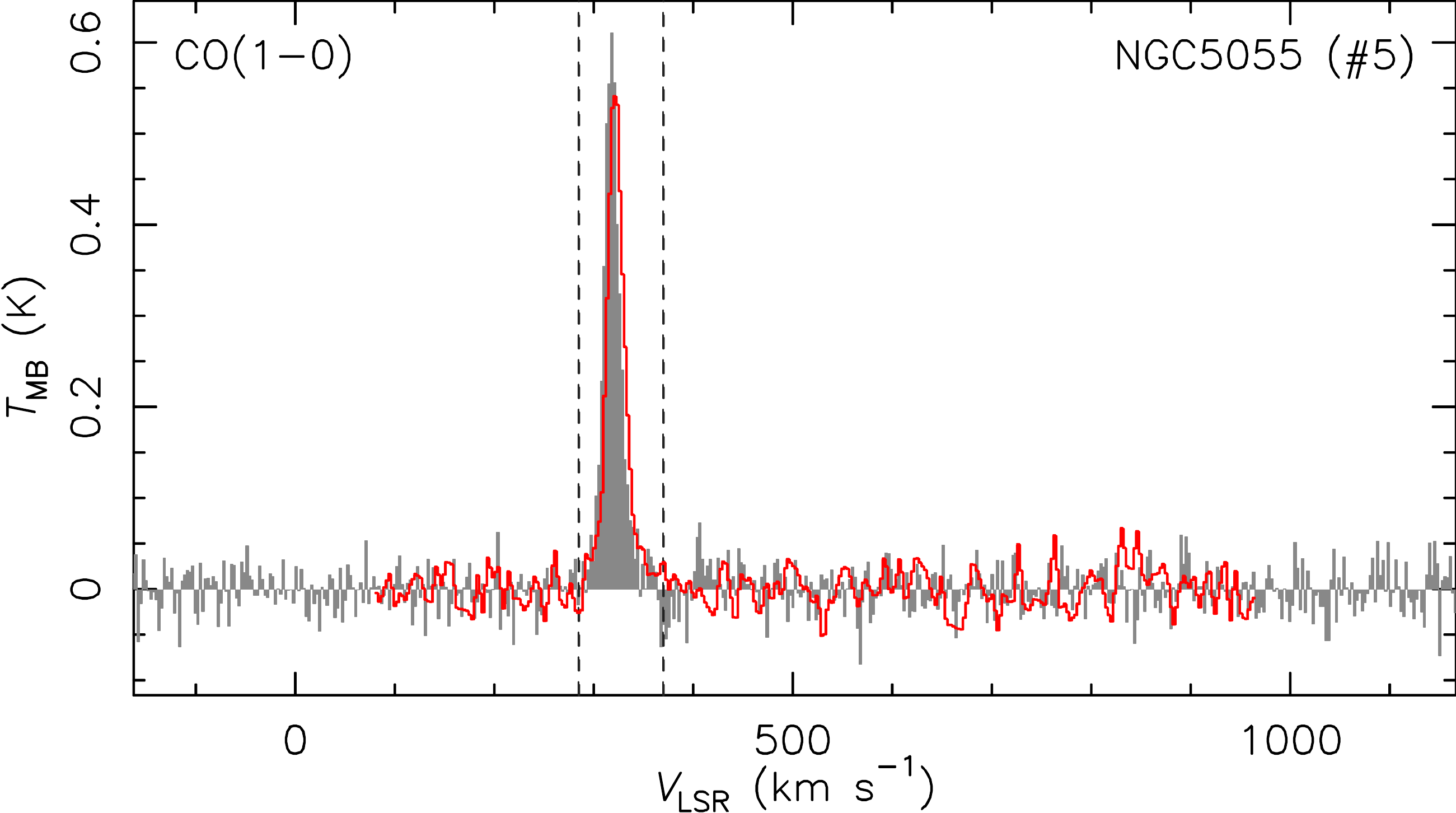} \\ 
\end{tabular}
\end{center}  
\caption{Same as Fig.~\ref{f-spec-1} for NGC~5055.} 
\end{figure} 
\newpage
\begin{figure}[h!]
\begin{center}  
\begin{tabular}{cc} 
\includegraphics[width=0.42\textwidth]{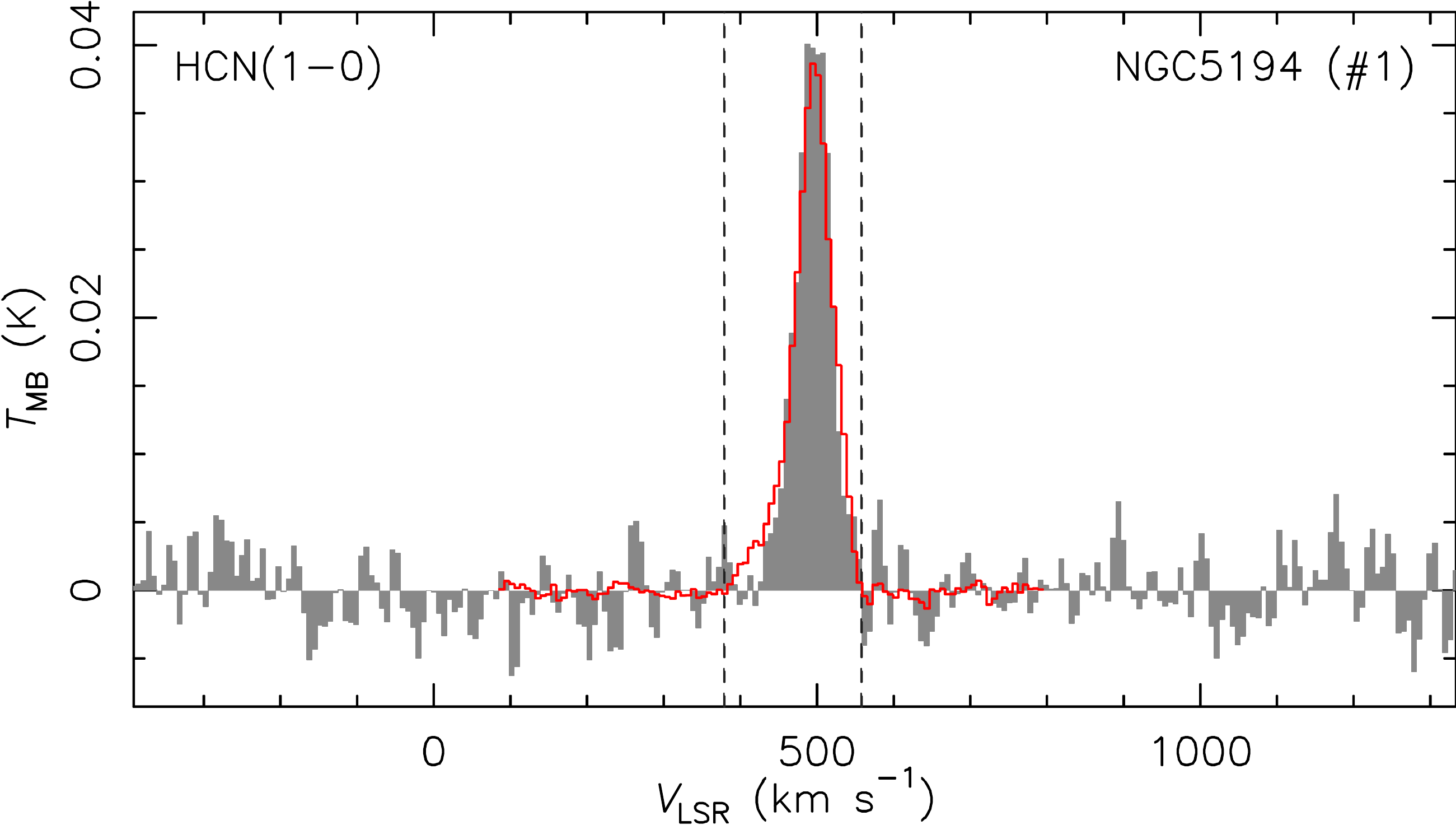} & 
\mybox{} \\ 
\includegraphics[width=0.42\textwidth]{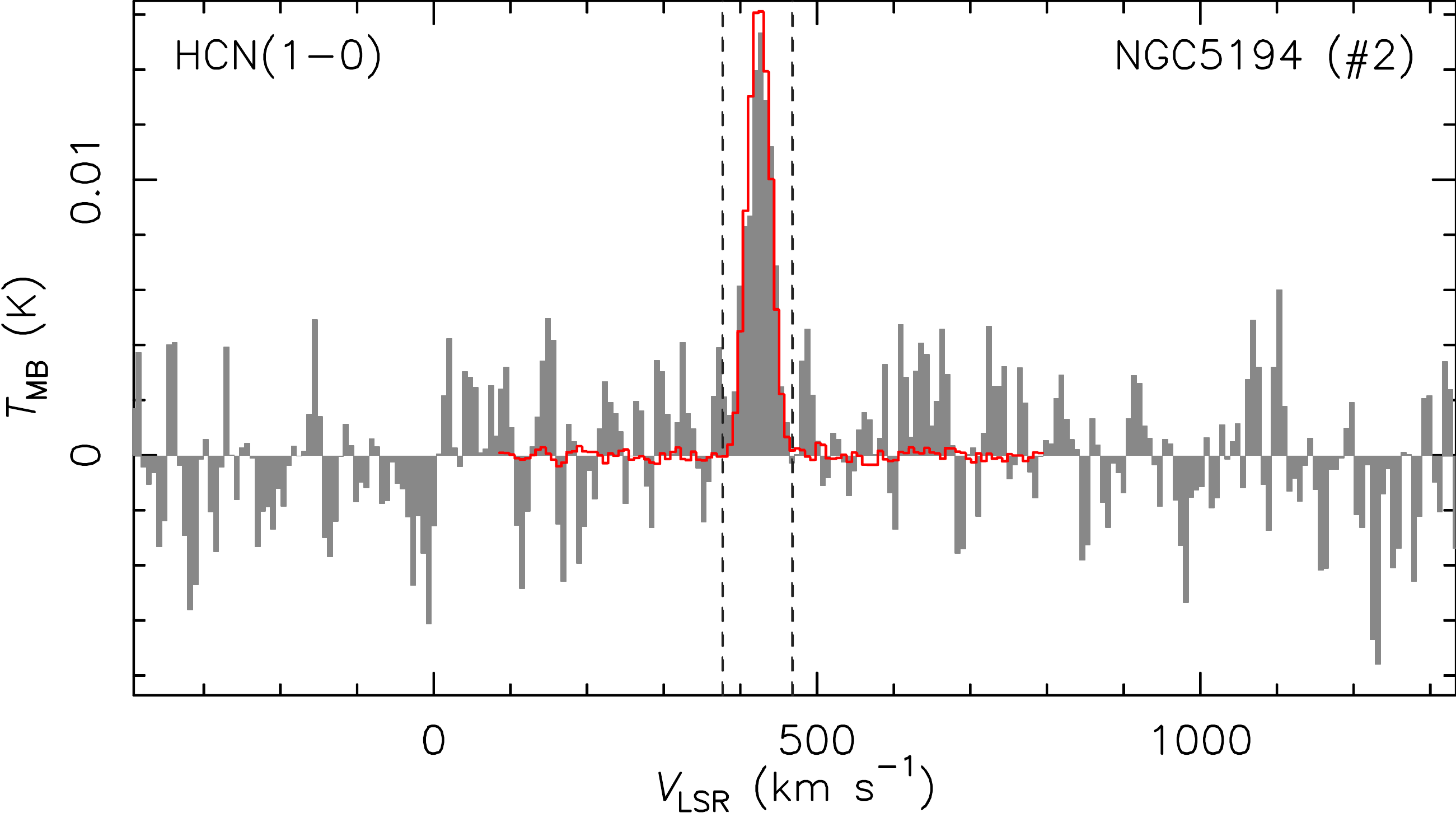} & 
\mybox{} \\ 
\includegraphics[width=0.42\textwidth]{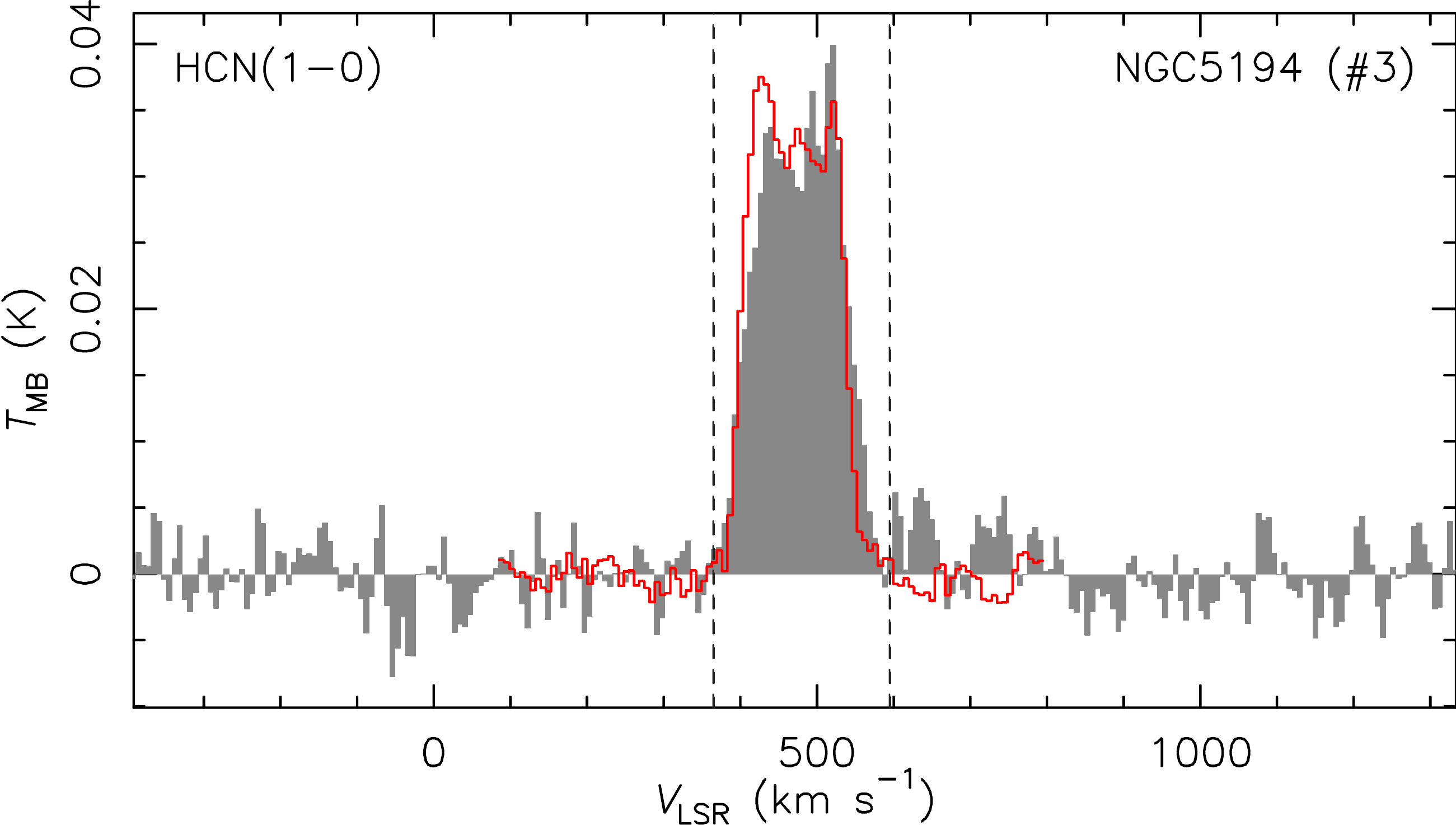} & 
\mybox{} \\ 
\includegraphics[width=0.42\textwidth]{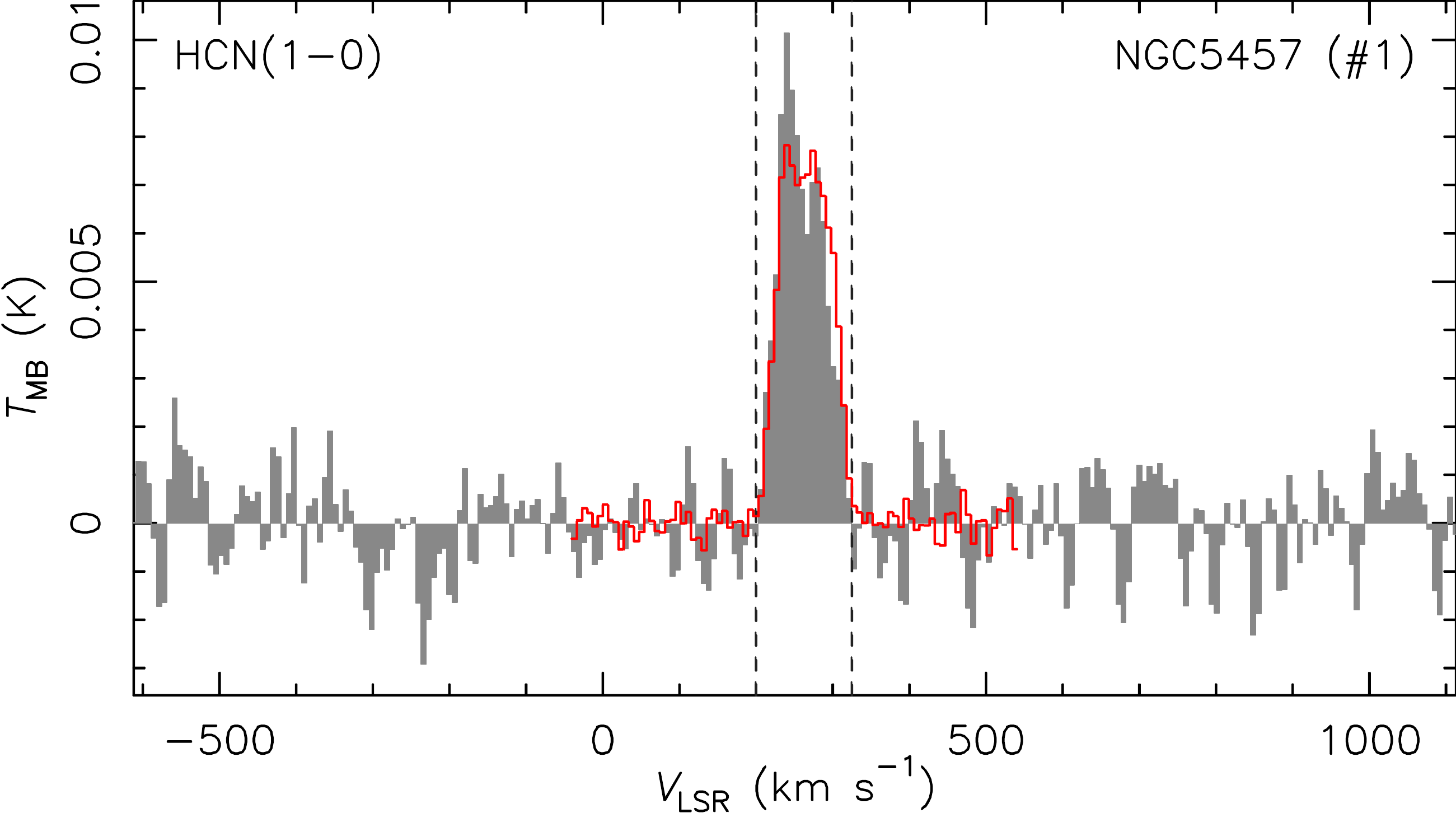} & 
\mybox{} \\ 
\includegraphics[width=0.42\textwidth]{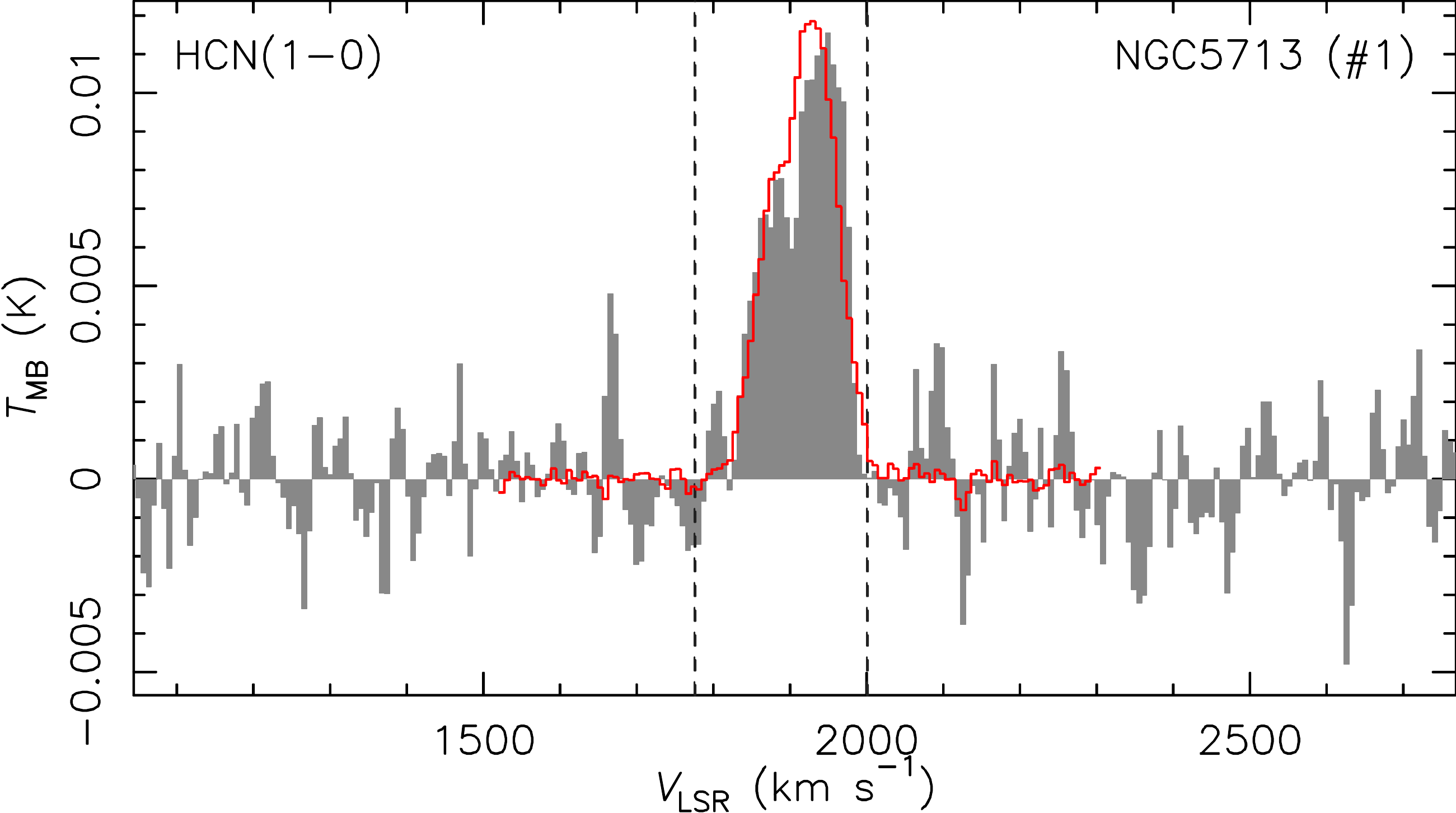} & 
\includegraphics[width=0.42\textwidth]{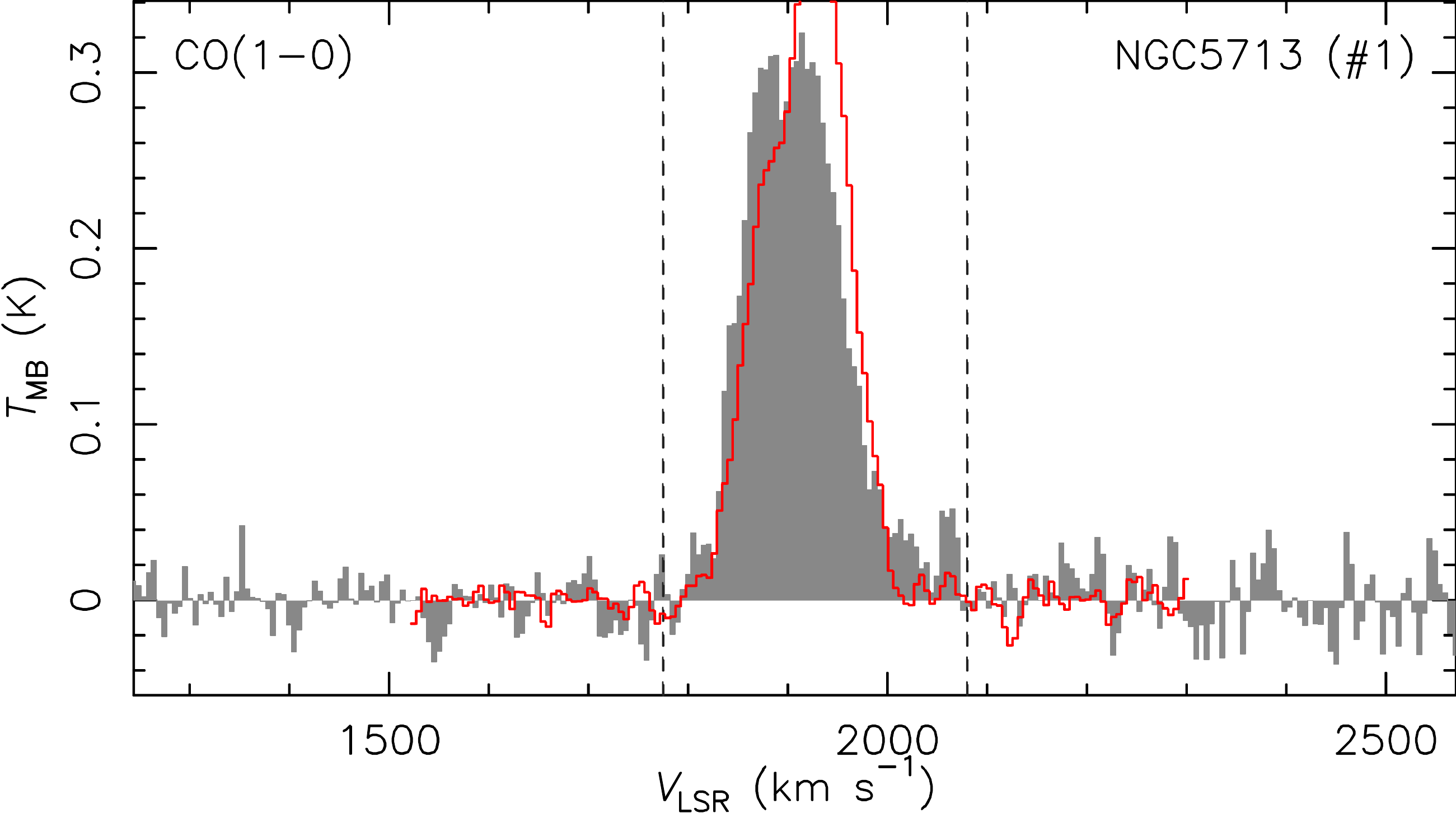} \\ 
\end{tabular}
\end{center}  
\caption{Same as Fig.~\ref{f-spec-1} for NGC~5194 and NGC~5457, and NGC~5713.} 
\end{figure} 
\newpage 
\begin{figure}[h!]
\begin{center}  
\begin{tabular}{cc} 
\includegraphics[width=0.42\textwidth]{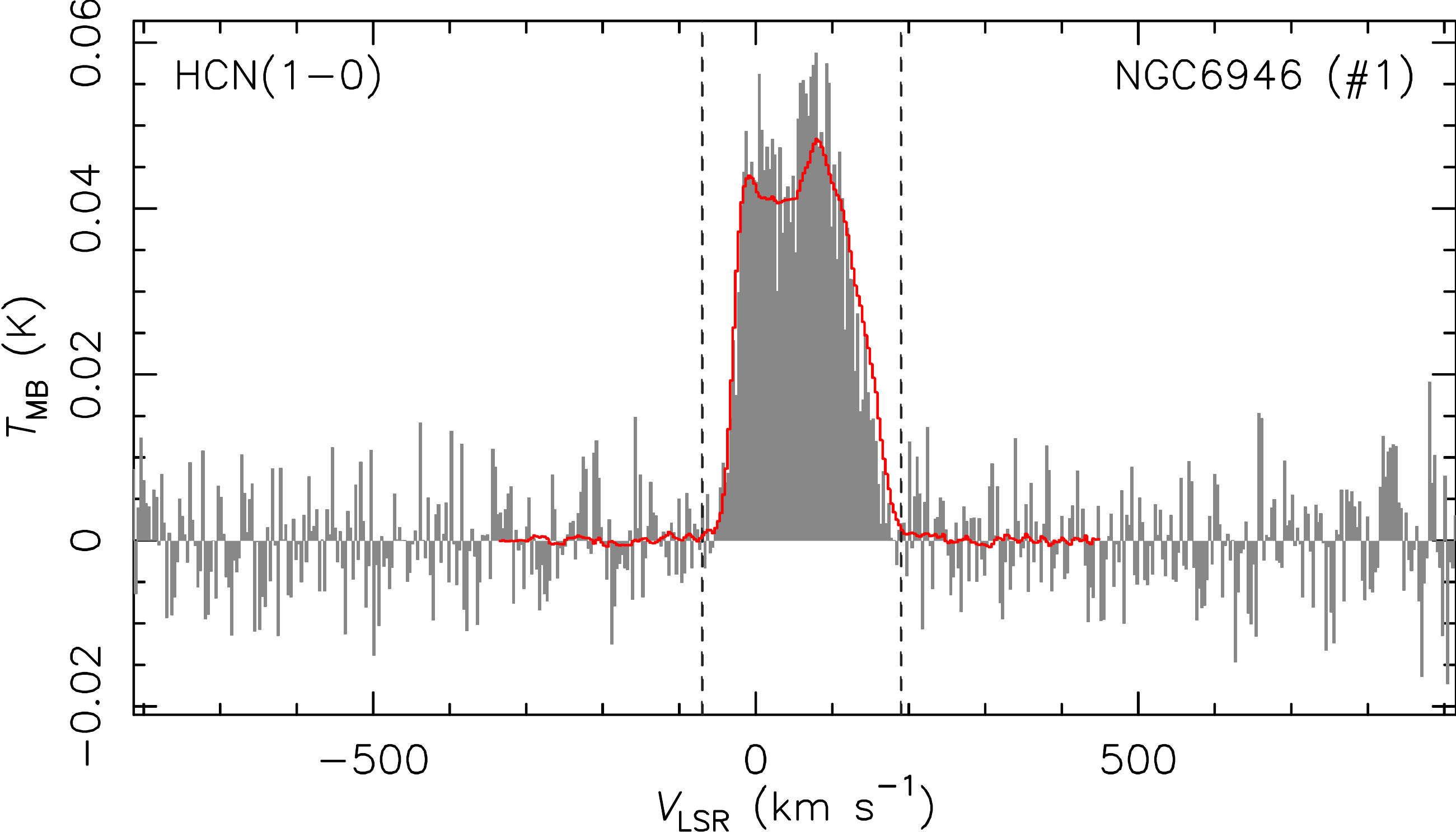} & 
\includegraphics[width=0.42\textwidth]{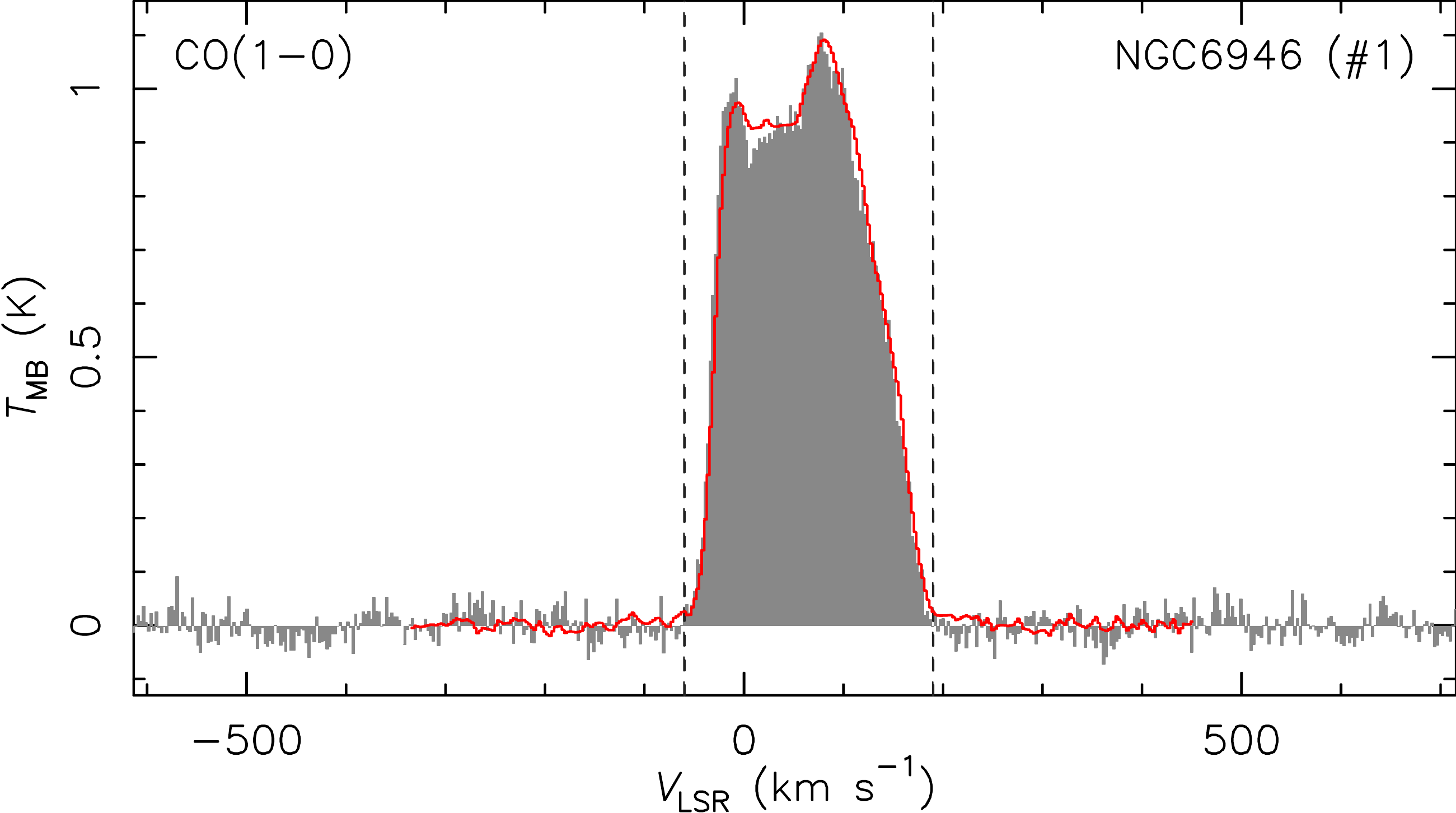} \\ 
\includegraphics[width=0.42\textwidth]{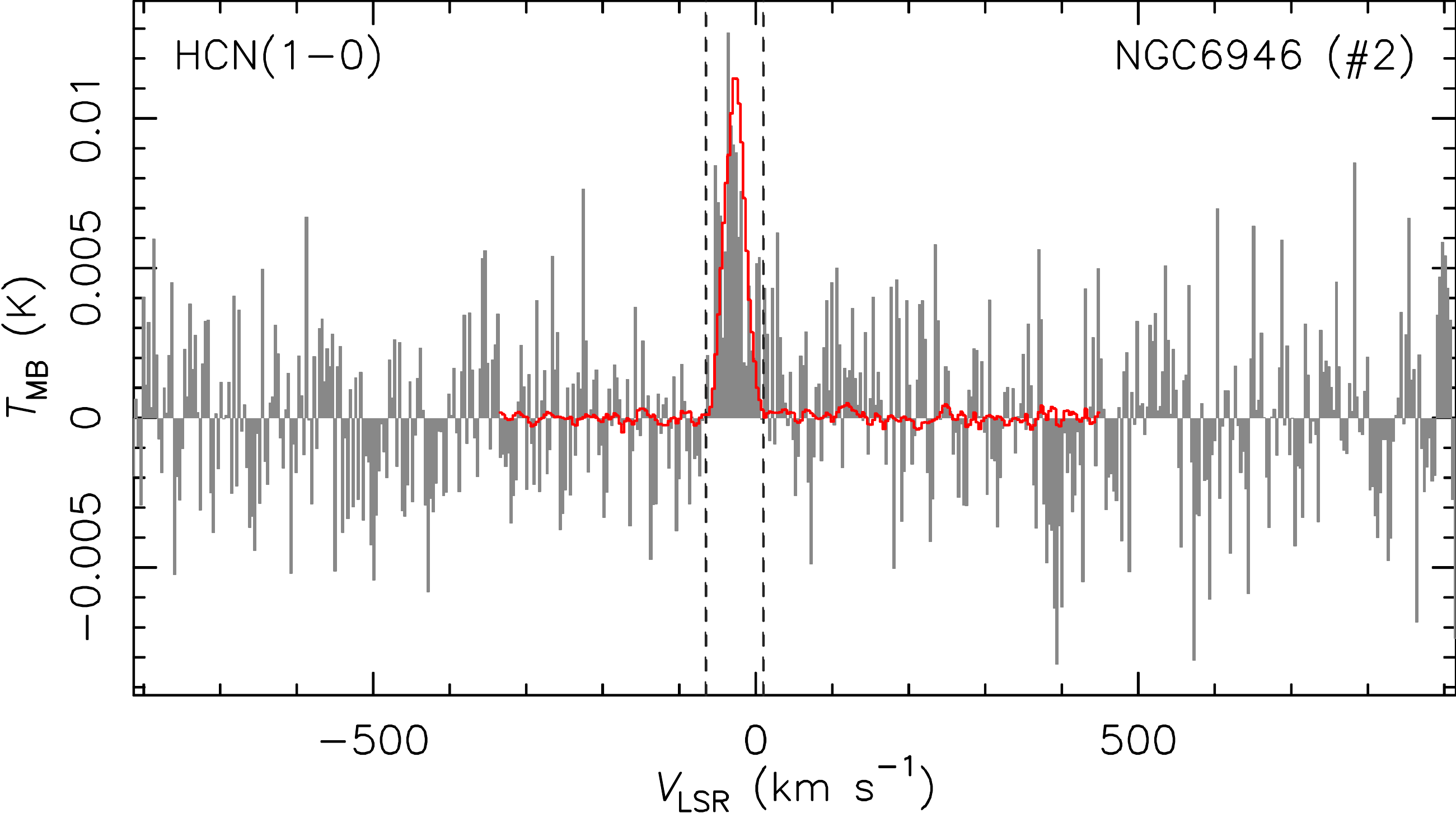} & 
\includegraphics[width=0.42\textwidth]{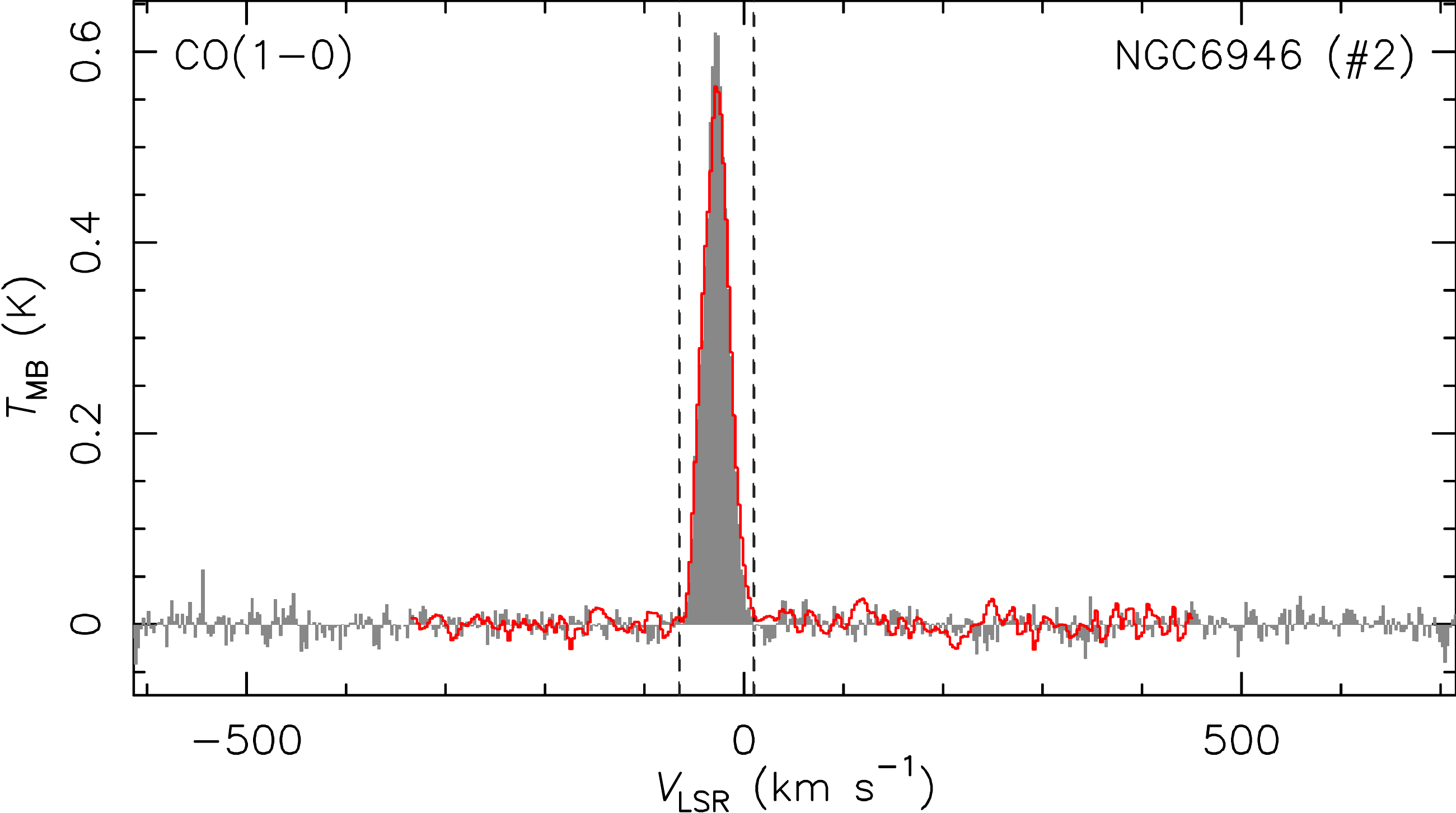} \\ 
\includegraphics[width=0.42\textwidth]{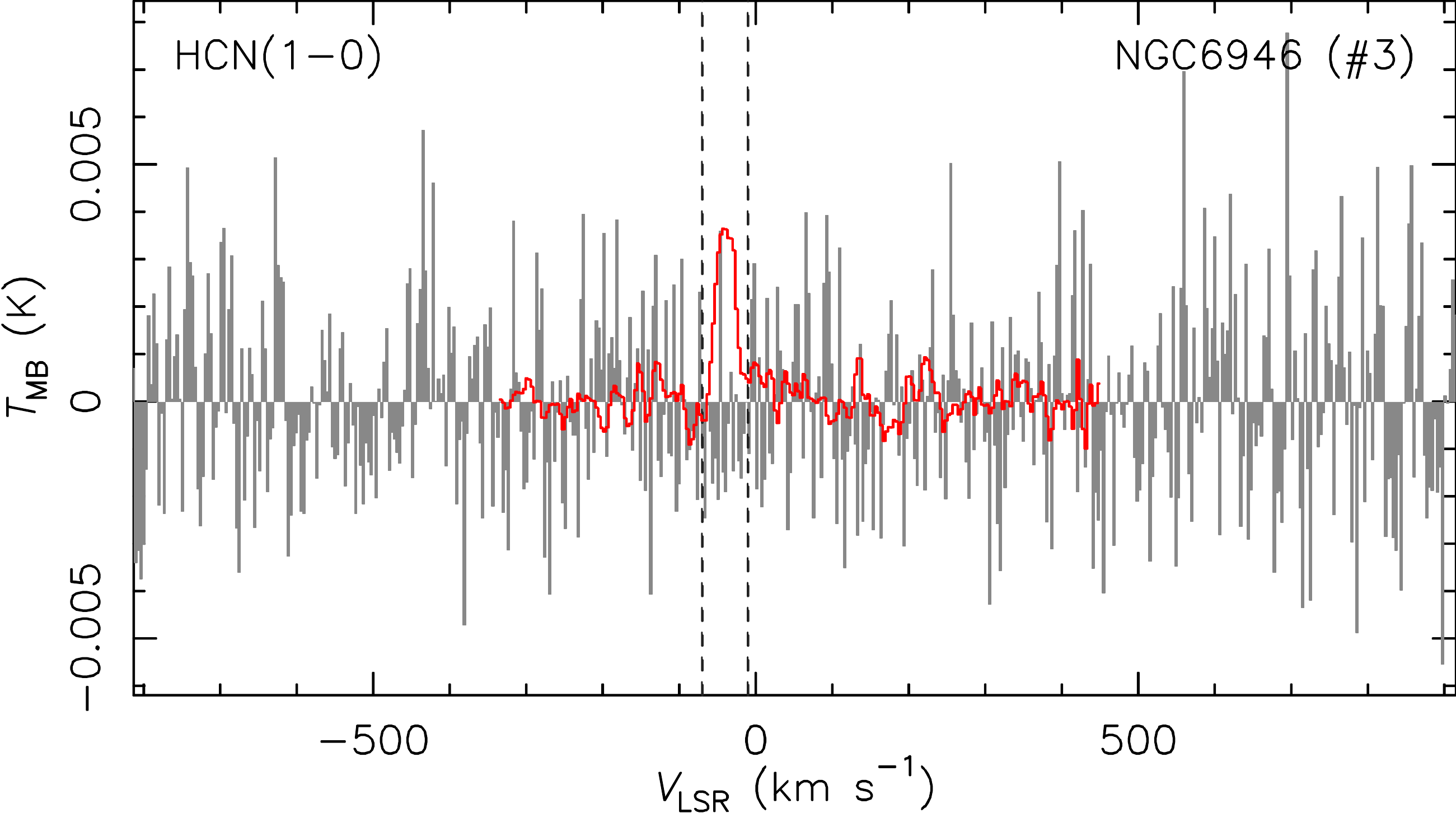} & 
\includegraphics[width=0.42\textwidth]{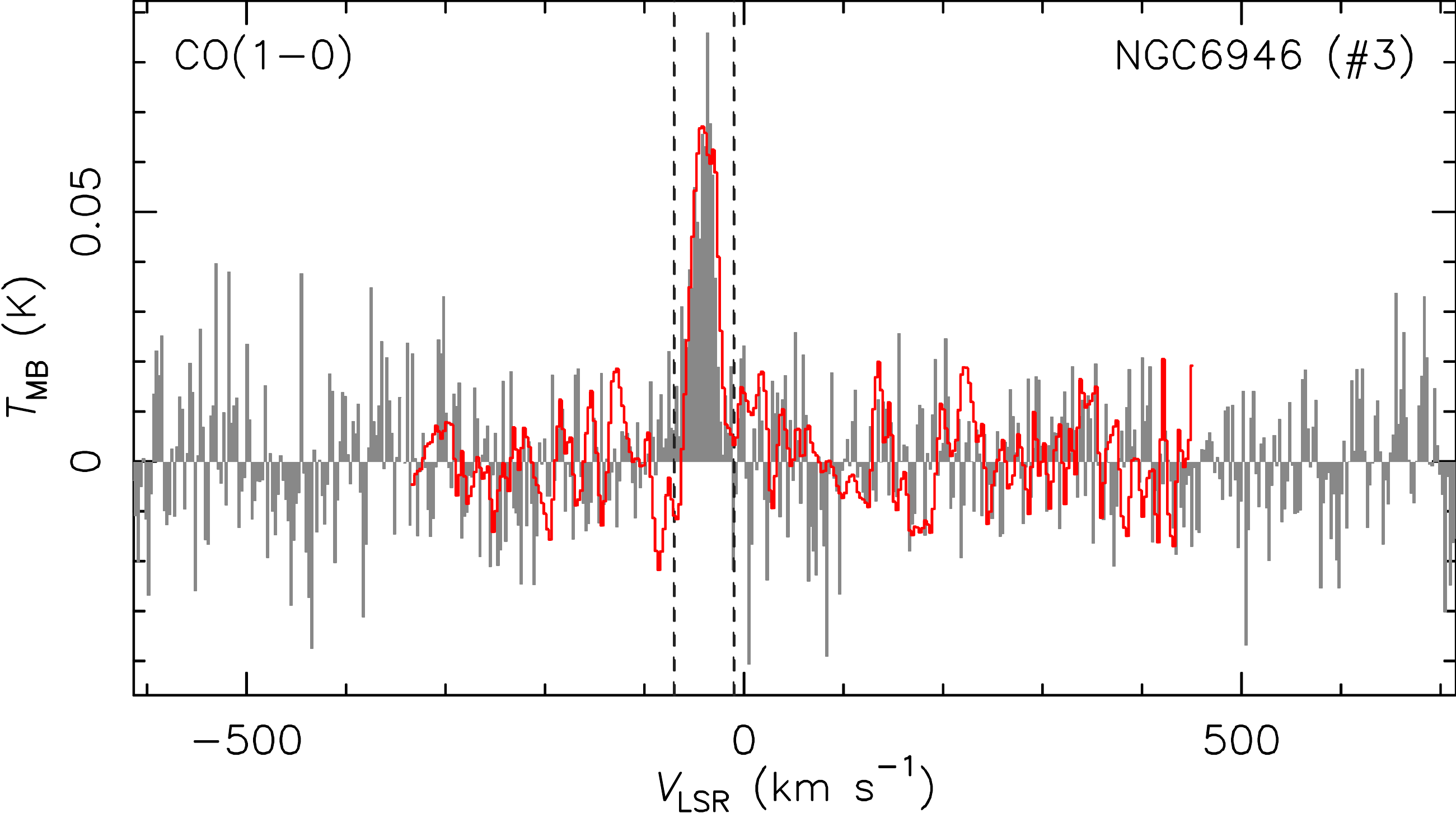} \\ 
\includegraphics[width=0.42\textwidth]{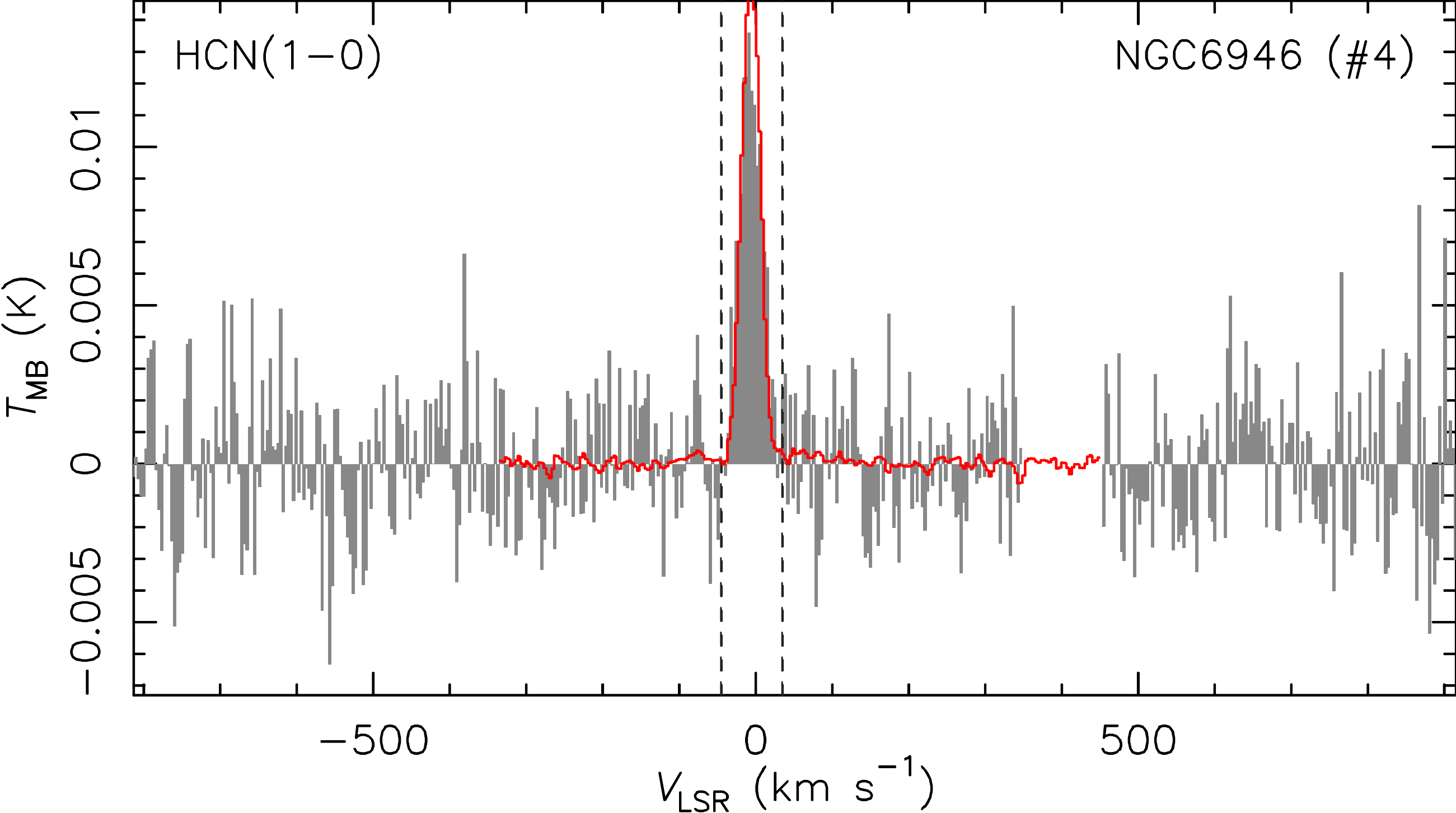} & 
\includegraphics[width=0.42\textwidth]{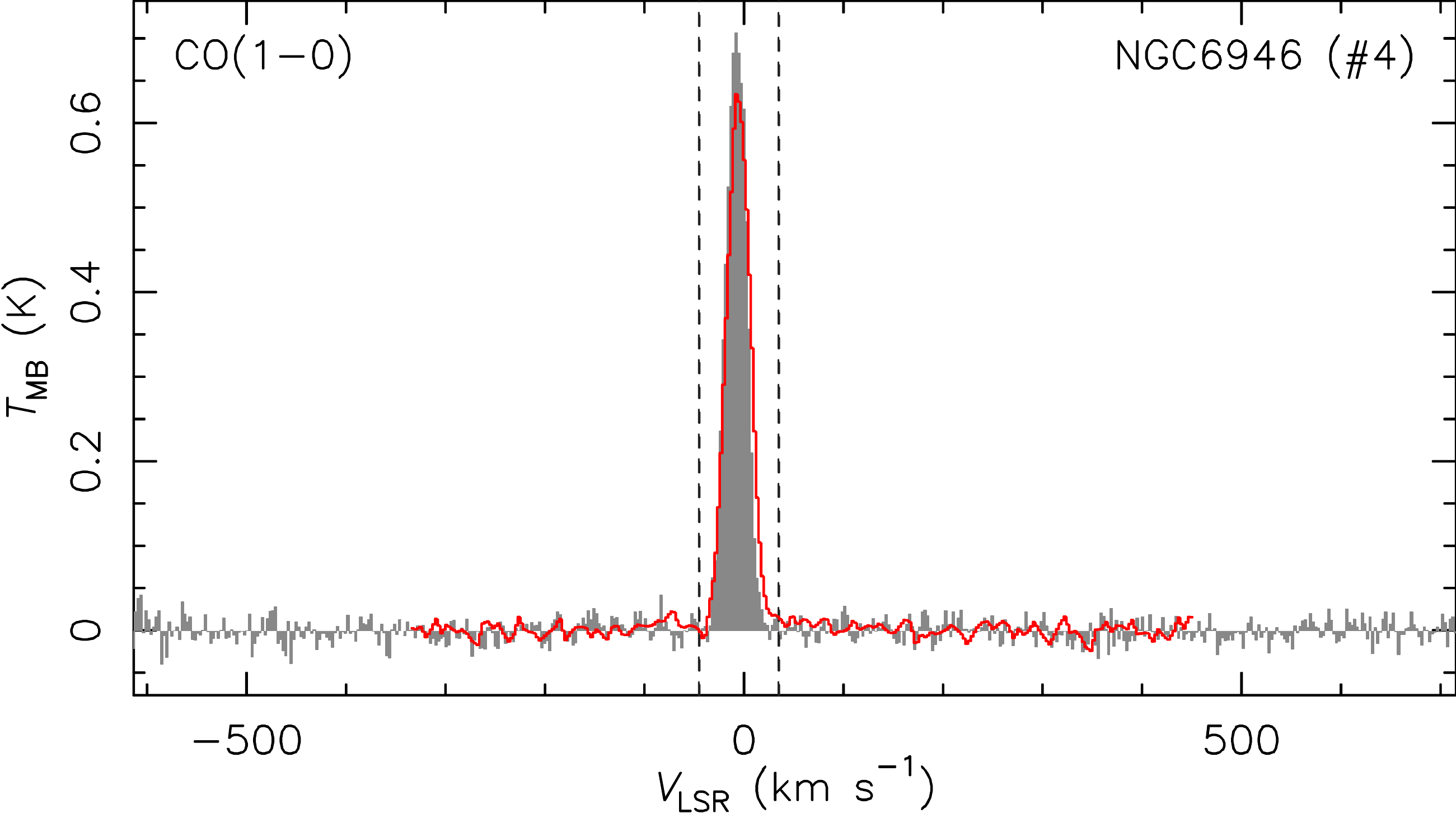} \\ 
\includegraphics[width=0.42\textwidth]{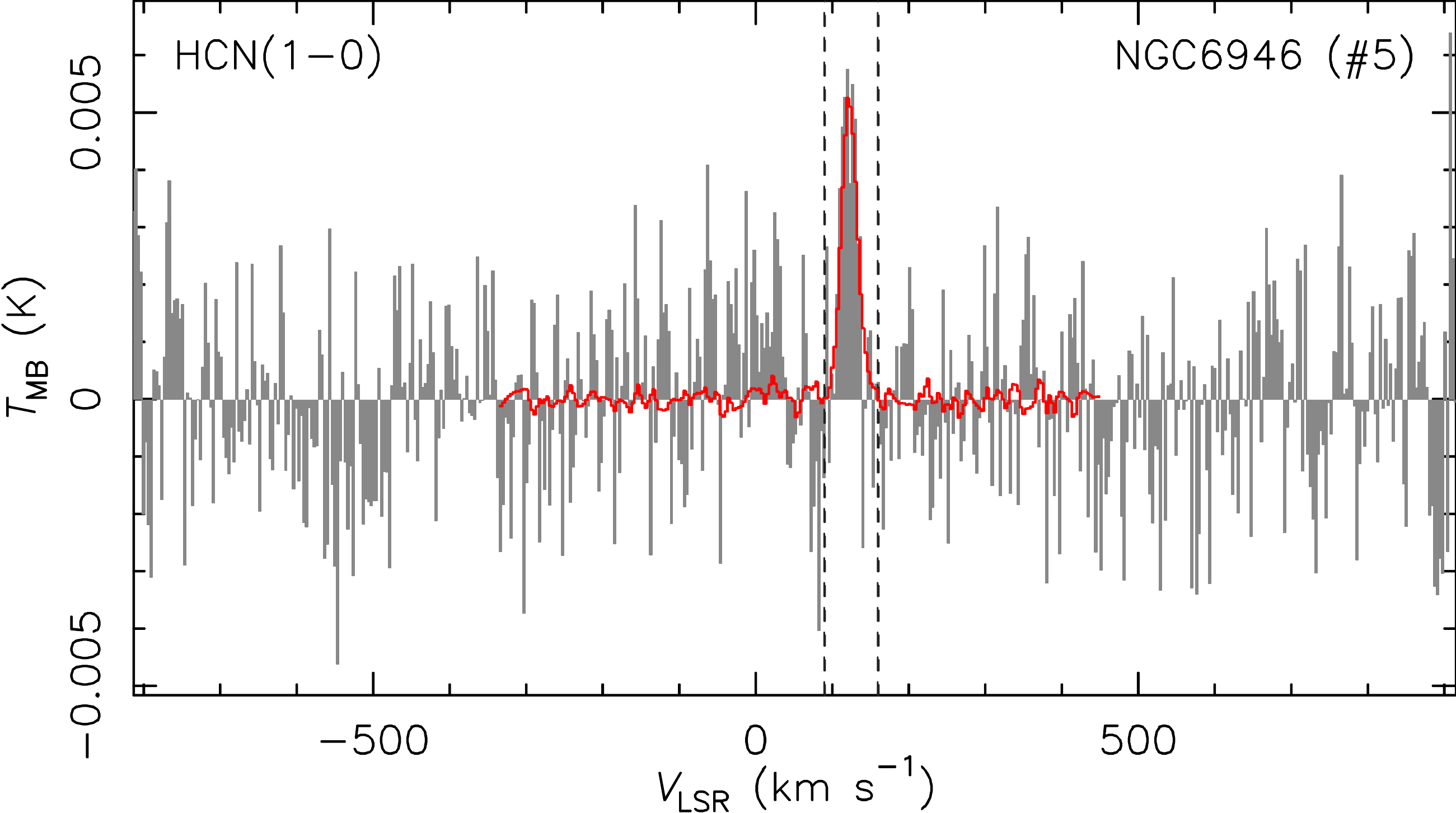} & 
\includegraphics[width=0.42\textwidth]{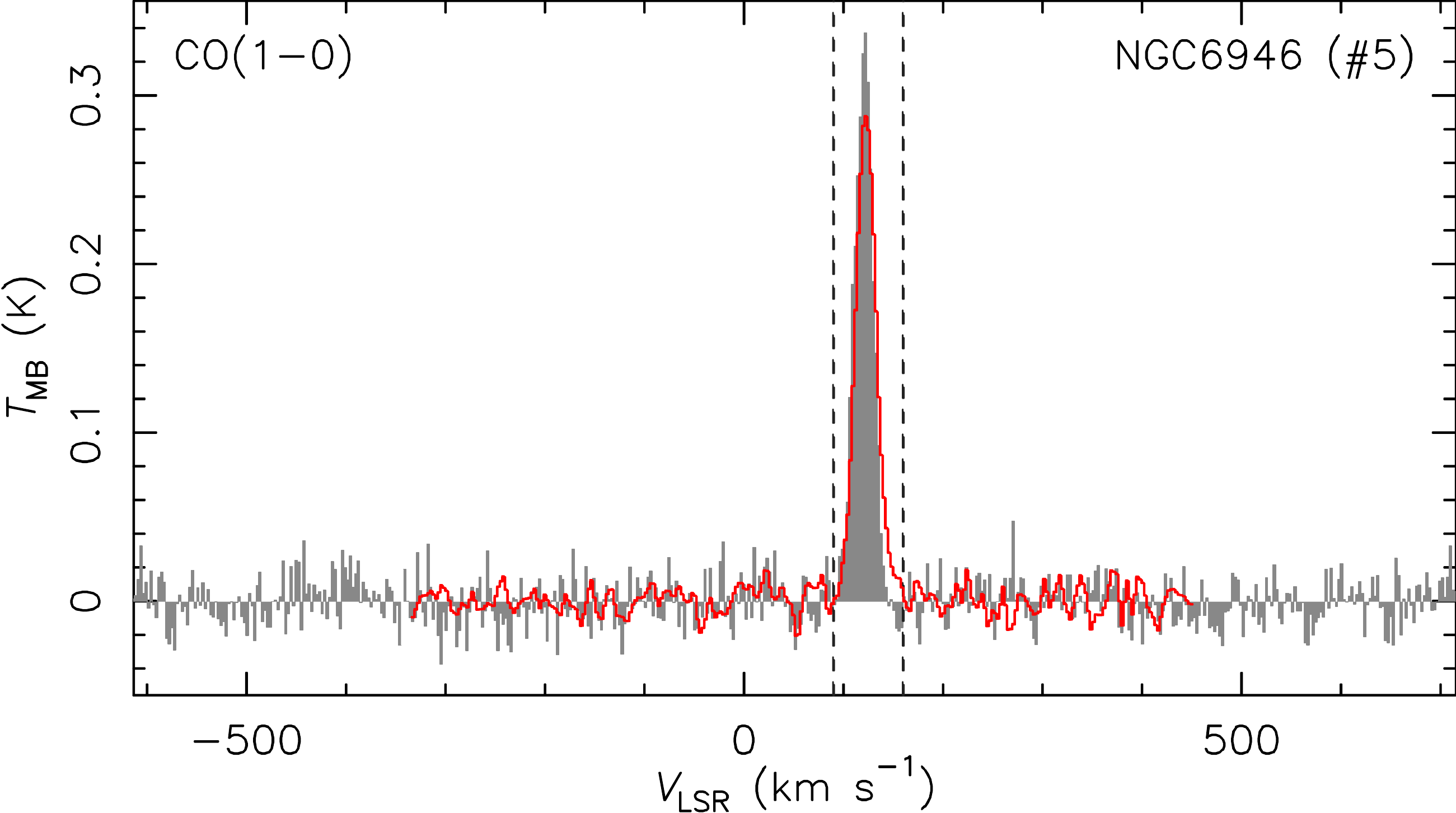} \\ 
\end{tabular}
\end{center}  
\caption{Same as Fig.~\ref{f-spec-1} for NGC~6946.} 
\end{figure} 
\newpage 
\begin{figure}[h!]
\begin{center}  
\begin{tabular}{cc} 
\includegraphics[width=0.42\textwidth]{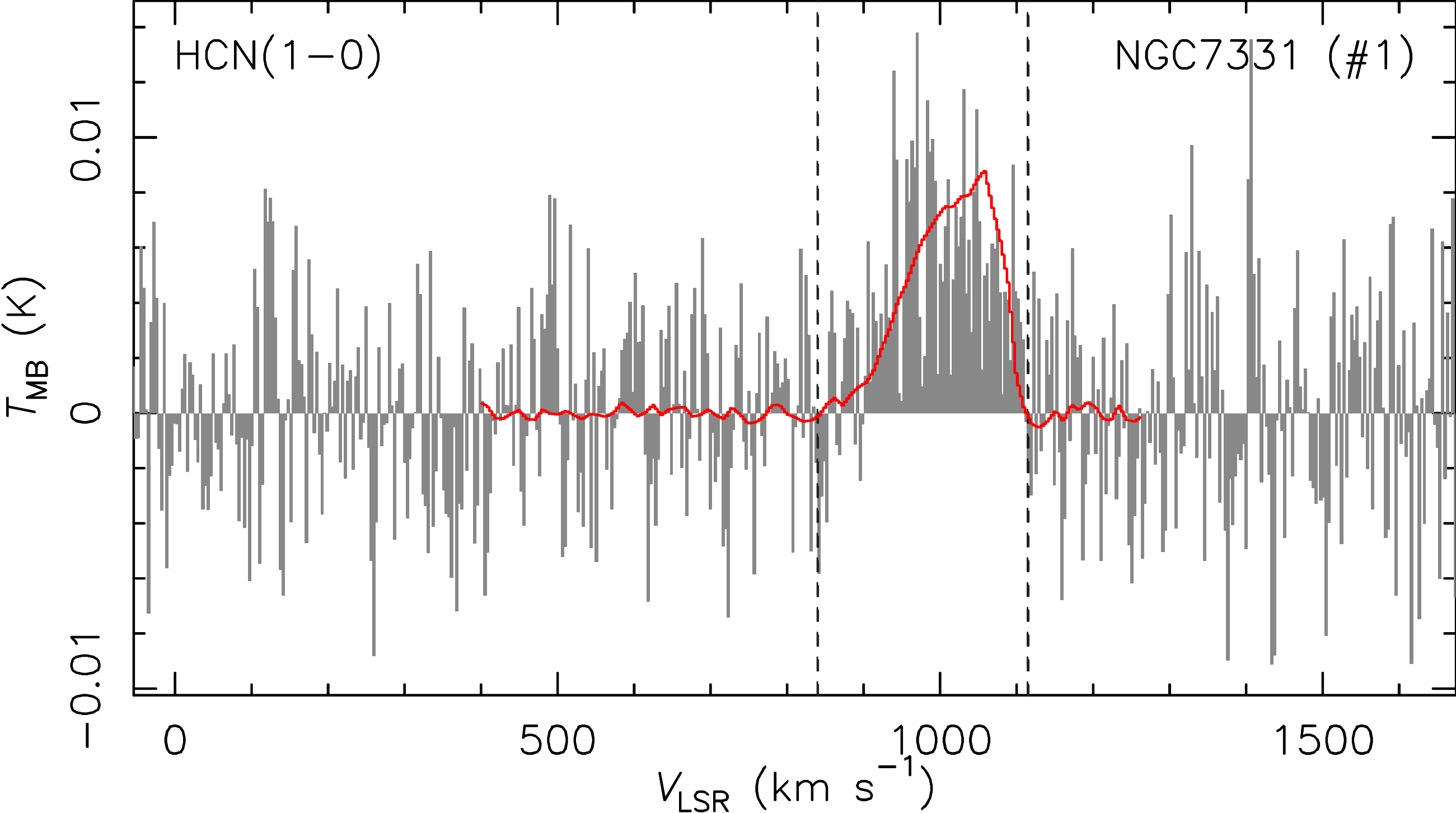} & 
\includegraphics[width=0.42\textwidth]{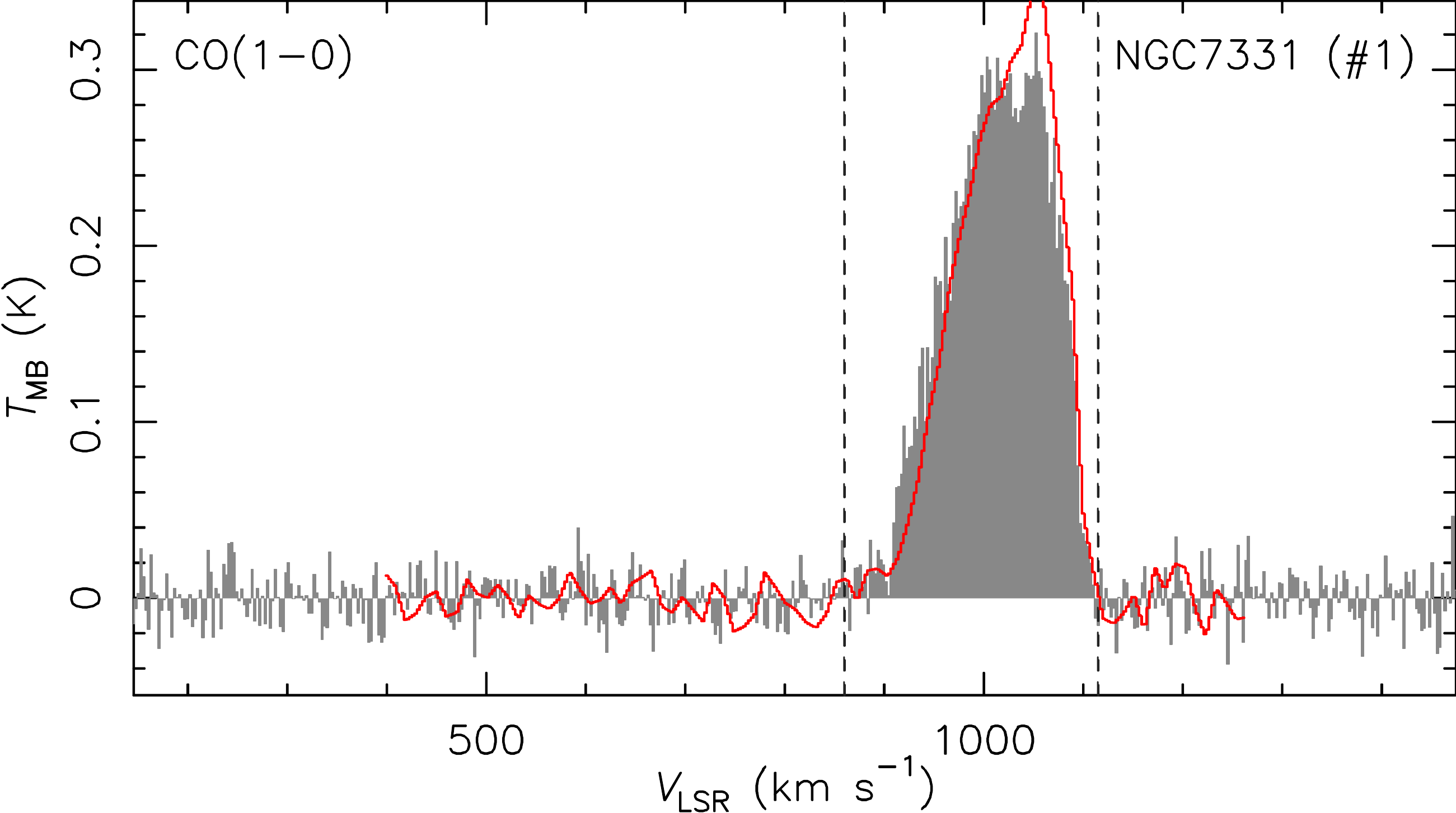} \\ 
\includegraphics[width=0.42\textwidth]{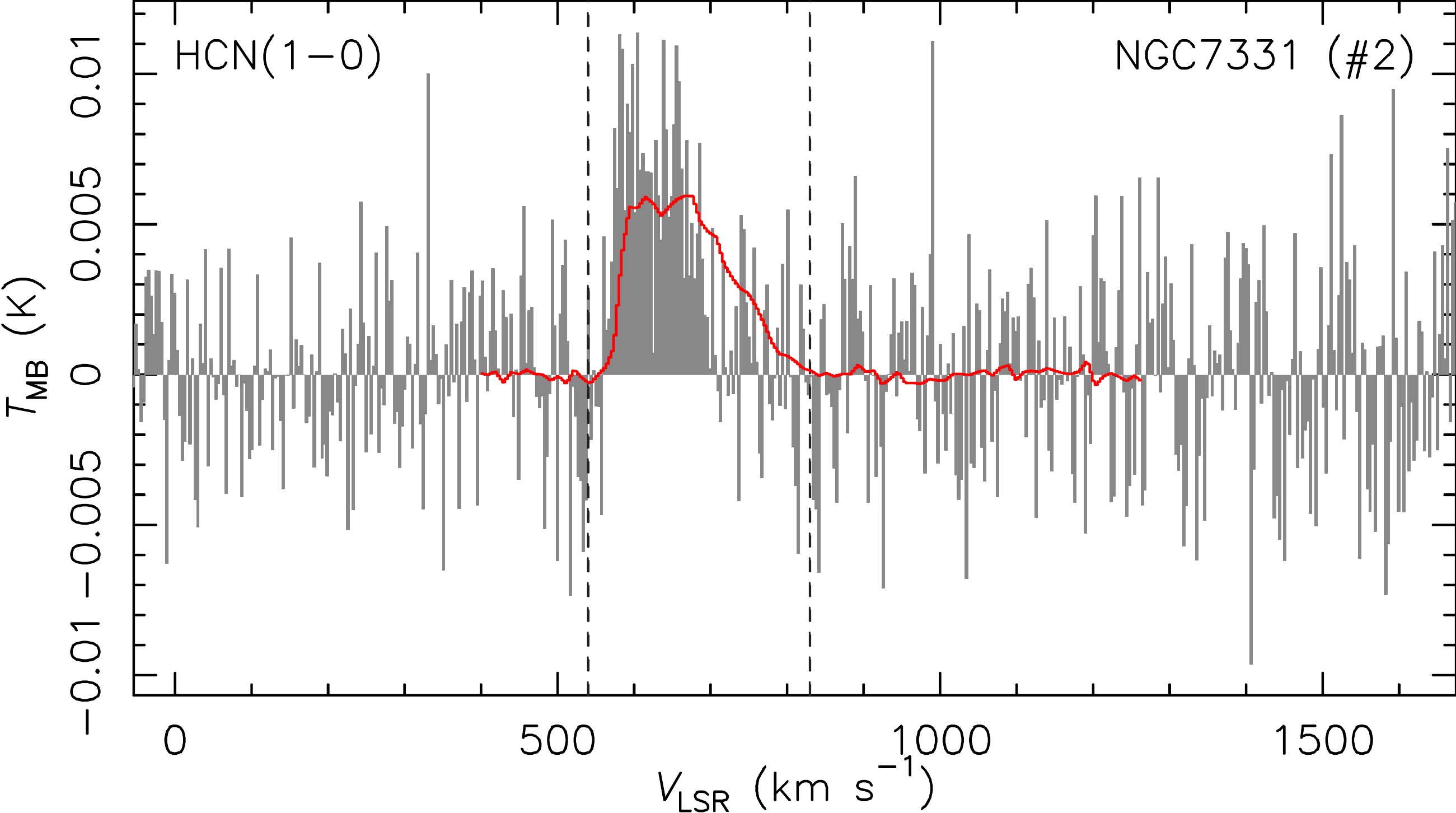} & 
\includegraphics[width=0.42\textwidth]{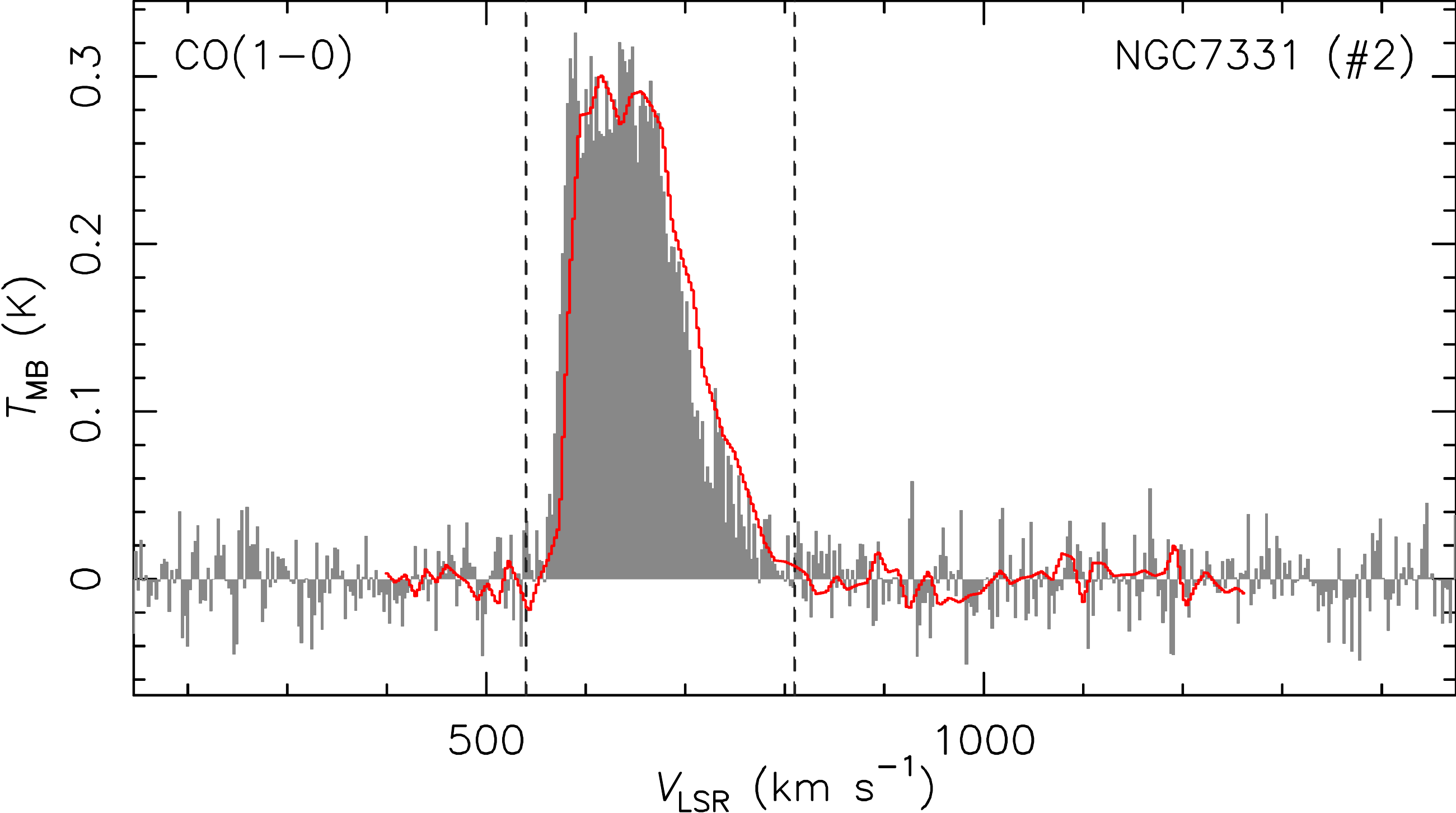} \\ 
\end{tabular}
\end{center}  
\caption{Same as Fig.~\ref{f-spec-1} for NGC~7331.} 
\label{f-spec-14}
\end{figure} 

\end{center}

\end{document}